Universidad Nacional de La Plata
Facultad de Ciencias Astronómicas y Geofísicas

Tesis para obtener el grado académico de
Doctor en Astronomía

# Formación de planetas gigantes en el marco del modelo de inestabilidad nucleada

Lic. Andrea Fortier

Director: Dr. Omar G. Benvenuto

Co-Director: Dr. Adrián Brunini

Marzo de 2009

# Agradecimientos

Esta Tesis Doctoral fue posible gracias al apoyo que recibí de muchísima gente, que durante estos cinco años estuvieron presentes en mi vida compartiendo alegrías y desencuentros, e impulsándome a seguir adelante. Hoy, el resultado se traduce en este manuscrito. Sin embargo, entre las citas que encontrarán en él generalmente no aparecen mencionadas las personas que, directa o indirectamente, colaboraron para la realización de este trabajo. Es por ésto que considero importante mencionarlas en este apartado. Y si bien seguramente no podré hacer justicia al intentar transmitir lo que ellos significaron para mí en este tiempo, espero al menos poder dejar en claro que una parte muy importante de este trabajo se las debo a todos ellos.

Por empezar, mi más profundo agradecimiento a mis directores, Omar Benvenuto y Adrián Brunini, por haberme dejado participar de este proyecto, ambicioso quizá en un principio, de intentar entender cómo se formaron los planetas gigantes. Me resulta increíble al mirar hacia atrás, cuando en los comienzos del año 2004, empecé a intentar "empaparme" en el tema, luchando contra un código numérico que nunca hacía lo que yo pretendía, y encontrar que hoy podemos presentar los resultados que conforman esta Tesis. Fueron Omar y Adrián quienes estuvieron presentes siempre, guiándome, enseñándome y, fundamentalmente, apoyándome para que el trabajo pudiera salir adelante. Pero además, junto a ellos aprendí otras cosas, que se dieron con el tiempo, a partir de charlas y discusiones, y que quizá sean tan importantes como estos resultados. Fundamentalmente, puedo decir que gracias a ellos pude comenzar a formarme un *criterio*. Un criterio para distinguir las cosas importantes de las que no lo son tanto, y que no solo se aplica al armado de un modelo para estudiar un problema particular, sino que se extiende a las relaciones humanas y a la vida en general. Por todo lo que compartimos y por todo lo que aprendí estando al lado de ellos, gracias.

Agradezco la Universidad Nacional de La Plata y, muy en particular, al personal docente y no docente de la Facultad de Ciencias Astronómicas y Geofísicas por la ayuda brindada en este tiempo. También, le agradezco al Instituto de Astrofísica de La Plata (IALP) y al Consejo Nacional de Investigaciones Científicas y Técnicas (CONICET).

A la Facultad de Ingeniería y, en particular, al Departamento de Ciencias Básicas les agradezco por permitir mi desarrollo docente y por el reconocimiento que en el transcurso de estos años me brindaron. Mi especial consideración para la Lic. Lilina Carboni (Jefa del Departamento de Ciencias Básicas), la Prof. Nélida Echebest (Coordinadora de la Cátedra de Matemática C) y para mis compañeros de cátedra. Mi agradecimiento también al Prof. Raúl Rossignoli por los años compartidos y la confianza que siempre tuvo en mi persona.

A mis compañeros de oficina, Alejandra De Vito y Rubén Martinez, todo mi cariño por haberme sabido acompañar, escuchar y ayudar. Los tres, junto también a Omar Benvenuto, hemos compartido muchas horas de nuestras vidas, que quedarán por siempre en mis recuerdos. Gracias por, mate mediante, haberme permitido conocerlos y crecer a su lado.

A los miembros del Grupo de Ciencias Planetarias: Romina Di Sisto, Lorena Dirani, Pablo Santamaría y Yamila Miguel por muchos años más compartiendo inquietudes, almuerzos y discusiones. Gracias por saber que puedo contar siempre con ustedes. A Héctor Viturro por el soporte técnico. Y a mi gran amigo, Gonzalo de Elía, compañero en las buenas y en las malas, mi "mellizo" de carrera, gracias, muchas, muchas gracias por todo. Este trayecto de mi vida podría haber sido muy distinto sin su amistad. Gracias, Gonza, por haber estado a mi lado.

En estos años tuve la suerte de asistir a reuniones científicas las cuales, no solo contribuyeron a mi formación profesional, sino que me permitieron conocer gente que, con su ejemplo, me impulsó a seguir en esta carrera. Por el aliento y por las enseñanzas que, probablemente sin proponérselo, me dieron quiero agradecer a: Prof. J. Fernandez, Prof. G. Wuchterl, Prof. M. Kürster, y Prof. S. Ida. Pero, sin duda, lo más provechoso de estas reuniones fue conocer a muchos de mis pares, algunos de los cuales se convirtieron en grandes amigos a pesar de las distancias. A Nancy Sosa, Pablo Pais, Christoph Mordasini, Arnaud Cassan, Andrés Carmona, Olga Zakhozhay, Marin Treselj y Francesca De Meo, gracias.

Por el compañerismo, por el haber compartido tantas cosas juntos, por el deseo de que esto siga siendo así , gracias a: Javier Vasquez, Nicolás Duronea, Andrea Torres, Marcelo Miller, Ileana Andruchow, Diego Bagú, Jorge Panei, Leandro Althaus, Fernando de la Corte, Ricardo Amorín, Guillermo Hägele, Germán Rubino, Juan Ignacio Sabbione, Luis Guarracino, Sofía Cora y Sixto Gimenez-Benitez. A Mariana Orellana por su amistad.

Durante este tiempo, siete amigas pudimos construir una relación que me llena de orgullo y que representa una de las cosas más valiosas que tengo. A Verónica Firpo, Cecilia Fariña, Anahí Granada, Anabella Araudo, Claudia Scoccola y Antonela Monachesi gracias por la amistad que cultivamos. Gracias por ser como son.

A Melina Bersten, gracias por todas las cosas compartidas. A pesar de la distancia, nuestra amistad dio sobradas pruebas de ser auténtica y robusta; espero saber honrarla.

A Juan Cappi y Lorena Praderio, por todas las cenas compartidas y por todas las que vendrán. Gracias por formar parte de mi vida y por permitirme formar parte de la suya.

Y por último, a los verdaderos responsables de que esta Tesis se haya podido escribir: mi mamá Beatriz, mi papá Eduardo, mi hermano Andrés, mi hermana "postiza" Florencia y a mi abuela Claire, gracias. Por el amor, la paciencia y el apoyo, gracias. Por todos los momentos que compartimos juntos, buenos y malos, gracias. Por su presencia incondicional, gracias. Por aceptarme y quererme como soy ... no tengo palabras.

*A mi mamá, a mi papá y a mi hermano.*

# Resumen

Los planetas del Sistema Solar han captado la atención de la humanidad por siglos. Sin embargo, a pesar de toda la información con la que se cuenta, ya sea teórica o proveniente de datos observacionales, son todavía muchas las incógnitas que involucran la evolución de los sistemas planetarios y de los diversos cuerpos que los componen. El objetivo de esta tesis es estudiar la formación de los objetos de masa subestelar más grandes que pueblan los sistemas planetarios: los planetas gigantes.

Existen dos modelos, cualitativamente muy diferentes, que explican el proceso de formación de los planetas gigantes: la hipótesis de inestabilidad gravitatoria del disco y la hipótesis de inestabilidad nucleada. Si bien no se puede descartar la plausibilidad de la primera como escenario de formación, es la hipótesis de inestabilidad nucleada la que cuenta con mayor consenso en la comunidad científica debido a que consigue explicar en forma sencilla la actualmente aceptada estructura interna de los planetas gigantes. La hipótesis de inestabilidad nucleada propone dos etapas para la formación de los planetas gigantes: en una primera instancia, por acreción de planetesimales, se forma un núcleo sólido. Cuando éste llega a aproximadamente unas 10 masas terrestres, comienza la segunda etapa, durante la cual el planeta alcanza su masa final a instancias de acretar grandes cantidades de gas en una corta escala de tiempo. El presente trabajo fue desarrollado adoptando la hipótesis de inestabilidad nucleada como modelo de formación.

Por su parte, el crecimiento de los embriones sólidos no se produce a tasa constante sino que atraviesa también varios estadíos. Al principio los planetesimales siguen un crecimiento retroalimentado conocido como crecimiento runaway, cuyo producto es la aparición de los primeros embriones planetarios en tan solo unas decenas de miles de años. Cuando estos embriones alcanzan aproximadamente la masa de la Luna perturban gravitatoriamente a los planetesimales vecinos, aumentando las velocidades relativas y disminuyendo así su tasa de crecimiento. Este régimen se conoce como crecimiento oligárquico ya que solo los embriones más grandes son capaces de continuar capturando planetesimales. La tasa de acreción de sólidos puede ser crucial para estimar el tiempo de formación total del planeta. Los datos observacionales provenientes de discos circumestelares limitan la escala de tiempo de formación de los planetas gaseosos, la cual no puede superar los 10 millones de años. Esta es una de las mayores dificultades asociadas a la hipótesis de inestabilidad nucleada, ya que en general los resultados de las simulaciones no consiguen encuadrarse satisfactoriamente dentro de esta cota. Es por esto que la mayoría de los trabajos que se encuentran en la bibliografía adoptan el crecimiento runaway como único régimen de crecimiento para el núcleo durante toda la formación planetaria.

Esta Tesis compila los resultados de nuestro estudio de la formación de planetas gigantes. El trabajo se realizó bajo la suposición de que el planeta se encuentra en órbita circular, no perturbada, alrededor del Sol, donde no se han tenido en cuenta la existencia de campos magnéticos, turbulencias o rotación del planeta. Los cálculos fueron realizados con un código de tipo Henyey, el cual resuelve las ecuaciones diferenciales de evolución estelar acopladas en forma autoconsistente a la tasa de de acreción de planetesimales. Se consideró, por primera vez en este tipo de cálculos, al crecimiento oligárquico como el régimen dominante para el crecimiento del núcleo durante toda la formación del planeta, incorporando así un tratamiento más realista para la acreción del material sólido. Como consecuencia de esto, el tiempo de formación resulta mucho más prolongado que lo encontrado en trabajos previos. Por otra parte, sin embargo, hemos observado una dependencia muy fuerte entre el tiempo de formación y el tamaño de los planetesimales acretados: cuanto más pequeños son, más rápido se completa el proceso. Consecuentemente, dado que los planetesimales que pueblan el disco no son todos iguales, hemos incluido en el cálculo una función que represente la distribución de tamaños. Las simulaciones realizadas contemplan una población con hasta nueve tamaños, entre los 100 metros y 100 kilómetros de radio, siendo los más pequeños los más abundantes. Bajo estas hipótesis hemos conseguido modelar la formación de Júpiter, Saturno, Urano y Neptuno en menos de 10 millones de años. Además, las masas obtenidas para los núcleos se encuentran dentro de las cotas aceptadas actualmente.

Otro de los resultados relevantes de esta tesis, fue la aparición, bajo ciertas circunstancias, de eventos cuasi-periódicos de pérdida de gas cuando el planeta llega a ciertos valores críticos de su masa. Este efecto no había sido observado previamente, y podría tener consecuencias determinantes en el proceso de formación de los planetas gigantes.

# Índice general

# Capítulo 1

# Introducción

## 1.1. Un poco de historia[1]

El concepto de "Sistema Solar" surge después de la aceptación de la teoría heliocéntrica, quedando definitivamente incorporado en la comunidad científica hacia fines del siglo XVII. El modelo actual de formación del Sistema Solar (*la hipótesis nebular*) se remonta al siglo XVIII y tiene sus orígenes en las ideas de E. Swedenborg, I. Kant y P. S. Laplace. Según esta teoría, el Sol y los planetas se habrían formado a partir de una nube molecular rotante en contracción. Asociado a este proceso, tendría lugar la formación de un disco alrededor de la estrella, donde a su vez se originarían los planetas y demás cuerpos menores. Este escenario explica naturalmente la coplanaridad y la cuasi circularidad de las órbitas de los planetas del Sistema Solar. Sin embargo, en el pasado esta teoría fue muy cuestionada debido a la distribución actual de momento angular entre el Sol y los planetas: mientras que el 99,86 % de la masa del Sistema Solar está en el Sol, el 99,5 % del momento angular total del sistema reside en los planetas. De la aplicación directa de la teoría de Swedenborg, Kant y Laplace, se desprende que la mayor parte de la masa y del momento angular del sistema deberían haber permanecido en el Sol. Este hecho hizo que se desestimara a la hipótesis nebular por casi dos siglos. Sin embargo, desde comienzos de 1980 las observaciones muestran que la mayoría de las estrellas jóvenes están rodeadas por un disco de gas y polvo, lo cual concuerda con las predicciones de la hipótesis nebular. Actualmente, la hipótesis nebular se encuentra completamente aceptada como el escenario estándar de formación del Sistema Solar.

Por su parte, Descartes, en 1644, propuso que los planetas eran el resultado último de un sistema de vórtices girando alrededor del Sol. Esta idea no incluía la presencia de un disco protoplanetario ni tampoco sugería la manera en que llegaron a formarse dichos vórtices. Tres siglos más tarde, en 1944, C.F. von Weizsäcker retomó esta idea proponiendo que entorno al Sol existió un disco turbulento, de composición solar, en el cual se generaron vórtices que permitieron la acreción de pequeñas partículas, lo cual finalmente habría dado lugar a los planetas. El aporte más importante de von Weizsäcker fue mostrar que en un disco turbulento el momento angular se transporta como consecuencia de la viscosidad, donde durante el proceso de acreción la masa fluye hacia el centro y el momento angular es transportado hacia afuera. Esta idea permitía resolver uno de los mayores problemas de la hipótesis nebular, según la cual el Sol debería rotar mucho más rápido de lo que lo hace en realidad. La propuesta de la formación de vórtices fue bastante criticada, sobre todo porque los vórtices tendrían una vida media muy corta, con lo cual no fue tomada en serio hasta 1995 cuando Barge & Sommeria mostraron que las partículas más pequeñas resultaban fácilmente capturadas por estos vórtices, permitiendo la formación de los núcleos planetarios en alrededor de $10^5$ años, acelerando así el proceso de formación y dando, de este modo, una posible solución al problema de las escalas de tiempo involucradas en la formación de los planetas gigantes (ver capítulo 6).

[1] Estas notas están basadas, fundamentalmente, en el artículo *Historical notes on planet formation* (Bodenheimer 2006).

Por su parte, Hoyle (1960) invocó la ruptura magnética para explicar la "lenta" rotación del Sol. Según sus cálculos, los efectos hidrodinámicos por sí mismos (como, por ejemplo, la viscosidad) no serían suficientes para explicar la transferencia de momento angular ya que, de haber sido así, el material del disco debería haber estado en contacto con el Sol y, en ese caso, la velocidad de rotación del Sol habría disminuido solo hasta encontrarse en co-rotación con la parte interior del disco. Hoyle propuso entonces que durante el proceso de formación del sistema Sol-disco se debió haber abierto una brecha entre ambos, pronunciada aún más por el campo magnético, lo cual permitió la transferencia de momento angular desde el Sol hacia la parte interior del disco, y desde allí hacia afuera. Entonces, durante su contracción, el Sol habría acelerado su velocidad de rotación pero, a su vez, el momento angular habría continuado siendo transferido a través del campo magnético. Para obtener el acoplamiento magnético, la temperatura debió ser del orden de los 1.000 K, y la intensidad del campo magnético de alrededor de 1 gauss. Más allá del borde interno del disco no se requiere de este acoplamiento ya que la viscosidad del disco permitiría la transferencia de momento angular hacia la región exterior del mismo.

Hoy en día la explicación básica para la pérdida de momento angular por parte de la estrella central radica en el transporte magnético y la turbulencia. Sin embargo, los cálculos actuales estiman que los campos magnéticos involucrados deberían ser del orden de los 1000 gauss.

## 1.2. El Sistema Solar: generalidades

La estructura básica del Sistema Solar involucra la presencia de objetos muy diversos, tanto en tamaño como en composición. El Sistema Solar está dominado por el Sol, una estrella enana de tipo G2V según su clasificación espectral, estable, que se encuentra atravesando la mitad de su existencia y que es la principal responsable de que en La Tierra se desarrolle la vida. La composición química del Sol es: un 70 % de hidrógeno, un 28 % de helio y un 2 % de elementos pesados. El Sol produce su energía por fusión nuclear en su centro, donde los átomos de hidrógeno se combinan para dar átomos de helio. La temperatura central del Sol, donde ocurre la fusión nuclear, es de $15,7 \times 10^6$ K, mientras que la temperatura promedio en la fotosfera es de 5800 K.

**Tabla 1.1.** Características principales de los planetas del Sistema Solar[a].

| Planeta | $a$ [UA] | Período orbital [años] | $e$ | $i$ | Masa [$M_\oplus$] | Radio [$R_\oplus$] | Densidad Media [g cm$^{-3}$] |
|---|---|---|---|---|---|---|---|
| Mercurio | 0,39 | 0,24 | 0,206 | 7,0° | 0,055 | 0,38 | 5,43 |
| Venus | 0,72 | 0,61 | 0,007 | 3,4° | 0,817 | 0,35 | 5,24 |
| Tierra | 1,00 | 1,00 | 0,017 | 0,0° | 1,000 | 1,00 | 5,52 |
| Marte | 1,52 | 1,88 | 0,093 | 1,8° | 0,107 | 0,53 | 3,90 |
| Júpiter | 5,20 | 11,86 | 0,048 | 1,3° | 317,8 | 11,2 | 1,30 |
| Saturno | 9,54 | 29,46 | 0,054 | 2,5° | 94,30 | 9,41 | 0,70 |
| Urano | 19,19 | 84,07 | 0,047 | 0,8° | 14,60 | 4,11 | 1,30 |
| Neptuno | 30,07 | 164,8 | 0,009 | 1,8° | 17,20 | 3,81 | 1,50 |

[a] El semieje orbital $a$ está dado en Unidades Astronómicas (UA), la masa del planeta en masas de la Tierra ($M_\oplus$) y el radio del planeta en radios de la Tierra ($R_\oplus$). La excentricidad se designa con $e$, y la inclinación del plano orbital respecto de la eclíptica con $i$.

Entorno al Sol orbitan numerosos cuerpos, siendo los planetas los más destacados. Los planetas orbitan alrededor del Sol en planos muy cercanos a la eclíptica, todos en la misma dirección, que corresponde al sentido contrario al de las agujas del reloj visto desde el polo norte ecliptical. La arquitectura de nuestro sistema planetario puede resumirse, según la distancia orbital creciente en: cuatro *planetas terrestres* (Mercurio, Venus, La Tierra y Marte), el *Cinturón de Asteroides*, dos *planetas gigantes gaseosos* (Júpiter y Saturno), dos *planetas gigantes de hielo* (Urano y Neptuno), el *Cinturón de Kuiper* (compuesto por asteroides y cometas) y, finalmente, en los confines del Sistema Solar, la *Nube de Oort*. En la tabla 1.1 se listan las características principales de los planetas del Sistema Solar.

La región comprendida entre Marte y Júpiter está ocupada por numerosos cuerpos irregulares y se la conoce como Cinturón de Asteroides. Más de la mitad de su masa está contenida en los cuatro objetos más grandes: Ceres (planeta enano), Vesta, Pallas e Hygiea. El resto de la población se distribuye entre asteroides más pequeños y partículas de polvo. Los asteroides son planetesimales[2] remanentes de la formación del Sistema Solar que no pudieron ser acretados para formar un planeta debido a las perturbaciones gravitatorias a las que se vieron sometidos por parte de los planetas gigantes. Las colisiones entre es-tos cuerpos habrían sido muy violentas, siendo mayormente destructivas, lo cual habría impedido la acreción en un cuerpo más masivo.

[2]La palabra *planetesimal* proviene del concepto matemático de infinitesimal y literalmente significa la fracción más pequeña que compone a un planeta. La denominación "planetesimal" se utiliza para referirse a los cuerpos rocosos más pequeños presentes en el Sistema Solar durante el proceso de formación planetaria.

El Cinturón de Kuiper está formado por asteroides y cometas, y es similar al Cinturón de Asteroides pero mucho más grande, tanto en tamaño como en masa. Esta región del Sistema Solar, que se extiende desde la órbita de Neptuno ($\sim$ 30 Unidades Astronómicas - UA) hasta aproximadamente unas 55 UA, está compuesta por los remanentes de la formación del Sistema Solar. A diferencia de los miembros del Cinturón de Asteroides, los cuerpos del Cinturón de Kuiper están compuestos mayormente por volátiles helados, como metano, amoníaco y agua. El cuerpo más grande del Cinturón de Kuiper es Eris. El primer objeto, descontando a Plutón, que confirmó la existencia del Cinturón de Kuiper se descubrió en 1992 (Jewitt & Luu, 1992), aunque su existencia fue propuesta durante décadas por varios investigadores. Actualmente se conocen más de mil objetos pertenecientes a esta región, y se estima que debe haber más de 70.000 con un diámetro mayor a los 100 km. En el Cinturón de Kuiper se distinguen cuatro poblaciones dinámicamente diferentes. El Cinturón de Kuiper clásico está formado por cometas de bajas inclinaciones y excentricidades ($40 < a < 48$ UA, $e < 0,2$). Una segunda población de objetos con órbitas excéntricas y más inclinadas, generalmente de mayor semieje, forman el *scattered disk* (o "disco disperso"), reservorio de los cometas de corto período. Estos objetos tienen perihelios cercanos a la órbita de Neptuno ($30 < q < 39$ UA, $e > 0,2$), e interactúan gravitatoriamente con él. Se presume que Tritón, el satélite más grande de Neptuno, sería un objeto capturado del Cinturón de Kuiper. Otra región es el Scattered Disk extendido, con $q > 39$ UA y $a > 50$ UA. Por último se distingue la Población Resonante, formada por objetos que tienen alguna relación de conmensurabilidad con Neptuno, siendo la resonancia 3:2 la más poblada. Debido a la controversia que genera la denominación "Cinturón de Kuiper", y las diferentes regiones que se pueden distinguir más allá de la órbita de Neptuno, actualmente los astrónomos se refieren a estos objetos en forma general como *objetos transneptunianos* (TNOs).

La masa del Cinturón de Kuiper contenida hasta 50 UA sería de aproximadamente $0,5\,\mathrm{M}_{\oplus}$. En función del número de cometas observados, el número de cometas del Cinturón de Kuiper entre 30 y 50 UA de diámetro mayor que 1 km está estimado en $10^9$. Del mismo orden sería el número de cometas que conforman el Scattered Disk.

Los cometas son planetesimales helados de la región externa de la nebulosa solar que orbitan más allá de Júpiter. Estos proto-cometas que originalmente tenían órbitas entre los planetas gigantes fueron eyectados por sus interacciones gravitatorias con ellos, la mayoría hacia el espacio interestelar. Sin embargo, una fracción de ellos quedó ligado gravitatoriamente al Sol, formando una esfera de aproximadamente 1 pc de radio. Las perturbaciones estelares y galácticas aumentaron los perihelios de los cometas llevándolos fuera de la región planetaria, generando una distribución aleatoria de sus órbitas lo cual le confirió una configuración esférica. La Nube de Oort sería entonces una aglomeración más o menos esférica de cometas que se encuentra a aproximadamente 50.000 UA del Sol. Su límite exterior define los confines de la influencia gravitatoria del Sistema Solar. La Nube de Oort se podría pensar como constituida por dos estructuras: una región interna y una externa, esta última de forma esférica. Los miembros de la Nube de Oort serían objetos que se formaron en el interior del Sistema Solar y que, debido a interacciones gravitatorias con los planetas gigantes, fueron dispersados hacia el exterior. Se cree que los cometas de largo período, y los de tipo Halley, provienen de la Nube de Oort. Hasta la fecha no se ha confirmado la observación de objetos pertenecientes a dicha región, aunque Sedna y otros dos objetos transneptunianos son fuertes candidatos a ser miembros de la Nube de Oort interior. Se estima que en la Nube de Oort hay alrededor de $10^{1}2$ cometas, con una masa total de algunas masas terrestres ($\sim 10\,M_{\oplus}$). Entre el 50-80 % de su población se encuentra en una región más densa, a unas 10.000 UA del Sol.

## 1.3. Clasificación de los objetos subestelares

Los objetos subestelares son cuerpos celestes cuya masa es inferior a la necesaria para que la fusión de hidrógeno se produzca o sea sostenida durante escalas de tiempo significativamente largas. La masa mínima necesaria estimada para que la fusión de hidrógeno sea posible estaría en el rango 0,07 - 0,092 $M_{\odot}$, dependiendo de la composición química de la estrella (0,07 $M_{\odot}$ para objetos de metalicidad solar y 0,092 $M_{\odot}$ para metalicidad cero). Los objetos subestelares pueden, a su vez, subclasificarse en enanas marrones, planetas, cuerpos de masa "planetaria", satélites y todo tipo de cuerpos pequeños (como asteroides y cometas).

En los últimos 20 años, con el progreso en las capacidades tecnológicas y en las técnicas observacionales, se han encontrado numerosos objetos de masa planetaria alrededor de otras estrellas (los llamados *planetas extrasolares*). Pero además se descubrieron nuevos cuerpos orbitando alrededor del Sol, fundamentalmente en la región del cinturón de Kuiper, algunos de los cuales tienen dimensiones comparables a la de Plutón. Todos estos descubrimientos pusieron bajo la lupa el concepto mismo de *planeta*. Por este motivo, la Unión Astronómica Internacional (IAU por sus siglas en inglés) designó un grupo de trabajo que tuvo como objetivo redefinir este término. Finalmente, luego de acalorados debates, la resolución B5 de la IAU, tomada en la Asamblea General celebrada en Praga el 24 de Agosto de 2006, define a un **planeta** como un cuerpo celeste que:

(a) está en órbita alrededor del Sol,

(b) tiene la masa suficiente como para que su autogravedad contrarreste las fuerzas de cuerpo rígido, de modo que su forma sea aquella correspondiente a la de equilibrio hidrostático (cuasi esférica), y

(c) que haya "limpiado" la vecindad de su órbita.

El punto (c) hace referencia a que el objeto en cuestión sea gravitatoriamente dominante en la región que comprende su órbita y adyacencias, con lo cual no debe haber en sus cercanías ningún otro objeto de tamaño comparable, exceptuando sus lunas si las tuviera.

Como se desprende inmediatamente de la definición, la palabra *planeta*, estrictamente hablando, puede ser usada solo para hacer referencia a objetos del Sistema Solar. De hecho, los únicos planetas son, por ahora, Mercurio, Venus, La Tierra, Marte, Júpiter, Saturno, Urano y Neptuno. Una definición más abarcativa, que tenga en cuenta a objetos extrasolares, sea que estos estén en órbita entorno a una estrella o no, está actualmente siendo estudiada. De todos modos, y por las reminiscencias históricas del término, a los cuerpos de masa planetaria que orbitan una estrella o una enana marrón se los sigue llamando planetas.

Los planetas extrasolares, o *exoplanetas*, son planetas que no pertenecen al Sistema Solar y que orbitan alrededor de una estrella. La mayoría de los detectados hasta el momento son planetas gigantes de características jovianas, aunque seguramente la detección de planetas de tipo terrestres será corriente en los próximos años, tal como lo predice la teoría (Ida & Lin 2008a, 2008b). En última instancia, lo que se persigue con esta intensa exploración de sistemas extrasolares es la búsqueda de vida. Claro que, probablemente, la detección de vida tal como la conocemos en nuestro planeta todavía esté muy lejos de ocurrir. Sin embargo, como veremos en el capítulo 2, en las últimas dos décadas la detección de exoplanetas ha aumentado de una manera extraordinaria, al punto que las teorías de formación y evolución planetaria quedaron varios pasos por detrás de las observaciones.

Pero no solo se han detectado planetas entorno a estrellas y enanas marrones, sino también "planetas sueltos" en el espacio (ver, por ejemplo, Zapatero Osorio et al. 2002). A estos objetos de masa planetaria se los denomina *planemos*, *rogue planets* o *free floating planets* ("planetas flotando a la deriva"). El descubrimiento de estos objetos que no están asociados a ninguna estrella (en general pertenecen a cúmulos abiertos muy jóvenes (Caballero et al. 2006)) fue otro hito en las Ciencias Planetarias, ya que despertó nuevos interrogantes, tanto sobre la formación estelar como sobre la formación de sistemas planetarios. Si estos "planetas" en sus inicios orbitaban una estrella y por algún encuentro fueron separados de ella, o si pudieron formarse en soledad en alguna región de formación estelar, es todavía una incógnita.

Si bien la definición de "planeta extrasolar" no ha sido oficialmente establecida todavía, el grupo de trabajo de la IAU trabajando en este tema propone que:

- Objetos con masas debajo de la fusión termonuclear del deuterio (alrededor de 13 masas de Júpiter para la metalicidad solar), que orbitan estrellas o remanentes estelares son considerados "planetas", sin importar cómo se formaron. El límite inferior de masa requerido para un planeta extrasolar es el mismo que el utilizado en el Sistema Solar.

- Objetos subestelares con masas superiores por encima del límite correspondiente a la fusión del hidrógeno son "enanas marrones", sin importar su mecanismo de formación.

- Free-floating objects en cúmulos estelares con masas por debajo de la necesaria para la fusión del deuterio no son "planetas" sino "sub-enanas marrones". Su existencia es de particular importancia para determinar los mecanismos de formación planetaria. Podrían ser planetas que fueron arrancados de sus órbitas por perturbaciones gravitatorias. Sin embargo, en la región de Orión no hay suficientes estrellas que pudieran albergar sistemas planetarios. Luego, estos planetas podrían provenir del colapso gravitatorio por fragmentación directa de la nube.

### 1.3.1. Enanas marrones

Con el término de *enana marrón* se denomina a los objetos subestelares que no tienen la masa necesaria para mantener la fusión de hidrógeno en su núcleo[3] pero que, sin embargo, pueden quemar deuterio. Las enanas marrones son completamente convectivas, tanto en su interior como en su superficie. El posible rango de valores para su masa está entre aquellas correspondientes a la de los planetas gigantes y a la de las estrellas más pequeñas de la Secuencia Principal, esto es, entre 13 y 80 masas de Júpiter ($M_J$). Es en este intervalo de masas donde se alcanzaría la temperatura de $0,5 \times 10^6$ K necesaria para la fusión de deuterio. Las enanas marrones de mayor masa (0,06 - 0,08 $M_\odot$) pueden también quemar litio transitoriamente durante algunas etapas de su evolución ya que la temperatura en su interior podría llegar a los $2,5 \times 10^6$ K. La presencia de litio en un espectro es lo que permite la identificación de las enanas marrones y su diferenciación de las estrellas de muy baja masa. Las estrellas queman su contenido de litio muy rápidamente, no dejando rastros de su presencia, mientras que en los objetos subestelares el litio no llega a ser completamente consumido. Entonces, la diferencia más importante entre las estrellas y las enanas marrones es que estas últimas no alcanzan la secuencia principal correspondiente a la quema de hidrógeno. Así, el proceso evolutivo de una enana marrón es simplemente su enfriamiento. En este aspecto son más cercanas a los planetas gaseosos, siendo la mayor diferencia entre ambos su temperatura para una determinada edad, la cual depende de la masa del objeto.

La primera enana marrón confirmada no estaba sola en el espacio sino en órbita alrededor de una estrella. A partir de ese momento, las enanas marrones se han encontrado en forma aislada o como miembros de sistemas múltiples, pudiéndose distinguir dos subtipos de sistemas: aquellos formados solo por enanas marrones y aquellos en los que la enana marrón tiene un compañero de masa estelar (Bate 2006). Este último caso (enana marrón-estrella) es de particular interés a la hora de comparar a las enanas marrones con los planetas. En el caso de sistemas binarios de enanas marrones se observan dos grandes diferencias respecto de los sistemas binarios estelares: por un lado la separación entre las enanas marrones es menor que la que ocurre en los pares estelares, pero además la relación de masa entre ambos miembros es siempre cercana a la unidad, mientras que la frecuencia de estrellas binarias tiene un pico cuando la relación de masas es del orden de 0,25. Por otro lado, tenemos a las enanas marrones como acompañantes de una estrella. Como mencionaremos más adelante, la mayor cantidad de exoplanetas fueron descubiertos por el método de velocidades radiales. Es esperable entonces que esta técnica de como resultado el descubrimiento de muchas enanas marrones. Sin embargo, eso no ha sido así. Muy pocas enanas marrones fueron descubiertas en órbitas cercanas a estrellas, dando lugar a lo que se conoce como "desierto de enanas marrones". Esta ausencia hace pensar en un mecanismo de formación diferente entre los planetas y las enanas marrones. Existen varios escenarios propuestos para la formación de las enanas marrones, dependiendo de si forman parte de un sistema (y de qué clase de sistema) o si son objetos aislados. Aquellas enanas marrones (y posiblemente también los planemos) que se encuentran aisladas han seguido, muy probablemente, un proceso de formación análogo al de las estrellas de masa solar, atravesando en sus orígenes una fase T-Tauri. Como evidencia de esto se tiene generalmente la presencia de un disco circumestelar, el cual se detecta en la mayoría de las enanas marrones jóvenes, además de líneas de emisión anchas y asimétricas (las cuales indican procesos de acreción) y líneas prohibidas (Jayawardhana & Ivanov 2006). Estas analogías entre las estrellas de baja masa y las enanas marrones sugieren un escenario de formación común. Sin embargo, debido a su baja masa, las reacciones termonucleares de fusión de hidrógeno que ocurren en el núcleo no producen la luminosidad suficiente para mantener su estructura. El colapso gravitatorio de la protoestrella no calienta en forma efectiva al núcleo y antes que se alcance la temperatura necesaria para el desarrollo eficiente de la fusión nuclear, la densidad se vuelve muy alta, llegando al estado de degeneración electrónica. Bajo estas condiciones, la contracción gravitatoria se detiene y el resultado es una "estrella fallida" o enana marrón. A partir de ahí, el objeto simplemente se enfría, radiando su energía interna.

[3] La temperatura necesaria para la quema de hidrógeno es aproximadamente $3 \times 10^6$ K.

Se ha sugerido que otro mecanismo posible para la formación de enanas marrones podría tener lugar durante el proceso de formación estelar en regiones donde hay varios objetos en formación simultánea. Allí se producirían eyecciones por interacciones dinámicas y las enanas marrones serían objetos que fueron eyectados antes de llegar a alcanzar la masa necesaria para convertirse en una estrella. Este escenario fue estudiado mediante simulaciones hidrodinámicas que mostraron su factibilidad (ver Bate 2006 y las referencias que allí se mencionan).

Otra posibilidad es que las enanas marrones se formen debido a una finalización abrupta de la acreción de gas debido a que la radiación de estrellas masivas cercanas produciría la evaporación del resto de gas necesario. Si bien este escenario es factible, no podría ser el único mecanismo de formación de enanas marrones ya que no podría explicar por sí solo la abundancia de estos objetos.

Una característica sobresaliente de las enanas marrones es que todas tienen más o menos el radio de Júpiter. En todo el rango de masas posibles, el tamaño del radio solo varía en un $10-15\,\%$. Distinguir una enana marrón de un planeta gigante no es una tarea sencilla, sobre todo cuando, como veremos más adelante, de las observaciones de planetas extrasolares generalmente solo se puede estimar su masa mínima. El primer objeto candidato a enana marrón fue descubierto en 1988, GD 165B, compañera de una enana blanca (Becklin & Zuckerman 1988). Por años existió la controversia de si era realmente una enana marrón. Hoy es el prototipo de las estrellas menos masivas, las de tipo espectral L, que están en la frontera de los objetos estelares. Fue recién en 1995 que se encontraron tres candidatos que no presentaban controversia. El más significativo fue el descubrimiento de Gliese 229B. Este objeto es lo suficientemente frío como para no ser una estrella pero demasiado luminoso y distante de su compañera como para ser un planeta. Su espectro no solo muestra que su temperatura y su luminosidad están por debajo de la estelar (Oppenheimer et al. 1995), sino que también presenta líneas de absorción de metano en 2 micrones (estas líneas no son esperables en los espectros estelares), una característica asociada, hasta ese momento, solo a nuestros planetas gigantes. Otra característica que distingue a los objetos subestelares son las líneas de litio. Si bien el litio también está presente en los espectros de las estrellas jóvenes, en esos casos se las puede distinguir de las enanas marrones por su tamaño. Las enanas marrones emiten la mayor parte de su flujo en las bandas del infrarrojo cercano. La temperatura superficial se encuentra entre los 2.200 y los 750 K.

### 1.3.2. Planetas gigantes

Planetas gigantes o jovianos es la denominación con la que se conoce colectivamente a Júpiter, Saturno, Urano y Neptuno, los planetas del Sistema Solar caracterizados por sus masivas envolturas gaseosas.[4]

Júpiter y Saturno son conocidos desde tiempos remotos y estuvieron asociados a la mitología y religión de numerosos pueblos. Fueron los romanos quienes les dieron los nombres con los que los conocemos en la actualidad. Júpiter (Zeus para los griegos) era el padre de todos los dioses y, entre otras cosas, era el protector de las actividades militares de los romanos fuera de los límites del Imperio. Saturno, por su parte, era considerado el dios de la cosecha y para los romanos era equivalente al dios Kronos de los griegos. Júpiter es el planeta de mayor masa del Sistema Solar. Su magnitud aparente es -2.8 en oposición y -1.6 en conjunción, lo cual lo hace el tercer objeto más brillante del cielo nocturno, después de La Luna y Venus (aunque en ciertos momentos del año Marte puede ser aún más brillante).

En tanto, el nombre de Urano proviene de la antigua Grecia. Urano era una deidad del cielo, padre de Kronos (Saturno) y abuelo de Zeus (Júpiter). Si bien Urano es visible a ojo desnudo, no fue reconocido como un planeta sino hasta fines del siglo XVIII ya que se lo confundía con una estrella. En 1781 William Herschel anunció el descubrimiento de lo que él pensaba era un cometa. Sin embargo, luego de sucesivas observaciones, no logró identificar su cola y, por otra parte, la estimación de sus elementos orbitales indicaban que la órbita era bastante circular. Estos datos sugerían que se trataba de un nuevo planeta, pero Herschel no se atrevía a darle esa clasificación. Sin embargo, rápidamente, el resto de la comunidad científica lo reconoció como tal y luego de varios debates se le dio finalmente el nombre de Urano.

Por su parte, Neptuno fue el primer objeto del Sistema Solar predicho matemáticamente. Apartamientos de su trayectoria real respecto de la órbita calculada de Urano indujeron a los astrónomos de la época a proponer la existencia de un octavo planeta. Así fue como gracias a los cálculos del francés Le Verrier, se descubrió Neptuno en septiembre de 1846, en la región del cielo en la cual se lo buscaba. El nombre de Neptuno (dios del mar en la mitología romana) lo propuso el astrónomo F. Struve a fines de ese mismo año y fue ampliamente aceptado por la comunidad científica internacional.

[4] Las características de estos planetas serán desarrolladas en detalle en los capítulos que siguen. Esta sección fue incluida aquí por completitud.

### 1.3.3. Planetas terrestres

Los planetas terrestres (rocosos, telúricos, o interiores) son cuerpos similares a La Tierra, compuestos fundamentalmente por silicatos. Según su radio orbital creciente, ellos son: Mercurio, Venus, La Tierra y Marte. Los planetas terrestres son los más cercanos al Sol. Todos los planetas terrestres tienen estructuras similares: un núcleo metálico compuesto fundamentalmente de hierro, rodeado por un manto de silicatos. Las atmósferas de estos planetas son secundarias, ya que fueron generadas por vulcanismos o impactos cometarios, y no como la de los planetas gigantes donde el gas proviene de la nebulosa protoplanetaria.

Si bien la mayoría de los planetas extrasolares descubiertos hasta el momento tienen características asimilables a las de nuestros planetas gigantes, el primer sistema de exoplanetas detectado (Wolszczan & Frail 1992) resultó estar compuesto por planetas de tipo terrestres, siendo sus masas 0,02, 4,3 y 3,9 masas terrestres ($M_\oplus$). La gran diferencia es que estos planetas no orbitan una estrella como el Sol sino un púlsar (PSR B1257+12). Su descubrimiento fue completamente accidental: durante la observación del púlsar se detectaron interrupciones en las emisiones de radio, que resultaron ser causadas por los tránsitos de estos planetas. Si no hubieran estado orbitando un púlsar, no habrían sido descubiertos.

Cuando en 1995 se descubrió el primer exoplaneta entorno a una estrella de tipo solar (Mayor & Queloz, 1995), 51 Peg b, muchos astrónomos pensaron que se trataba de un gigante terrestre puesto que no se pensaba que fuera posible que un planeta gaseoso pudiera estar tan cerca de su estrella ($a$ = 0,052 UA). Recién en junio de 2005 se detectó un planeta de tipo terrestre alrededor de una enana roja, Gliese 876 (Rivera et al. 2005). Este objeto tendría una masa de entre 5 y 7 $M_\oplus$ y un período orbital de tan solo 2 días. En agosto del mismo año se observó, con la técnica de microlensing, un planeta de 5,5 $M_\oplus$ en una ubicación respecto de su estrella parecida a la del cinturón de asteroides local (OGLE-05-390LB, Beaulien et al. 2006). En abril de 2007, con el telescopio de La Silla (ESO), se descubrió que alrededor de la estrella Gliese 581 (enana roja), en su región de habitabilidad[5], hay un planeta cuya masa sería 5 veces la de la Tierra (Udry et al. 2007, Selsis et al. 2007). Sin embargo, todavía no se tiene la certeza de que este planeta sea de tipo rocoso.

### 1.3.4. Planetas enanos y cuerpos pequeños

En la década del '90 comenzaron a confirmarse observacionalmente los primeros objetos del Cinturón de Kuiper (hasta ese momento solo se conocía el sistema Plutón - Caronte). A comienzos de este siglo se descubrieron los primeros objetos transneptunianos comparables con Plutón tanto en tamaño como en sus características orbitales: Quoar, Sedna y Eris. Se sabía entonces que en un corto tiempo aparecerían muchísimos objetos similares y que, aplicando el criterio aceptado hasta ese momento, todos deberían ser llamados *planetas* (situación parecida a la que ocurrió en el Siglo XIX luego del descubrimiento de Ceres). Esto fue lo que motivó que la IAU redefiniera la palabra *planeta*, hecho que tuvo como consecuencia que surgieran otras dos clasificaciones nuevas: *planetas enanos* y *cuerpos pequeños del Sistema Solar*.

[5]La región de habitabilidad de una estrella se define como el rango de distancias orbitales en las cuales podría haber potenciales reservorios de agua en estado líquido, ingrediente fundamental para el desarrollo de organismos biológicos complejos (Kasting et al.1993).

Un **planeta enano** es un cuerpo celeste que: (a) está en órbita alrededor del Sol, (b) tiene la masa suficiente como para que su autogravedad contrarreste las fuerzas de cuerpo rígido, de modo que su forma sea aquella correspondiente a la de equilibrio hidrostático (cuasi esférica), (c) no limpió la vecindad de su órbita, y (d) no es un satélite.

Cualquier otro objeto, exceptuando los satélites, que no sea un planeta o un planeta enano será clasificado como un **cuerpo pequeño del Sistema Solar**. Entre ellos encontramos fundamentalmente a los asteroides, la mayoría de los objetos transneptunianos y los cometas.

Hasta el momento, la IAU reconoce cinco planetas enanos: Ceres (en el Cinturón de Asteroides), Plutón, Haumea, Makemake y Eris (pertenecientes al Scattered Disk). Se sospecha que otros 40 objetos ya conocidos también clasificarían como planetas enanos, y se estima que se encontrarán muchos más cuando se explore mejor el Cinturón de Kuiper. De hecho, los candidatos corresponden a un survey del hemisferio norte. Por simetría se esperaría encontrar otros tantos en el hemisferio sur.

El 11 de Junio de 2008, la IAU anunció una categoría dentro de los planetas enanos, los *plutoides*. Se define como **plutoide** a todo cuerpo celeste en órbita alrededor del Sol a una distancia superior a la de Neptuno, con masa suficiente como para que su autogravedad sea capaz de contrarrestar la fuerza de cuerpo rígido de modo que su forma sea la correspondiente a la de equilibrio hidrostático (cuasi esférica) y que no haya limpiado la vecindad de su órbita. No hay que confundir el término plutoide con *plutino*. Un **plutino** es un objeto transneptuniano en resonancia 2:3 con Neptuno. La denominación plutino remite a Plutón, el cual está atrapado en dicha resonancia. Esta clasificación, que incluye al propio Plutón y a sus lunas, solo se refiere a las propiedades dinámicas de los objetos y no a sus características físicas.

## 1.4. ¿Por qué estudiar la formación de planetas gigantes?

El ser humano ha sido siempre curioso, y ha buscado incansablemente respuestas a todas sus preguntas. Exceptuando Urano y Neptuno, el resto de los planetas del Sistema Solar han sido familiares a los ojos de millones de espectadores del cielo desde el principio de la humanidad. En los finales del siglo XX se sumaron, a los planetas ya conocidos, centenares de exoplanetas orbitando otros soles, la mayoría de ellos en configuraciones completamente distintas a la de nuestro Sistema Solar. La búsqueda de otros mundos tiene como objetivo principal encontrar análogos a nuestra Tierra donde se pueda haber desarrollado la vida. El desafío que esto plantea para la ciencia está recién en sus comienzos. En su exploración del Universo, los astrónomos han encontrado sistemas planetarios en configuraciones inesperadas, las cuales seguramente nunca hubieran sido propuestas en una reunión científica de no mediar evidencias irrefutables: Tierras orbitando púlsares, Júpiters a distancias de su estrella central más pequeñas que la de Mercurio al Sol, o mucho mayores que la de Plutón; Uranos y Neptunos *calientes*... ¿Cómo se formaron estos planetas? La teoría todavía no ha podido encontrar una respuesta. Pero, quizá, lo más sorprendente, es lo poco que todavía sabemos de nuestros "propios planetas". En la actualidad, no podemos precisar la estructura interna de Júpiter, ni si los planetas gigantes se formaron donde se encuentran hoy. Tampoco sabemos exactamente cuánta de la masa de Urano y Neptuno está en estado gaseoso, ni por qué Urano es el único de los planetas gigantes que emite menos radiación que la esperada, en relación a la que recibe del Sol. De hecho, no contamos con una teoría unívocamente aceptada para explicar la formación de los planetas gaseosos.

Los datos observacionales nos muestran, por un lado, la gran variedad de sistemas planetarios que pueden existir en el Universo. Pero por otra parte, imponen serias restricciones a los modelos que intentan explicar su origen. Los planetas gigantes están presentes en la mayoría de los sistemas planetarios detectados hasta el momento. En lo que respecta al Sistema Solar, se cree que jugaron un papel fundamental para que en la Tierra pudiera desarrollarse la vida. Sin embargo, la teoría de formación de los planetas gigantes dista mucho de estar cerrada. Más allá de las bases del modelo, que podrían considerarse "simples", existen numerosos actores que intervienen directamente, y pueden modificar de manera sustancial, el escenario de formación de estos objetos. Para un estudio consistente de la formación de los planetas gigantes se necesita contar tanto con resultados precisos para la ecuación de estado de un gas compuesto por hidrógeno y helio, como de la dinámica a gran escala del Sistema Solar. A diferencia de otros problemas astrofísicos, es difícil en este caso aislar y, al mismo tiempo, estudiar en forma realista un determinado aspecto de la cuestión. Por otro lado, es imprescindible hacer simplificaciones, muchas de ellas necesariamente gruesas, si se quiere comenzar a comprender cómo ocurre este proceso.

Debido a la dificultad que plantea el problema de la formación de los planetas gigantes, existen todavía muchas incógnitas, y se hace necesario que los modelos adopten hipótesis tan realistas como sea posible si se quiere llegar a una descripción confiable de este proceso. En función de todas las preguntas sin respuesta que rodean al origen de los sistemas planetarios creemos que vale la pena estudiar, en particular, la formación de los planetas gigantes. Dada la diversidad de situaciones que rodean a este problema, hemos decidido circunscribir este estudio a los planetas del Sistema Solar, esperando que en el futuro nuestro modelo pueda ser extendido para atacar el problema de la formación de planetas extrasolares.

# Capítulo 2

# Datos observacionales

La naturaleza de los planetas gigantes es muy distinta a la de sus pares terrestres. En este capítulo presentaremos la información que aportan los datos observacionales, tanto en lo que respecta a los planetas del Sistema Solar como a los planetas extrasolares.

## 2.1. Características de los planetas gigantes del Sistema Solar

La característica fundamental que distingue a los planetas gigantes de los planetas terrestres es su masa y su composición. Júpiter y Saturno son dos órdenes de magnitud más masivos que la Tierra, y Urano y Neptuno, si bien relativamente mucho más pequeños , superan a la masa de la Tierra en un factor 10. Pero además, en el caso de Júpiter y Saturno, más del 90 % de su masa está compuesta por hidrógeno y helio.

A continuación describiremos los aspectos más relevantes que caracterizan al interior de los planetas gigantes. Es importante destacar que esta información se infiere de la construcción de modelos teóricos, en conjunto con los datos observacionales disponibles. Los aspectos generales de los párrafos que siguen corresponden a un curso dictado por T. Guillot (Guillot, 2001).

La masa de los planetas gigantes se puede calcular de la observación del movimiento de sus satélites naturales (sus valores se listan en la tabla 1.1). Del análisis de las trayectorias descriptas por los vuelos de las misiones espaciales, sobre todo de aquellas que estuvieron en órbita polar, se pueden obtener estimaciones más precisas de su campo gravitatorio. Debido a la rápida rotación que tienen los planetas, su campo gravitatorio se aparta del correspondiente a un cuerpo esférico. Si expandimos el potencial gravitatorio en polinomios de Legendre, $P_i(\cos\theta)$, obtenemos, bajo la hipótesis de simetría axial:

$$\phi(r,\theta) = -\frac{GM_\mathrm{p}}{r}\left(1 - \sum_{i=1}^{\infty}\left(\frac{R_\mathrm{ec}}{r}\right)^i J_i P_i(\cos\theta)\right) \tag{2.1}$$

siendo $r$ la coordenada radial, $\theta$ el ángulo polar, $M_\mathrm{p}$ la masa del planeta, $R_\mathrm{ec}$ su radio ecuatorial y $J_i$ los momentos gravitatorios. Dado que los planetas gigantes se encuentran casi en equilibrio hidrostático, solo los coeficientes correspondientes a los términos pares resultan no despreciables (los $J_i$ impares son nulos debido a la simetría que existe respecto del ecuador). La estructura interna de estos objetos se puede estimar a partir de su masa $M_\mathrm{p}$, su radio ecuatorial $R_\mathrm{ec}$ y de los momentos gravitatorios. Los momentos gravitatorios son medidos por las sondas y misiones espaciales que sobrevuelan ocasionalmente a los planetas. Una primera estimación de la densidad se puede obtener de dividir la masa total del planeta, $M_\mathrm{p}$, por el volumen de la esfera de radio $R_\mathrm{ec}$. De la tabla 1.1 se puede ver que las densidades de los planetas gigantes rondan la unidad, siendo muy bajas comparadas con las de los planetas terrestres. Saturno es el menos denso de todos, con una densidad promedio inferior a 1. Cálculos más detallados se obtienen utilizando el período de rotación y las mediciones de los momentos $J_2$, $J_4$ y $J_6$. Los momentos gravitatorios están directamente relacionados con los momentos de inercia, con lo cual dependen sensiblemente de la velocidad de rotación, de la distribución de la masa, etc. De este modo, el conocimiento de los momentos gravitatorios junto con una ecuación de estado para el gas permiten estimar la densidad $\rho_\mathrm{g} = \rho_\mathrm{g}(r,\theta)$, la cual es equivalente (vía la ecuación de estado) al perfil de densidad $\rho_\mathrm{g}(P,T)$, donde $P$ es la presión y $T$ la temperatura.

El procedimiento para obtener información del interior de los planetas es integrar la ecuación de equilibrio hidrostático (incluyendo el término de rotación) junto con una ecuación de estado y opacidades apropiadas. Las condiciones de borde para esta integración surgen de los datos observacionales (masa, temperatura superficial, gravedad superficial, etc.). De este modo, el conocimiento de la estructura interna es indirecto y depende fuertemente del modelo físico empleado. En este proceso, la ecuación de estado (EOS, por sus siglas del término en inglés Equation of State) juega un papel fundamental en la descripción de la estructura interna de los planetas gigantes. La EOS que se utiliza en la mayoría de los cálculos de formación, evolución y estructura de los planetas gigantes es la calculada por Saumon, Chabrier & van Horn (1995, en adelante SCVH; ver capítulo 4 para comentarios al respecto de la EOS). Estos autores calculan la ecuación de estado para el hidrógeno y el helio por separado y sugieren una determinada interpolación para el caso de una mezcla de ambos gases. Hay que notar que al hacer esto se desprecian las interacciones entre ambas especies. Por otra parte, la presencia de otros elementos que no sean el hidrógeno y el helio, se incorpora utilizando otras ecuaciones de estado, las cuales resultan aún menos confiables. Es importante subrayar que la ecuación de estado SCVH se calcula sobre la base de numerosas simplificaciones pero es, hasta el momento, la más detallada con la que se cuenta.

Las atmósferas de los planetas gigantes están compuestas, fundamentalmente, por hidrógeno molecular ($H_2$) y helio (He). Estas moléculas son difíciles de detectar puesto que su momento dipolar es nulo. En cuanto a las líneas producidas por las transiciones electrónicas, las mismas corresponden a altitudes atmosféricas altas y, por lo tanto, no aportan información sobre la estructura interior en regiones más profundas. La abundancia de elementos pesados, por otra parte, resulta fundamental para entender el proceso de formación de estos objetos. Sin embargo, exceptuando el caso de Júpiter, de donde se tiene información gracias a la sonda Galileo, las abundancias en el resto de los planetas gigantes ha sido muy pobremente estimada y solo la abundancia de metano ($CH_4$) pudo ser medida con confianza.

Se cree que tanto en Júpiter como en Saturno los elementos que componen sus envolturas están bien mezclados. En el caso de Júpiter, las mediciones *in situ* de la sonda Galileo dan como resultado que la abundancia de helio es $Y/(X+Y) = 0,238 \pm 0,007$, siendo $Y$ la fracción en masa de todas las especies de helio, y $X$ la correspondiente al hidrógeno. En cuanto a Saturno, las incertezas todavía son bastante grandes. Los datos con los que se cuenta hasta el momento son los correspondientes a la sonda Voyager 2, que dan como resultado $Y/(X+Y) = 0,06 \pm 0,05$. Según se cree, estos últimos valores son poco confiables y se piensa que en realidad serían mayores. Se espera que, en breve, los resultados de las mediciones de la misión Cassini-Huygens puedan aportar datos más precisos. De todos modos, la abundancia superficial de helio en estos planetas resulta siempre inferior a la solar[1]. Sin embargo, si aceptamos que la nebulosa primordial tenía abundancias solares, debe haber operado algún mecanismo para que las abundancias superficiales observadas en los planetas exteriores sean menores a las solares. Una posibilidad es que exista una separación de fases entre el hidrógeno y el helio, en la cual "gotitas" de helio puedan crecer lo suficientemente rápido como para precipitar hacia el interior por la acción de la gravedad y no ser afectadas por la convección. Una evidencia que apoya esta hipótesis es la deficiencia en la abundancia de neón, dado que el neón tiende a disolverse en el helio, con lo cual habría descendido al interior dentro de estas gotas.

En la atmósfera de Júpiter se detecta la presencia de metano, vapor de agua, amoníaco, carbono y oxígeno, entre otros "elementos pesados". En lo que se refiere a elementos pesados en general, el enriquecimiento sería de un factor entre 2 y 4 en relación a la abundancia solar. De todas formas, todavía hay elementos cuyas abundancias no han podido ser bien determinadas. Basados en datos espectroscópicos, la composición de Saturno podría ser muy similar a la de Júpiter, o incluso aún más rico en elementos pesados.

Utilizando la EOS de SCVH, Saumon & Guillot (2004) estimaron, dentro del rango de incertezas de la EOS, el estado del gas en los interiores de Júpiter y Saturno. Mientras que el interior de Saturno se encuentra dentro de una región bastante bien estudiada de la ecuación de estado de la mezcla hidrógeno-helio (donde el hidrógeno es molecular), en el caso de Júpiter la situación es diferente. Puesto que gran parte de su estructura se encuentra a presiones intermedias, donde la ecuación de estado es bastante menos conocida, las incertezas entorno a estas estimaciones son grandes (ver capítulo 4 y figura 4.4). El modelo teórico que se propone para el cálculo del interior de estos planetas distingue tres estructuras: un núcleo central, denso, presumiblemente formado por rocas (silicatos, hierro) e hielos (agua, amoníaco, metano); un interior rico en helio y una envoltura externa deficiente en helio (ésta última hipótesis basada en las observaciones). Una representación esquemática del interior de los planetas gigantes se muestras en la figura 2.1. La existencia del núcleo sólido resulta generalmente necesaria para reproducir los campos gravitatorios medidos. Es importante destacar, sin embargo, que para los modelos de formación, las cantidades más relevantes que se quieren estimar son: la masa del núcleo ($M_c$) y la masa de elementos pesados contenida en el interior del planeta ($M_Z$). La determinación de estos parámetros surge de aquellos modelos que mejor ajusten los valores del $R_{ec}$ y los momentos gravitatorios $J_2$ y $J_4$. En el caso de Júpiter, de acuerdo con este modelo, la masa del núcleo podría acotarse entre 0 y 11 $M_\oplus$, mientras que la masa de elementos pesados estaría entre 1 y 39 $M_\oplus$. Comparado con la abundancia de elementos pesados en el Sol, estos valores corresponden a un enriquecimiento de un factor entre 1,5 y 6. El caso de Saturno es menos problemático, ya que al ser menos masivo que Júpiter resulta menos sensible a las incertezas de la EOS en los regímenes correspondientes al interior de este planeta. Los valores estimados para el núcleo son $9\,M_\oplus < M_c < 22\,M_\oplus$, y para el resto de los elementos pesados en la envoltura $1\,M_\oplus < M_Z < 8\,M_\oplus$. Esto representa un enriquecimiento de un factor entre 6 y 14 por encima de la abundancia solar.

[1]Los modelos evolutivos estelares estiman que la relación del helio relativa al hidrógeno en el Sol es $Y/(X+Y) = 0,270 \pm 0,005$.

Recientemente, Militzer y colaboradores (2008) calcularon el interior de Júpiter utilizando una EOS derivada a partir de primeros principios. El modelo empleado no presupone la estructura del planeta dividida en tres regiones sino simplemente compuesta por una parte "sólida" (el núcleo) y una envoltura gaseosa. Según este modelo, la masa del núcleo sería de $16 \pm 2\,M_\oplus$, siendo la metalicidad de la envoltura mucho menor que la estimada por Saumon & Guillot (2004). A diferencia de los modelos calculados con la EOS de SCVH, donde la masa del núcleo era bastante pequeña y las abundancias en el manto eran grandes, en el modelo de Militzer et al. la situación se revierte. Según las mediciones de la sonda Galileo, si se extrapolan los valores superficiales (con un modelo adecuado) a toda la envoltura, la masa total de hielos fuera del núcleo no superaría las 6 $M_\oplus$. En el modelo de Militzer et al., la masa de hielos sería de 4 $M_\oplus$, en buen acuerdo con los datos observacionales.

Por otra parte, los valores de los momentos multipolares también reflejan la rotación del planeta; cuanto más altos sean más afectados se verán por rotaciones no uniformes. Para ajustar el valor de $J_4$, Militzer et al. no necesitan proponer la diferenciación de la envoltura como lo hacen Saumon & Guillot. Por las características de su modelo, ellos sugieren que el valor de $J_4$ está afectado por los vientos superficiales y, de hecho, bajo esta propuesta ellos logran ajustar este valor sin necesidad de modificar otros parámetros.

Los resultados de Militzer et al. (2008) son muy recientes y probablemente no hayan sido todavía discutidos en profundidad por la comunidad científica. Además, por el momento, la EOS que emplearon para calcular sus resultados no fue publicada en forma de tabla, por cuanto no es para nada trivial intentar hacer cálculos bajo las hipótesis de su modelo. Por su parte, Nettelman y colaboradores (2008) también estimaron, a partir de una EOS obtenida por primeros principios, el interior de Júpiter. Sin embargo, sus resultados no concuerdan con los de Militzer et al. De este modo, la masa del núcleo y la abundancia total de sólidos presente en el interior de de los planetas gigantes sigue siendo uno de los ejes principales de discusión en relación a la estructura interna de los planetas.

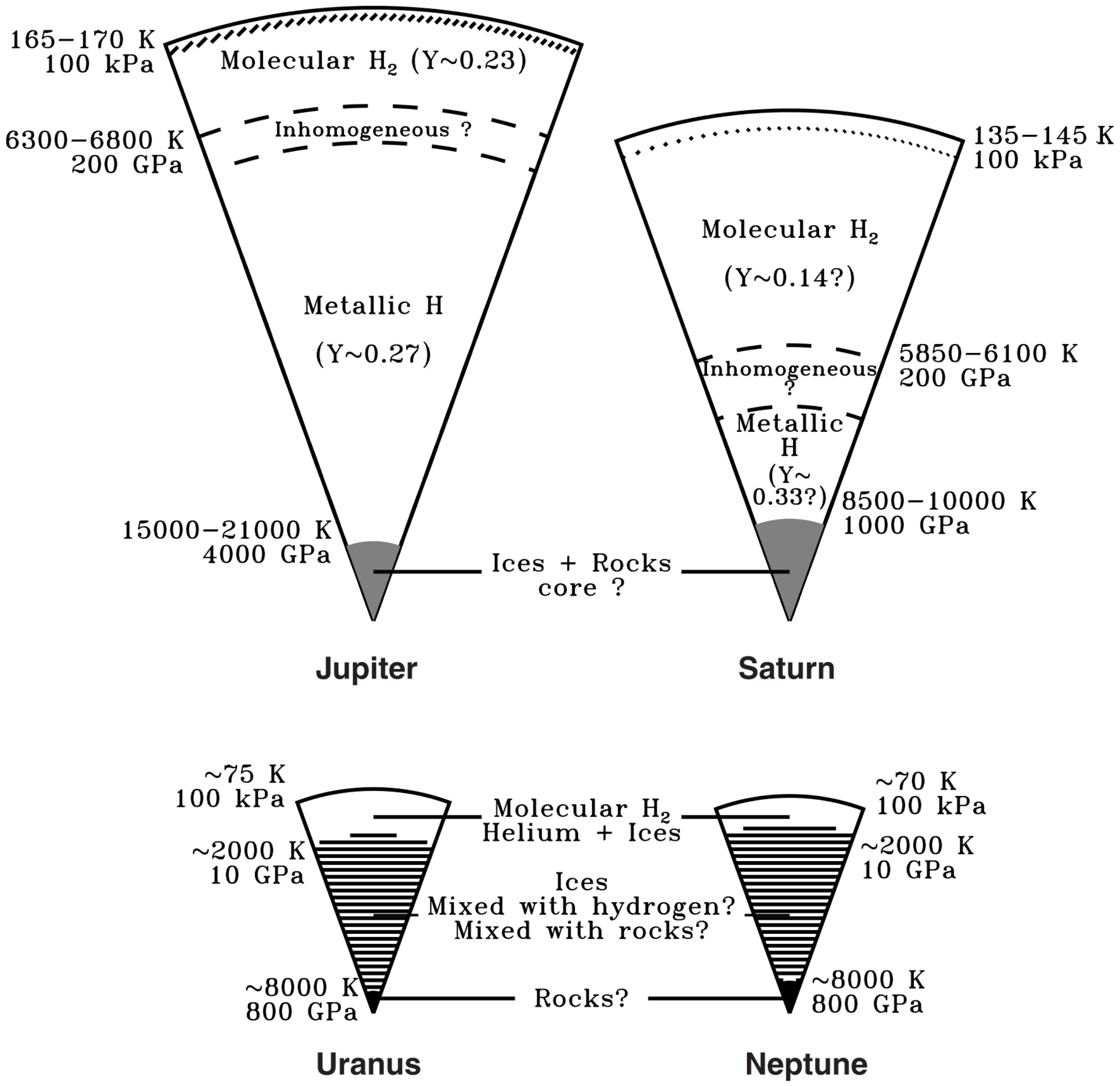

**Figura 2.1.** Representación esquemática del interior de Júpiter, Saturno, Urano y Neptuno (figura tomada del artículo de Guillot, 1999).

En lo que respecta a las atmósferas de Urano y Neptuno se observa un enriquecimiento en elementos pesados de hasta un factor 30 por encima de la composición solar, fundamentalmente de $CH_4$. Los perfiles de densidad que se pueden inferir de los momentos gravitatorios hacen pensar que estos planetas están compuestos fundamentalmente por hielos (mezcla de $H_2O$, $CH_4$ y $NH_3$), y que solo en su atmósfera predominan el hidrógeno y el helio. Los modelos del interior de Urano y Neptuno proponen también una estructura de tres capas: un núcleo central rocoso (compuesto fundamentalmente por silicatos e hierro), una capa de hielos y una atmósfera gaseosa. Cada una de estas capas no estaría homogéneamente mezclada. La existencia de estas regiones inhomogéneas confirmaría la presunción de que el hidrógeno se encuentra confinado solo a la atmósfera. Los modelos predicen que la relación entre hielos y rocas es del orden de 10, mucho mayor que el valor protosolar, que sería 2,5. Por otra parte, si se impone que la relación entre hielo y rocas sea acorde a la composición solar se obtiene que, en rasgos generales, ambos planetas están formados por un 25 % de rocas, un $60 - 70$ % de hielos y el restante $5 - 15$ % de hidrógeno y helio. Esto resulta en que la estimación de la masa de gas para Urano sea de 3 $M_\oplus$ y para Neptuno de 5 $M_\oplus$. La estructura característica de los planetas gigantes puede observarse en la figura 2.1.

Las opacidades son otro punto fundamental a la hora de explicar el interior de los planetas. A bajas temperaturas ($\sim 1500 - 2000$ K) la opacidad media de Rosseland está dominada por los granos de polvo. Las tablas de opacidad que se usan en astrofísica se calculan generalmente imponiendo la existencia de un equilibrio químico global donde las especies condensadas son retenidas y además, suponiendo que la distribución de tamaños de las partículas es la correspondiente al medio interestelar. Sin embargo, no está claro que esta última hipótesis sea válida para el caso de los planetas gigantes y las enanas marrones. Dado que el interior de los planetas sería mayormente convectivo, esto modificaría la distribución de tamaños de los granos, hecho que tendría profunda repercusión en los modelos de formación de los planetas gigantes (Podolak 2003). En lo que respecta a Urano y Neptuno, la opacidad de Rosseland es todavía incierta. De las estimaciones hechas al momento, ambos planetas tendrían una región radiativa a temperaturas de alrededor de 1500 K. Sin embargo, estas regiones podrían desaparecer de existir ciertas fuentes de opacidad como el sodio y el potasio, presentes en las enanas marrones. Sin embargo, vale la pena notar que si bien la existencia de una zona radiativa es bastante significativa a la hora de calcular la evolución de estos objetos, no es determinante para el cálculo de la estructura interior.

Júpiter, Saturno y Neptuno emiten mucha más energía de la que reciben del Sol. El caso de Urano es menos claro: si bien su flujo intrínseco es mucho menor que el del resto de los planetas gigantes, está en duda si emite o no más energía de la que recibe del Sol. En el caso de Júpiter se mostró que esto puede ser explicado por la continua contracción y enfriamiento del planeta. Una consecuencia importante de la existencia del flujo de calor intrínseco en un planeta es que requiere de temperaturas del orden de los $10^4$ K o más, lo cual implica que los planetas gigantes son fluidos, y que su interior resultaría ser esencialmente convectivo. Estimar las temperaturas superficiales es por demás complicado pero, asumiendo como superficie la correspondiente a 1 bar de presión se obtiene: para Júpiter, $T = 165 \pm 5$ K, para Saturno $T = 135 \pm 5$ K, para Urano $T = 76 \pm 2$ K y para Neptuno $T = 72 \pm 2$ K.

Las atmósferas de los planetas gigantes son de naturaleza compleja y turbulenta. Por ejemplo, los vientos son muy variables dependiendo de la latitud. En el caso de Júpiter y Saturno, los vientos en el ecuador tienen una velocidad mucho mayor que en los polos, mientras que para Urano y Neptuno se da a la inversa. Sin embargo, dado que los cuatro planetas tienen altas velocidades de rotación entorno a su propio eje, las velocidades de los vientos son más bajas que la velocidad de rotación (la velocidad típica de los vientos en Júpiter es de 360 km $h^{-1}$). Los vientos en Saturno tienen velocidades mucho más altas que en Júpiter, pudiendo alcanzar los 1.800 km $h^{-1}$. Sin embargo, es Neptuno el planeta con los vientos más intensos del Sistema Solar, llegando a velocidades de hasta 2.100 km $h^{-1}$. Además de los vientos, estos planetas presentan tormentas, que pueden tener la extensión de todo el planeta, y durar por semanas o por siglos. La gran mancha roja de Júpiter es un centro anticiclónico de 12.000 km de diámetro, situado 22° al sur del ecuador con, al menos, 300 años de edad. Los modelos indican que la tormenta es estable y que podría ser una característica permanente del planeta. Júpiter está cubierto por nubes compuestas de cristales de amoníaco y posiblemente bisulfuro de amonio. Las nubes están en la tropopausa y cubren su superficie en forma de bandas. Los colores marrones y anaranjados de las nubes provienen de compuestos que suben desde el interior a la superficie y que cambian de color cuando son expuestos a la luz ultravioleta solar. Si bien la composición exacta se desconoce, las posibles sustancias que la componen podrían ser fósforo, sulfuros o hidrocarburos. En lo que a Saturno respecta, se observaron tormentas desarrollándose en todo el planeta.

Por último, debemos hacer mención a los satélites y anillos de los planetas gigantes. De acuerdo a sus características orbitales, los satélites se pueden clasificar en regulares e irregulares. Los primeros generalmente se encuentran en órbitas casi circulares en el plano del ecuador, rotando en sentido directo. Los irregulares, por su parte, no tienen una forma definida y describen órbitas excéntricas, alargadas y retrógradas. Al momento, sabemos que Júpiter cuenta con 63 satélites, de los cuales ocho son regulares y los restantes son irregulares. Los satélites regulares se dividen en dos grupos: los cuatro interiores (Grupo Amaltea) y los cuatros satélites galileanos, descubiertos por Galileo hace 400 años. Estos últimos, en orden decreciente según su tamaño, son: Ganímides, Calisto, Io y Europa. Saturno, por su parte, tiene 60 satélites naturales, siendo el más grande Titán que, después de Ganímides, es el segundo en tamaño en el Sistema Solar. Urano y Neptuno cuentan con 27 y 13, respectivamente. Se cree que los satélites regulares se formaron en el disco de acreción del planeta, mientras el planeta se encontraba en sus últimas etapas del proceso de formación. Los irregulares, en cambio, serían planetesimales capturados por el campo gravitatorio del planeta. Los cuatro planetas gigantes también tienen anillos y se supone que los satélites son los que, constantemente, les proveen el material debido a las colisiones que se producen entre ellos. De entre todos, el sistema de anillos de Saturno es el más espectacular y el único que puede observarse con binoculares. De hecho, el área que ocupan los anillos es tan grande que al reflejar la luz que proviene del planeta resultan ser tan o más brillantes que el propio Saturno.

## 2.2. Planetas extrasolares

Si bien esta Tesis no trata el problema específico de los planetas extrasolares, es innegable que el interés por los planetas gigantes del Sistema Solar fue reavivado a partir del descubrimiento de estos objetos alrededor de otras estrellas. Creemos entonces adecuado hacer mención de las características más importantes de los planetas extrasolares, como así también de los mecanismo que permiten su detección, y de las preguntas y teorías que surgen en base a estos descubrimientos.

### 2.2.1. Métodos de detección[2]

Los métodos de detección de planetas extrasolares pueden clasificarse en dos grandes grupos: los indirectos y los directos. Se dice que un método es indirecto si a través de él se puede inferir la presencia de un planeta, lo cual significa que lo que se detecta son los efectos que tiene el planeta sobre la estrella y no una señal propia proveniente del planeta. Variaciones regulares en la posición de una estrella en el cielo, corrimientos por efecto Doppler de las líneas espectrales y cambios periódicos en la luminosidad son algunos de los efectos medibles que pueden indicar la presencia de un planeta orbitando entorno a una estrella. Veremos a continuación que hay numerosas técnicas de detección indirecta de planetas extrasolares. Por otra parte, cuando podemos observar la luz proveniente del planeta, cualquiera sea la longitud de onda involucrada, decimos que el método de detección es directo.

#### Métodos indirectos

Existen numerosas técnicas que permiten la detección indirecta de exoplanetas. En general, cada una de ellas resulta más efectiva que otra dependiendo de la configuración del sistema planetario, por cuanto cada técnica tiene asociado un sesgo intrínseco respecto de las características del objeto detectado. A continuación describiremos brevemente cuatro de las técnicas más utilizadas las cuales, a su vez, presentan marcadas diferencias conceptuales entre sí.

#### Velocidades radiales

El método de las velocidades radiales o de corrimiento Doppler es, hasta el momento, el método de detección de planetas extrasolares más exitoso, puesto que la mayoría de los descubrimientos han estado relacionados a esta técnica. De hecho, el primer exoplaneta orbitando alrededor de una estrella de tipo solar, 51 Peg b, fue descubierto en 1995 de esta manera (Mayor & Queloz 1995). Consideremos un sistema estrella-planeta. Tanto la estrella como el planeta describen una órbita en torno al centro de masa del sistema. Si nos focalizamos en la estrella, un observador en la Tierra podrá detectar el movimiento de la misma en la dirección de la visual a través, por ejemplo, de un estudio de su espectro. Las variaciones en la velocidad radial pueden inferirse a través mediciones altamente precisas de corrimientos en las líneas del espectro de la estrella. Así, mediante un detallado análisis espectroscópico, se podrá determinar el corrimiento periódico al rojo o al azul de las líneas espectrales, cuya magnitud será $\Delta\lambda/\lambda = v/c$, siendo $v$ la velocidad de la estrella alrededor del baricentro del sistema. Las variaciones de las líneas espectrales solo miden la componente del movimiento relativo que se aleja o se acerca del observador. Del análisis de los espectros obtenidos a lo largo de uno o varios períodos orbitales se puede inferir la curva de velocidad radial de la estrella, cuya amplitud máxima, que puede ser deducida de las ecuaciones de Newton y de la segunda ley de Kepler, será:

$$K = \frac{2\pi G}{P_{\rm orb}} \frac{M_{\rm p} \sin i}{(M_{\rm p} + M_*)^{2/3}} \frac{1}{(1-e^2)^{1/2}} \tag{2.2}$$

donde $P_{\rm orb}$ es el período orbital del planeta, $e$ es la excentricidad de la órbita, e $i$ es su inclinación entre el plano del cielo y el plano orbital. $K$, $P_{\rm orb}$ y $e$ se pueden determinar si se toman varios espectros durante una órbita completa del planeta. Notemos que como $M_{\rm p} \ll M_*$, $M_{\rm p} + M_* \sim M_*$, la ecuación anterior no depende de la masa del planeta sino de la *masa mínima*, $M_{\rm p} \sin i$. De este modo, estimando $M_*$ de la clasificación espectral, se puede despejar $M_{\rm p} \sin i$. Si bien la masa del planeta no podrá ser determinada ya que la geometría del problema es *a priori* desconocida, podemos estimar qué probabilidades hay de que la masa del planeta sea muy distinta a la masa mínima. Claramente, si $i \simeq 90\,^\circ$, $M_{\rm p} \sin i \simeq M_{\rm p}$, pero si $\sin i \simeq 0$ el valor estimado será muy distinto del valor real de la masa del objeto. Sin embargo, estos casos son poco probables: si suponemos que las orientaciones orbitales están distribuidas al azar, la probabilidad que $\sin i < 0{,}1$ será tan solo del 5 %.

[2] La bibliografía consultada para el armado conceptual de esta sección corresponde a los cursos dictados durante la XVI Canary Island Winter School por T. Brown, L. Doyle y S. Udry (ver bibliografía: Brown 2008, Doyle 2008, Udry 2008).

La detección de exoplanetas con esta técnica depende de la precisión con la que se puede medir el corrimiento Doppler de las líneas. Por ejemplo, para medir variaciones del orden de 10 m $s^{-1}$ en la velocidad radial de la estrella, la sensibilidad de la medición tiene que ser tal que permita detectar una variación de $10^{-3}$ en el ancho de una línea (un valor típico es $10^{-4}$ Å). El efecto de Júpiter sobre el Sol es de 12 m $s^{-1}$, el de Saturno 4 m $s^{-1}$ y el de la Tierra de tan solo 0,1 m $s^{-1}$. Por otra parte, existe un límite para este tipo de detección impuesto por las fluctuaciones intrínsecas de la superficie estelar. Este límite es de 1 m $s^{-1}$. Sin embargo, la detección de planetas como Júpiter resulta muy útil para comparar sistemas planetarios extrasolares con nuestro propio Sistema Solar, como así también para estimar la efectividad de planetas de tipo jovianos como escudos protectores de potenciales planetas terrestres, en órbitas interiores, que estén en condiciones de desarrollar alguna forma de vida.

### Tránsitos

Supongamos que el observador, el plano de la órbita del planeta y la estrella se encuentran los tres sobre la línea de la visual ($i \simeq 90°$). Cuando el planeta pasa por delante de la estrella, oculta parte del disco estelar, de modo que el observador detecta una disminución del brillo de la estrella mientras este evento ocurre. Si se acepta que la orientación de las órbitas sigue una distribución al azar, la probabilidad de observar un tránsito se puede estimar con tres parámetros: $R_*$, $R_\mathrm{p}$ y $a$, y resulta ser:

$$\text{Probabilidad} = \frac{R_* + R_\mathrm{p}}{a}. \tag{2.3}$$

Como $R_* \gg R_\mathrm{p}$, la probabilidad se reduce a $\frac{R_*}{a}$, lo cual indica que es más probable observar tránsitos cuanto menor sea la separación orbital entre el planeta y la estrella. En el caso del Sistema Solar, la probabilidad de que un observador externo detecte un tránsito de Mercurio es del orden del 1 %, mientras que para el caso de Júpiter cae a apenas 0,1 %. En el caso de los *hot Jupiters*[3] la probabilidad puede ascender a un 20 % debido a los pequeños semiejes de estos objetos. Dado que no hay una forma clara de mejorar la probabilidad de detección de un tránsito planetario, la única estrategia disponible es la observación de un gran número de estrellas, ya sea con programas automatizados de búsqueda de variaciones fotométricas en estrellas muy brillantes, o bien a través del estudio detallado de las estrellas más débiles confinadas a una determinada región del cielo.

Existen dos técnicas de detección de tránsitos: la fotométrica y la espectroscópica. En los tránsitos fotométricos lo que se detecta es la sombra que produce el planeta cuando pasa por delante del disco estelar. El área de cobertura del disco estelar dependerá del radio del planeta ya que cuanto mayor sea el tamaño del planeta, más significativa será la disminución del brillo estelar. De hecho, el cambio en la luminosidad $\Delta L$ será:

$$\frac{\Delta L}{L} \propto \left(\frac{R_\mathrm{p}}{R_*}\right)^2, \tag{2.4}$$

donde $L$ es la luminosidad total del sistema cuando no se encuentra en situación de tránsito. Dado que esta técnica depende del radio y no de la masa del planeta, permite la detección de planetas no necesariamente muy masivos. Además, para un dado límite de detección, la observación de estrellas pequeñas permitiría la detección de planetas pequeños. Si el sistema tiene observaciones espectroscópicas que permitan calcular el corrimiento Doppler de las líneas, se puede obtener la masa real de objeto, ya que ésta será aproximadamente igual a la masa mínima. De este modo, como conocemos el radio y la masa del objeto podemos estimar ciertos parámetros físicos muy relevantes, como ser su densidad media. Del análisis de la curva de luz se puede obtener información sobre la estructura del planeta, de posibles satélites en órbita como también la potencial presencia de otros compañeros planetarios.

[3]Los llamados *hot Jupiters* son planetas extrasolares con masa similar a la de Júpiter pero que se encuentran muy cerca de su estrella central, como en el caso de 51 Peg.

El análisis espectroscópico de un tránsito puede aportar muy valiosa información, pero es mucho más difícil de detectar que su contraparte fotométrica ya que el número de fotones que involucra la obtención de un espectro es menor y la precisión requerida es, por otra parte, mayor. Supongamos al planeta y a la estrella como cuerpos esféricos, el observador medirá radios diferentes dependiendo de la opacidad de la atmósfera del planeta. Así, el radio del objeto que provoca el ocultamiento depende de la longitud de onda en que se observe, la cual, según la región de la atmósfera que atraviese, será completa o parcialmente absorbida. Esto lleva directamente a la observación de la atmósfera de un planeta extrasolar. En el caso de la Tierra, por ejemplo, su atmósfera es opaca por debajo de los 3.000 Å con lo cual, si se la observara durante un tránsito en longitudes de onda cortas, se la vería 60 km más grande que en longitudes de onda largas.

**Microlensing**

Consideremos que tenemos en el campo de la visual dos estrellas, una de las cuales se encuentra mucho más lejos de nosotros que la otra. Supongamos que la más cercana, a la que llamaremos lente $L$, en su desplazamiento relativo, se alinea con la otra ($S$), ocultándola por un cierto intervalo de tiempo. Debido a efectos relativísticos relacionados con la curvatura del espacio-tiempo, la luz que proviene de la estrella más lejana $S$ se curvará al encontrarse con la estrella $L$. Este encuentro causa un incremento temporario del brillo combinado de ambas estrellas debido a la amplificación que sufre la estrella $S$. Este efecto fue observado por primera vez en galaxias y se conoce como *lentes gravitacionales*. Si la estrella lente $L$ tiene, a su vez, un planeta en órbita, el patrón de amplificación del brillo se desviará del patrón estándar debido a que el planeta modifica el campo gravitatorio de su estrella. Si bien la probabilidad de alineación de dos estrellas es de una en un millón, si esto ocurre, la probabilidad de que un planeta provoque una amplificación que exceda el 5 % de la propia amplificación producida por la estrella $L$ es de uno en cinco. La duración de estos eventos depende de la velocidad relativa con que se desplaza la lente, de la distancia, y de la masa del planeta. Por ejemplo, si la estrella lente se encuentra a 5 kiloparsecs, para un planeta de la masa de Júpiter, el evento durará 3 días y para un planeta como la Tierra tan solo 4 horas. En el primer caso, la amplificación del brillo será de 3 magnitudes y en el segundo de 1 magnitud. Como se ve, estos eventos son detectables con los instrumentos actuales. Sin embargo, no son previsibles ni repetibles, por cuanto se requiere la observación constante de un gran número de estrellas lejanas para tener la posibilidad de registrarlos. Este método permite obtener información sobre los datos orbitales del planeta como así también la razón entre la masa de la estrella y del planeta. Al momento, es el método que mejor funciona a grandes distancias y el que, seguramente, aportará información planetaria teniendo en cuenta poblaciones estelares diferentes, como pueden ser la del disco y del bulge galáctico.

**Astrometría**

La búsqueda astrométrica de exoplanetas es la más antigua de todas estas técnicas. De hecho, desde mediados del siglo XX se hacen anuncios de planetas extrasolares debido mediciones astrométricas pero, después de un análisis más profundo, estos objetos resultan ser demasiado masivos para ser considerados planetas. La astrometría mide variaciones periódicas de la posición de una estrella en el plano del cielo (eliminando previamente el movimiento aparente de la estrella debido al movimiento propio y la paralaje anual). Si la estrella tiene un planeta girando en torno a ella describirá un movimiento elíptico alrededor del baricentro del sistema estrella-planeta cuyo semieje mayor, medido en segundos de arco, será

$$\alpha = \frac{M_{\mathrm{p}}}{M_*}\frac{a_{\mathrm{p}}}{d}, \tag{2.5}$$

si $d$ (distancia entre el observador y la estrella) se mide en parsecs y $a_{\mathrm{p}}$ en UA. Para que esta técnica sea efectiva, se deben tomar numerosas imágenes de la estrella durante al menos una fracción importante del período orbital del planeta. Este método es complementario al de velocidades radiales ya que es más sensible a largos períodos (lo cual es equivalente a grandes semiejes), mientras que el método de velocidades radiales favorece la detección de exoplanetas en órbitas cercanas a la estrella. De la ecuación 2.5 se puede ver que la detección de exoplanetas con esta técnica depende de la distancia al sistema, por cuanto, por el momento, queda limitada a estrellas cercanas. Para tener una idea de las magnitudes relacionadas con esta técnica, si consideramos una estrella a 5 parsecs de distancia, el efecto que veríamos si un "Júpiter" estuviera en órbita alrededor de ella sería un desplazamiento de un *mili*segundo de arco en el plano del cielo, mientras que para una "Tierra" sería de tan solo 0,6 *micro*segundos de arco. En la actualidad, las mediciones astrométricas están en el límite de la precisión necesaria para poder detectar estas variaciones. Recién en junio de 2009 se anunció el primer exoplaneta descubierto mediante esta técnica[4] (Pravdo & Shakland 2009) y se espera que la astrometría aporte numerosos hallazgos en el futuro cercano. De hecho, la misión de la NASA *Kepler*, destinada a la búsqueda de tránsitos permitirá también la búsqueda astrométrica usando los mismos datos fotométricos.

**Métodos directos**

La obtención directa de la imagen de un planeta en la región del visible depende de la luz que éste refleja de la estrella central, lo cual, a su vez, depende de la separación orbital con la estrella, del tamaño del planeta y de la naturaleza de su atmósfera. Registrar esta luz requiere de telescopios muy poderosos. La razón entre el brillo del planeta y de la estrella es el factor determinante para la detección ya que de eso depende que el planeta quede enmascarado por el brillo de los anillos de difracción que provoca la estrella en el telescopio.

[4]Previamente, las mediciones astrométricas permitieron la confirmación de Gliese 876b como exoplaneta (Benedict et al. 2002), el cual fue descubierto espectroscópicamente (Delfosse et al. 1998, Marcy et al. 1998).

Por ejemplo, para una estrella a 5 parsecs y un brillo de 1 magnitud, el brillo de un planeta como Júpiter será de 23 magnitudes y el de una Tierra, de 25. La búsqueda de exoplanetas con esta técnica requiere de herramientas específicas como ópticas adaptativas para corregir los efectos atmosféricos que distorsionan la imagen y de coronógrafos para bloquear la luz que proviene de la estrella. El uso de un coronógrafo es la idea del proyecto *TPF–C*. Por su parte, *TPF–I* hará su búsqueda utilizando interferometría destructiva. La idea es que la luz proveniente de la estrella interfiera destructivamente, cancelando así el flujo proveniente de la estrella. El planeta, por su parte, como se encuentra en otra ubicación angular, no se verá afectado. El gran contraste que existe ente la estrella y el planeta se puede mejorar si en vez de observar en el visible se observa en el infrarrojo. En estas longitudes de onda el planeta no solo refleja parte de la luz que recibe de la estrella sino que también emite su propia energía térmica. Por otra parte, los planetas de tipo joviano que tengan asociado un gran campo magnético (por ejemplo, esto puede ocurrir si los planetas tienen un núcleo metálico y si, a su vez, rotan relativamente rápido) pueden emitir un flujo de radiación bastante importante en las longitudes de onda de radio. Si consideramos un planeta como Júpiter alrededor de una estrella de tipo solar, en la frecuencia de 10 MHz, el cociente de flujo planeta-estrella podría llegar a 4 (el planeta más brillante que la estrella), en una etapa de baja actividad estelar. Sin embargo, los electrones presentes en el medio interestelar añaden mucho ruido a la detección en este rango de longitudes de onda, por cuanto, por el momento, no es posible la detección en este rango de frecuencias.

### 2.2.2. Programas espaciales

#### Actualmente en funcionamiento

*Spitzer Space Telescope* El telescopio espacial Spitzer fue puesto en órbita alrededor del Sol en agosto de 2004 y se encuentra orbitando detrás de la Tierra. Trabaja en el infrarrojo. Entre las detecciones más importantes se encuentra un posible cinturón de asteroides alrededor de una estrella cercana y la primera detección de emisión térmica de varios planetas extrasolares.
Spitzer Homepage: http://www.spitzer.caltech.edu/

*CoRoT (Convection, Rotation and planetary Transits)* La misión espacial CoRoT fue ideada con el objeto de obtener información sobre la estructura estelar, mediante técnicas astrosismológicas, y, por otra parte, detectar planetas extrasolares utilizando el método de tránsitos. Además, es la primera misión diseñada para la búsqueda de planetas extrasolares de tipo terrestres. El satélite fue lanzado el 27 de diciembre de 2006 y se encuentra en una órbita polar circular a 896 km de altitud. Está previsto que la misión se desarrolle por el plazo de dos años y medio. Cuenta con un telescopio afocal de 27 cm y 4 cámaras CCD. CoRoT es una misión liderada por la French National Space Agency, aunque el consorcio está integrado también por ESA (Agencia Espacial Europea), Alemania, España, Bélgica, Austria y Brasil. Hasta el momento, los descubrimientos más significativos fueron: la detección de dos planetas extrasolares, la detección de un objeto que clasificaría entre enana marrón y planeta gigante (una masa de alrededor de 20 masas de Júpiter y su radio de aproximadamente 0,8 radios de Júpiter) y otro exoplaneta que presenta un período de translación igual al período de rotación de su estrella.
Corot Homepage: http://smsc.cnes.fr/COROT/

*EPOCh (Extrasolar Planet Observations and Characterization)* Es un proyecto que utiliza las cámaras de la estación espacial Deep Impact (puesta en órbita en 2005) para buscar tránsitos de exoplanetas, observar el movimiento en el plano del cielo de estrellas con planetas y analizar la luz reflejada por los planetas. Se encuentra funcionando desde comienzos de 2008.
Homepage: http://discovery.nasa.gov/epoxi.html

**Proyectos**

*TPF (Terrestrial Planet Finder)* TPF es un proyecto de la NASA para la búsqueda de planetas de tipo terrestres en regiones potencialmente habitables. TPF consta de dos instrumentos a bordo: un coronógrafo (TPF-C) y un interferómetro (TPF-I). Actualmente el proyecto está muy poco avanzado y si bien las estimaciones de lanzamiento son: TPF-C en 2014 y TPF-I en 2020, el proyecto siempre es postergado.
TPF Homepage: http://planetquest.jpl.nasa.gov/TPF

*Kepler* se encargará de monitorear fotométricamente, con un telescopio de un metro de diámetro, cientos de miles de estrellas en búsqueda de tránsitos de planetas tipo terrestres. Está previsto el lanzamiento de Kepler para el año 2009. De esta manera se podrá obtener una estimación estadística de cuál es la frecuencia de planetas como la Tierra en sistemas extrasolares.
Kepler Homepage: http://kepler.nasa.gov/

*SIM (Space Interferometry Mission)* se encuentra en desarrollo y su objetivo será determinar las posiciones y distancias a cientos de estrellas en forma más precisa que cualquier programa previo. SIM está siendo desarrollada por el Jet Propulsion Laboratory bajo contrato de la NASA. Tendrá como objetivo el primer censo de planetas tipo terrestres que orbiten estrellas cercanas. Este último hecho tiene que ver con que se hará detección directa de estos objetos de modo de poder obtener información sobre sus propiedades físicas, sus masas y los elementos orbitales. Esta misión está siendo específicamente diseñada para optimizar las mediciones astrométricas y será capaz de detectar el desplazamiento en el cielo de una estrella, con una precisión de un microsegundo de arco. Kepler y SIM son complementarias a TPF: Kepler brindará estadísticas de la frecuencia de planetas terrestres utilizando estrellas distantes y SIM hará censos en estrellas cercanas, con lo cual ambos proveerán una selección de objetos para ser luego observados por TPF.
SIM Homepage: http://planetquest.jpl.nasa.gov/SIM

*Darwin* Darwin está previsto que sea una flotilla de cuatro o cinco telescopios espaciales cuyo objetivo será la búsqueda de planetas terrestres extrasolares y, consecuentemente el análisis atmosférico en busca de señales químicas de vida. Dado que en el óptico sería casi imposible detectar este tipo de objetos debido a la luminosidad de la estrella central, Darwin hará su búsqueda en el infrarrojo. Por otra parte, es en el infrarrojo donde la vida tal como la conocemos en la Tierra deja señales detectables. Uno de los fundamentos para que el telescopio se encuentre en el espacio es la necesidad de que esté a muy bajas temperaturas, de modo que él mismo no emita radiación en la longitud de onda a observar, para poder así detectar señales muy tenues de planetas lejanos. Tres de las naves espaciales transportarán el telescopio que tendrá un diámetro de aproximadamente 3 metros. Darwin operará desde el punto lagrangiano L2, que se encuentra a 1,5 millones de km de la Tierra en dirección opuesta al Sol. Si bien todavía no hay fecha estimada para el lanzamiento, dada la similitud de objetivos y la complejidad de los proyectos Darwin y TPF, ESA y NASA están pensando en unificarlos y hacerlos operar en conjunto.
Darwin Homepage: http://www.esa.int/science/darwin

*GAIA* Gaia es un proyecto de la ESA. Entre otros, su objetivo es medir, con una precisión de 20 microsegundos de arco, las posiciones de mil millones de estrellas (que se encuentran hasta 200 parsecs del Sol), tanto de nuestra Galaxia como de otros miembros del Grupo Local (observaciones astrométricas). También hará observaciones espectroscópicas y fotométricas de todos los objetos. Su lanzamiento está previsto para fines del año 2011.
GAIA Homepage: gaia.esa.int/

### 2.2.3. Características de los planetas extrasolares detectados

Al día de hoy, y durante los 13 años desde la confirmación del primer planeta extrasolar descubierto alrededor de una estrella de tipo solar (hay registros previos de planetas orbitando alrededor de púlsares), se han detectado 340 candidatos a planetas extrasolares (ver La Enciclopedia de Exoplanetas http://exoplanet.eu). El hecho que se los mencione como *candidatos* (aunque, generalmente, se utilizará el término *planeta* o *exoplaneta*) está relacionado con que no teniéndose una definición de la palabra *planeta* afuera del Sistema Solar, se considerará *ad-hoc* que son planetas los objetos subestelares de masa inferior a 20 $M_J$ (no entran dentro de esta clasificación los llamados “free-floating planets” u objetos con masa planetaria que se encuentran solos en el espacio). Teniendo en cuenta que más del 90 % de los mismos fue detectado por el método de velocidades radiales (menos del 25 % presentan, a su vez, tránsitos), este límite para la masa es todavía más arbitrario si se tiene en cuenta que, en general, solo se puede estimar la masa mínima del objeto. Otro motivo para no hacer mucho hincapié en el número de candidatos detectados es que aumentan día a día debido a la alta tasa de descubrimientos, consecuencia del mejoramiento en la precisión instrumental. De hecho, el 36 % de los exoplanetas fueron detectados entre los años 2007–2009.

Del análisis de los datos disponibles se puede decir que, de los 340 exoplanetas, habría 16 ($\sim 5\,\%$) con masa menor a 10 $M_\oplus$. Claramente, esto no significa que los planetas más pequeños sean menos abundantes sino que las capacidades tecnológicas no han permitido todavía su detección en un gran número. Como veremos a continuación, las limitaciones tecnológicas traen aparejado un sesgo en la detección, viéndose naturalmente favorecido el descubrimiento de planetas alrededor de estrellas cercanas ($d < 400$ parsecs, con un máximo entorno a los 30 parsecs). El método de velocidades radiales es el que ha tenido más éxito en la búsqueda de exoplanetas y, por sus características, favorece el descubrimiento de planetas de gran masa y con semiejes pequeños. De hecho, más del 50 % de los exoplanetas tiene un período orbital inferior a un año y una masa superior a 0,1 $M_J$. Sin embargo, si bien las técnicas utilizadas favorecen este tipo de descubrimientos, fue una gran sorpresa que sistemas planetarios tan distintos al Sistema Solar existieran. Otros datos que despertaron el asombro son el amplio abanico de excentricidades (alrededor del 80 % de los exoplanetas tiene una excentricidad superior a 0,1), la presencia de planetas en estrellas binarias y la alta metalicidad de muchas de las estrellas con planetas.

Al momento, alrededor del 60 % de las estrellas que albergan planetas muestran una metalicidad mayor a la solar. La detección de planetas entorno a estrellas de metalicidad alta provocó que en los subsiguientes programas de búsqueda de exoplanetas se favorecieran ciertos blancos (Santos et al. 2005, Wuchterl 2008). Por otra parte, por las dificultades intrínsecas de los métodos de detección se suelen descartar cierto tipo de estrellas, como aquellas que presentan altas velocidades de rotación, alta actividad cromosférica, estrellas sin una clase de luminosidad bien definida, etc. Existe además un sesgo en la búsqueda espectroscópica que favorece la detección de planetas entorno de estrellas más metálicas, ya que estas presentan líneas de absorción metálicas muy fuertes. La variación en la velocidad aumenta desde algunos m seg$^{-1}$ para estrellas de metalicidad solar hasta 16 m seg$^{-1}$ para estrellas poco metálicas (1/4 o menos). Este hecho podría llevar a que la frecuencia de exoplanetas en relación a la metalicidad de las estrellas que los albergan esté sobrestimada en hasta un factor 2 (Durisen et al. 2007). Todo esto sugiere que la correlación entre la metalicidad de las estrellas y la frecuencia de planetas deba, por ahora, ser analizada con cautela. Es notable, además, que en los últimos dos años el número de descubrimientos de exoplanetas entorno a estrellas más y menos metálicas que el Sol se ha ido balanceando. Parecería, entonces, necesario esperar a tener muestras estadísticamente más representativas de los sistemas extrasolares antes de precipitar conclusiones respecto de las características que presentan los sistemas planetarios en general.

Luego del descubrimiento de los primeros exoplanetas, y debido a que estos resultaron ser muy diferentes de los planetas del Sistema Solar, se pensó que el Sistema Solar era un caso atípico y no representativo de un sistema planetario “normal”. Ahora bien, estas conclusiones suenan al menos un poco prematuras si tenemos en cuenta lo mencionado en los párrafos anteriores respecto de los efectos de selección que son intrínsecos a la disponibilidad tecnológica actual. Además, no hay que perder de vista que el Sistema Solar es el sistema planetario más estudiado y mejor conocido por los científicos. Y que el acceso que se tiene a diversos datos, como ser su estructura, su composición, la distribución de masa y de momento angular, la información detallada del Sol, las propiedades superficiales de los planetas, los datos que aportan los cometas, asteroides y satélites; en fin, toneladas de información que la humanidad lleva siglos recopilando y analizando, y que demorará mucho tiempo, si es que alguna vez es posible, obtenerla con este nivel de detalle para sistemas extrasolares. Es así que la sugerencia de la violación del principio copernicano aplicado a la cosmogonía, que ha surgido en debates frente a la aparición de tan inesperados sistemas extrasolares, parece, como mínimo, muy apresurada. Es innegable la existencia de sistemas planetarios muy diferentes al nuestro, que presentan condiciones extremas que no eran imaginables hace 20 años atrás. Pero entonces, una teoría general de formación y evolución de los sistemas planetarios tiene que poder explicar en forma autoconsistente las diversas configuraciones observadas, incluida la local. Y éste es, sin duda, el desafío más grande de las Ciencias Planetarias.

## 2.3. Teorías de migración

La presencia de exoplanetas en órbitas muy cercanas a la estrella central provocó consternación en la comunidad astronómica ¿Pudieron estos objetos formarse *in situ*? ¿Pueden haber sobrevivido en esa ubicación el tiempo estimado de vida de su estrella? ¿Qué otras alternativas hay para explicar la existencia de estos sistemas?

Poco tiempo después del descubrimiento de 51 Peg b (Mayor & Queloz 1995), Guillot et al. (1996) demostraron que el planeta pudo, efectivamente, permanecer tan cerca de su estrella durante todo su tiempo estimado de vida. Por su parte, Wuchterl (1996) probó que este planeta se pudo haber formado *in situ* si en su zona de alimentación había suficiente material disponible para ser acretado. Pero, a su vez, otras alternativas fueron propuestas. Basadas en la idea generalizada de que los planetas gigantes deben formarse más allá de la línea de hielo (ver 4.1), puesto que en la región interior no habría la cantidad de sólidos necesarios para construir planetas gigantes, estas teorías sugieren que debe haber algún mecanismo capaz de llevar a los protoplanetas desde órbitas exteriores a órbitas muy cercanas a la estrella como las observadas (Wuchterl 2008). Los mecanismos propuestos para el cambio en los elementos orbitales fueron dos:

- *jumping Jupiters*, para lo cual se necesita de un sistema planetario que en sus orígenes haya contado con varios planetas gigantes,
- *migración orbital*, donde se propone un decrecimiento gradual del radio orbital debido a la interacción del protoplaneta con el disco de gas y de planetesimales.

La primera alternativa, propuesta por Weidenschilling & Marzari (1996), necesita de la formación de varios planetas gigantes, en un corto lapso de tiempo que permita que se generen perturbaciones destructivas que den por resultado la supervivencia de solo algunos de ellos en órbitas excéntricas y cercanas a la estrella central. En estos casos, los planetas de menos masa son expulsados fuera del sistema. Más allá de que se requiere de configuraciones muy particulares para que esto ocurra, las distancias orbitales resultantes serían todavía mayores a las observadas, por cuanto se requeriría de algún otro mecanismo para continuar el acercamiento entre la estrella y los planetas.

La migración orbital, por su parte, ha sido favorecida por muchos especialistas. La idea básica que engloba este concepto es el cambio sistemático del semieje mayor orbital de un planeta (Papaloizou et al. 2007). Históricamente, la teoría de migración estuvo asociada al desplazamiento hacia afuera de Urano y Neptuno desde su supuesto lugar de formación hasta su ubicación actual, lo cual dio una alternativa para acortar los tiempos de crecimiento de estos planetas. La migración se habría producido por un intercambio de momento angular entre los planetas y el disco de planetesimales remanentes. Después del descubrimiento de los primeros exoplanetas surgieron otras teorías de naturaleza muy distinta que se conocen como migración de tipo I, de tipo II y de tipo III, las dos primeras sugeridas por Ward en 1997 y la tercera así bautizada por analogía. En estos casos, la migración sería consecuencia del intercambio de momento angular entre el planeta y el disco de gas, siendo importante tanto para los planetas terrestres como para los gigantes. Esta propuesta ofrece una alternativa para explicar la existencia de los "hot Jupiters".

Para un planeta inmerso en un disco de gas, en órbita circular entorno a una estrella, el intercambio de momento angular entre el planeta y el gas ocurre en las ubicaciones de las resonancias de corrotación y de Lindblad (Armitage 2007). De su interacción con las resonancias de Lindblad interiores el planeta gana momento angular, lo cual hace que se desplace hacia afuera. Por su parte, el gas pierde momento angular y se mueve hacia adentro. De la interacción con las resonancias de Lindblad exteriores, el planeta pierde momento angular (lo que hace que se desplace hacia adentro) y el gas, que gana momento angular, se mueve hacia afuera. De este modo, el planeta repele al gas entorno a su órbita. El flujo de momento angular intercambiado en cada resonancia es proporcional a la masa del planeta al cuadrado y a la densidad superficial de gas en ese punto. Cuando la masa del planeta es baja, digamos de alrededor de 1 $M_\oplus$ aunque esto depende de las propiedades del disco, el flujo de momento angular es despreciable comparado con el transporte de momento angular por la viscosidad del disco, con lo cual el perfil de densidad de gas no se altera, y el torque neto que actúa sobre el planeta es la suma de los torques de las resonancias. En este caso la escala de tiempo de migración va con el inverso de la masa del planeta con lo cual, mientras se mantenga válida la hipótesis de que el perfil de gas no cambia, los planetas más grandes son los más afectados. El sentido en el cual se produce la migración no resulta evidente de la suma de los torques. Sin embargo, se encuentra invariablemente que las resonancias de Lindblad exteriores son más importantes que las interiores provocando que el planeta migre hacia el interior. El torque en la resonancia de corrotación no invierte el sentido de la migración.

En una primera aproximación, la eficiencia del transporte de momento angular tiene poco impacto en la estimación de la tasa de migración de tipo I. Sin embargo, cuando se hacen simulaciones más realistas, considerando discos turbulentos, se producen fluctuaciones en la densidad del disco protoplanetario, las cuales ejercen torques aleatorios sobre el planeta. Esto provoca que la variación del semieje no sea en un único sentido, produciéndose una migración estocástica o *random walk*. Simulaciones recientes muestran que para planetas de masa menor a 10 $M_\oplus$ domina la componente de *random walk* sobre la migración de tipo I "clásica". Si consideramos planetas de mucha masa, el flujo de momento angular domina localmente al debido a la viscosidad. En este caso se produce la "repulsión" entre el planeta y el gas que lo rodea generándose una *brecha* de baja densidad entre ambos. Para que se forme la brecha se deben satisfacer dos condiciones: por un lado, el radio de Hill del planeta debe ser comparable a la altura de escala del disco; por el otro, los torques deben ser capaces de remover el gas más rápido de lo que los efectos viscosos tardan en llenarlo. Un planeta de la masa de Saturno estaría en el límite para la apertura de la brecha. Cuando un planeta tiene la masa necesaria para abrir una brecha, su evolución orbital ocurre en una escala de tiempo comparable a la del disco protoplanetario. Esta es la migración de tipo II. Para el caso de un Júpiter, el tiempo característico de migración sería de 0,75 millones de años aproximadamente.

La migración de tipo III se refiere a una variación violenta del semieje del protoplaneta por inestabilidades en la interacción entre el planeta y el disco. Este tipo de migración surge de la transferencia de material a través de la resonancia de corrotación. Este modo de migración se manifestaría en aquellos planetas que hayan vaciado su región de corrotación y que se encuentren inmersos en discos masivos. La migración puede ser hacia adentro o hacia afuera dependiendo de las condiciones iniciales. En algunas simulaciones se vio que el semieje del planeta puede variar considerablemente en menos de 100 órbitas. Hasta el momento no está claro cuánto tiempo puede operar este mecanismo ni de que manera un planeta que sufre esta migración puede ser frenado.

El estudio de la interacción planeta disco lejos está de ser sencillo. El estudio analítico de Ward (1997) toma como modelo las ecuaciones de la mecánica de fluidos linealizadas, una ley de potencias para la densidad nebular y un cierto potencial gravitatorio que refleje el problema. Esta aproximación, sin embargo, no tiene en cuenta la perturbación en la densidad del disco como consecuencia de la presencia del protoplaneta (Wuchterl 2008). Las partes más densas del protoplaneta, en la región más interior de su esfera de Hill, difieren enormemente del resto del disco. De este modo, la presión en el interior de la esfera de Hill estaría siendo altamente subestimada. El protoplaneta entra entonces en los cálculos como una fuente gravitatoria pero la presión que ejerce el gas en el sentido inverso está siendo omitida. Otra cuestión es la consideración, necesaria para un análisis lineal, de un disco kepleriano sin perturbar. Sin embargo, la mera presencia de un cuerpo masivo hace que la idea de un disco kepleriano sea, en estos casos, artificial. Por su parte, los cálculos hidrodinámicos bidimensionales (2D) y tridimensionales (3D) son terriblemente costosos y requieren, a su vez, de numerosas simplificaciones para funcionar. Sin embargo, son este tipo de simulaciones las que pueden aportar una aproximación más realista de cómo es la interacción planeta-disco. Masset, D'Angelo & Kley (2006) estudiaron la migración de tipo I con simulaciones tridimensionales para un disco isotermo y mostraron que los efectos no lineales se hacen importantes cuando el objeto tiene una masa de, aproximadamente, 5 $M_\oplus$, reduciendo el torque neto, lo que lleva a tasas de migración más lentas. Debemos notar que, en modelos unidimensionales de protoplanetas en discos isotermos (Pečnik & Wuchterl 2007) la masa de cruce[5] es de apenas 0,1 $M_\oplus$, un factor 100 por debajo del valor estimado para situaciones más realistas. Entonces, los regímenes típicos que se estudian en modelos tridimensionales son supercríticos en un factor 100. Por ejemplo, la interacción dinámica planeta-disco, la acreción y la migración calculada en un disco isotermo para un objeto de 5 $M_\oplus$ de algún modo se corresponden con un objeto de 500 $M_\oplus$ (aproximadamente $1\frac{1}{2}$ $M_J$). En realidad, escalear correctamente ambos casos no es trivial y requiere relacionar ambos cálculos con mucho cuidado. Por otra parte, la dificultad intrínseca de los modelos 3D limita todavía la realización de cálculos con diversas condiciones iniciales que permitan una comparación directa con los modelos unidimensionales (1D). Sin embargo, seguramente estas simulaciones serán posibles en un futuro cercano, arrojando luz sobre valores más realistas de la tasa de migración, como también sobre otras cuestiones de la formación de los planetas gigantes que hasta ahora solo se han podido estudiar en forma lineal.

[5]Masa de cruce: corresponde a la masa del núcleo (o de la envoltura de gas) de un planeta cuando, en el proceso de formación, la masa del núcleo es igual a la de la envoltura. (Ver capítulo 6, Ec. (6.8)).

# Capítulo 3

# Escenarios de formación

## 3.1. Formación de un sistema planetario

### 3.1.1. Nubes moleculares gigantes: la nursery

Las nubes moleculares son las progenitoras de las estrellas, y por consiguiente, de los sistemas planetarios. Estas nubes están formadas fundamentalmente por moléculas de hidrógeno, aunque también se observan CO, $^{13}CO$ y $NH_3$ (por ejemplo, Armitage 2007). Algunas de ellas se ven en el cielo como regiones oscuras; ésto se debe a que el polvo que contienen oscurece la luz de las estrellas de la Galaxia que se encuentran detrás de ellas. Generalmente, dentro de las nubes moleculares se detectan estrellas muy jóvenes (de pocos millones de años de edad), las cuales corresponden a objetos de pre-secuencia principal, lo cual significa que están aún en contracción y que todavía no alcanzaron las condiciones necesarias como para comenzar la fusión de hidrógeno.

La formación estelar ocurre en los brazos espirales de la Galaxia (por ejemplo, Wuchterl 2008). Las nubes moleculares se fragmentan en subestructuras que a su vez colapsan para dar lugar a las estrellas. El resultado de la fragmentación de la nube y del colapso de sus componentes es un cúmulo estelar. El proceso de formación estelar involucra pasar por múltiples escalas tanto de tamaño como de masa. La masa típica de una nube molecular es de alrededor de $10^6$ $M_\odot$ y su extensión de 100 pc, siendo su velocidad de rotación muy baja. Por su parte, los fragmentos que dan lugar a las estrellas tienen un tamaño de aproximadamente 0,1 pc y masas típicamente estelares de, aproximadamente, 1 $M_\odot$. La fragmentación de la nube tiene su origen en las mareas provocadas por la Galaxia, los campos magnéticos y la turbulencia, los cuales juegan un rol fundamental para que se formen "grumos" de alta densidad que luego colapsan para dar origen a las estrellas. Sin embargo, la física que domina a los fragmentos no es la misma que provoca los fragmentos, y la formación estelar estará gobernada por la competencia entre la fuerza de gravedad y la presión. En las primeras etapas de colapso la velocidad de rotación es dinámicamente despreciable. Sin embargo, el momento angular de cada grumo es grande.

De este modo, durante el colapso de los fragmentos, la rotación se vuelve cada vez más importante. Mientras se produce la contracción, la conservación del momento angular lleva a que estas estructuras entren en una rotación rápida. Sin embargo, en el caso del Sistema Solar el momento angular actual es mucho menor que el momento angular estimado para el estado inicial del colapso. Esto se conoce como “el problema del momento angular durante la formación estelar” (ver Armitage 2007). La solución a este problema sería que probablemente exista también una redistribución del momento angular entre las distintas componentes que surgen del fragmento (sistemas binarios o múltiples, formación de discos circumestelares, pérdida de momento angular en forma de jets), lo cual a su vez impediría que cada uno de los fragmentos alcance las velocidades de ruptura.

Durante el colapso de una nube se pueden distinguir dos etapas: la fragmentación, y el posterior colapso y acreción de los fragmentos. La fragmentación de la nube daría origen a núcleos de masas superiores a 0,01-0,007 $M_{\odot}$. Por debajo de este límite la fragmentación es poco probable ya que, en ese caso, los fragmentos serían tan densos que se volverían lo suficientemente opacos como para impedir el transporte de energía, llevando a que la temperatura aumente sin límites. Los fragmentos pueden caracterizarse básicamente de la siguiente manera: son cuasi-homogéneos, fríos ($T \sim 10$ K), opacos para las longitudes de onda correspondientes al visible pero transparentes para aquellas encargadas del transporte de energía y, consecuentemente, isotermos.

### 3.1.2. Formación del disco protoplanetario

El colapso gravitacional de una nube protoestelar se detiene cuando su núcleo se vuelve opaco y es entonces capaz de calentarse lo suficiente como para que la presión contrarreste a la gravedad (Wuchterl 2008). Dado que tanto la presión como la gravedad actúan en forma isotrópica, las estrellas resultantes son esféricas. Sin embargo, dado que el momento angular no es nulo, hay un segundo agente que puede llevar a detener el colapso: la fuerza centrífuga. Durante el colapso de la nube, la conservación del momento angular lleva a que la velocidad de rotación aumente en varios órdenes de magnitud. A diferencia de la presión, la fuerza centrífuga no es isotrópica y actúa en forma perpendicular al eje de rotación. De este modo el colapso, que se produce paralelo al eje de rotación, queda prácticamente inalterado. A partir del momento en que la nube alcanza un determinado radio, la fuerza centrífuga se vuelve lo suficientemente importante como para que el gas que caiga hacia el centro lo haga en forma paralela al eje de rotación. El material llega entonces al plano ecuatorial de la estrella, donde la componente vertical de la fuerza gravitatoria es cero y la componente radial se equilibra con la fuerza centrífuga. Se forma así un disco circumestelar en el plano ecuatorial. El ancho del disco queda determinado por el equilibrio entre la presión del gas y la componente vertical de la fuerza gravitatoria que ejerce la estrella. El balance de fuerzas en la dirección radial se da entre la fuerza gravitatoria y la fuerza centrífuga más el gradiente de presión del gas. Luego, a diferencia de un problema de dos cuerpos donde solo actúa la gravedad, en este caso el balance de fuerzas involucra también a la presión del gas.

Entonces, que exista una fuerza debida a la presión del gas es lo que hace que la velocidad de rotación del disco sea un poco más baja que en el modelo de un disco kepleriano.

El resultado global del colapso de una condensación de gas es una protoestrella con un disco de acreción, el cual eventualmente puede convertirse en un disco protoplanetario. Es importante notar que tanto a nivel observacional como teórico es poco lo que se sabe respecto de los discos protoplanetarios y su origen. En el primer caso el problema radica en que la sensibilidad y la resolución necesaria para este tipo de estudios debe ser muy grande. En cuanto a los estudios teóricos, existen numerosas dificultades a nivel computacional para calcular un disco protoplanetario como resultado del colapso de una nube, y además el transporte de momento angular es algo que todavía no está satisfactoriamente resuelto.

A grandes rasgos, la información que se tiene de las observaciones muestra que la mayoría de las estrellas jóvenes presentan un disco circumestelar cuya masa máxima no superaría el 10 % de la masa de la estrella central. Los discos observados tienen una extensión que en general no supera las 1.000 UA (Millan-Gabet et al. 2007). Debido a los límites actuales de resolución, la información sobre propiedades físicas como la temperatura y la densidad, solo pudo obtenerse a partir de las 50 UA en el caso de estrellas cercanas (Wuchterl 2008). Para radios menores a las 50 UA, el disco no puede ser resuelto espacialmente, y su estructura puede inferirse al combinar la distribución espectral de energía con los modelos teóricos. Además, las observaciones muestran una disminución en la emisión infrarroja de discos con diferencias de edad de algunos millones de años, lo cual indicaría que se encuentran en distintos estadios evolutivos. Es probable que en aquellos discos con baja emisión esté ocurriendo la formación planetaria o que los propios discos se encuentren en proceso de disipación.

### 3.1.3. Del polvo a los planetesimales

El escenario estándar de la formación de un sistema planetario se puede pensar en cuatro etapas (Nagasawa et al. 2007). Primero, el polvo sedimenta hacia el plano de la nebulosa protoplanetaria formando un disco. Luego, la formación de un sistema planetario comprende las tres etapas restantes, que están caracterizadas por la aparición en el disco de: 1) los planetesimales, 2) los protoplanetas, y finalmente 3) los planetas. El disco circumestelar está formado originalmente por gas y polvo. La formación de los planetas requiere entonces del crecimiento de las partículas de polvo en, al menos, 12 órdenes de magnitud (Armitage 2007). Se considera *polvo* a las partículas con tamaños por debajo del micrón y hasta aproximadamente el centímetro. El movimiento de estas partículas está acoplado al del gas y su crecimiento ocurre por colisiones inelásticas. A medida que estas partículas van aumentando su tamaño se van desacoplando dinámicamente del gas, estando su movimiento bien descripto como una combinación entre la correspondiente a una órbita kepleriana y la fricción aerodinámica producto de que estos objetos se desplazan más rápido que el gas que los rodea. El crecimiento de las partículas en este régimen debería ser rápido, aunque todavía no está bien entendido. Los planetesimales son cuerpos del orden del kilómetro, su-ficientemente masivos como para que se los pueda considerar dinámicamente desacoplados del gas. Los planetesimales son los cuerpos más pequeños que se tienen en cuenta cuando se estudia la formación de los planetas dado que sus interacciones pueden ser consideradas puramente gravitatorias. Cuando la acreción de planetesimales da lugar a cuerpos de tamaño terrestre, estos vuelven a acoplarse al disco de gas, pero en este caso predomina la interacción gravitatoria: aparecen los fenómenos de migración (ver Capítulo 2, § 2.3). Su subsecuente crecimiento dará lugar a los núcleos de los planetas gaseosos: cuando un embrión alcanza aproximadamente 10 $M_{\oplus}$ es capaz de acretar grandes cantidades del gas circundante dando origen a las extensas atmósferas de los planetas gigantes (Mizuno 1980, Stevenson 1982).

En síntesis, los primeros cuerpos masivos en aparecer en el disco son los planetesimales, que se forman por la coagulación de granos de polvo. Luego, los protoplanetas se forman por acreción de planetesimales. Finalmente, la tercera etapa es la que distingue la formación de los planetas terrestres de los planetas gigantes: mientras que los primeros surgen de las colisiones de protoplanetas, los segundos son protoplanetas capaces de acretar grandes cantidades de gas de la nebulosa. Este escenario, conocido como *escenario estándar*, es el más aceptado actualmente, pero hay que remarcar que su plausibilidad está ligada a la formación de los planetesimales, hecho que todavía no pudo ser explicado satisfactoriamente.

En el proceso de formación de un sistema planetario hay dos cuestiones que no están bien entendidas: la formación de los planetesimales y la formación de los planetas gigantes. Esta Tesis tiene como objetivo estudiar lo segundo, dando por sentado la existencia de los planetesimales. De todos modos, en esta sección daremos un breve resumen de lo que se sabe hasta el momento del proceso de formación de los planetesimales. Lo que sigue está basado en el curso dictado por A. Youdin durante la escuela de invierno Les Houches Winter School 2008.

La estimación del tiempo de formación de los planetesimales puede hacerse de dos maneras completamente independientes. Por un lado, una cota superior proviene de la escala de tiempo de vida de los discos observados alrededor de las estrellas T Tauri. Por otra parte, se pueden hacer estimaciones a partir del estudio de los meteoritos. Afortunadamente, ambos procedimientos arrojan resultados concordantes.

Dado que los planetesimales se forman en regiones donde hay abundante polvo, es importante estimar la vida media de los discos óptimamente gruesos. Las observaciones de objetos estelares jóvenes en el infrarrojo cercano revelan la existencia de un disco interior que se encuentra a altas temperaturas. Si bien la parte gaseosa del disco es más difícil de estimar que su componente de polvo, existe gran evidencia de que las zonas de altas temperaturas en el disco correlacionan con abundantes cantidades de gas. De este hecho surge que en el cálculo de la formación de los planetesimales se debe tener en cuenta la presencia del gas, que implica que no se puede despreciar el frenado gaseoso. Uno de los motivos por los cuales los planetesimales deben formarse mientras todavía hay gas en el disco, tiene que ver con que los planetas gigantes deben a su vez formarse mientras el gas esté presente, y como dijimos, los planetesimales son los bloques fundamentales para la construcción de un planeta. Dadas las estimaciones actuales de la vida media de los discos circumestelares, este proceso no puede durar más que algunos millones de años. Se concluye entonces que los planetesimales deben formarse durante los primeros millones de años de evolución de la estrella en la pre-secuencia principal, cuando los discos tienen todavía abundante cantidad de polvo. Sin embargo, dado que las partículas centimétricas no emiten radiación en forma eficiente, no se puede observar a los planetesimales durante su proceso de formación.

Por otra parte, de los meteoritos del Sistema Solar se puede obtener información en ciertos aspectos más detallada. Los meteoritos primitivos se conocen como *condritas* (no están diferenciados en un núcleo rico en hierro y un manto rocoso como los planetas y los asteroides más grandes) y se mantienen relativamente inalterados desde su origen. Las condritas están formadas mayoritariamente por condrulos. Sin embargo, algunas presentan inclusiones de calcio y aluminio, las cuales son altamente refractarias. Se conoce a estas inclusiones como CAIs (Calcium Aluminun Inclusions) y representan el material más antiguo conocido del Sistema Solar, generalmente utilizado para datar su edad. Si las condritas forman parte de la primera generación de planetesimales, la diferencia de edad entre las condritas y las CAIs da un límite inferior a la escala de tiempo de formación de los meteoritos primitivos. Del decaimiento radiactivo del $^{26}$Al se puede estimar que las CAIs son más viejas que los condrulos por algunos millones de años. Esto significa que la formación de los planetesimales en el Sistema Solar fue un proceso que se extendió por varios millones de años, en acuerdo con las estimaciones de vida de los discos protoplanetarios.

Sin embargo, las partículas están inmersas en un disco de gas que no rota con velocidad kepleriana, lo cual a efectos prácticos resulta en un "viento" que las frena y las hace espiralar hacia la estrella central. La escala de tiempo involucrada en este proceso para objetos del tamaño del metro es tan solo de ¡100 años!. Luego, los planetesimales en crecimiento deberían superar este tamaño crítico con mucha rapidez o, alternativamente, debería existir algún mecanismo que retenga al material y le impida caer hacia la estrella central.

Actualmente existen dos mecanismos propuestos para la formación de los planetesimales: (1) colisiones sucesivas entre las partículas de polvo que terminen en la aglomeración de las mismas y (2) el colapso gravitatorio de pequeñas partículas sólidas en cuerpos más grandes. Estos dos mecanismos no son mutuamente excluyentes ya que podría ocurrir que las partículas de polvo crezcan primero como producto de colisiones inelásticas hasta que los cuerpos que formen se desacoplen del gas y luego colapsen como consecuencia de su autogravedad dando como resultado la aparición de los planetesimales.

Las colisiones entre partículas podrían llevar a la formación rápida de los planetesimales si el "pegoteo" entre las partículas fuera altamente efectivo. La eficiencia de la acreción por colisiones inelásticas no se conoce bien todavía, aunque actualmente se están llevando a cabo numerosos ensayos de laboratorio para ver bajo qué condiciones estas colisiones provocan el crecimiento de las partículas. Las velocidades relativas entre las partículas en el momento de una colisión determinan si la colisión dará lugar a acumulación o no. La velocidad de colisión crece monótonamente con el tamaño de las partículas hasta que el desacoplamiento entre ambas cantidades ocurre para cuerpos del orden del metro, cuando las velocidades relativas alcanzan los 100 m seg$^{-1}$. Sin embargo, los experimentos muestran que partículas del tamaño del milímetro tendrían ya dificultad en aglomerarse en forma efectiva, independientemente de las velocidades involucradas. Por otra parte, la idea del crecimiento de un planetesimal por colisiones con partículas pequeñas parece también difícil debido a las altas velocidades a las cuales estos eventos deberían producirse.

La otra posibilidad para la formación de los planetesimales es la inestabilidad gravitatoria. Este mecanismo es responsable de la formación de muchos de los objetos del Universo, en las más diversas escalas. En el caso de los planetesimales, la idea es que estos se formarían en el disco protoplanetario, donde los sólidos forman un disco *frío* (debido a que pierden rápidamente su energía cinética por el frenado gaseoso y las colisiones inelásticas), el cual se fragmentaría debido a su autogravedad y provocaría de esta forma el colapso de pequeñas partículas sólidas en planetesimales. El problema de este modelo es la presencia de turbulencia. En ausencia de turbulencia, las partículas se acumularían en el plano central del disco hasta llegar a densidades suficientemente altas como para que se produzca el colapso gravitatorio. Pero aún el propio decantamiento de los sólidos provoca turbulencia en el disco, con lo cual ésta no puede ser despreciada. Recientemente, Johansen et al. (2007), hicieron simulaciones del colapso gravitatorio de los sólidos con un código híbrido MHD/partícula-malla donde se incluyeron diversas fuentes de turbulencia en el disco. En estas simulaciones las partículas se consideraron con un tamaño inicial de entre 15 y 60 cm. Si bien es incierto todavía cómo alcanzarían este tamaño, las simulaciones muestran la viabilidad de la formación de planetesimales por inestabilidad gravitatoria, ya que las partículas colapsan rápidamente hasta formar objetos del tamaño del kilómetro en tan solo un período orbital.

La formación de planetesimales presenta muchos interrogantes y quedan todavía serias cuestiones por resolver. Los recientes resultados de simulaciones numéricas muestran que el colapso gravitatorio es una solución posible, aunque el crecimiento de los granos de polvo hasta el tamaño del centímetro ocurra mediante mecanismos por ahora desconocidos. Las capacidades tecnológicas actuales no permiten tener registros observacionales del proceso de formación de los planetesimales, quedando limitada la información disponible a los registros obtenidos de estrellas T Tauri y a aquella que aportan lo meteoritos encontrados en La Tierra.

### 3.1.4. Planetas rocosos y planetas gaseosos

Los protoplanetas se forman por acreción entre los planetesimales. Clásicamente, se distinguen dos regímenes de crecimiento entre los planetesimales: el *ordenado* y el *runaway*. En el crecimiento ordenado todos los cuerpos que conforman la población de planetesimales tienen aproximadamente la misma tasa de crecimiento, y la razón entre sus masas resulta ser del orden de la unidad. En cambio, en el crecimiento runaway los objetos más grandes crecen más rápidamente que los pequeños, con lo cual la razón entre sus masas crece monótonamente.

El escenario clásico de la formación planetaria, planteado en forma teórica por Safronov en la década del '60, asumía que los embriones planetarios surgían por el crecimiento ordenado de los planetesimales. Sin embargo, si esto fuera así, los núcleos de los planetas gigantes no se podrían formar lo suficientemente rápido como para capturar el gas necesario para su envoltura antes que la nebulosa se disipe (Safronov 1969). A partir de la década del '80, el crecimiento de los planetesimales comenzó a estudiarse en más detalle a través de simulaciones numéricas. Estas simulaciones plantearon un escenario diferente, ya que mostraron que si un objeto del disco de planetesimales se distingue de los demás por ser más masivo que la media seguirá un crecimiento runaway y no un crecimiento ordenado como se pensaba antes (Kokubo & Ida 1998, 2000, 2002). En el crecimiento runaway, el planetesimal que, por cuestiones probabilísticas, surge más masivo que el resto continúa acretando planetesimales de su vecindad, aumentando así su masa, diferenciándose cada vez más del resto de los planetesimales del disco e impidiendo que éstos crezcan en forma significativa. Es destacable, sin embargo, que la mayor parte de la masa del sistema continúa contenida en los planetesimales más pequeños. El crecimiento runaway no implica que el tiempo de crecimiento de un planetesimal decrezca con la masa sino que, la razón entre la masa del cuerpo en crecimiento runaway y la del resto de los planetesimales crece con el tiempo. Una vez que el embrión alcanza la masa necesaria como para excitar gravitatoriamente a los planetesimales que lo rodean, el crecimiento runaway se autolimita. Los embriones resultantes tienen, aproximadamente, la masa de la Luna y se encuentran separados por una distancia de $\sim 10\,\mathrm{R_H}$ (con $\mathrm{R_H}$ nos referimos al radio de Hill de un objeto[1]). Estos embriones continúan creciendo, pero no ya dentro de un régimen runaway sino siguiendo lo que se conoce como *crecimiento oligárquico*, ya que solo estos objetos siguen acretando planetesimales. Entonces, el crecimiento oligárquico es el régimen que corresponde a los embriones que surgieron por el crecimiento runaway de los planetesimales y que siguen su crecimiento alimentándose de los planetesimales que pueblan el disco pero que se vieron impedidos de crecer. La tasa de crecimiento correspondiente al régimen oligárquico es menor que durante el crecimiento runaway, pero mayor que en un crecimiento ordenado.

El resultado del crecimiento oligárquico es un sistema protoplanetario: el crecimiento oligárquico es el que, de alguna manera, establece cuáles son las configuraciones orbitales finales, como también las masas resultantes y las escalas de tiempo involucradas (Kokubo 2001). En la tabla 3.1 se muestran los resultados de simulaciones numéricas realizadas para distintos radios orbitales en el Sistema Solar. Llegado a este estado, la fase final de la formación planetaria ocurre dependiendo de la región del disco donde se encuentran los protoplanetas. En la región de los planetas terrestres, la masa final y la separación orbital estimadas de los embriones resultantes es mucho más pequeña que la de los planetas actuales. Luego, para alcanzar las masas planetarias el proceso de acreción debe continuar con colisiones entre los protoplanetas. El sistema protoplanetario resultante del crecimiento oligárquico se puede volver dinámicamente inestable en escalas de tiempo largas. Las excentricidades aumentarían entonces por perturbaciones gravitatorias de otros protoplanetas o de los planetas jovianos. De esta manera se daría el cruce orbital necesario para que se produzcan las colisiones entre los protoplanetas. Sin embargo, las simulaciones de $N$-cuerpos muestran que las excentricidades finales de los planetas resultarían mucho más elevadas que las actuales. Una posibilidad para bajar las excentricidades provendría de la fricción dinámica de los planetesimales residuales en el disco; otra posibilidad serían las interacciones con el disco de gas.

[1] Ver definición del radio de Hill en la sección "Símbolos y Unidades" o en el capítulo 4, ecuación (4.13).

Por otra parte, para los embriones que se encuentran en la región de los planetas exteriores la abundancia de material gaseoso permite que la gravedad de los cuerpos sólidos atraiga al gas circundante del disco, formando de esta manera la característica estructura de los planetas gaseosos.

## 3.2. Teorías de formación de planetas gigantes

Se han propuesto varios modelos para explicar la formación de los planetas gigantes. Entre otros argumentos, el hecho observacional de la muy baja presencia de objetos con masas entre 5 y 50 $M_J$ en órbita alrededor de estrellas de tipo solar (conocido como "desierto de enanas marrones"), lo cual representa una discontinuidad en la función inicial de masa, sugiere que el mecanismo de formación de los planetas es diferente al de las estrellas. En la sección 1.3.1 mencionamos algunos escenarios de formación de enanas marrones, fundamentalmente aquellos que podrían explicar la existencia de estos objetos en forma aislada. Para el caso de las enanas marrones con compañeros de masa estelar, la escasa población de estos objetos con períodos menores a cinco años podría estar relacionada con las dificultades del *modelo de inestabilidad del disco* como escenario de formación de objetos subestelares, como veremos a continuación. Por otra parte, en el Sistema Solar, el hecho que los planetas terrestres estén formados fundamentalmente por sólidos y que, a medida que la masa de los planetas crece, aumenta su componente gaseosa pero evidenciando siempre un enriquecimiento respecto de la abundancia solar, lleva a pensar también en un escenario de formación común para todos los planetas. Además, las propiedades observadas en otros sistemas planetarios, como por ejemplo el caso de HD 149026b que tendría un núcleo sólido de 67 $M_\oplus$ (Sato et al. 2005), sugiere que los exoplanetas se habrían formado en una manera análoga a la de los planetas del Sistema Solar.

**Tabla 3.1.** Masa final, $m_f$, y tiempo requerido, $t_f$, para el crecimiento oligárquico de embriones planetarios ubicados en las actuales posiciones de La Tierra, Júpiter, Saturno, Urano y Neptuno (Kokubo 2001). $\Sigma_0$ es la densidad superficial de sólidos y $\Delta a$ la separación orbital de los embriones ($\Delta a = 10\,R_H$).

| $a$ [UA] | $\Sigma_0$ [g cm$^{-2}$] | $\Delta a$ [UA] | $m_f$ [$M_\oplus$] | $t_f$ [$10^6$ años] |
|---|---|---|---|---|
| 1 | 10 | 0,07 | 0,16 | 0,7 |
| 5 | 4 | 1 | 5 | 40 |
| 10 | 1,4 | 3 | 9 | 300 |
| 20 | 0,5 | 6 | 14 | 2000 |
| 30 | 0,3 | 10 | 20 | 7000 |

Actualmente, existen dos modelos que intentan explicar la formación de los planetas gigantes: el modelo de inestabilidad nucleada y el modelo de inestabilidad gravitatoria. Ambos modelos son conceptualmente diferentes. Si bien el modelo de inestabilidad nucleada es el que cuenta con mayor aceptación por parte de la comunidad científica en general (y es el modelo en el cual está basado el desarrollo de esta Tesis), la falta de argumentos irrefutables contra el modelo de inestabilidad gravitatoria hace que éste se encuentre todavía en consideración.

### 3.2.1. Modelo de inestabilidad nucleada

Siguiendo a Lissauer & Stevenson (2007), el modelo de inestabilidad nucleada puede resumirse de la siguiente manera:

- Las partículas de polvo presentes en la nebulosa solar se acumulan para formar planetesimales, los cuales a su vez son acretados mutuamente dando origen al núcleo sólido del planeta (este proceso es completamente análogo al de la formación de los planetas terrestres). Mientras esto ocurre, el núcleo comienza a rodearse de una envoltura de gas que en principio es poco masiva. Por su parte, la tasa de acreción de sólidos es mucho más alta que la de gas. A medida que la zona de alimentación[2] del planeta se va vaciando, la acreción de sólidos disminuye mientras que la de gas aumenta.

- En el transcurso del proceso de acreción, se llega a que, en algún momento, la masa del núcleo y la masa de la envoltura gaseosa se hacen iguales. Cuando esto ocurre, la tasa de acreción de gas cambia, haciéndose cada vez más alta. Esta etapa se conoce como crecimiento *runaway* de la envoltura, ya que el protoplaneta incorpora grandes cantidades de gas en un corto período de tiempo.

  Estas dos primeras etapas se conocen como *estadío nebular* ya que el borde externo del planeta permanece en contacto con la nebulosa protoplanetaria, con lo cual la densidad y la temperatura en la interfase son las correspondientes a la de la nebulosa. Estas dos etapas son las que se estudiarán en detalle en el presente trabajo.

  Siguiendo en el proceso de formación, se tiene que:

- El crecimiento runaway de la envoltura queda luego limitado a la tasa a la cual la nebulosa es capaz de transportar gas a la vecindad del planeta. Una vez alcanzado este estado, la región de la envoltura que se encuentra en equilibrio se contrae y continúa la acreción de gas en forma hidrodinámica.

[2]Región adyacente a la órbita del planeta de donde éste obtiene el material para su formación.

- Finalmente, la acreción de gas termina, ya sea por la apertura de una brecha en el disco (*gap*) como consecuencia de los efectos de marea que provoca el planeta, o por la propia disipación de la nebulosa. Una vez finalizada la acreción, el planeta evoluciona aislado, contrayéndose y enfriándose a masa constante.

El modelo de inestabilidad nucleada fue investigado por primera vez por Perri & Cameron (1974) y Cameron (1978) quienes, proponiendo una envoltura completamente adiabática, estudiaron las soluciones de equilibrio para la envoltura. Más tarde, Mizuno et al. (1978), perfeccionaron el modelo aproximando la estructura de la envoltura por capas isotermas y adiabáticas. Finalmente, en 1980, Mizuno publicó el trabajo que es considerado referente para los cálculos que se realizaron a posteriori.

Mizuno (1980) estudió la formación de los planetas gigantes en base a modelos completamente hidrostáticos y en equilibrio térmico. El procedimiento del que se valió fue la construcción de una secuencia de modelos donde se incrementaba la masa del núcleo conforme transcurría el tiempo y, para cada modelo resolvía el estado de la masa de gas ligado. El límite externo de la envoltura correspondía al radio tidal (o radio de Hill) del objeto, en donde las condiciones de borde eran las que caracterizan a la nebulosa protoplanetaria para el dado radio orbital. Mizuno encontró que existía un valor máximo para la masa del núcleo a partir del cual no existía una solución estática para la envoltura. De acuerdo a sus cálculos, la "masa crítica" del núcleo a partir de la cual se perdía el estado de equilibrio hidrostático era de aproximadamente 12 $M_\oplus$. Esto concordaba con las estimaciones para los núcleos de los planetas gigantes del Sistema Solar. Y dado que este valor resultaba ser insensible a la posición del planeta en la nebulosa, esto es, las condiciones de borde no afectaban notoriamente la masa final del núcleo, se daba así una explicación lógica para el hecho de que las masas de los núcleos de los planetas gigantes del Sistema Solar fueran similares.

Seis años más tarde, Bodenheimer & Pollack (1986) construyeron el primer código capaz de resolver numéricamente y en forma autoconsistente las cuatro ecuaciones diferenciales características de la evolución estelar aplicadas a la envoltura del planeta (ver capítulo 4, § 4.4.1), proponiendo así un modelo más realista que el de Mizuno para estudiar la formación de los planetas gigantes. Adoptando una tasa de acreción de sólidos constante para la formación del núcleo, la resolución de estas ecuaciones en conjunto con las condiciones de borde, dan lugar a una secuencia de modelos donde el crecimiento del núcleo y de la envoltura se calculan en forma acoplada y consistente. Una vez que el planeta llegaba a la masa final, Bodenheimer & Pollack estudiaron además la evolución del mismo durante su proceso de enfriamiento.

En la primera fase de la formación del planeta, el núcleo crece a una tasa mucho más alta que a la que lo hace la envoltura. En esta etapa, la energía de la envoltura proviene principalmente de la energía gravitatoria de los planetesimales ingresantes. Cuando la masa de la envoltura se hace igual a la masa del núcleo, la energía proveniente de la acreción de sólidos se vuelve insuficiente para contrarrestar "el peso" de las capas de gas, y la luminosidad de la envoltura comienza a estar dominada por la contracción de las capas de gas. Esta transición, si bien bastante rápida, se produce de manera suave y continua. El valor de la masa del núcleo para el cual ella ocurre puede identificarse con la "masa crítica" definida por Mizuno. Sin embargo, es importante notar que, a diferencia de lo que se pensaba con anterioridad, la envoltura nunca deja de estar en equilibrio hidrostático. El concepto de "masa crítica" puede reformularse como la masa del núcleo a partir de la cual el crecimiento del planeta se debe fundamentalmente a la acreción de gas. En la práctica, este punto se alcanza cuando el núcleo y la envoltura tienen masas aproximadamente iguales. Bodenheimer & Pollack encontraron que la masa del núcleo para la cual esto ocurría era de entre 10 y 30 $M_\oplus$.

El trabajo de Bodenheimer & Pollack (1986) fue pionero en lo que respecta al cálculo autoconsistente de la formación de planetas gigantes en el marco del modelo de inestabilidad nucleada. Si bien presenta numerosas simplificaciones (como considerar la tasa de acreción de sólidos constante), sus resultados continúan siendo conceptualmente válidos. El estudio de Bodenheimer & Pollack incorporó por primera vez, bajo la hipótesis de un modelo unidimensional, toda la física relevante del problema. Probablemente uno de los resultados más importantes de este trabajo, más allá de la confirmación de la plausibilidad del modelo de inestabilidad nucleada, fue el hecho de que todo el proceso de acreción de gas ocurre con la envoltura en equilibrio hidrostático, aún después de alcanzar la "masa crítica" para el núcleo. Las simulaciones más detalladas que continuaron con el estudio de este problema confirmaron también este hecho. Cabe, sin embargo, mencionar que Wuchterl (1990, 1991, 1995), con un tratamiento completamente hidrodinámico, encuentra que cuando el núcleo alcanza la masa crítica se desencadena una inestabilidad en la envoltura que provoca la eyección de grandes cantidades de gas. Mientras que estos eventos están presentes cuando la densidad nebular es baja (como podría atribuirse a las regiones de Urano y Neptuno), desaparecen para densidades más altas. Wuchterl propone que este mecanismo fue el que le impidió a los gigantes helados continuar la acreción de gas y por este motivo presentan envolturas menos masivas que Júpiter y Saturno. Las inestabilidades encontradas por Wuchterl no pudieron ser reproducidas por otros autores. Sin embargo, el tratamiento que Wuchterl hace del transporte de la energía (considerando que la convección depende del tiempo), no fue tampoco incluido en ningún otro modelo de formación de planetas gigantes.

### 3.2.2. Modelo de inestabilidad del disco

El modelo de inestabilidad del disco propone un escenario completamente distinto al de la inestabilidad nucleada para la formación de los planetas gigantes. Según este modelo, los planetas gigantes se forman por la contracción de "grumos" de gas que surgen debido a inestabilidades gravitatorias en el disco protoplanetario. Las simulaciones numéricas (Boss 1997, 1998) muestran que condensaciones de $\sim$ 1$M_J$ se podrían formar en un disco de gas gravitatoriamente inestable. Sin embargo, debido al costo computacional que estas simulaciones requieren la evolución de estas condensaciones solo se siguen por algunos períodos orbitales y resulta incierto si ellas son capaces de sobrevivir lo suficiente como para que luego contraerse hasta alcanzar el tamaño de un planeta. De hecho, las inestabilidades gravitatorias que provocan las condensaciones generan también ondas de densidad, las cuales transportan momento angular lo cual haría que el disco se ensanche y, por lo tanto, disminuya su densidad en la región. Esto provocaría que los grumos finalmente se desarmen y el disco vuelva a ser estable. Raficov (2005) mostró que los discos que podrían dar lugar a planetas gaseosos por inestabilidad gravitatoria, a una distancia de algunas decenas de UA tendrían que ser muy calientes ($T > 10^3$ K), luminosos y de gran masa, propiedades que, de ser frecuentes, serían fácilmente detectables en discos circumestelares. Para discos con propiedades más o menos estándar, la formación por inestabilidad gravitatoria podría producirse más allá de las 100 UA. En estos casos no solo habría que invocar mecanismos de migración para posicionarlos en las ubicaciones actuales sino que, además, las masas de estos objetos no podrían ser inferiores a las 10 $M_J$. Este hecho estaría en concordancia con el "desierto de enanas marrones", dado que si el mecanismo de inestabilidad del disco fuera eficiente para semiejes pequeños deberían detectarse muchos objetos en el rango de masas de las enanas marrones. Sin embargo, la frecuencia de enanas marrones alrededor de estrellas de tipo solar aumenta con la separación orbital. Dado que, en cambio, sí se observan cuerpos de masa planetaria para pequeños semiejes, se considera a este hecho como un argumento a favor de que la formación de planetas gigante y enanas marrones correrían por caminos diferentes. Por otra parte, otro argumento en contra del modelo de inestabilidad del disco es que, para impedir que las condensaciones se desarmen bajo circunstancias plausibles para el disco, se necesitaría que el disco se enfríe demasiado rápido y/o que haya acreción de masa, de modo que se mantenga inestable. Aún si esto fuera posible, la composición del planeta sería la misma de la nebulosa, con lo cual se necesitaría un proceso independiente para explicar el enriquecimiento respecto de la composición solar que tienen Júpiter y Saturno. Además, en este escenario la formación de planetas como Urano y Neptuno (que, de algún modo, se puede pensar que están a mitad de camino entre los planetas terrestres y los gigantes gaseosos) resulta bastante difícil de explicar.

El estudio de la evolución de los grumos requiere que las simulaciones sean integradas por largos períodos de tiempo, con lo cual habría que considerar a su vez los procesos dinámicos que ocurren en el disco. Por otra parte, la capacidad computacional actual es insuficiente para hacer un seguimiento detallado de la evolución de estos grumos por un lapso de tiempo mayor a algunos miles de años. Además, si se quiere estudiar en profundidad la capacidad de supervivencia de los grumos es vital incluir en el modelo una descripción termodinámica más detallada de la que se ha estado empleando hasta ahora. En una reciente revisión sobre la formación de los planetas gigantes por inestabilidades gravitatorias en el disco, Durisen et al. (2007) plantean que en simulaciones donde la ecuación de estado considerada es isoterma, los grumos crecen hasta aproximadamente 10 $M_J$ en algunos miles de años. Sin embargo, al usar la EOS del gas ideal (que no es isoterma) estos grumos apenas llegan a 1 $M_J$. Este tipo de resultados pone en evidencia la importancia de un tratamiento termodinámico adecuado para analizar tanto la supervivencia de los grumos como también su crecimiento.

### 3.2.3. ¿Un modelo híbrido?

En los últimos años, varios autores han sugerido que la formación de planetas gigantes por inestabilidad nucleada podría verse favorecida si se considerara la presencia de inestabilidades gravitatorias en el disco de gas. La complejidad de este tipo de simulaciones hace que los resultados obtenidos hasta el momento estén lejos de ser concluyentes; sin embargo son sugestivos y proponen nuevos caminos para la investigación de este problema.

En base a simulaciones 3D hidrodinámicas, Durisen et al. (2007) especulan con que en los discos protoplanetarios las inestabilidades gravitatorias lleven a que se produzcan, en una escala de tiempo de algunos miles de años, arcos o anillos donde la densidad del disco se ve incrementada en algunos factores respecto del valor original. Si bien no necesariamente estos anillos llevarían a la formación de planetas por inestabilidades gravitatorias en el gas, podrían acelerar el proceso de la inestabilidad nucleada. La formación de los anillos no solo implica un aumento en la densidad sino también un pico de presión en esa región. En la vecindad de regiones de alta presión la velocidad del gas puede ser super o sub-kepleriana dependiendo del gradiente de presión. Los planetesimales, que sufren la fricción producida por el gas, se desplazan hacia afuera o hacia adentro respectivamente, siendo el efecto neto una acumulación de sólidos en las regiones de alta presión. Esta acumulación podría llevar a la formación, más rápida y en gran cantidad, de planetesimales en esta zona y, consecuentemente, se aceleraría la formación del núcleo de los planetas gigantes, haciendo que todo el proceso sea más eficiente. Además, la formación de cuerpos sólidos por aglomeración de partículas de polvo disminuye la opacidad del gas, lo cual favorece la acreción y la contracción de las capas de gas. Si bien no se han realizado simulaciones que demuestren que realmente estas sean las consecuencias de la formación de anillos en el disco de gas, ponen de manifiesto que es necesario estudiar estos escenarios en más profundidad.

Otro mecanismo "híbrido" de formación de planetas gigantes es el propuesto por Klahr & Bodenheimer (2003). La idea de estos autores es que el núcleo sólido se forme en el centro de un vórtice anticiclónico, dado que los vórtices colectan material del disco. Los vórtices se forman por perturbaciones que produce la materia cuando ésta cae hacia el disco o bien como resultado de inestabilidades hidrodinámicas. De simulaciones numéricas se obtiene que, una vez capturadas, las partículas de hasta el tamaño del metro, quedan dentro del vórtice por varios cientos de períodos orbitales, aún si el disco es turbulento. En el caso de cuerpos del orden del kilómetro o más grandes ya la situación no es tan clara. Es posible que los planetesimales sean también eyectados del vórtice, con lo cual sería poco probable que llegue a formarse un cuerpo lo suficientemente masivo como para ser considerado un embrión. Sin embargo, este mecanismo podría dar origen a objetos de hasta un kilómetro de tamaño que, como se mencionó anteriormente, es algo que todavía no está bien resuelto. Por otra parte, los planetesimales que se formen quedan radialmente confinados, lo cual facilita la acreción mutua para formar el núcleo, y un enriquecimiento en la zona de alimentación para que el proceso sea rápido. Con lo cual, la presencia de vórtices en el disco aceleraría la formación de un planeta, aún en discos de baja masa. Además, pone de manifiesto un mecanismo alternativo para la formación de planetesimales. El mayor inconveniente que hasta el momento tiene esta hipótesis es que no existen observaciones que confirmen la formación de vórtices en los discos protoplanetarios. Está entre los objetivos de ALMA[3] la búsqueda de estas estructuras.

Es importante entonces que se continúe el estudio de los procesos de inestabilidad gravitatoria en los discos protoplanetarios ya que, si bien puede que éstos no den lugar a la formación directa de planetas, pueden favorecer la formación de los planetesimales y de los núcleos planetarios. De hecho, aún cuando no sea posible que se formen cuerpos de tamaños superiores al metro, las inestabilidades gravitatorias impedirían la migración abrupta de las partículas sólidas hacia la estrella central. La estructura compleja y variable en el tiempo de las inestabilidades gravitatorias aumentaría el tiempo de permanencia de las partículas en el disco dándoles el tiempo suficiente como para crecer hasta el tamaño necesario para desacoplarse del gas.

[3] Atacama Large Millimiter/submillimiter Array. Proyecto astronómico internacional que consiste en un conjunto de radio telescopios. Estudiará la formación estelar en el universo temprano e intentará detectar en forma directa planetas extrasolares.

# Capítulo 4

# Modelo teórico de formación de planetas gigantes

En los capítulos anteriores se resumió el estado actual del conocimiento de los procesos involucrados en la formación de los planetas gigantes. También, se hizo referencia a los últimos datos observacionales disponibles, tanto en lo que respecta a los planetas del Sistema Solar como a los planetas extrasolares. En lo que sigue, se presentarán en detalle las componentes del modelo adoptado para la elaboración de esta Tesis sobre la formación de los planetas gigantes.

Como ya mencionamos, el objetivo de esta Tesis es calcular la formación de planetas gigantes en el marco de la hipótesis de inestabilidad nucleada. Este estudio se hará utilizando un código numérico, cuya versión original fue desarrollada por Benvenuto & Brunini (2005), y posteriormente actualizada por Fortier, Benvenuto & Brunini (2007). Para este tipo de cálculo deben elegirse modelos de crecimiento, tanto para el núcleo sólido como para la envoltura gaseosa. El estudio detallado del crecimiento de un embrión sólido requiere de simulaciones de $N$-cuerpos, lo cual está fuera del alcance de los objetivos de este trabajo. Sin embargo, de este tipo de simulaciones, y en conjunto con modelos semi-analíticos (Ida & Makino 1993, Kokubo & Ida 1996) pueden obtenerse expresiones analíticas aproximadas, las cuales nos permiten incorporar al código un modelo del régimen de acreción de sólidos. En cuanto a la formación de la envoltura gaseosa, ésta se calcula en forma autoconsistente resolviendo las ecuaciones estándar de evolución estelar junto con las condiciones de borde adecuadas. Es importante notar que la tasa de acreción de sólidos está acoplada al cálculo de la formación de la envoltura. De este modo, el régimen de acreción de los planetesimales y la interacción de éstos con la envoltura tendrán una incidencia importante en la forma en que ocurra la acreción de gas. Por este motivo, describiremos los dos regímenes involucrados en la formación del núcleo de los planetas gigantes, el crecimiento de tipo runaway y, en más profundidad, el crecimiento oligárquico (post-runaway); siendo éste último el que adoptaremos para nuestro modelo. Por otra parte, dado que durante su formación el planeta está inmerso en un disco de gas y sólidos, debemos considerar un modelo que lo describa. Utilizaremos uno sencillo, el modelo de la nebulosa solar estándar (Hayashi 1981), el cual determinará parte de las condiciones iniciales y de borde necesarias a la hora de resolver las ecuaciones diferenciales.

## 4.1. El disco protoplanetario

El disco protoplanetario será caracterizado según el modelo propuesto por Hayashi (1981). Consideraremos que la nebulosa solar se encuentra en equilibrio, en lo cual está implícito que ha transcurrido el tiempo suficiente como para que los granos de polvo hayan sedimentado hacia el plano ecuatorial del disco, de modo que la estructura gaseosa estará compuesta fundamentalmente por moléculas de hidrógeno y átomos de helio. Se supone que cualquier efecto magnético es despreciable y, por lo tanto, no serán tenidos en cuenta en el modelo.

La masa mínima de la nebulosa solar corresponde a la mínima cantidad de materia que debe contener el disco protoplanetario para dar cuenta de todos los cuerpos presentes actualmente en el Sistema Solar. Para estimar su valor se calcula la masa total de material sólido que se encuentra hoy en día presente en los planetas, asteroides, cometas, satélites, etc., y se calcula la cantidad de gas necesario para obtener la abundancia solar. El perfil de densidad surge de dividir la masa total de sólidos que contiene cada planeta por la región del disco que ocupa. La masa total se estima en $10^{-2}\,\mathrm{M}_{\odot}$, en un rango de distancias entre 0,35 y 36 UA. A este modelo del disco protoplanetario se lo conoce como Nebulosa Solar de Masa Mínima (NSMM).

Para el desarrollo de esta Tesis supondremos que el disco protoplanetario está bien descripto por los siguientes tres parámetros: el perfil de temperatura, la densidad superficial de sólidos y la densidad volumétrica de gas.

Aceptando la hipótesis de que los granos de polvo presentes en el disco se comportan como un cuerpo negro, con lo cual toda la energía que absorben es reemitida, y planteando la ley de Steffan-Boltzmann se obtiene para el perfil de temperatura una ley de potencias dependiente de la distancia:

$$T\left(a\right) = \alpha T_{\mathrm{eff}}\left(1\ \mathrm{UA}\right) T_{\mathrm{eff}}^{\star}\, a^{-q}, \tag{4.1}$$

donde $a$ es la distancia a la estrella central, $T_{\mathrm{eff}}^{\star}$ es la temperatura efectiva relativa a la del Sol (es una cantidad adimensional y para los casos considerados en este trabajo $T_{\mathrm{eff}}^{\star} = 1$), $T_{\mathrm{eff}}(1\ \mathrm{UA})$ es la temperatura efectiva del disco a 1 UA, fijada en 280 K, que es la temperatura actual en el Sistema Solar en esa ubicación, y $\alpha$ es una constante que se introduce para tener en cuenta que en sus inicios el Sol era más luminoso que en la actualidad ($\alpha$ fue fijado arbitrariamente en 1,08; vale destacar que esta consideración no tiene ninguna incidencia en los resultados). La línea de hielo, $a_{\mathrm{snow}}$, es la ubicación en el disco donde la temperatura es tal que el agua se encuentra en estado sólido. Este cambio de fase en el estado del agua provoca un aumento en la densidad de sólidos lo cual, según se cree, sería fundamental para la formación de planetas de gran masa. La línea de hielo se sitúa donde la temperatura es $T = 170$ K, que en este modelo corresponde a $a = 3{,}16$ UA.

La definición de la NSMM lleva a que el ajuste del perfil de densidad de sólidos siga una ley de potencias:

$$\Sigma\,(a) = \eta_{\rm ice}\,\sigma_{\rm z}\,(1\ {\rm UA})\,a^{-p}, \tag{4.2}$$

donde consideramos que $\sigma_{\rm z}\,(1\ {\rm UA}) = 10$ g cm$^{-2}$, $p = 3/2$ y $\eta_{\rm ice}$ es el factor de enriquecimiento más allá de la línea de hielo dado por

$$\eta_{\rm ice} = \begin{cases} 1 & \text{si } a < a_{\rm snow} \\ 4 & \text{si } a > a_{\rm snow}. \end{cases}$$

Si bien en esta ecuación está implícita la hipótesis de que los sólidos se encuentran distribuidos uniformemente, para el cálculo de la formación de planetas gigantes es también necesario adoptar un valor para la masa de los planetesimales del disco, como así también un valor para su radio. En la mayoría de las simulaciones, consideraremos que la densidad en masa de los planetesimales es para todos la misma e igual a 1,5 g cm$^{-3}$. En cuanto al tamaño, se considerarán poblaciones de planetesimales con radio único y poblaciones donde se adopta una función para representar la distribución de tamaños. Como veremos en los próximos capítulos, los resultados dependen fuertemente de estas hipótesis.

En cuanto a la densidad volumétrica de gas, el perfil está dado por:

$$\rho\,(a) = \sigma_{\rm g}(1\ {\rm UA})\,\frac{a^{-b}}{2H}, \tag{4.3}$$

donde $\sigma_{\rm g}$ (1 UA) es la densidad superficial de gas 1 UA, siendo ésta igual a $2 \times 10^3$ g cm$^{-2}$, $b = 3/2$ y $H$ es la altura de escala del disco de gas,

$$H\,(a) = 0{,}05\ a^{5/4} \times (1{,}5 \times 10^{13}{\rm cm}). \tag{4.4}$$

Notemos que $a$ está expresada en UA pero $H$ debe estar en cm para que $\rho\,(a)$ esté en g cm$^{-3}$, con lo cual aparece un factor de conversión $1{,}5 \times 10^{13}$ cm. Los valores de las constantes $\sigma_{\rm g}\,(1\ {\rm UA})$ y $\sigma_{\rm z}\,(1\ {\rm UA})$ no son los propuestos en el artículo original de Hayashi (1981) sino que han sido actualizados y son un poco más elevados, correspondiendo en este caso al modelo propuesto por Kokubo (2001).

Las características más importantes de la NSMM en la posición de Júpiter ($a = 5{,}2$ UA) se listan en la Tabla 4.1.

## 4.2. Sobre el crecimiento de los planetesimales

A la hora de hablar del crecimiento de los planetesimales es necesario mencionar brevemente cómo operan los efectos de enfocamiento gravitatorio y fricción dinámica. Sea $m$ la masa de un planetesimal, $r_m$ su radio, $v_{\rm rel}$ la velocidad relativa promedio entre este planetesimal y los que lo rodean, y sea $\rho_{\rm Pl}$ la densidad de planetesimales (masa por unidad de volumen del disco). Si el planetesimal acreta todos los planetesimales con los que tiene encuentros en un tiempo $dt$, su tasa de crecimiento será:

$$\frac{dm}{dt} = \pi r_m^2 \rho_{\rm Pl} v_{\rm rel}. \tag{4.5}$$

**Tabla 4.1.** Características de la NSMM en la ubicación actual de Júpiter ($a$ = 5,2 UA).

| | |
|---|---|
| Temperatura | 133 K |
| Densidad superficial de sólidos | 3,37 g cm$^{-2}$ |
| Densidad volumétrica de gas | $1,5 \times 10^{-11}$g cm$^{-3}$ |

Sin embargo, debido al efecto de atracción gravitatoria mutua entre los planetesimales, su trayectoria relativa no es lineal sino hiperbólica. Entonces, los planetesimales acretados por $m$ no serán solo aquellos cuyo parámetro de impacto sea menor o igual a $r_m$ (o sea, todos aquellos que directamente chocan con él), sino también aquellos que, por acción gravitatoria, sean deflectados de su trayectoria original y ésta termine en una colisión (ver figura 4.1). El valor máximo del parámetro de impacto para el cual se satisface esta condición puede obtenerse imponiendo que durante un encuentro debe conservarse la energía y el momento angular. De este modo, la acción de la gravedad produce un aumento en la sección eficaz de colisión que se denomina *enfocamiento gravitatorio*. El radio efectivo del planetesimal satisface la ecuación:

$$r_{\rm ef}^2 = r_m^2(1 + 2\theta), \tag{4.6}$$

donde $\theta$ se conoce como número de Safronov y está definido como:

$$\theta \equiv \frac{1}{2}\frac{v_{\rm esc}^2}{v_{\rm rel}^2}, \tag{4.7}$$

donde $v_{\rm esc}$ es la velocidad de escape de la superficie del planetesimal. Luego, cuanto más pequeña es la velocidad relativa o cuanto más masivo es el objeto, mayor es la sección eficaz de colisión. Entonces, los planetesimales más masivos o que se encuentren en bajos regímenes de velocidades relativas tendrán mayores probabilidades de crecimiento.

Otro efecto que favorece el crecimiento de los planetesimales de mayor masa es la fricción dinámica. Consideremos un cuerpo de masa $M$ que se desplaza a través de un mar infinito y homogéneo de partículas de masa $m$. El efecto neto, luego de sucesivos encuentros entre $M$ y los pequeños cuerpos que lo rodean, es similar al que experimenta un objeto moviéndose en un fluido, solo que en el primer caso la naturaleza de la "fricción" que sufre $M$ es de carácter gravitatorio. El cuerpo $M$ experimenta una desaceleración, cuya componente actúa en dirección paralela a su velocidad, que provoca una reducción de la misma. Se dice que esta continua desaceleración es debida a la *fricción dinámica*. La fricción dinámica, al oponerse a la velocidad orbital, provoca el decaimiento orbital del objeto, pero también afecta la excentricidad del mismo, tendiendo a circularizar las órbitas. El frenado es proporcional a la masa del objeto: cuanto mayor es su masa, mayor es la desaceleración. El efecto neto resulta ser una transferencia de energía cinética desde los objetos de mayor tamaño hacia los más pequeños. Ésto es de crucial importancia, pues gracias al enfocamiento gravitatorio los cuerpos más masivos conservan su status en la medida en que las velocidades evolucionan, mientras que los objetos más pequeños pierden ese privilegio debido a que la fricción dinámica no es operativa sobre ellos.

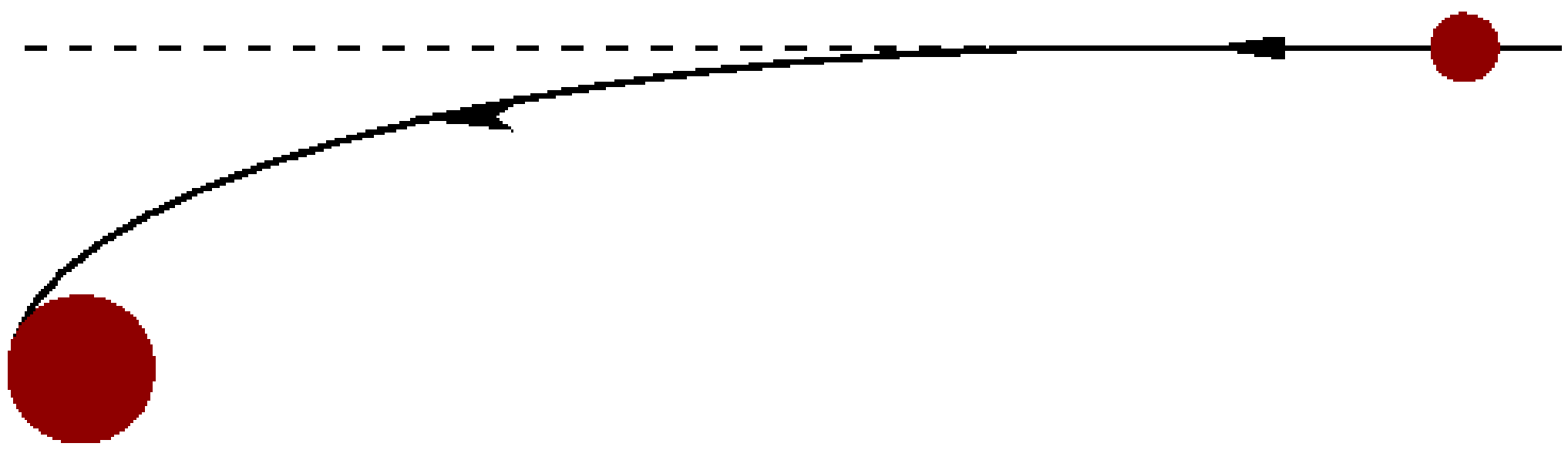

**Figura 4.1.** Representación esquemática del enfocamiento gravitatorio que produce un cuerpo masivo sobre los planetesimales más pequeños que se encuentran en su zona de influencia y presentan una velocidad relativa no nula respecto de éste. Los planetesimales, que de seguir su trayectoria original no colisionarían con el embrión, sufren una deflección debido al efecto gravitatorio del mismo, resultando finalmente acretados por el embrión.

En lo que respecta al crecimiento de los planetesimales para dar origen a los protoplanetas se puede decir que la formación de los planetas terrestres atraviesa tres etapas, mientras que la formación de los núcleos de los planetas gigantes se completa en tan solo las dos primeras (Thommes & Duncan 2006). En la primera etapa, los planetesimales crecen siguiendo un régimen runaway, lo cual significa que los más grandes pobladores del disco serán favorecidos en su crecimiento y rápidamente ganarán la masa suficiente para distinguirse del resto de los planetesimales. Cuando aparecen los primeros embriones protoplanetarios (cuerpos del orden de $10^{24}$ g, aunque este valor depende de la ubicación en el disco), el crecimiento se vuelve autorregulado puesto que, en estos casos, los objetos más grandes perturban gravitatoriamente a los planetesimales de su adyacencia lo cual restringe la probabilidad de acreción dado que aumenta el régimen de velocidades relativas. Esta etapa se conoce como crecimiento oligárquico ya que solo los cuerpos de mayor masa alcanzan su “masa de aislación”, lo cual significa que llegan a consumir todos los planetesimales que se encuentran bajo su influencia gravitatoria. En la región de los planetas terrestres, la masa de aislación típica resulta ser aproximadamente igual a la masa de Marte. La tercera etapa de crecimiento se produce por la colisión de estos embriones, en un crecimiento típicamente ordenado. Los modelos de formación de planetas terrestres estiman un período de acreción del orden de $10^8$ años cuando se consideran escenarios en ausencia de gas (Chambers 2001), o de solo algunas decenas de millones de años si se tienen en cuenta los efectos dinámicos de las resonancias con los planetas gigantes (Nagasawa, Lin & Thommes 2005). Este último hecho requiere de la pre-existencia de los planetas gigantes, con lo cual éstos deben formarse antes que los planetas terrestres. Para que ésto ocurra, y teniendo en cuenta el tiempo que involucra el crecimiento por colisión entre embriones, que corresponde a un régimen de crecimiento ordenado, la tercera etapa no debería ser significativa en la formación de los núcleos de los planetas gigantes, los cuales serían entonces el producto final del crecimiento oligárquico. O sea que, el crecimiento runaway y el oligárquico deben ser capaces de dar origen a embriones sólidos de 10 $M_\oplus$ en la región exterior del Sistema Solar para poder formar los núcleos de los planetas gigantes sin necesidad de invocar además colisiones entre embriones para llegar a masas de ese orden.

A continuación describiremos en detalle los regímenes de crecimiento runaway y oligárquico, y estableceremos las características del modelo utilizado en nuestras simulaciones.

### 4.2.1. Régimen de crecimiento runaway

Aceptando que, por algún mecanismo, planetesimales de tamaño del orden del kilómetro se forman en el disco protoplanetario, existe un acuerdo general en que las primeras semillas de los embriones planetarios surgen debido a la acreción mutua entre los planetesimales. En el escenario clásico de la formación planetaria (Safronov 1969), el crecimiento de los planetesimales se considera *ordenado*. Sin embargo, bajo este régimen, el tiempo de crecimiento de los núcleos de Júpiter y Saturno excedería el tiempo de vida estimado para la nebulosa protoplanetaria, con lo cual estos objetos no tendrían la posibilidad de capturar el gas necesario para formar la envoltura. Además, la escala de tiempo de formación de Urano y Neptuno superaría, en ese caso, la edad del Sistema Solar. A partir de la década del '80, el crecimiento de los planetesimales comenzó a ser profundamente investigado bajo la motivación de resolver este problema. Mediante simulaciones numéricas se encontró que el crecimiento de los planetesimales en los primeros estadíos de la acreción era el correspondiente al de un régimen de crecimiento en fuga (Greenberg et al. 1978, Kokubo & Ida 1996). De acuerdo con Ida & Makino (1993), este tipo de crecimiento se puede resumir de la siguiente manera: al principio, dado que las excentricidades e inclinaciones de los planetesimales se mantienen bajas (las velocidades aleatorias de los planetesimales resultan proporcionales a su excentricidad e inclinación), las velocidades aleatorias de los planetesimales más chicos son pequeñas debido a que están reguladas por la dispersión mutua entre los planetesimales. Esto favorece al crecimiento del protoplaneta, el cual se acelera a medida que aumenta su masa debido al enfocamiento gravitatorio. Se desencadena entonces un proceso retroalimentado: los planetesimales más grandes crecen más rápido porque son gravitatoriamente dominantes entre sus vecinos, comenzando así el crecimiento en fuga

o *crecimiento runaway*. Los cuerpos en crecimiento runaway rápidamente se vuelven más masivos que el resto, siendo este el mecanismo que da origen a los primeros embriones planetarios en el disco. Durante la etapa de crecimiento runaway, tanto la tasa de crecimiento del protoplaneta como la proporción de masa entre el protoplaneta y el resto de los planetesimales se incrementa con el tiempo.

### 4.2.2. Régimen de crecimiento oligárquico

Ida & Makino (1993) investigaron la evolución post-runaway de las excentricidades e inclinaciones de los planetesimales debido a la dispersión producida por la acción del campo gravitatorio del protoplaneta. Cuando los embriones en crecimiento runaway adquieren la masa suficiente como para afectar gravitatoriamente a los planetesimales que pueblan su zona de influencia, las velocidades aleatorias de los planetesimales aumentan y el crecimiento del embrión se vuelve más lento. Ida & Makino (1993) bautizaron a esta etapa como "etapa dominada por el protoplaneta". Para estudiarla, los autores realizaron simulaciones tridimensionales de $N$-cuerpos, con las cuales analizaron la evolución de las excentricidades e inclinaciones de un sistema de planetesimales, todos de ellos de igual masa, frente a la presencia de un protoplaneta. Tuvieron en cuenta las interacciones protoplaneta-planetesimal y planetesimal-planetesimal, pero no consideraron que hubiera acreción entre ellos. Ida & Makino (1993) encontraron que los planetesimales son fuertemente dispersados por el protoplaneta y que parte de la energía que adquieren en la dispersión se distribuye luego entre el resto de los planetesimales. Los planetesimales excitados de esta manera quedan confinados en una región alrededor del protoplaneta que es aproximadamente el doble del ancho de su zona de alimentación. En esta región, las excentricidades ($e_m$) y las inclinaciones ($i_m$) de los planetesimales son fuertemente perturbadas, y las velocidades aleatorias pueden aproximarse mediante la ecuación:

$$v_{\mathrm{rel}} \simeq \sqrt{\langle e_m^2 \rangle + \langle i_m^2 \rangle}\, v_{\mathrm{k}}. \tag{4.8}$$

Luego, es claro que un incremento en $e_m$ e $i_m$ se traduce en un aumento de las velocidades aleatorias. Aquí $v_{\mathrm{k}}$ es la velocidad kepleriana correspondiente a la posición del protoplaneta ($v_{\mathrm{k}} = a\,\Omega_{\mathrm{k}}$ donde $\Omega_{\mathrm{k}}$ es la velocidad angular kepleriana de un objeto que gira en torno al Sol en una órbita circular de radio $a$), y $\langle e_m^2 \rangle^{1/2}$ ($\langle i_m^2 \rangle^{1/2}$) es la excentricidad (inclinación) cuadrática media de los planetesimales en el disco.

En base a los resultados de sus simulaciones, Ida & Makino (1993) proponen un crecimiento de tipo escalonado para los protoplanetas. En una primera etapa, los protoplanetas siguen un crecimiento runaway en el sentido usual: la perturbación que sufren los planetesimales está dominada por los propios planetesimales, siendo las velocidades relativas generalmente bajas, lo cual favorece el rápido crecimiento de los embriones. Un crecimiento runaway no significa necesariamente que el tiempo de crecimiento del embrión disminuya a medida que aumenta su masa, sino que la razón de la masa entre dos cuerpos (embrión - planetesimal) aumenta con el tiempo. Además, durante el crecimiento runaway, la mayor parte de la masa del sistema se encuentra en los cuerpos pequeños, cuya distribución de masa puede ajustarse con una ley de potencias, mientras que la masa del embrión se separa de esta distribución. Además, debido a la fricción dinámica, la excentricidad e inclinación del embrión se mantienen bajas, lo cual facilita también el proceso de acreción. Cuando un protoplaneta es suficientemente masivo como para perturbar a los planetesimales de su vecindad, el sistema entra en una etapa en la cual su dinámica está dominada por el protoplaneta. Las simulaciones numéricas de $N$-cuerpos muestran que esta etapa ocurre en el régimen dominado por la dispersión, donde la relación entre los valores medios de las excentricidades y las inclinaciones es:

$$\langle e_m^2 \rangle^{1/2} / \langle i_m^2 \rangle^{1/2} \simeq 2. \tag{4.9}$$

En esta segunda etapa de crecimiento post-runaway, la tasa de crecimiento del protoplaneta decrece como consecuencia del aumento en las velocidades relativas entre los planetesimales y el protoplaneta. Sin embargo, el cociente de masas entre el protoplaneta y los planetesimales sigue aumentando con el tiempo.

Ida & Makino (1993) derivaron, en base a un modelo semi-anlítico la condición para el dominio de la dispersión protoplaneta-planetesimal sobre la dispersión planetesimal-planetesimal:

$$2\,\Sigma_M\, M > \Sigma\, m, \tag{4.10}$$

donde $M$ es la masa del embrión sólido, $m$ es el valor promedio de la masa de los planetesimales, $\Sigma$ es la densidad superficial de masa debida a los planetesimales del disco y $\Sigma_M$ se define como:

$$\Sigma_M = \frac{M}{2\pi a \Delta a}, \tag{4.11}$$

donde $2\pi a\Delta a$ es el área de la región que ocupan los planetesimales excitados. Para una nebulosa solar estándar, esta condición puede ser trasladada a una relación entre $M$ y $m$, la cual depende del perfil de densidad superficial del disco, del semieje mayor $a$ y de la masa de los planetesimales. Para un modelo de nebulosa de baja masa, la transición entre los dos regímenes de crecimiento, ocurre cuando $M/m \simeq 50 - 100$. Esta condición, derivada en forma semi-analítica, fue corroborada por los autores con simulaciones de $N$-cuerpos. La segunda etapa en el régimen de crecimiento del protoplaneta comienza cuando el protoplaneta es todavía mucho menos masivo que su correspondiente masa de aislación. De este modo, cuando la masa del protoplaneta excede en, aproximadamente, 100 veces la masa típica de un planetesimal, el protoplaneta excita las velocidades de los planetesimales que lo rodean y las velocidades aleatorias comienzan a depender de su masa. Ida & Makino (1993) estimaron el tiempo característico de crecimiento del protoplaneta $T_{grow}$ ("the mass–doubling time"), bajo la suposición de una densidad superficial de sólidos constante, y obtuvieron que:

$$T_{grow} \equiv \left(\frac{1}{M}\frac{dM}{dt}\right)^{-1} \propto M^{-1/3}\langle e_m^2 \rangle. \tag{4.12}$$

Para estimar $\langle e_m^2 \rangle$, consideraron un estado de equilibrio donde la dispersión producida por el embrión se encuentra compensada por el amortiguamiento debido a la viscosidad con el gas nebular. Como resultado, en la primera etapa (correspondiente al crecimiento runaway, donde las perturbaciones están dominadas por los planetesimales), $\langle e_m^2 \rangle \propto m^{8/15}$ (notar que no depende de la masa del protoplaneta), mientras que en la etapa post-runaway, $\langle e_m^2 \rangle \propto m^{1/9}M^{2/3}$. Así, en la primera etapa $T_{grow} \propto M^{-1/3}$, de modo que los protoplanetas crecen lo suficientemente rápido como para que la relación $M/m \simeq 50 - 100$ se alcance en una escala de tiempo despreciable. En la segunda etapa, es el protoplaneta el perturbador dominante y $T_{grow} \propto M^{1/3}$, con lo cual el crecimiento del protoplaneta se hace más lento a medida que su masa aumenta. Este tipo de dependencia temporal es característica del crecimiento de tipo ordenado (que es el régimen correspondiente a la última etapa de formación de los planetas terrestres). Sin embargo, la etapa dominada por el protoplaneta es mucho más corta que la correspondiente al crecimiento ordenado porque los protoplanetas crecen por la acreción de planetesimales y no por la colisión entre cuerpos de tamaños similares, donde la fricción dinámica está ausente.

Continuando la investigación de Ida & Makino (1993), Kokubo & Ida (1998) hicieron simulaciones tridimensionales de $N$-cuerpos para estudiar el escenario de acreción post-runaway. Como condición inicial consideraron la presencia de dos protoplanetas, ahora sí, creciendo por la acreción de planetesimales. En la primera etapa de acreción, confirmaron que los protoplanetas crecen según el régimen runaway. El crecimiento runaway se interrumpe cuando $M \sim 50\,m$, tal como lo predecía el modelo semi-analítico de Ida & Makino (1993). Luego, los protoplanetas continúan su crecimiento manteniendo una separación orbital típica de $10\,R_{\mathrm{H}}$, donde $R_{\mathrm{H}}$ es el Radio de Hill del protoplaneta,

$$R_{\mathrm{H}} = a\left(\frac{M}{3M_\star}\right)^{1/3}. \tag{4.13}$$

Durante su crecimiento se manifiesta una repulsión orbital entre los protoplanetas, la cual es consecuencia del acoplamiento entre la dispersión provocada por los cuerpos de mayor masa y la fricción dinámica en la población de planetesimales. Los protoplanetas crecen manteniendo el cociente entre sus masas cercano a uno, pero la relación de las masas entre los protoplanetas y los planetesimales se incrementa en el tiempo. De este modo, se dice que los protoplanetas siguen un crecimiento de tipo "oligárquico" ya que no se produce acreción entre los planetesimales y solo crecen los cuerpos de mayor masa. La distribución de masa en el disco se vuelve bimodal: hay un pequeño grupo de embriones, embebidos en un disco poblado por un gran número de planetesimales. Kokubo & Ida (1998) definieron formalmente a este tipo de crecimiento como *crecimiento oligárquico* en el sentido de que no solo uno sino varios son los protoplanetas que dominan al sistema de planetesimales.

Kokubo & Ida (2000) investigaron, también con simulaciones tridimensionales de $N$-cuerpos la evolución de un enjambre de planetesimales hasta que algunos alcanzan la masa necesaria para ser considerados embriones, pero ahora incluyendo el efecto de la fricción nebular. Confirmaron la existencia de un crecimiento runaway en la primera etapa de la formación protoplanetaria y una segunda etapa de crecimiento oligárquico de los embriones, en el mismo sentido que lo encontrado por Kokubo & Ida (1998). Así mismo, al analizar la evolución de la excentricidad y la inclinación cuadrática media de los planetesimales durante el crecimiento oligárquico de los embriones, encontraron que sus resultados concordaban con aquellos predichos por la teoría semi-analítica de Ida & Makino (1993).

En función de estos resultados, tanto numéricos como analíticos, es ampliamente aceptado que el crecimiento de los embriones sólidos atraviesa dos etapas: la primera de tipo runaway, cuya duración es corta comparada con las escalas de tiempo involucradas en la formación de los planetas gigantes; y la segunda, de tipo oligárquico, que es la que domina durante el proceso de formación.

## 4.3. La aparición de los embriones planetarios

Cuando la dinámica de la población de planetesimales está dominada por la dispersión, la tasa de crecimiento de un embrión sólido está bien descripta por la aproximación de partículas en una caja (Safronov 1969),

$$\frac{dM_{\rm c}}{dt} \simeq F \frac{\Sigma}{2h} \pi \, R_{\rm eff}^2 \, v_{\rm rel}, \tag{4.14}$$

donde $M_{\rm c}$ es la masa de la componente sólida del protoplaneta (que en nuestro caso corresponde al núcleo de un planeta gigante), $h$ es la altura de escala del disco de planetesimales, $R_{\rm eff}$ es el radio efectivo de captura del embrión y $F$ es un factor que se introduce para compensar la tasa de crecimiento producto de utilizar la aproximación de dos cuerpos en el cálculo de la velocidad de dispersión de los planetesimales, la cual está modelada por un único valor para la excentricidad y la inclinación, correspondiente al valor cuadrático medio. $F$ está estimado en $\sim 3$ (Greenzweig & Lissauer 1992).

Debido al enfocamiento gravitatorio, el radio efectivo de captura del protoplaneta, $R_{\rm eff}$, es mayor que el radio geométrico del embrión sólido, $R_{\rm c}$,

$$R_{\rm eff}^2 = R_{\rm c}^2 \left( 1 + \left( \frac{v_{\rm esc}}{v_{\rm rel}} \right)^2 \right), \tag{4.15}$$

con $v_{\rm esc}$ la velocidad de escape de su superficie y $v_{\rm rel}$ la velocidad relativa entre el protoplaneta y los planetesimales,

$$v_{\rm rel} \simeq \sqrt{e^2+i^2}\, a\, \Omega_{\rm k}, \tag{4.16}$$

donde, a partir de ahora, definimos $e \equiv \langle e_m^2 \rangle^{1/2}$ ($i \equiv \langle i_m^2 \rangle^{1/2}$ ); $e$ ($i$) es el valor cuadrático medio de la excentricidad (inclinación) con respecto al plano del disco protoplanetario ($e$, $i$ $<<$ 1) y $\Omega_{\rm k}$ es la velocidad angular kepleriana. En lo que sigue utilizaremos la aproximación $i \simeq e/2$ y $h \simeq ai$ ($h$ es la altura de escala de los planetesimales). Siguiendo a Thommes, Duncan & Levison (2003), adoptamos para $e$ la expresión de equilibrio que se deduce para el caso donde las perturbaciones gravitatorias debidas al protoplaneta están balanceadas por el arrastre gaseoso. Sea $T_{\rm VS}$ la escala de tiempo en la cual un cuerpo de masa $M_{\rm p}$ (en nuestro caso, $M_{\rm p}$ es la masa total del protoplaneta, ésto es, la masa del núcleo más la masa de la envoltura) excita a los planetesimales que lo rodean (Ida 1990),

$$T_{\rm VS} \simeq \frac{1}{40}\left(\frac{\Omega_{\rm k}^2 a^3}{GM}\right)^2 \frac{e^4 M_{\rm p}}{\Sigma_M a^2 \Omega_{\rm k}}, \tag{4.17}$$

y sea la escala de tiempo asociada al amortiguamiento del gas (Adachi, Hayashi & Nakazawa 1976):

$$T_{gas} \simeq \frac{1}{e}\frac{m}{(C_{\rm D}/2)\pi r_m^2 \rho a \Omega_{\rm k}}, \tag{4.18}$$

donde $C_{\rm D}$ es el coeficiente de amortiguamiento de la componente gaseosa del disco (adimensional y de orden uno para estos casos), y $r_m$ el radio de un planetesimal de masa $m$. Igualando las ecuaciones (4.17) y (4.18) y despejando $e$, se obtiene para la excentricidad de equilibrio de los planetesimales:

$$e = \frac{1{,}7 m^{1/15} M_{\rm p}^{1/3} \rho_m^{2/15}}{\beta^{1/5} C_{\rm D}^{1/5} \rho^{1/5} M_\star^{1/3} a^{1/5}}, \tag{4.19}$$

donde $\rho_m$ es la densidad en masa de un planetesimal, $\rho$ es la densidad volumétrica de gas en el disco protoplanetario, y $\beta$ es al ancho, en unidades del radio de Hill, de la región que ocupan los planetesimales excitados por el protoplaneta (considerando que, potencialmente, hay otros embriones creciendo en la vecindad, $\beta$ es del orden de 10). Según Chambers (2006), utilizar los valores de equilibrio para $e$ e $i$ es una aproximación aceptable cuando se consideran embriones $m > 10^{-2}$ $M_\oplus$, consistente con las masas de los núcleos iniciales de todas nuestras simulaciones (ver capítulos 6 y 7).

Sin embargo, cuando un embrión sólido alcanza la masa necesaria como para ligar gravitatoriamente gas de la nebulosa en la cual está inmerso, la presencia de esta envoltura debe ser considerada en el cálculo del radio efectivo de captura, $R_{\rm eff}$. La envoltura gaseosa modifica la trayectoria de los planetesimales ingresantes, ya que estos se ven afectados por la fricción del gas, lo cual aumenta el radio de captura del protoplaneta. De este modo, además del enfocamiento gravitatorio, tendremos un "enfocamiento viscoso" debido a la fricción producida por la atmósfera.

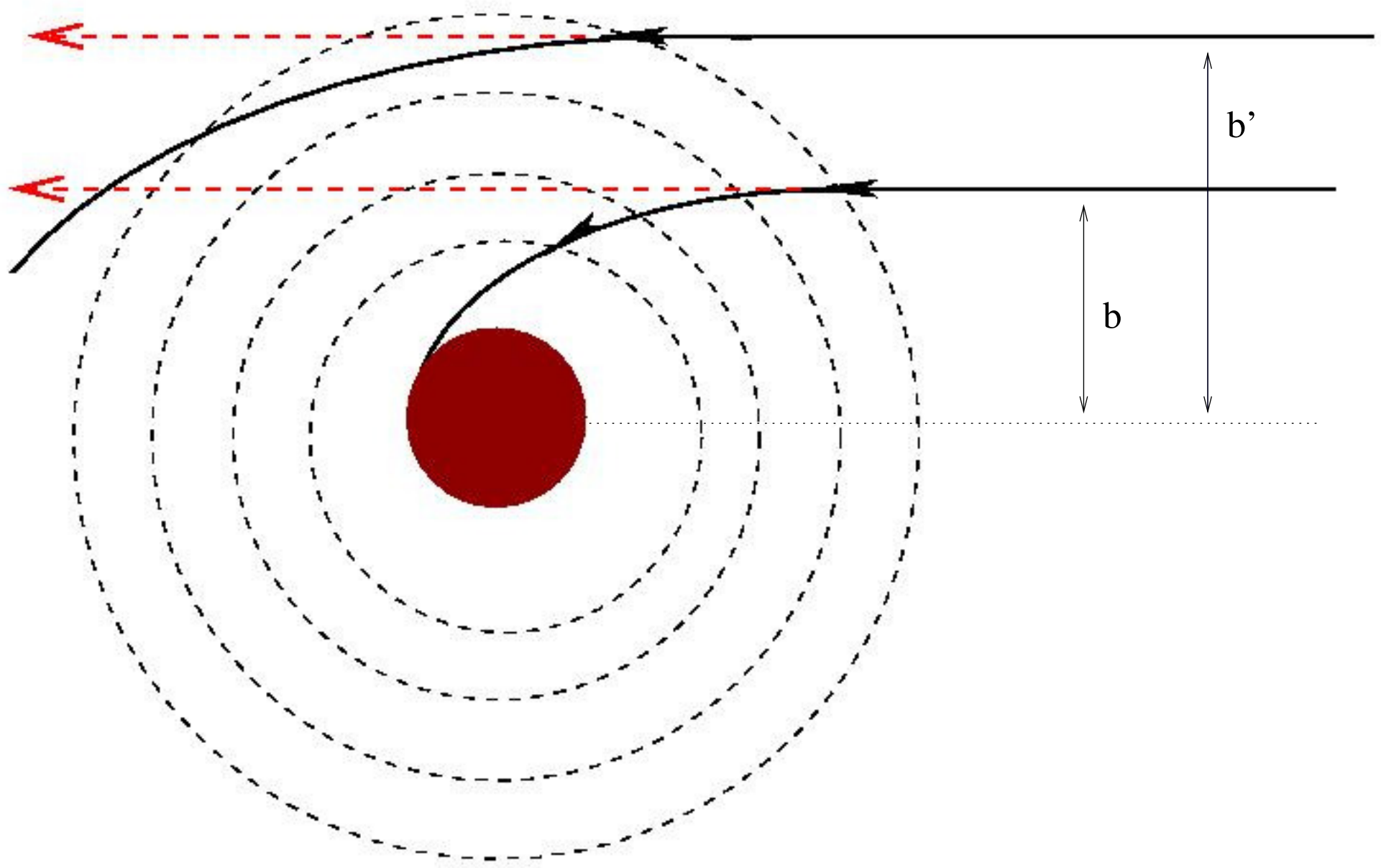

**Figura 4.2.** Esquema del enfocamiento gravitatorio y viscoso de un protoplaneta. Debido a la gravedad del planeta y a la viscosidad del gas la sección eficaz de captura de un protoplaneta es mayor que su sección eficaz geométrica. La gravedad y el frenado viscoso deflectan la trayectoria de un planetesimal. Cuando el parámetro de impacto se va reduciendo, de b' a b, llegará un momento en que el planetesimal seguirá una trayectoria que lo hace colisionar con el núcleo. El valor de b para el cual esto ocurre se define como el radio efectivo del planeta.

El radio efectivo $R_{\rm eff}$, en la forma dada en la Ec. (4.15) domina cuando la masa de gas ligado es despreciable. Pero cuando un embrión adquiere la cantidad de gas suficiente para formar apenas una delgada atmósfera, su radio efectivo aumenta, separándose considerablemente del calculado anteriormente (ver figura 4.2). Para calcular el radio efectivo del protoplaneta en presencia de una atmósfera hay que considerar que sobre los planetesimales ingresantes actúan dos fuerzas: la gravedad y el arrastre viscoso. Consideremos un planetesimal que ingresa a la atmósfera del protoplaneta con una velocidad $v_{\rm rel}$ (Ec. (4.16)). Su ecuación de movimiento se obtiene de considerar, entonces, la acción de la gravedad,

$$\vec{f}_{\rm G} = -\frac{GM_r m}{r^2}\hat{r}, \tag{4.20}$$

junto con la acción de la fricción gaseosa. En la Ec. (4.20), $r$ es la coordenada radial con origen en el centro del protoplaneta y $M_r$ es la masa contenida hasta un radio $r$. La fuerza de arrastre viscoso actuando sobre un cuerpo esférico de radio $r_m$ que viaja a una velocidad $v$ es:

$$\vec{f}_{\rm D} = -\frac{1}{2}C_{\rm D}\,\pi\, r_m^2\, \rho_{\rm g} v^2\, \hat{v}, \tag{4.21}$$

donde aquí $\rho_{\rm g}$ es la densidad de la envoltura y $C_{\rm D} = 1$ (Adachi, Hayashi & Nakazawa 1976). Luego, la ecuación de movimiento surge de combinar ambas ecuaciones,

$$m\frac{d\vec{v}}{{\rm d}t} = \vec{f}_{\rm G} + \vec{f}_{\rm D} = -\frac{GM_r m}{r^2}\hat{r} - \frac{1}{2}C_{\rm D}\,\pi\, r_m^2\, \rho_{\rm g} v^2\, \hat{v}. \tag{4.22}$$

Para comprender la importancia de la viscosidad en el cálculo del radio efectivo, la figura 4.3 muestra, para una de nuestras simulaciones, los valores que toman, a lo largo de todo el proceso de formación, el radio efectivo considerando solo el enfocamiento gravitatorio ($R^*_{\rm eff}$) y el que surge de considerar también el efecto viscoso de la envoltura ($R_{\rm eff}$). Los detalles de la simulación que caracterizan este ejemplo se discutirán en el capítulo 6. Como se puede apreciar, aún cuando el gas ligado al embrión es despreciable (corresponde a cuando el radio del planeta es inferior al radio efectivo), la viscosidad del gas igualmente contribuye a aumentar el radio efectivo (en este caso, el gas no pertenece a la envoltura del planeta pero sí pertenece a su esfera de Hill). El apartamiento entre ambos radios se acentúa cuando la masa de la envoltura domina la masa total del planeta. De hecho, la diferencia entre $R_{\rm eff}$ y $R^*_{\rm eff}$ es de más de un orden de magnitud. Cuando el planeta es muy masivo, notamos que $R^*_{\rm eff}$ decrece. Ésto se debe a que la dispersión de los planetesimales es muy grande (la velocidad relativa, $v_{\rm rel}$, es alta), lo cual atenúa el enfocamiento gravitatorio (Ec. (4.15)). Sin embargo, como la viscosidad domina el cálculo de $R_{\rm eff}$, este hecho no tiene un efecto apreciable.

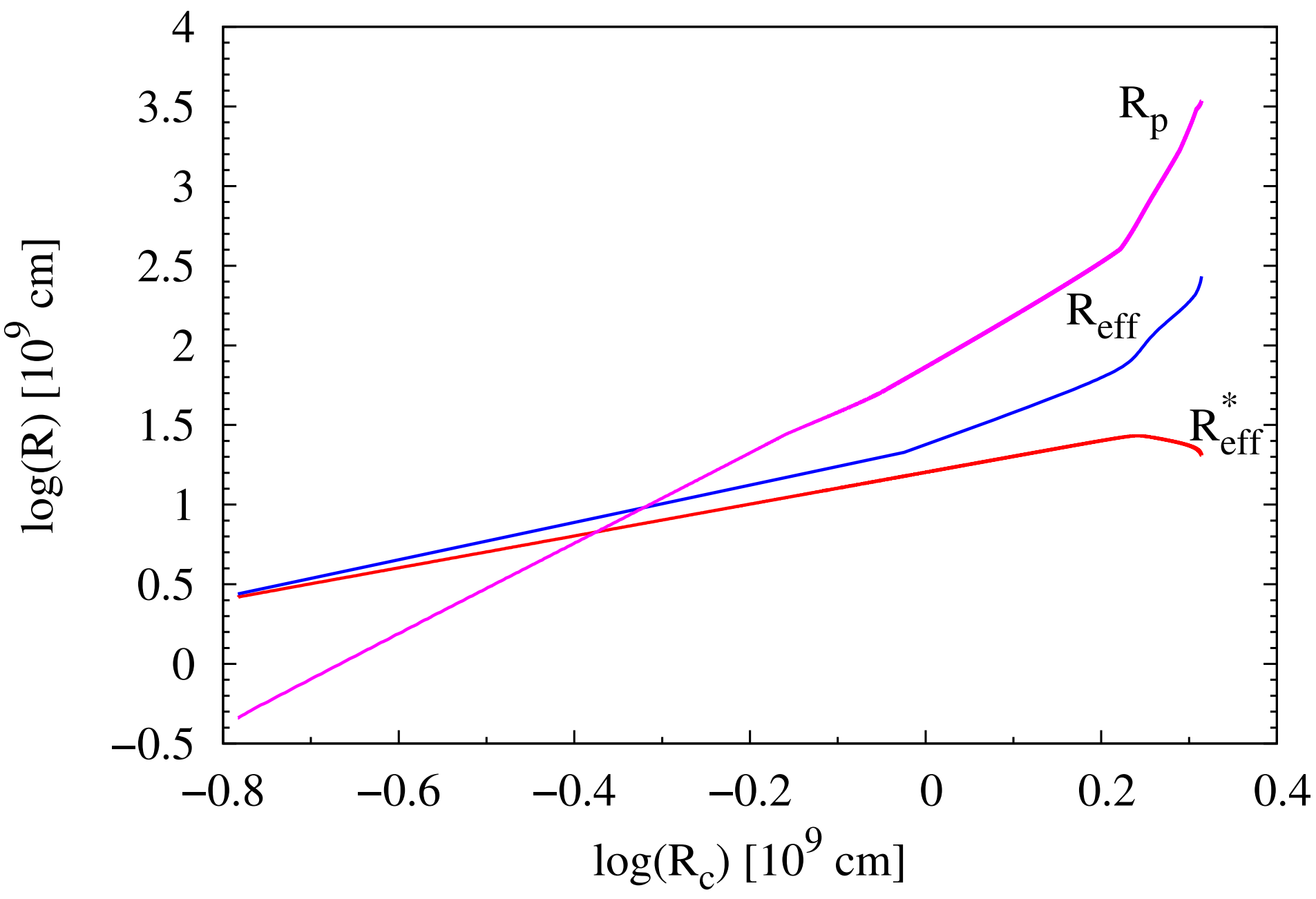

**Figura 4.3.** En esta figura se muestra la evolución del radio efectivo del planeta, $R_{\rm eff}$, y del radio efectivo por enfocamiento gravitatorio sin considerar la viscosidad del gas, $R^{*}_{\rm eff}$. La abscisa corresponde al logaritmo del radio del núcleo. Se incluyó también la evolución del radio del planeta a modo de referencia (el radio del planeta se calcula teniendo en cuenta la masa del núcleo y la masa de gas). Debido, justamente, a la naturaleza de la escala logarítmica, los estadíos iniciales de la formación del planeta, que corresponden a un embrión sólido rodeado de una pequeña atmósfera de gas, se pueden observar en detalle ($\log \mathrm{R_c} \lesssim 0,2$). Notemos que, muy al principio de la formación, el radio efectivo es mayor que el radio del planeta ($\log \mathrm{R_c} \lesssim -0,3$). Esto se debe a que el valor del radio efectivo está dominado por el enfocamiento gravitatorio. Cuando la masa de la envoltura crece, la viscosidad se vuelve cada vez más importante y el $R_{\rm eff}$ se aparta de $R^{*}_{\rm eff}$. El radio del planeta es generalmente mayor que el radio efectivo debido a que las capas de gas que conforman la parte exterior del planeta tienen una densidad muy baja y forman una envoltura muy extendida.

## 4.4. La envoltura gaseosa

### 4.4.1. Ecuaciones constitutivas

Las ecuaciones que gobiernan la estructura del planeta son las ecuaciones diferenciales estándar de la evolución estelar (ver, por ejemplo, Kippenhahn & Weigert 1990). Si despreciamos la rotación del planeta y la presencia de campos magnéticos, y adoptamos la formulación lagrangiana considerando simetría esférica, donde las variables independientes son la masa $M_r$ y el tiempo $t$, tenemos que la estructura gaseosa del planeta está bien descripta por cuatro ecuaciones diferenciales acopladas:

$$\frac{\partial r}{\partial M_r} = \frac{1}{4\pi r^2 \rho_{\mathrm{g}}} \qquad \text{Ecuación de conservación de la masa} \tag{4.23}$$

$$\frac{\partial P}{\partial M_r} = -\frac{G M_r}{4\pi r^4} \qquad \text{Ecuación de equilibrio hidrostático} \tag{4.24}$$

$$\frac{\partial L}{\partial M_r} = \epsilon_{\mathrm{ac}} - T\frac{\partial S}{\partial t} \qquad \text{Ecuación de balance energético} \tag{4.25}$$

$$\frac{\partial T}{\partial M_r} = -\frac{G M_r T}{4\pi r^4 P}\nabla \qquad \text{Ecuación de transporte} \tag{4.26}$$

donde $r$ es la coordenada radial, $\rho_{\mathrm{g}}$ es la densidad, $P$ es la presión, $G$ es la constante de gravitación universal, $L$ es la luminosidad, $\epsilon_{ac}$ es luminosidad debido a la acreción de planetesimales, $T$ es la temperatura, $S$ es la entropía por unidad de masa y $\nabla$ es el gradiente de temperatura adimensional,

$$\nabla \equiv \frac{d\ln T}{d\ln P}, \tag{4.27}$$

que depende del tipo de transporte de energía. Si el transporte es radiativo, $\nabla$ está dado por:

$$\nabla_{\mathrm{rad}} = \frac{3}{16\pi a_{\mathrm{r}} c G}\frac{\kappa L P}{M_r T^4} \tag{4.28}$$

con $a_{\mathrm{r}}$ la constante de densidad de radiación, $c$ la velocidad de la luz en el vacío y $\kappa$ la opacidad media de Rosseland. Si el transporte es convectivo se considerará que el gradiente es el correspondiente al caso adiabático, $\nabla = \nabla_{\mathrm{ad}}$. El tipo de transporte (radiativo o convectivo) estará determinado por el Criterio de Estabilidad de Schwarzschild,

$$\nabla = \begin{cases} \nabla_{\mathrm{rad}} & \text{si} \quad \nabla_{\mathrm{rad}} < \nabla_{\mathrm{ad}} \\ \nabla_{\mathrm{ad}} & \text{si} \quad \nabla_{\mathrm{rad}} \geq \nabla_{\mathrm{ad}}. \end{cases}$$

En lo que respecta a la energía, notemos que la ecuación (4.25) está caracterizada por dos términos:

- $\epsilon_{ac}$ representa el calor, por unidad de tiempo y unidad de masa de la envoltura, que es suministrado por los planetesimales cuando estos son capturados por el planeta. Supondremos que los planetesimales son suficientemente grandes como para no desintegrarse antes de llegar al núcleo. De hecho, supondremos que en su trayectoria no sufren ningún proceso de ablación (lo cual es una hipótesis simplificadora importante), con lo cual llegan al núcleo con la misma masa con la que ingresaron. El intercambio de energía que tienen con la envoltura será solo como consecuencia de la fricción con el gas. El procedimiento para el cálculo de $\epsilon_{ac}$ será detallado en § 4.4.5.

- el segundo término de la ecuación es $-T\frac{\partial S}{\partial t}$ que, según la primera ley de la termodinámica, resulta

$$T\frac{\partial S}{\partial t} = \frac{dQ}{dt} = \frac{\partial U}{\partial t} + P\,\frac{\partial V}{\partial t} \tag{4.29}$$

donde $Q$ es el calor, $U$ es la energía interna y $V$ es el volumen, todos por unidad de masa. De esta expresión se puede ver que hay otros dos mecanismos que controlan la energía: la contracción gravitatoria (reflejada en el término $P\frac{\partial V}{\partial t}$) y la variación de la energía interna, $\frac{\partial U}{\partial t}$. En la práctica, el término $T\,dS$ será evaluado según la relación termodinámica:

$$T dS = C_{\rm P}\, dT - \frac{\delta}{\rho_{\rm g}}\, dP, \tag{4.30}$$

donde $C_{\rm P}$ es el calor específico a presión constante (por unidad de masa) y $\delta$ se define como:

$$\delta \equiv \left.\frac{\partial \ln \rho_{\rm g}}{\partial \ln T}\right|_P. \tag{4.31}$$

Dependiendo de la fase de formación en la que se encuentre el planeta, dominará el primero o el segundo término de la ecuación (4.25). Cuando el proceso está controlado fundamentalmente por la acreción de sólidos (la primera etapa), la fuente primordial de luminosidad son los planetesimales ingresantes. Como veremos más adelante, la energía que liberan los planetesimales es la que "sostiene" a las capas de gas, impidiendo que se desplomen contra el núcleo, regulando así la escala de tiempo de la formación. Cuando el protoplaneta incorporó a la mayoría de los sólidos de su vecindad, y la masa de gas ligada alcanza valores aproximadamente iguales a los de la masa del núcleo, la energía que aportan los planetesimales a la presión del gas ya no es suficiente para contrarrestar la fuerza de la gravedad y las capas de gas comienzan a contraerse, entrando el planeta en la segunda etapa. En este caso, la ecuación (4.25) está gobernada por el segundo término. Esta etapa es mucho más rápida que la primera y se la conoce como etapa del crecimiento runaway de gas o simplemente *etapa runaway*.

Por otra parte, la ecuación de equilibrio hidrostático en simetría esférica (y tomando como variables a $r$ y a $t$) tiene la siguiente forma:

$$\frac{\partial P}{\partial r} = -\rho \frac{\partial \phi}{\partial r} \tag{4.32}$$

donde, para el problema de dos cuerpos, el potencial gravitatorio $\phi$ es:

$$\phi = -\frac{GM_r}{r}, \tag{4.33}$$

que es el caso considerado en la ecuación (4.24) (los "dos cuerpos" son el protoplaneta por un lado y un elemento de gas por el otro). Sin embargo, si analizamos el problema a resolver, nos damos cuenta que no tenemos exactamente un problema de dos cuerpos ya que hay una influencia gravitatoria importante: la estrella central. Estrictamente hablando, el tratamiento debería involucrar el potencial gravitatorio para el problema restringido de los tres cuerpos. Por simplicidad para el tratamiento numérico, en cambio, conviene utilizar un potencial aproximado (Benvenuto & Brunini 2005). La idea es considerar una corrección al potencial de los dos cuerpos que establezca que el límite para la atracción gravitatoria del planeta termina en el radio de Hill[1] ($R_{\rm H}$), con lo cual $\frac{\partial \phi}{\partial r}$ deberá tender a cero cuando $r$ tienda a $R_{\rm H}$. Para ello proponemos:

$$\frac{\partial \phi}{\partial r} = \frac{GM_r}{r^2} f(x) \qquad \text{con } x = \frac{r}{R_{\rm H}}, \tag{4.34}$$

donde $f(x) = 1 - x^3$. Que $f(x)$ tenga esta forma no es arbitrario sino que surge como la primera corrección al potencial de los dos cuerpos cuando se hace un promedio sobre el potencial correspondiente al problema restringido de los tres cuerpos (o sea, cuando se calcula $\frac{1}{4\pi}\int_{4\pi} \phi_R \, d\Omega$ con $\phi_R$ el potencial del problema restringido). Si bien considerar esta corrección es analíticamente correcto, esta expresión no resulta conveniente en el tratamiento numérico cuando $x \to 1$. Por las características del problema que se quiere abordar, cuando las ecuaciones diferenciales se llevan a su forma en diferencias finitas (ver § 5.1) resulta conveniente extender el grillado más allá del borde del planeta ($R_{\rm p}$). Ahora bien, como hemos dicho, el potencial corregido tiende a cero cuando $r$ tiende a $R_{\rm H}$ y, de este modo, cuando $r > R_{\rm H}$ tendremos que $\dfrac{\partial \phi}{\partial r} < 0$. Pero, de la ecuación (4.24), vemos que éste hecho implicaría un gradiente de presión positivo en la nebulosa. Dado que el gradiente de presión está conectado con el gradiente de temperatura, ésto entraría en contradicción con nuestra hipótesis de temperatura constante para la nebulosa. Para que esto no ocurra, y para que las funciones sean bien comportadas, será conveniente que $f(x)$ tienda suavemente a cero cuando $x \to \infty$. Resulta entonces adecuado empalmar $1 - x^3$ con la función de Fermi $\dfrac{A}{e^{B(x-x_0)} + 1}$, ajustando los parámetros $A$ y $B$ para que $f(x)$ resulte continua en $x_0$ (que

[1]El radio de Hill corresponde al radio equivalente de una esfera de volumen igual al lóbulo de Roche del planeta.

es un número cercano a 1). De este modo, obtenemos:

$$f(x) = \begin{cases} 1 - x^3 & \text{si} \quad x \leq x_0 \\ \\ \dfrac{A}{e^{B(x-x_0)} + 1} & \text{si} \quad x > x_0 \end{cases}$$

con $A = 2(1 - x_0^3)$ y $B = \dfrac{6x_0^2}{1 - x_0^3}$. En nuestras simulaciones hemos considerado generalmente $x_0 = 0{,}9$, aunque es importante destacar que el valor que tome $x_0$ no repercute significativamente en los resultados, siempre y cuando, $x_0$ sea cercano a 1.

### 4.4.2. La ecuación de estado del gas

Para resolver las ecuaciones (4.23)-(4.26) es necesario establecer relaciones constitutivas, esto es, dar los valores de $\rho_{\rm g}$, $\nabla_{\rm ad}$, $S$ y $\kappa$ en función de la temperatura y la presión. Las propiedades termodinámicas de un fluido están caracterizadas por la ecuación de estado (EOS), la cual determina unívocamente la densidad y el resto de las cantidades termodinámicas en función de la presión y la temperatura. En esta sección discutiremos la EOS considerada para el gas y en la próxima sección discutiremos sobre las tablas utilizadas para la opacidad.

La EOS más ampliamente utilizada en los cálculos de estructura de la componente gaseosa de los planetas gigantes es la de Saumon, Chabrier & Van Horn (1995, de ahora en más SCVH) y es la que adoptaremos para nuestro modelo. SCVH calcularon numéricamente la EOS para gases de hidrógeno y helio puros, considerando detalladamente los efectos no ideales, bajo las condiciones usuales de presión y temperatura presentes en los objetos subestelares (estrellas de masa inferior a $1\,{\rm M}_\odot$, enanas marrones y planetas gigantes). Los resultados se agrupan en tablas que cubren el rango de temperatura de $2,10 \leq \log T \leq 7,06\,[{\rm K}]$ y de presión $4 \leq \log P \leq 19\,[\text{dinas cm}^{-2}]$.

Un esquema del diagrama de fases del hidrógeno de esta EOS se esquematiza en la figura 4.4. En la región de baja densidad y baja temperatura, el hidrógeno se encuentra en estado neutro en forma de átomos y moléculas. Las moléculas dominan a bajas temperaturas ($\log T \leq 3,5$), las cuales comienzan a disociarse a medida que la temperatura aumenta. A temperaturas más altas, los átomos se ionizan formando un plasma de protones y electrones. En la figura 4.4, la curva de rayas cortas delimita las tres regiones. Para densidades superiores a $\log \rho \sim -2$, los átomos y las moléculas interactúan fuertemente y forman un fluido no-ideal. A densidades cercanas a $\log \rho = 0$, la distancia entre los átomos de hidrógeno es comparable al radio de Bohr, con lo cual las funciones de onda de dos electrones cercanos se superponen. Los electrones son entonces forzados a pasar a estados no-ligados y el fluido se ioniza por presión. Según los cálculos de SCVH, esta ionización no es gradual, sino que ocurriría en forma discontinua a través de una transición de fase de primer orden, conocida como "Plasma Phase-Transition" (PPT). En una transición de fase de primer orden todas las cantidades termodinámicas son discontinuas, exceptuando la presión, la temperatura y los potenciales químicos. El valor de la temperatura a partir del cual la ionización comienza a producirse en forma continua corresponde al punto crítico, $\log T_c = 4{,}185$, cuyos valores de la presión y la densidad son: $\log P_c = 11,788$ y $\log \rho_c = -0{,}456$. La PPT separa la fase no ionizada (constituida por moléculas de $H_2$) de la fase ionizada a densidades más altas formada por hidrógeno metálico ($H^+$). Respecto de la PPT es importante aclarar que esta transición de fase no está muy comprendida y que no existen evidencias en el laboratorio que sostengan su existencia. Este hecho es extremadamente relevante puesto que, como vemos en la figura 4.4, el interior de Júpiter estaría afectado por la PPT. De hecho, una de las principales fuentes de incerteza a la hora de estimar el núcleo sólido de los planetas gigantes son los problemas que existen en torno al cálculo de la EOS. Afortunadamente, en los cálculos de formación de un planeta de la masa de Júpiter, la estructura del protoplaneta no estaría afectada por esta discontinuidad (si es que, de existir la PPT, ésta se ubica donde actualmente se cree que está), ya que durante la formación la densidad de las capas de la envoltura es mucho menor que en el caso de un planeta totalmente formado, que tuvo que sufrir, para llegar a ese estado, la contracción abrupta de sus capas de gas (Benvenuto & Brunini 2005).

En cuanto a la ecuación de estado para el helio, se identifican regímenes dominados por el He atómico y por sus iones, $He^+$ y $H^{++}$ (ver figura 4.5). En el caso del helio, los autores no resuelven el problema de la ionización por presión. Téngase en cuenta, sin embargo que en un planeta joviano, la envoltura está compuesta mayormente por hidrógeno.

Tanto en las envolturas de los planetas gigantes como en la mayoría de los problemas astrofísicos, el hidrógeno y el helio no se encuentran como sustancias puras sino mezclados entre sí. De este modo, resulta necesario tener una EOS para este tipo de mezclas. Lo ideal sería poder calcular esta EOS para una composición cualquiera en forma autoconsistente. Sin embargo, la complejidad del problema hace que, en la actualidad, deban utilizarse ciertas simplificaciones. SCVH sugieren que la EOS de la mezcla se puede obtener interpolando adecuadamente las tablas de las EOS para el hidrógeno y el helio. De entre los modelos de interpolación posibles, SCVH proponen que la interpolación se haga haciendo uso de la regla de adición volumétrica ("additive-volume rule"). Esta regla está basada en las propiedades termodinámicas de las variables. Las variables como la temperatura y la presión son variables intensivas, lo cual significa que son uniformes en un sistema en equilibrio. Las variables extensivas, como el volumen, la entropía y la energía interna, son aditivas en sistemas idénticos, en el límite del régimen ideal. En esencia, la regla de adición lo que considera es que dado un sistema formado por dos o más subsistemas, tendrá como ecuación de estado a aquella que surja como combinación de las EOS de cada subsistema. Claramente, esta regla no es exacta puesto que ignora las interacciones entre las diferentes especies de hidrógeno y helio. Además, el estado de equilibrio de ionización de una mezcla de hidrógeno y helio estará acoplado a la densidad electrónica. Esto no está tenido en cuenta en la regla de adición volumétrica. Este punto es especialmente importante en las regiones de ionización por presión, donde las interacciones mutuas son fuertes. De este

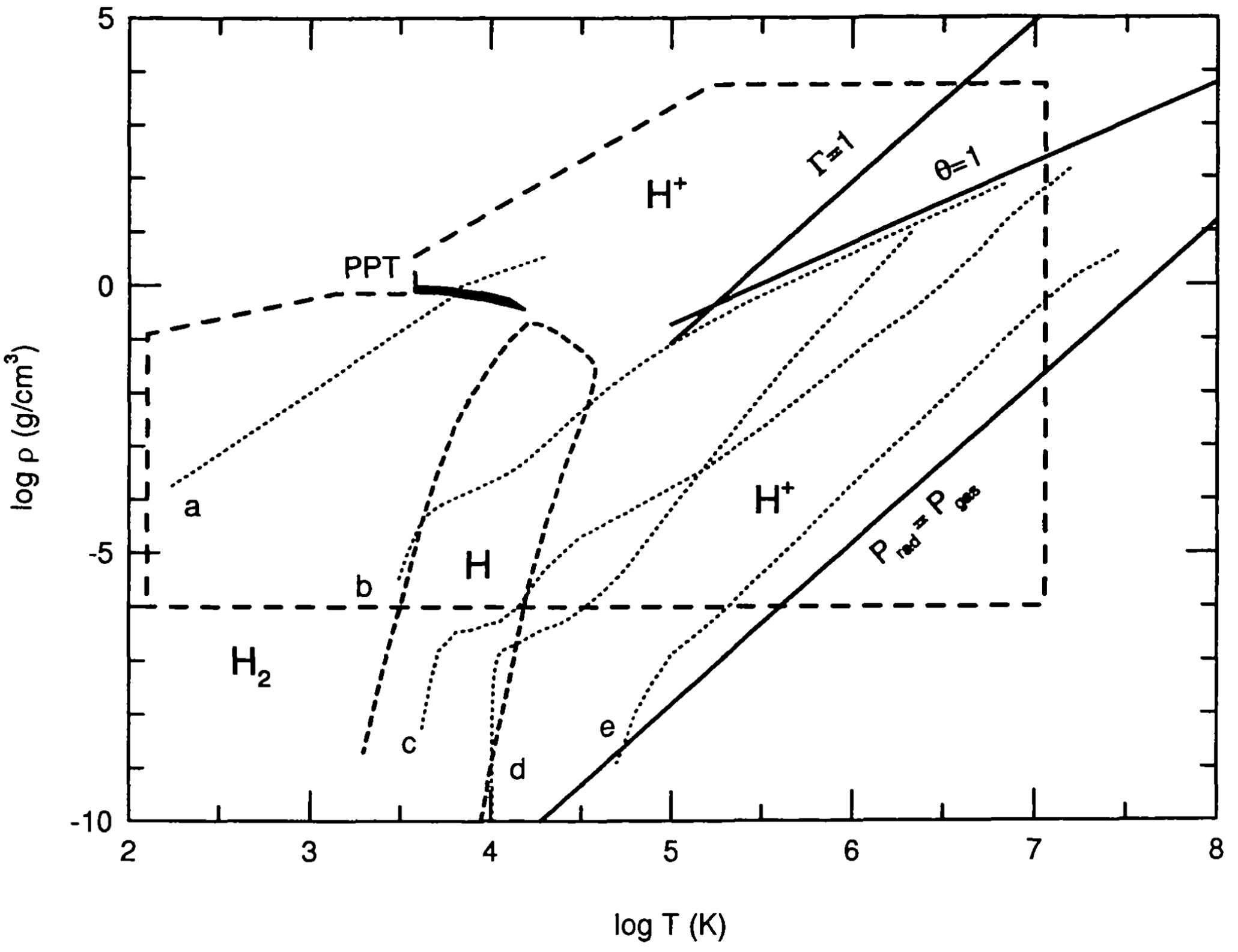

**Figura 4.4.** Diagrama de fases para el hidrógeno. La región limitada por la línea a rayas es la que cubre la EOS de SCVH. Las curvas etiquetadas con **a**, **b**, **c**, **d**, y **e** corresponden a los interiores de diversos objetos ricos en hidrógeno. La curva **a** representa el interior de Júpiter. La envoltura está prácticamente dominada en su totalidad por el hidrógeno molecular. Como se ve, su interior más profundo atraviesa la PPT. Justo debajo de la PPT, el fluido se aparta fuertemente del régimen ideal debido a las fuerzas intermoleculares repulsivas. Las curvas **b**, **c**, y **e** representan el interior de estrellas de secuencia principal, cuyas masas son 0,3, 1, y 15 $M_{\odot}$ respectivamente. La curva **d** corresponde a una enana blanca DA estratificada en una capa rica en hidrógeno y otra rica en helio, alrededor de un núcleo de carbono. (Figura perteneciente al artículo de Saumon, Chabrier & Van Horn (1995))

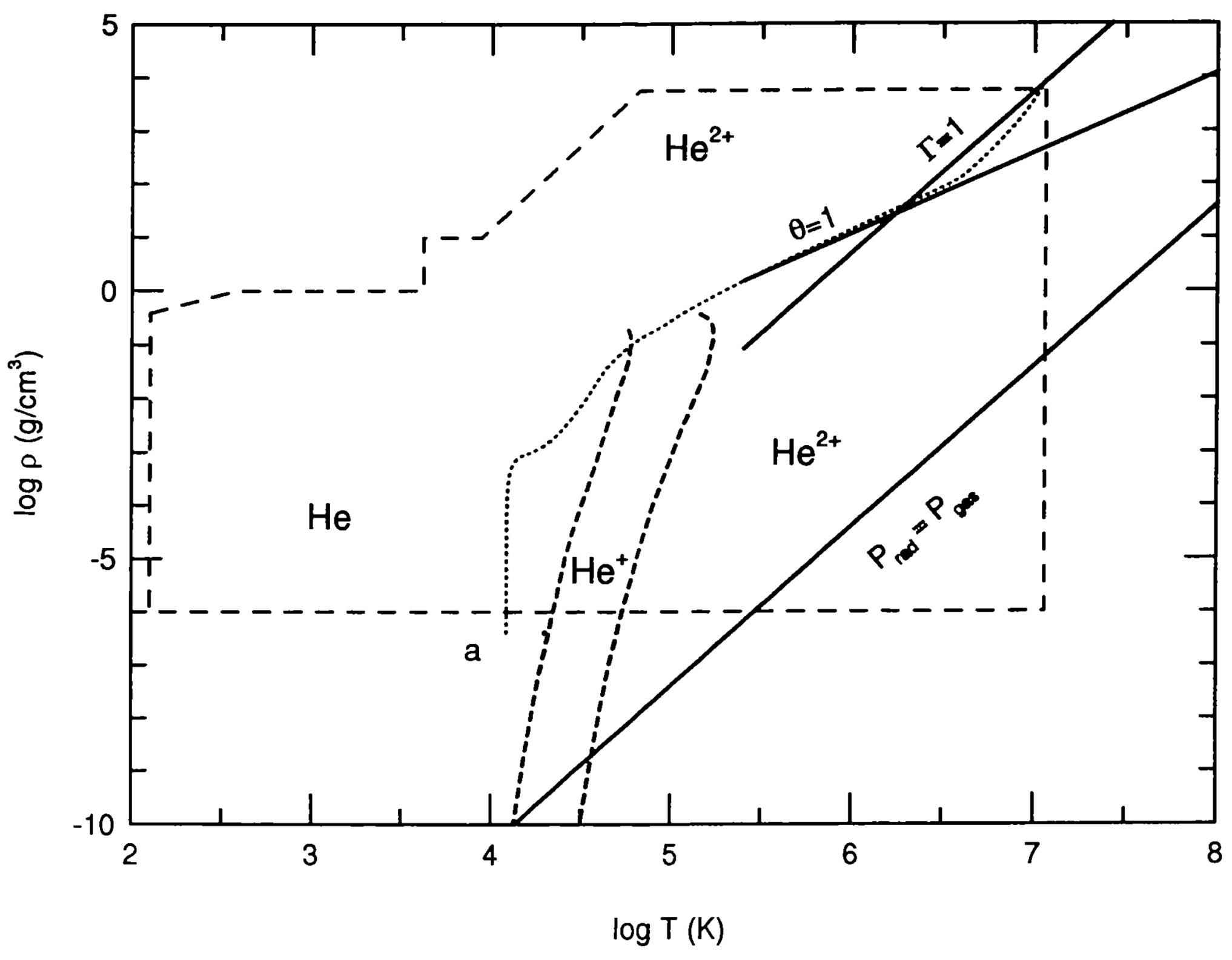

**Figura 4.5.** Diagrama de fases para el helio correspondiente a la EOS de SCVH. Aquí la línea de puntos, etiquetada por la letra **a**, representa el estado termodinámico de la envoltura de una enana blanca DB compuesta solamente por helio. (Figura perteneciente al artículo de Saumon, Chabrier & Van Horn (1995))

modo, la naturaleza, la localización y la mera existencia de la PPT estará afectada por la presencia del helio. En este régimen, la regla de adición volumétrica preserva a la PPT y a su curva de coexistencia.

En su artículo, SCVH derivan las fórmulas para la obtención de los valores interpolados de las cantidades termodinámicas. Desafortunadamente, en el artículo original de SCVH existen errores de tipeo en las fórmulas finales de interpolación, hecho que nos llevó a calcular erróneamente la EOS de la mezcla en nuestros primeros trabajo, y que tuvo importantes consecuencias en nuestros resultados (volveremos sobre este tema en el capítulo 6). Por este motivo, derivaremos a continuación las ecuaciones para la interpolación, escribiendo en negrita los términos con errores de tipeo del manuscrito original.

Dada una variable extensiva general, $W(P,T)$, la regla de volúmenes aditivos establece que, si $W_i(P,T)$ caracteriza a esta cantidad en el sistema $i$, para un sistema constituido por todos los subsistemas $i$ tendremos que:

$$W(P,T) = \sum_i X_i \, W^i(P,T), \tag{4.35}$$

donde $X_i$ es la fracción del sistema total ocupado por el subsistema $i$. En el caso de una mezcla de hidrógeno y helio, esto estará representado por la fracción en masa, $X$, para el hidrógeno, e $Y$, para el helio, tal que se satisface que:

$$X = 1 - Y. \tag{4.36}$$

Para el volumen específico $(1/\rho_{\rm g})$, y la energía interna tendremos, respectivamente:

$$\frac{1}{\rho_{\rm g}(P,T)} = \frac{1-Y}{\rho_{\rm g}^{\rm H}(P,T)} + \frac{Y}{\rho_{\rm g}^{\rm He}(P,T)} \tag{4.37}$$

$$U(P,T) = (1-Y)\, U^{\rm H}(P,T) + Y\, U^{\rm He}(P,T). \tag{4.38}$$

En el caso de la entropía, la interpolación se hace de la misma manera pero se agrega un término que tiene en cuenta la corrección para la entropía de mezcla del gas ideal. Con este término se recupera el límite en el caso del gas ideal. Escribimos entonces para la entropía por unidad de masa:

$$S\,(P,T) = (1-Y)\, S^{\rm H}(P,T) + Y\, S^{\rm He}(P,T) + S_{\rm mix}(P,T). \tag{4.39}$$

Para calcular las derivadas de $\rho_{\rm g}(P,T)$ y de $S$, diferenciamos las ecuaciones (4.37) y (4.39). Para ello resulta conveniente definir las derivadas logarítmicas:

$$\rho_T = \left.\frac{\partial \log \rho_{\rm g}}{\partial \log T}\right|_P, \quad \rho_P = \left.\frac{\partial \log \rho_{\rm g}}{\partial \log P}\right|_T. \tag{4.40}$$

Similarmente,

$$S_T = \left.\frac{\partial \log S}{\partial \log T}\right|_P, \quad S_P = \left.\frac{\partial \log S}{\partial \log P}\right|_T, \tag{4.41}$$

con lo cual se obtiene

$$\rho_T = (1-Y)\frac{\rho_g}{\rho_g^H}\rho_T^H + Y\frac{\rho_g}{\rho_g^{He}}\rho_T^{He}, \tag{4.42}$$

$$\rho_P = (1-Y)\frac{\rho_g}{\rho_g^H}\rho_P^H + Y\frac{\rho_g}{\rho_g^{He}}\rho_P^{He}, \tag{4.43}$$

En el caso de la entropía[2],

$$S_T = (1-Y)\,\frac{\mathbf{S}^H}{\mathbf{S}}\,S_T^H + Y\,\frac{\mathbf{S}^{He}}{\mathbf{S}}\,S_T^{He} + \frac{S_{mix}}{S}\left.\frac{\partial \log S_{mix}}{\partial \log T}\right|_P, \tag{4.44}$$

$$S_P = (1-Y)\,\frac{\mathbf{S}^H}{\mathbf{S}}\,S_P^H + Y\,\frac{\mathbf{S}^{He}}{\mathbf{S}}\,S_P^{He} + \frac{S_{mix}}{S}\left.\frac{\partial \log S_{mix}}{\partial \log P}\right|_T. \tag{4.45}$$

Las derivadas de la entropía son importantes puesto que definen el gradiente adiabático,

$$\nabla_{ad} = \left.\frac{\partial \log T}{\partial \log P}\right|_S = -\frac{S_P}{S_T}. \tag{4.46}$$

Para evaluar las derivadas hay que calcular la entropía de mezcla, la cual satisface

$$\begin{aligned} S_{mix} = k_B\,\frac{1-Y}{m_H}\frac{2}{1+X_H+3X_{H_2}} \times \{ &\ln(1+\beta\gamma) + \\ &- X_e^H \ln(1+\delta) + \beta\gamma\,[\ln(1+1/\beta\gamma) + \\ &- X_e^{He}\ln(1+1/\delta)]\} \end{aligned} \tag{4.47}$$

donde $m_H$ es la masa del hidrógeno y $X_A$ es la concentración de partículas A. Finalmente, las constantes $\beta$, $\gamma$ y $\delta$[3] se definen como:

$$\beta = \frac{m_H}{m_{He}}\,\frac{Y}{1-Y}, \tag{4.48}$$

$$\gamma = \frac{3}{2}\,\frac{(1+X_H+3X_{H_2})}{(1+2X_{He}+X_{He^+})}, \tag{4.49}$$

$$\delta = \frac{\mathbf{2}}{\mathbf{3}}\,\frac{(2-2X_{He}-X_{He^+})}{(1-X_{H_2}-X_H)}\,\beta\gamma. \tag{4.50}$$

Dadas estas fórmulas, se pueden interpolar las tablas de las ecuaciones de estado del hidrógeno y del helio para obtener la EOS de la mezcla para una dada composición. En los cálculos de esta Tesis se utilizó que la abundancia de helio era $Y = 0{,}3$.

[2] Los cocientes en negrita están invertidos en el artículo de SCVH.

[3] En lugar del factor 2/3, la ecuación para $\delta$ tenía en factor 3/2.

### 4.4.3. Opacidades

La opacidad, al igual que la EOS, juega un papel determinante en el cálculo de la formación de planetas gaseosos. Fundamentalmente, es la opacidad media de Rosseland de los granos un factor importantísimo a la hora de estimar las escalas de tiempo y la masa del núcleo para la cual comienza la etapa del runaway gaseoso. En nuestras simulaciones hemos considerado las tablas de opacidades de los granos de Pollack, McKay & Christofferson (1985), las cuales dominan para bajas densidades y temperaturas, mientras que para temperaturas superiores a los 1000 K utilizamos las tablas de Alexander & Ferguson (1994), las cuales están calculadas hasta temperaturas del orden de los $10^4$ K. Eventualmente, si algún cálculo lo requiriera, para temperaturas superiores se consideran las opacidades de Rogers & Iglesias (1992).

Respecto de la opacidad de los granos, es relevante hacer algunos comentarios sobre su influencia en los cálculos de formación de planetas gigantes, debido a que la tasa de contracción de las capas de gas de la envoltura es fuertemente dependiente de la opacidad. Como mencionamos en el párrafo anterior, la opacidad media de Rosseland utilizada es la publicada por Pollack, McKay & Christofferson (1985), la cual fue calculada para un conjunto de diferentes especies de granos de polvo que estarían presentes en la nebulosa solar primordial. Los resultados abarcan un rango de temperatura de entre 10 y 2.500 K (la opacidad debida a los granos de polvo domina hasta, aproximadamente, 1.500 K), con la densidad del gas nebular variando entre $10^{-14}$ y 1 g cm$^{-3}$. Las integraciones se hicieron considerando numerosas distribuciones de tamaños, incluyendo la correspondiente al medio interestelar la cual se consideró como caso nominal. Los autores encontraron que la opacidad no es demasiado sensible a la distribución considerada, siempre y cuando haya partículas cuyo tamaño supere las decenas de micrones.

Los sencillos modelos analíticos de Stevenson (1982) para el cálculo de la formación de los planetas gigantes mostraban ya la dependencia entre el comienzo de la etapa runaway de gas y la opacidad. De todas formas, las simplificaciones utilizadas para elaborar este modelo requieren que sus conclusiones sean tomadas con mucho cuidado. Sin embargo, las simulaciones numéricas de Pollack et al. (1996), donde se considera un modelo mucho más realista y detallado, mostraron que la dependencia con la opacidad era, efectivamente, muy importante. Pollack et al. encontraron que si la opacidad se reduce en un factor 50 (respecto de la correspondiente a la del gas interestelar), el tiempo de formación decrece en forma considerable. Si bien la composición de los granos en la atmósfera protoplanetaria debería ser similar a la presente en la nebulosa solar, la distribución del tamaño de los granos está influenciada por la microfísica asociada a los procesos atmosféricos, con lo cual no necesariamente sería igual a la del gas interestelar. Para evaluar la plausibilidad de una reducción en la opacidad en las atmósferas de los protoplanetas, Podolak (2003) calculó la opacidad de los granos utilizando los modelos de envoltura obtenidos por Pollack et al. (de estos modelos obtuvo la distribución de presión, densidad, temperatura y regiones convectivas que caracterizarían la estructura de la atmósfera de un planeta en formación). Notemos que, en este sentido, los cálculos de Podolak no son autoconsistentes puesto que los perfiles atmosféricos de Pollack et al. fueron obtenidos con las tablas de opacidad para la distribución de tamaños de la nebulosa primordial. Por otra parte, Podolak tuvo en cuenta que los planetesimales ingresantes sufren la ablación en su trayectoria hacia el núcleo, con lo cual fueron también considerados como una fuente de opacidad. De sus cálculos Podolak concluye que la opacidad debida a los granos puede ser, como máximo, del orden de $10^{-1}$ cm$^2$ g$^{-1}$, muy inferior a los valores estándares. El motivo de esta disminución sería la presencia de corrientes convectivas en la atmósfera, que llevarían a los granos desde regiones frías hasta regiones donde la temperatura es lo suficientemente alta como para que los granos se vaporicen.

Basados en estos resultados, Hubickyj, Bodenheimer & Lissauer (2005), hicieron varias simulaciones para estudiar en mayor detalle, entre otras cosas, la dependencia del proceso de formación de Júpiter con la opacidad. Dos de sus simulaciones consideran que los valores de la opacidad de los granos son tan solo el 2 % de la opacidad total correspondiente al gas interestelar. En uno de los casos (10L$^\infty$) la reducción se aplica en todo el rango del dominio de los granos, mientras que en el otro caso (10V$^\infty$) la reducción es del 2 % para $T < 350$ K, aumenta linealmente en el rango $350 \leq T \leq 500$ K, y toma el valor interestelar para $T > 500$ K. Respecto del caso en el que no se aplica ninguna reducción (10H$^\infty$), el tiempo de formación se reduce en un factor 3 (de, aproximadamente $6 \times 10^6$ años para el caso 10H$^\infty$ a $\sim 2 \times 10^6$ para 10L$^\infty$). Por otra parte, la diferencia en el tiempo de formación entre 10L$^\infty$ y 10V$^\infty$ es de tan solo $5 \times 10^5$ años. Estos resultados muestran la sensibilidad de este tipo de cálculos a la opacidad, y, además, que son las capas más externas de la atmósfera del protoplaneta las que dominan la contracción de la envoltura.

Sin embargo, recientemente, Dodson-Robinson et al. (2008) realizaron simulaciones de la formación de Saturno, con y sin reducción de la opacidad. A diferencia de los resultados anteriores, estos autores no encuentran que la reducción en la opacidad afecte significativamente la escala de tiempo de formación estimada para Saturno.

En función de las incertezas en torno al impacto real de las variaciones de la opacidad de los granos, y en vistas de que no existen tablas de opacidades calculadas en forma autoconsistente que avalen determinados valores para la opacidad, en nuestras simulaciones se utilizarán las tablas mencionadas más arriba en su forma original. Es, sin embargo, importante tener presente que es probable que los valores de la opacidad involucrados en este tipo de cálculos sean menores que los tabulados para el medio interestelar. El efecto de una reducción en la opacidad iría en la dirección de acortar los tiempos de formación, lo cual afectaría positivamente a nuestros resultados.

### 4.4.4. Condiciones de borde

Para resolver las ecuaciones diferenciales que describen el problema (ecuaciones (4.23) a (4.26)) se deben aplicar condiciones de contorno adecuadas. Éstas estarán dadas tanto en el límite interior de la envoltura gaseosa como en el exterior.

El borde interno de la envoltura está en contacto con el núcleo sólido, por cuanto allí tendremos: $M_r = M_c$, $r = R_c$ y $L_r = L_c = 0$. En nuestro trabajo, el núcleo se considera inerte, estando solo caracterizado por la densidad, la cual se fija constante ($\rho_c \simeq 3\,\mathrm{g\,cm^{-3}}$). De este modo, a medida que el núcleo aumenta su masa por la acreción de planetesimales, su radio crecerá según:

$$R_c = \sqrt[3]{\frac{3M_c}{4\pi\rho_c}}. \tag{4.51}$$

Durante la formación, el planeta se encuentra acretando material de la nebulosa solar, por cuanto las condiciones de contorno externas dependerán del estado de la misma. Ellas estarán dadas por la presión y la temperatura, $P = P_{nebulosa}$ y $T = T_{nebulosa}$. Estas magnitudes dependen, en principio, del tiempo y del semieje del planeta. En nuestro modelo consideraremos que tanto la temperatura como la densidad de la nebulosa se mantienen constantes durante todo el proceso de formación, donde los valores correspondientes a la ubicación del planeta se obtienen de las ecuaciones (4.1) y (4.2). Consideraremos solamente que la densidad de sólidos en el disco disminuye debido a la acreción por parte del protoplaneta (ver §5.3).

La tasa de acreción de gas para formar la envoltura, $\dot{M}_g$, queda determinada por la evolución del límite externo del planeta. El borde externo del planeta deberá encontrarse en el radio de acreción o en el radio tidal, según cuál sea el menor de ellos. El radio de acreción para un objeto de masa $M_r$ se define como

$$R_a = \frac{GM_r}{c_s^2}, \tag{4.52}$$

donde $c_s$ es la velocidad del sonido en la nebulosa. El radio de acreción marca el límite a partir del cual las partículas tienen una energía térmica mayor que la energía gravitatoria que las liga al planeta, con lo cual no formarían parte del mismo. El radio tidal (o radio de Hill), se define según la ecuación (4.13) y representa el límite para el cual la influencia gravitatoria del planeta es dominante. Más allá del radio de Hill, una partícula estará dominada por el campo gravitatorio de la estrella central. En general, las capas de gas de la envoltura se encuentran en contracción. Con lo cual, cuando el punto más externo del planeta, correspondiente a la masa $M_R$, se contrae hasta un radio menor a $R_a$ (o a $R_H$), se adiciona masa para satisfacer la condición de que $R = R_a$ (o $R = R_H$, según cuál sea el mínimo). De este modo transcurre la acreción de gas, la cual se produce en forma contínua. Los radios $R_a$ y $R_H$ se incrementan debido a que el planeta aumenta su masa, tanto de gas como de sólidos, lo cual agranda la esfera de influencia del planeta.

### 4.4.5. Interacción entre las capas de gas y los planetesimales ingresantes

En su trayectoria hacia el núcleo, los planetesimales acretados interactúan con la envoltura gaseosa del protoplaneta intercambiando energía. Para esta Tesis hemos implementado un modelo sencillo que contempla la interacción entre los planetesimales y la envoltura (para modelos más detallados ver Podolak et al. 1988; Pollack et al. 1996; Alibert et al. 2005). Para simplificar la situación, aceptaremos que los planetesimales que se acercan al protoplaneta lo hacen desde el infinito e ingresan a la envoltura con una velocidad $v_{\rm rel}$ (Ec. (4.16)), describiendo luego una trayectoria recta hacia el núcleo. El considerar una trayectoria recta parece inconsistente con el hecho que para definir $R_{\rm eff}$ calculamos las órbitas de los planetesimales ingresantes. Sin embargo, vale la pena mencionar que el cálculo orbital anteriormente citado se realiza solo hasta que los planetesimales acretados completan la primera revolución dentro de la esfera de Hill. Para el cálculo completo de la interacción energética sería necesario integrar toda la trayectoria hacia el núcleo, en conjunto con una descripción autoconsistente de la energía depositada por los planetesimales en las capas de la envoltura. Esto requiere de modelos mucho más complejos que, por ejemplo, consideren la ablación de los planetesimales. Si bien es cierto que la distribución de la energía es importante en los cálculos de la formación planetaria (Benvenuto & Brunini 2008), incorporar este tipo de modelos requiere de idear una buena estrategia numérica para que los tiempos computacionales no se tornen demasiado largos. Como el objetivo de esta Tesis no está focalizado en este punto, nos hemos conformado con un tratamiento simplificado que tiene en cuenta la existencia del intercambio energético entre los planetesimales y el gas, pero a sabiendas que este modelo debe ser profundizado en el futuro.

La energía de un planetesimal de masa $m$ en el borde externo de la envoltura es:

$$E = \frac{1}{2} m \, v_{\rm rel}^2. \tag{4.53}$$

El arrastre gaseoso y la aceleración de la gravedad son, según nuestro modelo, las dos fuerzas principales responsables de cambiar la energía de los planetesimales una vez que ingresan a la envoltura (ver figura 4.6). La fuerza viscosa que actúa sobre un planetesimal esférico de radio $r_m$, desplazándose a una velocidad constante $v$, está dada por la ecuación Ec. (4.21). La fuerza gravitatoria se calcula de la manera usual (Ec. (4.20)). Ambas fuerzas actúan simultáneamente y modifican la velocidad de los planetesimales capturados. Entonces, la fuerza total sobre un planetesimal es:

$$\vec{f} = \vec{f}_{\rm D} + \vec{f}_{\rm G} = -m \, \frac{dv}{dt} \, \hat{r} \tag{4.54}$$

y

$$\frac{dv}{dt} = v \, \frac{dv}{dr}, \tag{4.55}$$

con lo cual, la variación de la velocidad de un planetesimal es:

$$\frac{dv}{dr} = -\frac{C_{\rm D} \, \pi \, r_m^2 \, \rho_{\rm g} \, v}{2m} + \frac{G M_r}{r^2 \, v}. \tag{4.56}$$

Ahora bien, queremos calcular la energía que intercambian los planetesimales con el gas que atraviesan. La energía que las capas de gas absorben, en forma de calor, corresponde a la pérdida de energía cinética de los planetesimales por fricción. Entonces, si la variación de la velocidad debido a la viscosidad del medio es:

$$\frac{dv}{dt} = -\frac{1}{2}\frac{C_{\rm D}\pi\, r_m^2}{m}\rho_{\rm g}\, v^2, \tag{4.57}$$

la variación de energía cinética será:

$$\frac{dE_{{\rm k},m}}{dt} = mv\frac{dv}{dt} = -\frac{1}{2}C_{\rm D}\pi\, r_m^2\, \rho_{\rm g}\, v^3. \tag{4.58}$$

Pero más que la variación por unidad de tiempo, nos interesa cuantificar la pérdida de energía en las capas de gas:

$$\frac{dE_{{\rm k},m}}{dt}\frac{dt}{dr}\frac{dr}{dt} = \frac{dE_{{\rm k},m}}{dr}v, \tag{4.59}$$

entonces,

$$\frac{dE_{{\rm k},m}}{dr} = \frac{dE_{{\rm k},m}}{dt}\frac{1}{v}. \tag{4.60}$$

Con lo cual, teniendo en cuenta que la energía cinética perdida por un planetesimal debido a la fricción con el gas se transforma en calor, tenemos:

$$\frac{dE}{dr} = -\frac{dE_{{\rm k},m}}{dr} = \frac{1}{2}C_{\rm D}\, \pi r_m^2\, \rho_{\rm g}\, v^2, \tag{4.61}$$

donde $\frac{dE}{dr}$ es el calor absorbido por las capas de la envoltura. Esta cantidad ingresa, transformada en las unidades adecuadas, en las ecuaciones de estructura a través de $\epsilon_{\rm ac}$. El tratamiento numérico de este problema se detallará en el siguiente capítulo.

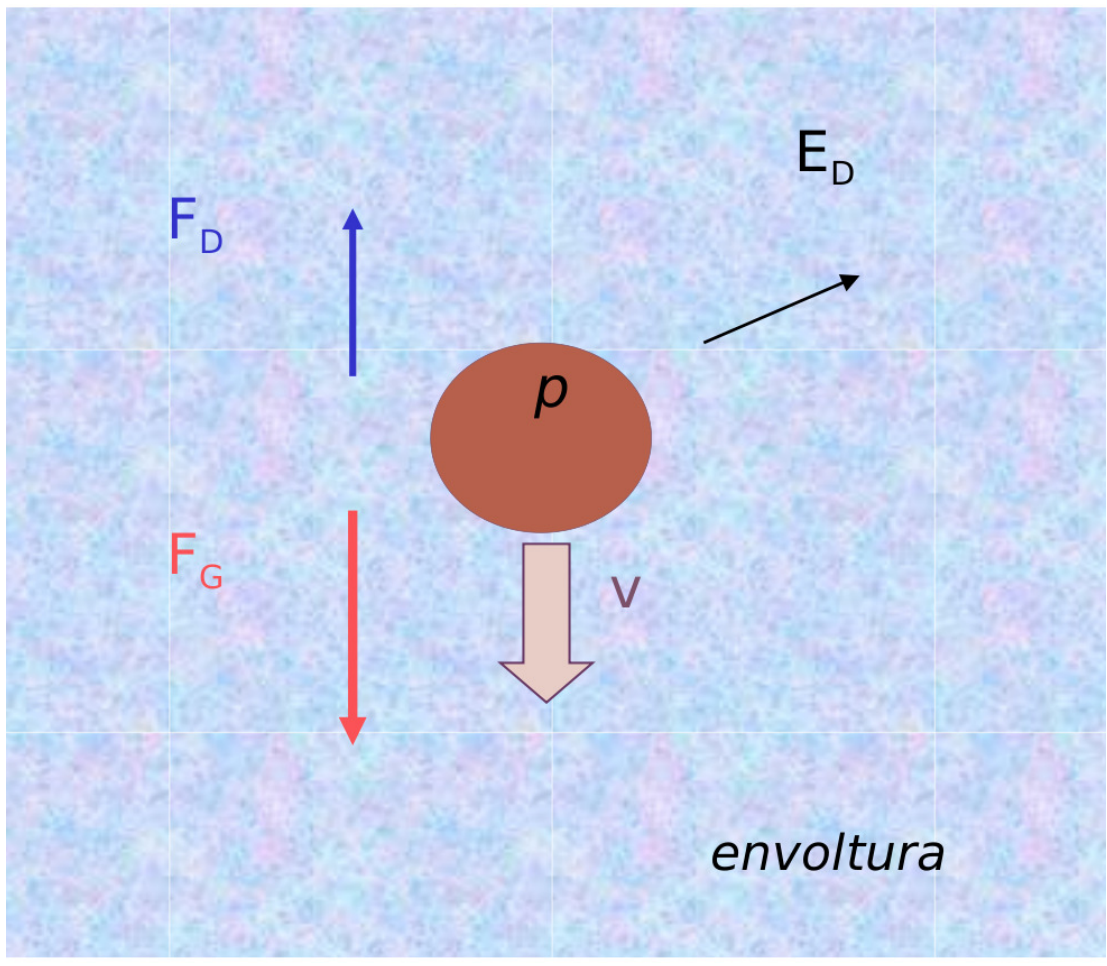

**Figura 4.6.** Esquema de un planetesimal que interactúa con la envoltura gaseosa del planeta. En nuestro modelo consideramos que la velocidad del planetesimal se ve afectada por la acción de la gravedad y la viscosidad del gas. La energía que pierde el planetesimal se transforma en calor absorbido por la envoltura.

# Capítulo 5

# Tratamiento numérico del modelo de formación de planetas gigantes

Las ecuaciones diferenciales introducidas en el capítulo anterior no tienen solución analítica, por cuanto hay que recurrir a métodos numéricos para resolverlas. La idea básica al hacer ésto es generar una secuencia de modelos: cada uno de ellos representa el estado de la estructura del planeta para un determinado instante de tiempo $t$. En la práctica, la estructura gaseosa se representa dividida en cáscaras concéntricas, donde se calcula para un tiempo $t$ y para cada una de estas cáscaras, las cantidades termodinámicas $P$, $T$, $L$, etc. Finalizado el cálculo del modelo, nos valemos de estos resultados para estimar el próximo modelo correspondiente a un $\Delta t$ posterior.

La implementación de este tipo de resolución está basada en *el método de Henyey* (Henyey et al. 1959, 1964) según la propuesta de Kippenhahn, Weigert y Hofmeister (1967).

## 5.1. Llevando las ecuaciones diferenciales a ecuaciones en diferencias

Benvenuto & Brunini (2005) desarrollaron y publicaron las ideas básicas concernientes al código utilizado en esta Tesis para el cálculo de la formación de planetas gigantes. Respecto de su versión original, el código fue varias veces actualizado, en la medida en que se fueron incorporando ingredientes nuevos al modelo físico. Sin embargo, el esquema básico y la metodología utilizada siguen siendo los originales.

El planteo del problema puede ser visualizado de la siguiente manera: el planeta está formado por un núcleo sólido y una envoltura gaseosa, y circundado por la nebulosa solar. Cerca del núcleo, la densidad del gas estará afectada por el campo gravitatorio del núcleo y por la energía liberada por los planetesimales acretados, con lo cual la densidad de la envoltura será distinta a la de la nebulosa. Además, el límite exterior del planeta no existe físicamente, sino que se lo define como el mínimo entre el radio de acreción y el radio de Hill para "separar" lo que se considera como parte del planeta de lo que forma parte de la nebulosa, ya que la componente gaseosa del planeta y el gas nebular forman un continuo. Debido a la hipótesis de simetría esférica, el problema puede ser descripto en forma unidimensional solo con la coordenada radial $r$ (o, en realidad, con $M_r$ ya que estamos haciendo un tratamiento lagrangiano). La coordenada $r$ tiene por origen al radio del núcleo, $R_{\rm c}$, que es donde comienza la envoltura del planeta y se extiende hasta el infinito, o sea que $r$ supera el límite del planeta. Es conveniente que esto sea así porque el límite externo del planeta se modifica modelo a modelo, en general aumentando, lo que significa que el planeta crece porque incorpora gas nebular. Para que el tratamiento sea suave, es conveniente que las variables termodinámicas especifican el estado del gas incorporado (gas que en el modelo anterior "pertenecía" a la nebulosa y ahora "pertenece" al planeta) sean continuas y diferenciables cerca del límite del planeta. Es por esto que las ecuaciones diferenciales (Ecs. 4.23–4.26) se integran más allá del borde del planeta.

Por razones de estabilidad numérico, la variable independiente $M_r$ y las variables dependientes $r$, $P$, y $T$ son inconvenientes para la integración de las ecuaciones diferenciales que describen la estructura gaseosa y resulta necesario realizar un cambio de variables. El cambio de variables crucial para la resolución numérica de las ecuaciones resulta de reemplazar a la variable independiente $M_r$ por:

$$\xi \equiv \ln\left(\frac{M_r}{M_{\rm c}} - 1\right) = \ln\left(\frac{M_{\rm g}}{M_{\rm c}}\right), \tag{5.1}$$

donde $M_{\rm g}$ es la masa de gas del planeta. Esta transformación resulta vital para lograr la convergencia del código en todo momento dado que elimina la necesidad de hacer interpolaciones para considerar la migración del gas desde la nebulosa hacia el interior del planeta. Es importante destacar que, a diferencia del caso estelar donde usualmente la masa es constante y la estructura de gas se extiende desde el centro de la estrella hasta el borde, la situación de un planeta en formación es bien diferente. Por empezar, la componente gaseosa comienza en la superficie del núcleo y no en el centro del planeta, la cual esta condición de borde no es estática sino que se desplaza hacia afuera a medida que el núcleo aumenta de tamaño, empujando con ella a las capas de gas. Pero además, la masa del planeta no es constante sino que aumenta (aunque pueden ocurrir eventos de expulsión de masa, como veremos en el capítulo 6) conforme pasa el tiempo. Estas características del problema (borde interno no fijo más masa total no constante) hacen que el cambio de variables adecuado para la masa no sea trivial.

Notemos que, por su definición, $\xi$ no es independiente de $t$. La coordenada $\xi_i$ representa dos elementos de masa diferentes en dos modelos consecutivos dado que la masa del núcleo creció en ese intervalo del tiempo. De esta manera, las derivadas temporales (como la que aparece en la ecuación de balance energético), que son las que conectan el estado termodinámico del gas entre dos modelos, deben "seguir" adecuadamente a un mismo elemento de masa de gas. Entonces, al reescribir las ecuaciones diferenciales hay que tener

en cuenta que:

$$\left.\frac{\partial}{\partial t}\right|_{M_{\rm g}} = \left.\frac{\partial}{\partial t}\right|_{\xi} + \left.\frac{\partial \xi}{\partial t}\right|_{M_{\rm g}} \left.\frac{\partial}{\partial \xi}\right|_{t},$$

donde de la ecuación 5.1 se obtiene que:

$$\left.\frac{\partial \xi}{\partial t}\right|_{M_{\rm g}} = -\frac{\partial}{\partial t} \ln M_{\rm c} = -\frac{\dot{M}_{\rm c}}{M_{\rm c}}.$$

Si bien es cierto que al hacer este cambio de variables se gana una singularidad en $M_r = M_{\rm c}$, $(\xi \rightarrow -\infty)$, ésta puede ser salvada imponiendo que la integración llegue sólo hasta un valor pequeño para $\xi$ que nos permita acercarnos a $M_c$ tanto como queramos, siendo el error cometido en esta aproximación despreciable (en nuestras simulaciones hemos adoptado $\xi = -19$). Consideraremos que $\xi$ se extiende hacia afuera hasta, aproximadamente, $\xi = 20$, lo cual supera ampliamente el valor de $\xi$ correspondiente a la masa final del planeta para todos los casos considerados en esta Tesis (observemos que el borde externo del planeta yace en algún punto $\xi < 20$).

Los otros cambios de variables utilizados son:

$$\begin{aligned} x &\equiv \ln r \\ \theta &\equiv \ln T \\ p &\equiv \ln P \end{aligned}$$

$L$ se sigue considerando una variable lineal.

Para hacer la integración debemos reemplazar las ecuaciones diferenciales por ecuaciones en diferencias. Para ello dividimos a la componente gaseosa en $m$ cáscaras concéntricas de espesor variable, que van desde $M_{\rm c}$ hasta algún valor $M$ (o desde $\xi_m$ hasta $\xi_0$). Tendremos $m-2$ capas interiores (cuyos bordes corresponden a $\xi_1, \xi_2 \cdots \xi_{m-1}$) y las dos capas restantes en los extremos: una linda con el borde interior del planeta (la capa que va desde $\xi_m$ hasta $\xi_{m-1}$), y la otra corresponde a algún punto exterior al planeta, en la nebulosa solar (la que va desde $\xi_1$ hasta $\xi_0$). Los puntos en los cuales calculamos las soluciones de las ecuaciones constituyen la grilla de integración. Luego, podemos expresar las ecuaciones (4.23–4.26) como ecuaciones en diferencias para estas capas o puntos de la grilla.

Mostraremos un ejemplo de cómo se hace esto, utilizando para ello la ecuación de continuidad. Primero expresamos a la Ec. (4.23) en función de las nuevas variables:

$$\frac{\partial x}{\partial \xi} = \frac{1}{4\pi} \frac{M_{\rm c}}{\rho_{\rm g}} e^{-3x} e^{\xi}.$$

Luego, esta ecuación escrita en diferencias, resulta en:

$$\frac{x_{j+1} - x_j}{\xi_{j+1} - \xi_j} = \frac{1}{4\pi} \frac{M_{\rm c}}{\rho_{{\rm g}\, j+\frac{1}{2}}} e^{-3x_{j+\frac{1}{2}}} e^{\xi_{j+\frac{1}{2}}} \qquad 1 \leq j \leq m-2$$

donde $j$ representa un punto de la grilla y las magnitudes con subíndice $j+\frac{1}{2}$ son calculadas de la siguiente manera:

$$b_{j+\frac{1}{2}} = \frac{1}{2}(b_{j+1} + b_j),$$

con $b_{j+1}$ y $b_j$ correspondientes al modelo que se está iterando (vemos que los $b_{j+\frac{1}{2}}$ resultan ser un promedio de la cantidad $b$ para la capa considerada).

La evolución temporal se calcula de la siguiente manera. Sea $Y$ alguna de las magnitudes termodinámicas, definimos:

- para magnitudes logarítmicas ($y \equiv \ln Y$, por ejemplo $x \equiv \ln r$),

$$y \equiv \ln Y = \ln(Y_{\text{prev}} + \delta Y) = \ln Y_{\text{prev}} + \ln\left(1 + \frac{\delta Y}{Y_{\text{prev}}}\right) = y_{\text{prev}} + \ln(1 + \delta y)$$

- y, para magnitudes lineales (como $L$)

$$y \equiv Y = Y_{\text{prev}} + \delta Y,$$

donde el subíndice "prev" se refiere al valor de la variable en el paso temporal previo. Se tiene, entonces, el modelo correspondiente al tiempo $t$ y se quiere calcular el modelo para $t + \Delta t$. El modelo al tiempo $t$ estará caracterizado por las magnitudes $y_{\text{prev}}$. Con estos valores y las actualizaciones de las condiciones de contorno, y de otros parámetros del modelo que no dependen de las ecuaciones de estructura, se calcula $\delta y$. El valor de $\delta y$ se mejora iterando las ecuaciones tantas veces como sea necesario hasta alcanzar una dada precisión. Concluido este proceso, se define el modelo que caracteriza el estado en $t + \Delta t$ como $y_{\text{prev}} + \delta y$ (o, $y_{\text{prev}} + \ln(1 + \delta y)$, según corresponda). De este modo, las cantidades que se iteran son $\delta x$, $\delta \theta$, $\delta p$ y $\delta L$, y no $x$, $\theta$, $p$ y $L$ ya que esto facilita el cálculo preciso de las derivadas $\frac{\partial}{\partial t}|_\xi$ y provee estabilidad numérica en todo momento.

Repitiendo el procedimiento para las cuatro ecuaciones tendremos, para cada uno de los $m - 2$ puntos de la grilla, cuatro ecuaciones en diferencias.

Matemáticamente, podemos escribir lo anterior de la siguiente manera:

$$G_i^j(p_j, \theta_j, x_j, L_j, \xi_j; p_{j+1}, \theta_{j+1}, x_{j+1}, L_{j+1}, \xi_{j+1}) = 0 \tag{5.2}$$

$$i = 1, \cdots, 4$$
$$j = 1, \cdots, m - 2.$$

Ahora bien, las incógnitas son $p_j, \theta_j, x_j, L_j$ para $1 \leq j \leq m - 1$, y $p_m, \theta_m$ (ya que $L_m$ y $x_m$ son las condiciones de borde internas), con lo cual tenemos $4(m - 1) + 2$ incógnitas y $4(m - 2)$ ecuaciones; luego faltan seis ecuaciones más.

Dos de ellas provienen de las condiciones de borde externas ($p_1 = \ln(P_{nebulosa})$ y $\theta_1 = \ln(T_{nebulosa})$). Simbólicamente escribimos esto como:

$$B_1(p_1, \theta_1, x_1, L_1) = 0 \quad (5.3)$$

$$B_2(p_1, \theta_1, x_1, L_1) = 0. \quad (5.4)$$

las cuatro ecuaciones restantes (que se aplican en la capa delimitada por $\xi_m$ y $\xi_{m-1}$) son las mismas ecuaciones $G_i$ pero que ahora solo dependen de las incógnitas $p_{m-1}$, $\theta_{m-1}$, $x_{m-1}$, $L_{m-1}$, $p_m$, $\theta_m$, y que renombramos:

$$C_i(p_{m-1}, \theta_{m-1}, x_{m-1}, L_{m-1}; p_m, \theta_m) = 0 \quad (5.5)$$

$$i = 1, \cdots, 4.$$

Queda entonces planteado el sistema de $4(m-1)+2$ ecuaciones con $4(m-1)+2$ incógnitas.

Notemos que hemos definido originalmente una grilla con $m$ capas pero solo hemos dado las condiciones que deben satisfacer $m$ de sus puntos (con lo cual faltaría especificar uno más, que corresponde al límite exterior). Sin embargo dado que las ecuaciones que caracterizan las condiciones de borde externas son por definición ecuaciones en diferencias (y no ecuaciones diferenciales discretizadas), las condiciones de borde pueden aplicarse a los puntos de la grilla correspondientes a $\xi_1$, ya que en $\xi_0$ tomarán el mismo valor.

**Condiciones de borde externas**

Como ya mencionamos, las condiciones de borde externas ($P = P_{nebulosa}$, $T = T_{nebulosa}$) no se aplican en el límite del planeta sino en algún punto de la nebulosa protoplanetaria exterior al planeta. Desarrollaremos con un poco más de detalle este punto.

El estado de la nebulosa solar está perturbado por la presencia del planeta hasta, aproximadamente, su radio de Hill (el cual, durante gran parte del proceso de formación, define el borde del planeta). Más allá del radio de Hill consideraremos que el gas nebular está caracterizado por dos parámetros constantes en el tiempo: la presión y la temperatura. Por otra parte, como mencionamos en § 4.4.1, el potencial gravitatorio considerado no es el correspondiente al problema de dos cuerpos sino que se encuentra ablandado por el término $1 - \left(\frac{r}{R_{\rm H}}\right)^3$, con el objetivo de considerar la influencia solar en el campo gravitatorio del planeta. La introducción de este ablandamiento hace que, para $r > R_{\rm H}$, el potencial cambie de signo y entonces deje de tener sentido dentro de las hipótesis que estamos considerando. Por esto es que para radios que superan $R_{\rm H}$ el potencial se hace tender suavemente a cero (Ec. 4.4.1). De la ecuación de equilibrio hidrostático (Ec. 4.24, o en su forma correspondiente a la ecuación 4.32) vemos que si el potencial es estrictamente cero, el gradiente de presión también lo es, con lo cual más allá del radio de Hill la presión es constante. De combinar las ecuaciones de equilibrio hidrostático y de transporte vemos, además, que el gradiente de presión es proporcional al gradiente de temperatura,

$$\frac{1}{T}\frac{\partial \ln T}{\partial M_r} = \frac{1}{P}\frac{\partial \ln P}{\partial M_r}\nabla. \quad (5.6)$$

De este modo, el gradiente de temperatura será cero cuando lo sea el gradiente de presión, y la temperatura también será constante más allá del radio de Hill. Así, las condiciones de borde externas se pueden imponer en cualquier punto de la grilla bien entrada la nebulosa y serán satisfechas automáticamente en la posición del radio de Hill. Este procedimiento facilita la integración numérica, ya que a medida que el planeta crece y aumenta el radio de Hill las variables termodinámicas no experimentan ningún cambio abrupto, sino que lo hacen suavemente y con continuidad.

## 5.2. Resolución de las ecuaciones

Una vez que las ecuaciones diferenciales son transformadas en ecuaciones en diferencias deben ser resueltas mediante algún método numérico. Es importante notar que las ecuaciones no son lineales. Para la resolución, el programa de computadora que calcula la formación de planetas gigantes utiliza el *método de Henyey*, de amplio uso en el cálculo de evolución estelar. El método de Henyey utiliza el procedimiento de resolución del método de Newton y, en función de la estructura final del sistema de ecuaciones, propone una metodología para encontrar la solución haciendo uso de la forma particular que tiene la matriz que representa al sistema. A continuación daremos una explicación esquemática del mismo siguiendo a Kippenhahn, Weigert & Hofmeister (1967).

### El método de Newton

Sean $y_1, y_2, \cdots, y_n$ $n$ incógnitas que satisfacen un sistema de $n$ ecuaciones no lineales:

$$\begin{aligned} E_1(y_1, \cdots, y_n) &= 0 \\ &\vdots \\ E_n(y_1, \cdots, y_n) &= 0. \end{aligned}$$

Supongamos que conocemos una solución aproximada que llamaremos $y_1^0, y_2^0, \cdots, y_n^0$. Dado que, justamente, los $y_i^0$ son aproximaciones tendremos, en general, $E_k(y_1^0, \cdots, y_n^0) \neq 0$, con $k = 1, \cdots, n$. Se quiere mejorar la aproximación inicial, obteniendo una corrección de la misma, que llamaremos $\delta y_i^0$. De este modo, a primer orden,

$$E_k(y_1^0 + \delta y_1^0, \cdots, y_n^0 + \delta y_n^0) = E_k(y_1, \cdots, y_n) + \delta E_k^0, \tag{5.7}$$

con

$$\delta E_k^0 = \sum_i \left( \frac{\partial E_k}{\partial y_i} \right)^0 \delta y_i^0 \tag{5.8}$$

y, como se busca perfeccionar la solución, se pide que ella satisfaga el sistema original, con lo cual:

$$E_k^0 + \delta E_k^0 = 0 \tag{5.9}$$

entonces,

$$\delta E_k^0 = -E_k^0, \tag{5.10}$$

y, usando la expresión para $\delta E_k^0$,

$$\sum_i \left(\frac{\partial E_k}{\partial y_i}\right)^0 \delta y_i^0 = -E_k^0 \qquad k = 1, \cdots, n. \tag{5.11}$$

De este modo, como $E_k^0$ y $(\frac{\partial E_k}{\partial y_i})^0$ son cantidades conocidas, se puede despejar $\delta y_i^0$. Luego, la solución mejorada es:

$$y_i^1 = y_i^0 + \delta y_i^0 \qquad i = 1, \cdots, n. \tag{5.12}$$

Este procedimiento se repite nuevamente con $y_i^1$ (que pasa a tomar el lugar de $y_i^0$), y así sucesivamente, hasta que $\delta E_k^s$ (para todo $k$) sea menor que una tolerancia dada.

**Aplicación del método de Newton a las ecuaciones de estructura.**

El sistema que tenemos que despejar es:

$$B_1(p_1, \theta_1, x_1, L_1) = 0 \tag{5.13}$$

$$B_2(p_1, \theta_1, x_1, L_1) = 0 \tag{5.14}$$

$$G_i^j(p_j, \theta_j, x_j, L_j \xi_j; p_{j+1}, \theta_{j+1}, x_{j+1}, L_{j+1}, \xi_{j+1}) = 0 \tag{5.15}$$
$$i = 1, \cdots, 4$$
$$j = 1, \cdots, m-2$$

$$C_i(p_{m-1}, \theta_{m-1}, x_{m-1}, L_{m-1}; p_m, \theta_m) = 0 \tag{5.16}$$
$$i = 1, \cdots, 4.$$

Como mencionamos, para poder resolverlo, hay que dar una aproximación inicial a la solución $p_j^0, \theta_j^0, x_j^0, L_j^0$, a partir de la cual se buscarán las correcciones $\delta p_j^0, \delta\theta_j^0, \delta x_j^0, \delta L_j^0$. En la práctica, se genera un modelo inicial muy sencillo que surge de las condiciones de borde que impone la nebulosa, y del modelo del gas ideal[1]. El sistema lineal de ecuaciones queda, entonces, planteado de esta manera:

$$\left(\frac{\partial B_i}{\partial L_1}\right)^0 \delta L_1^0 + \left(\frac{\partial B_i}{\partial p_1}\right)^0 \delta p_1^0 + \left(\frac{\partial B_i}{\partial x_1}\right)^0 \delta x_1^0 + \left(\frac{\partial B_i}{\partial \theta_1}\right)^0 \delta\theta_1^0 = -(B_i)^0 \tag{5.17}$$
$$i = 1, 2$$

[1]No es necesario, en la mayoría de los casos, que el modelo inicial sea muy preciso ya que el algoritmo empleado no es muy sensible a las condiciones iniciales, y al cabo del cálculo de algunos modelos se consigue la independencia de las mismas. Es mucho más importante generar un modelo inicial que garantice la convergencia del código.

$$
\begin{aligned}
\left(\frac{\partial G_i^j}{\partial L_j}\right)^0 \delta L_j^0 + \left(\frac{\partial G_i^j}{\partial p_j}\right)^0 \delta p_j^0 + \left(\frac{\partial G_i^j}{\partial x_j}\right)^0 \delta x_j^0 + & \\
+ \left(\frac{\partial G_i^j}{\partial \theta_j}\right)^0 \delta \theta_j^0 + \left(\frac{\partial G_i^j}{\partial L_{j+1}}\right)^0 \delta L_{j+1} + \left(\frac{\partial G_i^j}{\partial p_{j+1}}\right)^0 \delta p_{j+1}^0 + & \\
+ \left(\frac{\partial G_i^j}{\partial x_{j+1}}\right)^0 \delta x_{j+1}^0 + \left(\frac{\partial G_i^j}{\partial \theta_{j+1}}\right)^0 \delta \theta_{j+1}^0 & = -(G_i^j)^0 \qquad (5.18)
\end{aligned}
$$

$$
i = 1, ..., 4 \text{ y } j = 1, ..., m-2
$$

$$
\begin{aligned}
\left(\frac{\partial C_i}{\partial L_{m-1}}\right)^0 \delta L_{m-1}^0 + \left(\frac{\partial C_i}{\partial p_{m-1}}\right)^0 \delta p_{m-1}^0 + \left(\frac{\partial C_i}{\partial x_{m-1}}\right)^0 \delta x_{m-1}^0 + \left(\frac{\partial C_i}{\partial \theta_{m-1}}\right)^0 \delta \theta_{m-1}^0 + & \\
+ \left(\frac{\partial C_i}{\partial p_m}\right)^0 \delta p_m^0 + \left(\frac{\partial C_i}{\partial \theta_m}\right)^0 \delta \theta_m^0 = -(C_i)^0 \quad (5.19) &
\end{aligned}
$$

$$
i = 1, ..., 4.
$$

Podemos escribir este sistema en forma matricial:

$$
\begin{pmatrix}
(\frac{\partial B_1}{\partial x_1})^0 & \cdots & \\
\vdots & & \\
& & (\frac{\partial C_4}{\partial \theta_m})^0
\end{pmatrix}
\begin{pmatrix}
\delta x_1^0 \\ \vdots \\ \delta \theta_m^0
\end{pmatrix}
=
\begin{pmatrix}
-B_1 \\ \vdots \\ -C_4
\end{pmatrix}
\qquad (5.20)
$$

La matriz de coeficientes tiene, esquemáticamente, la siguiente forma:

$$
\begin{bmatrix}
b_1 & b_1 & b_1 & b_1 & & & & & & & & & & \\
b_2 & b_2 & b_2 & b_2 & & & & & & & & & & \\
g_1 & g_1 & g_1 & g_1 & g_1 & g_1 & g_1 & g_1 & & & & & & \\
g_2 & g_2 & g_2 & g_2 & g_2 & g_2 & g_2 & g_2 & & & & & & \\
g_3 & g_3 & g_3 & g_3 & g_3 & g_3 & g_3 & g_3 & & & & & & \\
g_4 & g_4 & g_4 & g_4 & g_4 & g_4 & g_4 & g_4 & & & & & & \\
 & & & & & \cdot & \cdot & \cdot & & & & & & \\
 & & & & & & \cdot & \cdot & \cdot & & & & & \\
 & & & & & & & \cdot & \cdot & \cdot & & & & \\
 & & & & & & & & \cdot & \cdot & \cdot & & & \\
 & & & & & & & & c_1 & c_1 & c_1 & c_1 & c_1 & c_1 \\
 & & & & & & & & c_2 & c_2 & c_2 & c_2 & c_2 & c_2 \\
 & & & & & & & & c_3 & c_3 & c_3 & c_3 & c_3 & c_3 \\
 & & & & & & & & c_4 & c_4 & c_4 & c_4 & c_4 & c_4
\end{bmatrix}
$$

Básicamente, la matriz es una matriz diagonal por bloques. Se observan tres estructuras de bloques diferentes: la primera formada por un bloque de seis filas constituido por dos subestructuras (de elementos representados esquemáticamente por $b_i$ y $g_i$); la segunda estructura se compone por $m-3$ bloques como el correspondiente a $g_i$ solamente, y la tercera es el bloque de los elementos representados por $c_i$. Hemos representado con $b_i$ las derivadas de $B_i$, con $g_i$ las derivadas de $G_i$, y con $c_i$ a las de $C_i$.

Henyey propuso un método para resolver el sistema, aprovechando la forma particular de la matriz. El detalle del método puede encontrarse en el artículo de Kippenhahn, Weigert & Hofmeister (1967). En resumen, podemos decir que las correcciones se obtienen de atrás hacia adelante, o sea, primero se encuentra $\delta p_m^0$, $\delta\theta_m^0$, $\delta p_{m-1}^0 \cdots$ hasta obtener, por último, $\delta p_1^0, \cdots, \delta x_1^0$. Este procedimiento se repite hasta que las correcciones sean menores que una tolerancia dada o hasta que lo sea $\delta E_k$ para todo $k$.

## 5.3. Tratamiento de la acreción de sólidos

En las secciones previas mostramos el método de resolución de las ecuaciones diferenciales que describen el estado de la estructura gaseosa del planeta. Como ya mencionamos, el núcleo sólido se considera inerte, solo caracterizado por su densidad constante. Obviamente, con esta simplificación se están despreciando, entre otras cosas, cualquier proceso de liberación de energía que pueda ocurrir en el propio núcleo. Luego, la presencia del núcleo tiene solo un efecto gravitatorio en este problema. Sin embargo, los planetesimales acretados, que incrementarán la masa del núcleo a través del tiempo, juegan un papel fundamental en la energética del problema. Por este motivo, describiremos aquí el tratamiento numérico de la acreción de sólidos.

El protoplaneta incorpora material de la nebulosa solar. La región de la cual el protoplaneta puede acretar material se conoce como *zona de alimentación* y corresponde a un anillo centrado en la órbita del planeta. La extensión considerada para la zona de alimentación varía dependiendo de los autores, ya que no se calcula analíticamente sino que se estima en forma numérica. En general suele considerarse que se extiende entre 3 y 5 $R_{\rm H}$ a cada uno de los costados de la órbita del planeta. Para este trabajo hemos considerado un valor de 4 $R_{\rm H}$, por cuanto la zona de alimentación forma un anillo cuyo ancho total es de 8 $R_{\rm H}$, centrado en la circunferencia de radio igual al radio orbital del planeta.

La tasa de acreción de sólidos está dada por la ecuación (4.14). Esta ecuación depende de varios factores, entre ellos de la densidad superficial de sólidos. La densidad en la nebulosa sigue una ley de potencias con la distancia, por cuanto calcularemos su valor medio en la zona de alimentación para poder obtener la tasa de acreción. El valor medio de la densidad superficial de sólidos es:

$$\langle \Sigma \rangle = \frac{\int_{a_{\rm out}}^{a_{\rm int}} \Sigma(a) 2\pi a da}{\pi(a_{\rm out}^2 - a_{\rm int}^2)}, \tag{5.21}$$

donde $a_{\rm out}$ y $a_{\rm int}$ representan los límites externo e interno de la zona de alimentación. Como tenemos que

$$\Sigma(a) = \Sigma_0 \, a^{-p}, \tag{5.22}$$

el valor medio para la densidad superficial de sólidos es[2]:

$$\langle \Sigma \rangle = \frac{2\,\Sigma_0}{2-p} \left( \frac{a_{\rm out}^{2-p} - a_{\rm int}^{2-p}}{a_{\rm out}^2 - a_{\rm int}^2} \right). \tag{5.23}$$

Notemos que estamos considerando que, en cada instante, la nebulosa está suficientemente relajada como para que el material disponible en la zona de alimentación esté distribuido de manera uniforme. Ahora bien, el núcleo del planeta crece a expensas de acretar material de su zona de alimentación, por cuanto esto modifica el valor de $\langle \Sigma \rangle$. Llamemos $\Delta A$ al área de la superficie de la zona de alimentación en el modelo correspondiente al tiempo $t$,

$$\Delta A = \pi(a_{\rm out}^2 - a_{\rm int}^2), \tag{5.24}$$

y sea $\Delta\tilde{A}$ el área correspondiente al modelo anterior (en $t - \Delta t$). Entre ambos modelos el planeta incorporó masa (tanto en forma de sólidos como de gas), por cuanto el radio de Hill creció y los límites de la zona de alimentación se expandieron. Sea $\delta a$ el incremento en el ancho del anillo, tanto hacia el interior como hacia el exterior (ver figura 5.1. Notemos que lo que estamos llamando $\delta a$ en el texto se representa con $\delta$r en la figura). Asociado a $\delta a$ hay un aumento en el área del anillo que llamaremos $\delta A_{\rm out}$ y $\delta A_{\rm int}$, según corresponda al incremento exterior o interior. En cada uno de estos pequeños anillos que se incorporan a la zona de alimentación, la densidad de la nebulosa es la original para esa posición. Notar que no se considera la migración de los planetesimales por fricción con el gas nebular. Consideramos que el material en la zona de alimentación del planeta se modifica por la acreción y fuera de ella se mantiene intacto ya que no consideramos ningún tipo de evolución nebular. Luego, si la densidad en $t - \Delta t$ era $\langle \Sigma(t - \Delta t) \rangle$, la densidad superficial de sólidos media en la zona de alimentación al tiempo $t$ será:

$$\langle \Sigma \rangle = \frac{\Delta \tilde{A}}{\Delta A} \langle \Sigma(t - \Delta t) \rangle + \frac{\delta A_{\rm out}}{\Delta A} \langle \Sigma_{\rm out} \rangle + \frac{\delta A_{\rm int}}{\Delta A} \langle \Sigma_{\rm int} \rangle. \tag{5.25}$$

[2]Notemos que acá estamos suponiendo que $\Sigma$ es una función continua. Esto no es cierto si la zona de alimentación cruza la línea de hielo. De todos modos, en las simulaciones consideradas para este trabajo, siempre nos mantendremos bastante lejos de esta situación.

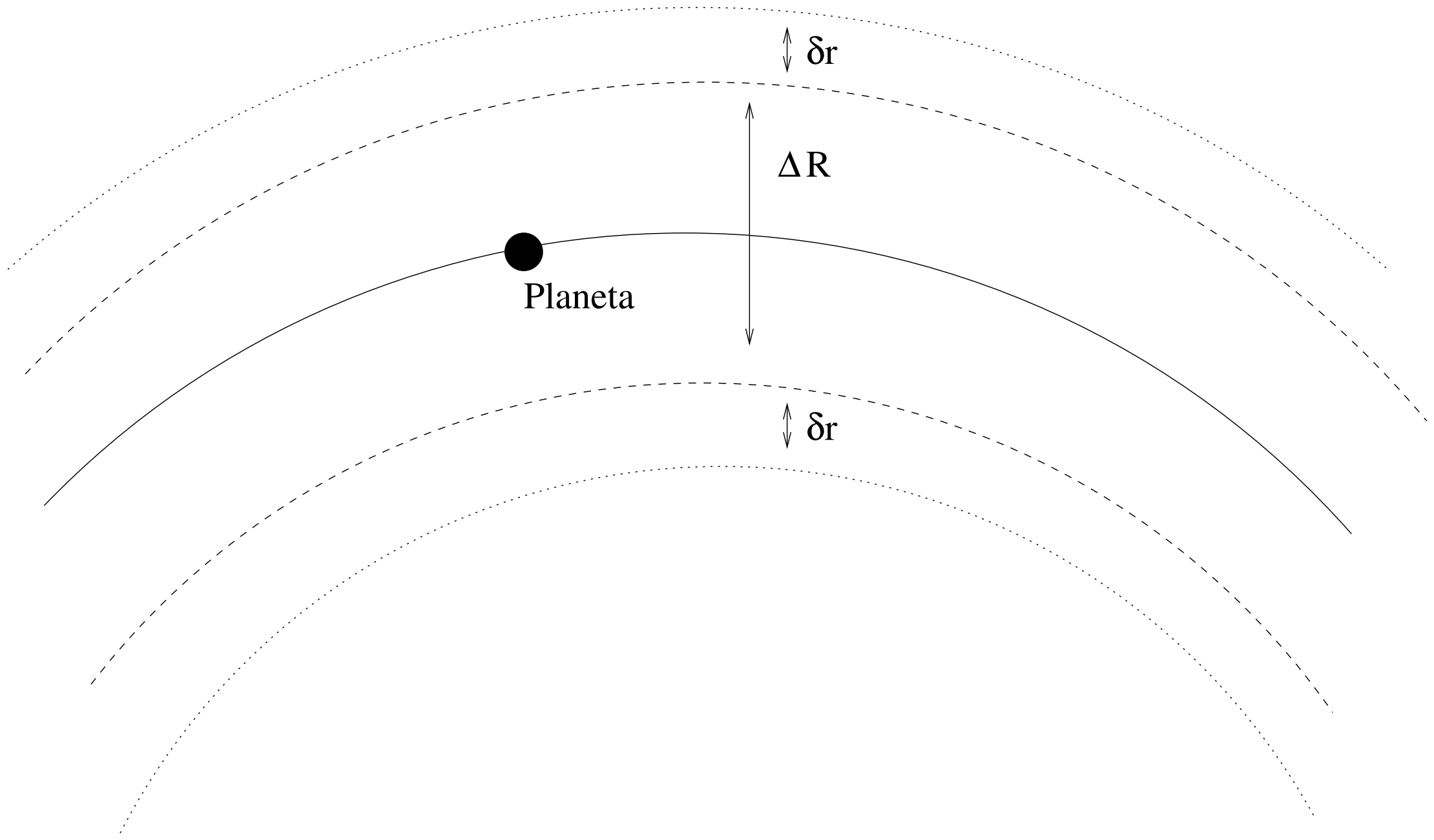

**Figura 5.1.** Representación esquemática de la zona de alimentación del planeta. En el instante $t - \Delta t$ el ancho de la zona de alimentación es ΔR. Para un instante posterior $t$, dado que el planeta incrementó su masa y que la zona de alimentación es proporcional a $R_{\rm H}$, sus límites se incrementan en $\delta$r.

Si llamamos $\Delta M$ a la masa de sólidos que acretó el planeta en el intervalo $\Delta t$, el valor final de la densidad media en la zona de alimentación para el modelo al tiempo $t$ será:

$$\langle \Sigma(t) \rangle = \langle \Sigma \rangle - \frac{\Delta M}{\Delta A}. \tag{5.26}$$

Repitiendo este procedimiento con cada nuevo modelo se va modificando la densidad superficial de sólidos, tanto por la acreción de material por parte del planeta, lo cual provoca una disminución en la densidad, como por la expansión de la zona de alimentación, que permite la incorporación de sólidos disponibles para ser acretados. Si, por algún motivo, el planeta perdiera material (que solo podría ser gas de la envoltura), el esquema anterior dejaría de ser válido. Para evitar esta inconsistencia, supondremos que si el planeta pierde masa, la zona de alimentación no modifica su tamaño, con lo cual estamos suponiendo que la zona de alimentación no puede decrecer. Esto es, si la masa del planeta disminuye, si bien su radio de Hill se reduce, consideraremos que la zona de alimentación mantendrá el tamaño previo. Esto se puede justificar de la siguiente manera: los planetesimales que se encuentran en la zona de alimentación son objetos cuyas órbitas sufren las perturbaciones debidas a la presencia del planeta de modo que, tarde o temprano, terminarán cruzando la órbita del planeta, lo cual les da la posibilidad de ser acretados. Si el $R_{\mathrm{H}}$ se reduce, eso no implica que la zona de alimentación deba estrecharse puesto que los planetesimales que en algún momento formaron parte de la zona de alimentación ya fueron excitados por el protoplaneta y seguirán teniendo probabilidad de ser acretados por el mismo.

Otro factor que influye en el cálculo de la tasa de acreción es el radio efectivo del protoplaneta. El radio efectivo es el máximo valor que puede tomar el parámetro de impacto de un planetesimal para ser capturado por el protoplaneta. Por otra parte, el valor mínimo posible para $R_{\mathrm{eff}}$ es el debido al enfocamiento gravitatorio (Ec. 4.15), donde no se considera la viscosidad del gas. Sea, entonces, un planetesimal que viene desde el infinito en una trayectoria recta, de la cual se ve deflectado por la presencia del protoplaneta. Supongamos que, en su trayectoria, el planetesimal ingresa en la atmósfera del protoplaneta. Para determinar el radio efectivo de captura debemos encontrar cuál es el parámetro de impacto más grande para el cual ésto ocurre. En el tratamiento numérico, la búsqueda del radio efectivo se hace de la siguiente manera: dado el $R_{\mathrm{eff}}^{\mathrm{prev}}$ del modelo anterior, proponemos como valor inicial para la búsqueda del radio efectivo del modelo que queremos calcular ($R_{\mathrm{eff}}$) una fracción del valor anterior ($R_{\mathrm{eff}} = \alpha\, R_{\mathrm{eff}}^{\mathrm{prev}}$). Si bien lo correcto sería comenzar la búsqueda a partir del radio mínimo, dado solo por el enfocamiento gravitatorio ($R_{\mathrm{eff}}^{*}$), estos dos valores son tan diferentes ($R_{\mathrm{eff}}^{*} \ll R_{\mathrm{eff}}$, ver figura 4.3), que el costo computacional sería enorme. Por otra parte, a menos que haya algún cambio abrupto en la densidad del gas (cosa que en general no ocurre), el radio efectivo siempre aumenta o, si disminuye (como en el caso de los planetas que pierden masa), el decrecimiento siempre es menor que el propuesto como valor inicial para el cálculo. Entonces, tomamos el valor previo del radio efectivo, $R_{\mathrm{eff}}^{\mathrm{prev}}$, lo reducimos en una fracción (generalmente entre un 5 y un 10 %), y buscamos el nuevo valor de $R_{\mathrm{eff}}$ para el modelo que estamos calculando. Para eso, integramos la trayectoria de un planetesimal desde el radio de Hill del planeta, moviéndose con una velocidad inicial dada por la ecuación (4.16). La integración de la ecuación de movimiento, considerando la aceleración de la gravedad y la fricción producida por el gas, se hace con un integrador Runge Kutta Fehlberg (Press et al. 1992). Si el planetesimal describe una trayectoria cerrada dentro de la esfera de Hill del planeta, se interrumpe la integración y se lo considera acretado. Se aumenta el valor del parámetro de impacto (en un factor que, usualmente, tomamos en $5 \times 10^{-4}$ del radio efectivo) y se procede nuevamente. Esto se repite hasta que las trayectorias que describen los planetesimales dejen de estar contenidas en la esfera de Hill del planeta. El mayor valor que toma el parámetro de impacto es el que se asigna como radio efectivo. Por la naturaleza del problema, y dado que el cálculo del radio efectivo se basa en considerar el frenado producido por el gas, cuyo estado está dominado por la evolución del planeta, el parámetro de impacto máximo posible es el dado por el radio de Hill. Si estamos en una fase de la formación en la cual el radio del planeta está definido por el radio de acreción ($R_{\rm p} = R_{\rm a}$) puede ocurrir que $R_{\rm eff} > R_{\rm p}$. Dado que estos planetesimales terminan cayendo siempre hacia el centro del planeta no hay contradicción en este punto, siendo perfectamente posible que en estos casos el radio efectivo sea mayor que el del planeta.

Una vez que queda establecida la tasa de acreción de sólidos, debemos calcular cuánta energía depositan los planetesimales en las capas de gas de la envoltura. Como dijimos en § 4.4.5, consideramos que los planetesimales acretados describen una trayectoria recta entre el borde del planeta y el núcleo. A lo largo de este camino, la velocidad de los planetesimales varía por la acción de la viscosidad del gas, que los frena, y de la fuerza gravitatoria que los acelera hacia el centro. La variación de la velocidad con el radio está dada por la ecuación (4.56) que, si la multiplicamos por la velocidad $v$, nos da:

$$v\frac{dv}{dr} = -\frac{C_{\rm D}\,\pi\, r_m^2\,\rho_{\rm g}\, v^2}{2m} + \frac{GM_r}{r^2}. \tag{5.27}$$

Sean $r_i$ y $r_{i+1}$ los valores de los radios correspondientes a dos puntos adyacentes de la grilla y supongamos que, entre estos puntos pueden considerarse constantes todas las variables de la ecuación anterior que no involucren a la velocidad. Esto es, si

$$C_1 = \frac{C_{\rm D}\,\pi\, r_m^2\,\rho_{\rm g}}{2m} \tag{5.28}$$

y

$$C_2 = \frac{GM_r}{r^2} \tag{5.29}$$

son consideradas constantes, podemos escribir,

$$v\frac{dv}{dr} = -C_1\, v^2 + C_2. \tag{5.30}$$

Esta ecuación tiene solución analítica, lo cual nos permite expresar

$$v_{i+1}^2 = v_i^2\, e^{-2\,C_1\,\Delta r_i} + \frac{C_2}{C_1}\left(1 - e^{-2\,C_1\,\Delta r_i}\right), \tag{5.31}$$

donde definimos $v_i \equiv v(r_i)$ y $\Delta r_i \equiv r_{i+1} - r_i$. Así, obtenemos la expresión para la velocidad con la cual un planetesimal atraviesa una capa de gas. Por efecto de la fricción con el gas de la envoltura, el planetesimal pierde parte de su energía cinética en forma de calor. La ecuación (4.61) corresponde al intercambio de energía entre un planetesimal con el gas que lo rodea. Si tenemos en cuenta todos los planetesimales ingresantes por unidad de tiempo, $\dot{M}_{\rm c}$, y transformamos las ecuaciones diferenciales en ecuaciones en diferencias, la variación total de la energía $\Delta E_i$ de la capa de gas $i$ en el intervalo de tiempo $\Delta t$ es:

$$\Delta E_i = \frac{1}{2} C_{\rm D}\, \pi\, r_m^2\, \rho_g \frac{\dot{M}_c}{m}\, v_i^2\, \Delta r_i\, \Delta t. \tag{5.32}$$

Finalmente, cuando un planetesimal llega al núcleo, toda su energía cinética remanente se deposita en la capa de gas adyacente a él.

## 5.4. Cálculo de la acreción de gas

La tasa de acreción de gas es una consecuencia de la condición de borde que define el límite del planeta. Como mencionamos en § 4.4.4, no existe una separación física entre el planeta y la nebulosa, por cuanto el radio del planeta debe ser definido. Es una norma general tomar como definición:

$$R_{\rm p} = \min[R_{\rm H}, R_{\rm a}]. \tag{5.33}$$

La búsqueda de $R_{\rm p}$ no es trivial ya que tanto $R_{\rm H}$ como $R_{\rm a}$ dependen de la masa del planeta, la cual no se conoce hasta tanto no se sabe donde está el borde externo del mismo.

De la resolución de las ecuaciones de estructura se obtiene, para cada punto de la grilla, el valor del radio $r$ asociado a la masa contenida hasta $r$, $M_r$. Para explicar el procedimiento, supongamos que $R$ generaliza el concepto del radio del planeta (con lo cual $R$ puede ser $R_{\rm a}$ o $R_{\rm H}$). Entonces, comenzando por el borde externo de la grilla, calculamos para cada punto de la grilla (que tiene asociado un valor de la masa $M_r$), el valor de $R$. Recordemos que tanto $R_{\rm a}$ como $R_{\rm H}$dependen de la masa total del planeta. En general, el valor de $R$ será distinto del de $r$ correspondiente a ese punto. Se calcula para cada punto la diferencia entre ambos radios. Dado que tanto $r$ como $R$ son monótonamente crecientes con la masa, la cantidad $r - R$ tendrá una única raíz en toda la grilla. Esa raíz estará entre dos puntos consecutivos de la grilla para los cuales $r - R$ cambie de signo. Luego, el valor de $R$ se obtiene de interpolar al punto medio entre ambos. Este procedimiento se realiza tanto para el radio de acreción, $R_{\rm a}$, como para el radio de Hill, $R_{\rm H}$. El mínimo entre ambos definirá el radio del planeta, $R_{\rm p}$, para el modelo que estemos calculando. La variación, modelo a modelo, del borde del planeta es lo que determina la cantidad de gas por unidad de tiempo que es acretado por el planeta.

El paso de tiempo de integración y la cantidad de puntos de la grilla no son constantes sino que varían a lo largo de todo el proceso, dependiendo de la precisión temporal o espacial que requiera el cálculo para un determinado estado del planeta. En las regiones en que las cantidades termodinámicas tienen un comportamiento suave muchas veces no es necesario que el grillado sea muy fino. En esos casos, se eliminan capas y se reestructura el grillado, reasignando las cantidades. Si, por el contrario, deben agregarse puntos, se calculan los valores de las cantidades termodinámicas en los puntos nuevos haciendo una interpolación lineal. En cuanto al paso de tiempo, su valor depende de cuán dificultosa sea la integración de un modelo. El algoritmo que determina cómo debe actualizarse el paso de tiempo depende del cambio máximo permitido para la luminosidad, la temperatura, la presión y el radio entre dos modelos, y el cambio que efectivamente fue calculado. Si el cambio ocurrido es menor que el permitido, el paso de tiempo aumenta; en el otro caso disminuye.

# Capítulo 6

# Resultados

En este capítulo presentaremos los resultados obtenidos con nuestro programa de formación de planetas gigantes. Durante el desarrollo de esta Tesis se realizaron numerosas simulaciones, considerando diversas situaciones que involucran: la posición del planeta en el disco, la densidad superficial de sólidos en la zona de alimentación del planeta y el tamaño de los planetesimales acretados. Los resultados obtenidos al variar estos parámetros motivaron que el modelo siguiera evolucionando en el sentido de la incorporación de una ley de potencias para la distribución de tamaños de los planetesimales acretados. El considerar que la población de planetesimales no sea uniforme hace que el abordaje del problema sea más realista pero, a la vez, el tratamiento numérico se vuelve más complejo y el tiempo de cálculo aumenta considerablemente. Esto reduce la posibilidad de hacer simulaciones que consideren una gran diversidad de casos. Sin embargo, los resultados pusieron de manifiesto que introducir una distribución de tamaños era un ingrediente fundamental en el cálculo de formación de planetas gigantes, algo que hasta el momento no se había tenido en cuenta. Finalmente, se consideró la arquitectura original del Sistema Solar según lo propuesto por el modelo de Niza (Tsiganis et al. 2005), y se calculó la formación de Júpiter, Saturno, Urano y Neptuno. Los resultados muestran una concordancia muy buena entre las masas de los núcleo obtenidas y las estimadas a partir de los datos observacionales, como así también entre los tiempos de formación y las cotas observacionales dadas por las escalas de vida de los discos circumestelares. Es importante destacar que, en algunas simulaciones, se encontraron eventos cuasi-periódicos de expulsión de masa gaseosa. Este fenómeno no tiene antecedentes en la bibliografía y es por demás interesante puesto que su aparición puede retrasar o, directamente interrumpir la formación de un planeta.

## 6.1. Simulaciones considerando una población de planetesimales de tamaño único

Todas las simulaciones que presentaremos en este capítulo tienen en común que fueron realizadas bajo las hipótesis del modelo que presentamos en los capítulos anteriores. A continuación sintetizaremos sus características principales. El embrión planetario se encuentra en órbita circular alrededor del Sol, siendo el Sol el único cuerpo que tiene influencia gravitatoria sobre el planeta en formación. No se consideran efectos magnéticos, de rotación o la presencia de otros embriones creciendo en la vecindad. El disco protoplanetario es estático, así que no se contempla la posibilidad de migración de los planetesimales, ni los procesos que operan para la disipación del disco (como, por ejemplo, la fotoevaporación). Tampoco se consideró la interacción gravitatoria entre el disco y el protoplaneta, por cuanto el protoplaneta permanece fijo en su órbita (el planeta no migra). La densidad superficial de sólidos en la zona de alimentación del planeta solo varía por la acreción por parte del propio planeta, mientras que la densidad de gas queda inalterada (esto último no tiene consecuencias relevantes puesto que la cantidad de gas que rodea al planeta es varias veces la masa del mismo). La tasa de crecimiento del núcleo sólido se calcula de acuerdo al modelo de crecimiento oligárquico: dado que el crecimiento runaway del embrión ocurre en una escala de tiempo despreciable respecto de la correspondiente a la formación del planeta (siendo la primera del orden de $10^4$ años, mientras que la segunda es del orden de $10^7$ años), ignoraremos la etapa runaway y comenzaremos las simulaciones con un embrión de masa lunar (aproximadamente del orden de $10^{-2}\,M_\oplus$). La masa de gas inicial corresponde a la masa de gas ligada gravitatoriamente a ese embrión, que es del orden de $10^{-10}\,M_\oplus$. La tasa de acreción de gas se calcula en forma autoconsistente resolviendo las ecuaciones diferenciales que describen la estructura del interior del planeta, las cuales están acopladas al proceso de acreción de sólidos. La finalización del proceso de formación se estipula en el momento en que el planeta alcanza la masa final, la cual es introducida como parámetro. Este criterio es arbitrario, puesto que no se están considerando los mecanismos reales que regulan la finalización de la acreción de gas, como la disipación del gas nebular o la apertura de una brecha en el disco. De hecho, el estadío final de la formación de un planeta gigante debería ser objeto de un estudio profundo, lo cual está fuera del alcance de esta Tesis. De todos modos, el estudio que presentaremos no se focaliza en esta última etapa, sino en el análisis de la formación hasta una vez iniciado el runaway de gas.

Para los resultados que presentaremos en esta sección, las simulaciones se hicieron considerando que los planetesimales acretados tienen un único tamaño. Los parámetros que se tuvieron en cuenta para estudiar la respuesta del proceso de formación frente a una variación de los mismos fueron: el radio de la órbita del embrión, la densidad del disco protoplanetario y el radio de los planetesimales acretados. La elección de estos tres parámetros tiene que ver con que ellos afectan directamente a la tasa de acreción de sólidos (Ec. (4.14) y siguientes) y esto repercute fuertemente en las dos cantidades que se considerarán fundamentales a la hora de evaluar globalmente el proceso que estamos estudiando: la masa del núcleo y la escala de tiempo de formación.

Dado que en lo que sigue utilizaremos en repetidas oportunidades el concepto de "masa de aislación", resulta oportuno aquí mostrar cómo puede calcularse esta cantidad. Recordemos que la masa de aislación es la masa total de sólidos en la zona de alimentación de un protoplaneta.

Sea $a$ el semieje orbital del planeta y sea $2\,\Delta a$ el ancho de la zona de alimentación. Definimos la *masa de aislación*, $\mathrm{M_{iso}}$, como la masa total de sólidos presente en la zona de alimentación. Entones tendremos que:

$$\mathrm{M_{iso}} = \int_{a-\Delta a}^{a+\Delta a} \Sigma(a)\, 2\,\pi\, a da. \tag{6.1}$$

Dado que la densidad superficial de sólidos satisface una ley de potencias con la distancia ($\Sigma(a) \propto a^{-p}$) la masa de aislación satisface:

$$\mathrm{M_{iso}} \propto a^{2-p}\Big|_{a-\Delta a}^{a+\Delta a}. \tag{6.2}$$

Ahora bien,

$$\Delta a = 4\, R_{\mathrm{H}} \propto a\, M_{\mathrm{p}}^{1/3}. \tag{6.3}$$

Haciendo un desarrollo a primer orden se obtiene para la masa de aislación:

$$\mathrm{M_{iso}} \propto a^{2-p} M_{\mathrm{p}}^{1/3}. \tag{6.4}$$

Dado que en nuestro modelo todos los planetesimales acretados son incorporados a la masa del núcleo estamos interesados en conocer la masa del núcleo, cuando se alcanza la masa de aislación para una dada masa total del planeta. Como $M_{\mathrm{p}} = M_{\mathrm{c}} + M_{\mathrm{g}}$ tendremos que:

$$M_{\mathrm{c}} \propto a^{2-p}\, M_{\mathrm{c}}^{1/3} \left(1 + \frac{M_{\mathrm{g}}}{M_{\mathrm{c}}}\right)^{1/3}, \tag{6.5}$$

con lo cual,

$$M_{\mathrm{c}} \propto a^{\frac{3\,(2-p)}{2}} \left(1 + \frac{M_{\mathrm{g}}}{M_{\mathrm{c}}}\right)^{1/2}. \tag{6.6}$$

Para el caso de un perfil de densidad como el de la nebulosa estándar, $p = 3/2$ y la masa de aislación resulta ser:

$$\mathrm{M_{iso}} = 0,65\, f_d^{3/2} \left(\frac{a}{1\,\mathrm{UA}}\right)^{3/4} \left(1 + \frac{M_{\mathrm{g}}}{M_{\mathrm{c}}}\right)^{1/2} \mathrm{M_{\oplus}}. \tag{6.7}$$

donde $f_d$ representa el aumento en la densidad de sólidos respecto de los valores estándar ($f_d = 1$ si se considera la nebulosa solar estándar).

### 6.1.1. Ejemplo y análisis de la formación de un planeta gigante

Dado que, frente a la variación de los parámetros que mencionamos antes (radio de los planetesimales, masa del disco, semieje del planeta, etc.), el proceso de formación solo cambia cuantitativamente, tomaremos aquí los resultados de una de las simulaciones para hacer el análisis detallado del proceso completo de formación. Los otros casos que presentaremos en esta sección son cualitativamente similares al siguiente ejemplo.

Consideremos el caso de un planeta ubicado en la posición actual de Júpiter ($a = 5{,}2$ UA), siendo la densidad superficial de sólidos en su zona de alimentación $\Sigma = 10$ g cm$^{-2}$ (correspondiente a, aproximadamente, 3 veces la Nebulosa Solar de Masa Mínima, NSMM), la densidad del gas nebular $\rho = 1,5 \times 10^{-11}$ g cm$^{-3}$, y la temperatura $T = 150$ K. El radio de los planetesimales acretados es $\sqrt{10}$ km ($r_m = 3,16$ km), siendo la densidad de los mismos $1,39$ g cm$^{-3}$ (según Pollack et al. 1996). La densidad del núcleo se fijó en $3,2$ g cm$^{-3}$. El hecho que la densidad del núcleo se considere mucho mayor a la de los planetesimales acretados tiene que ver con que el núcleo está sometido a la alta presión que ejerce la masa del planeta, lo cual hace que sea mucho más compacto. Denominaremos a este caso que estamos analizando a modo de ejemplo como $J1\sqrt{10}$.

De acuerdo con los resultados de nuestra simulación, para el caso $J1\sqrt{10}$, la formación del planeta se extiende por $7 \times 10^6$ años, como se puede observar de la figura 6.1. En alrededor de la mitad de este tiempo el núcleo alcanza una masa de, aproximadamente, $10\,M_\oplus$, mientras que en este proceso, la masa de gas ligada puede considerarse despreciable. Vemos, entonces, que el tiempo que insume la formación del núcleo regula la escala de tiempo que demanda la primera etapa de la formación de un planeta. En el régimen oligárquico, el tiempo característico de crecimiento de un embrión aumenta con su masa (Ec. 4.12), como se muestra en la figura 6.2. En el transcurso de este intervalo de tiempo, para su formación, el núcleo consumió gran parte de los sólidos del disco correspondiente a su región de influencia. Cuando el material en la zona de alimentación comienza a agotarse (último panel de la figura 6.1), la tasa de acreción de sólidos disminuye (tercer panel de la misma figura). En lo que resta para terminar la formación, el núcleo crecerá poco y muy lentamente. En cambio, la tasa de acreción de gas es la que comienza a dominar el proceso. Esto se debe fundamentalmente a dos cosas: por un lado, el embrión ya tiene una masa lo suficientemente grande como para que su influencia gravitatoria sea importante, lo cual le permite mantener ligado mayores cantidades de gas; pero por otro lado, y quizá más importante aún, son las consecuencias del vaciamiento de la zona de alimentación. En el tiempo transcurrido, el embrión utilizó casi todo el material sólido de su vecindad para la formación del núcleo, lo que hace que la acreción de planetesimales disminuya, lo cual, a su vez, significa que el aporte energético que estos cuerpos hacen al ingresar al planeta sea mucho menor. La energía que depositan los planetesimales en las capas de gas de la envoltura aumenta la presión del gas, que tiende a poner resistencia al colapso gravitatorio. Cuando el número de planetesimales acretados es pequeño, esta energía disminuye y la compresión de las capas de gas aumenta, lo cual da lugar a que la acreción de gas sea cada vez más alta. De hecho, durante la primera parte de la formación, la luminosidad del planeta proviene casi exclusivamente de la acreción de planetesimales, como se puede ver del segundo panel de la figura 6.1. Pero cuando los planetesimales ingresantes empiezan a menguar, la luminosidad proveniente de la contracción de las capas de gas comienza a ser del mismo orden que la debida a la de los planetesimales. Luego, el núcleo masivo y el decrecimiento de la energía proveniente de la acreción de los planetesimales, quienes aportan la resistencia a la contracción de las capas de la envoltura, le dan paso a un aumento en la acreción de gas. En un intervalo de tiempo comparable al que tomó la formación del núcleo, la masa de la envoltura alcanza un valor crítico: llega a ser igual a la masa del núcleo. Diremos que en este momento se alcanzó la *masa de cruce*, $M_{\rm cross}$, que se define como el valor de la masa del núcleo (o de la envoltura) en el momento en que la masa de la envoltura iguala a la masa del núcleo. El *tiempo de cruce* corresponde al valor de la coordenada temporal cuando se alcanza la masa de cruce. Matemáticamente, esto es:

$$M_{\rm cross} \equiv M_{\rm c} \qquad \text{cuando} \qquad M_{\rm g} = M_{\rm c} \tag{6.8}$$

$$t_{\rm cross} \equiv t(M_{\rm cross}). \tag{6.9}$$

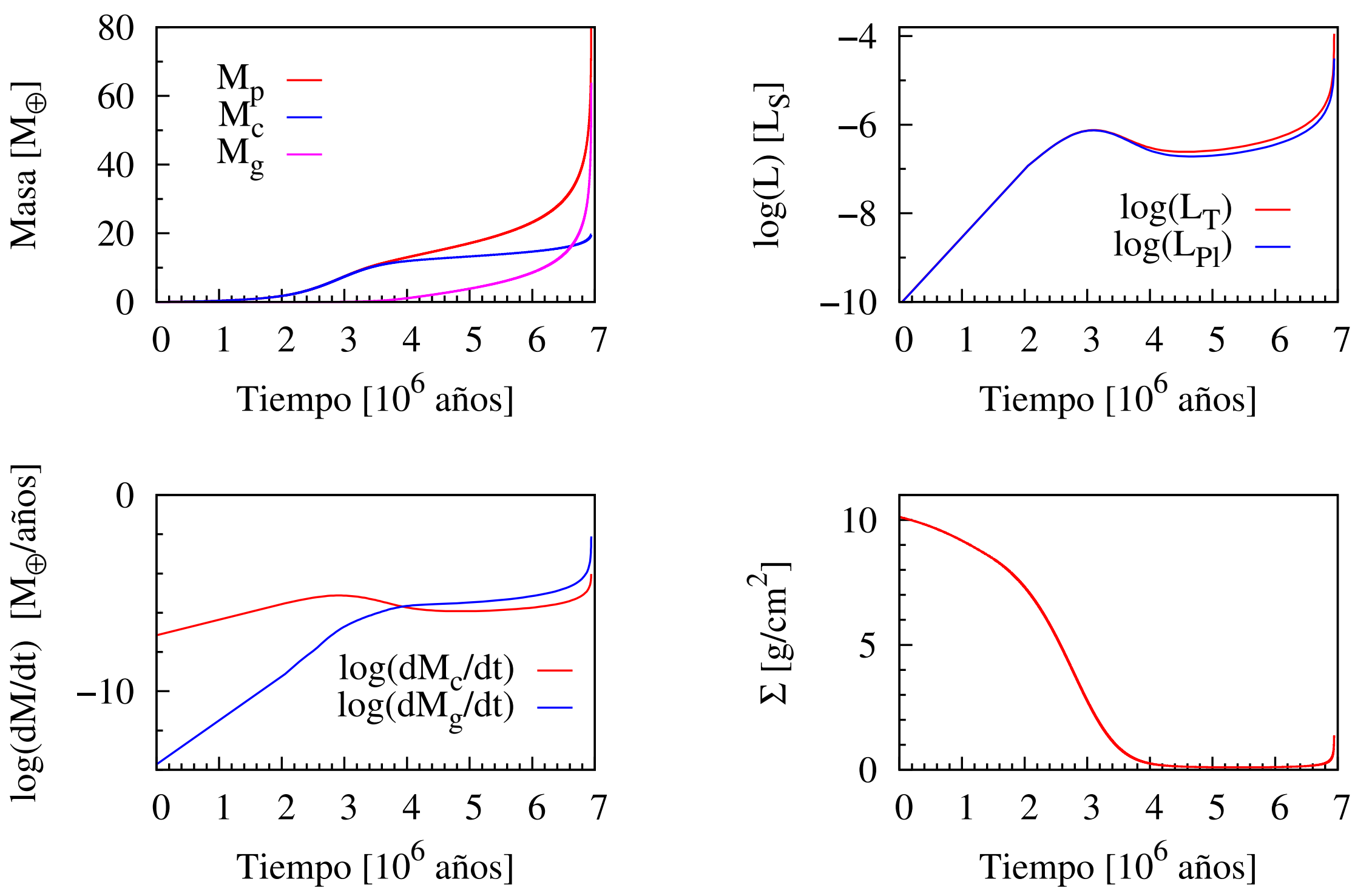

**Figura 6.1.** Los paneles de esta figura muestran la evolución de distintas cantidades en función del tiempo durante el proceso de formación de un planeta gigante, cuyo radio orbital es $a = 5{,}2$ UA. El disco protoplanetario es tres veces más masivo que la NSMM, y $r_m = \sqrt{10}$ km. De izquierda a derecha y de arriba hacia abajo, el primer panel muestra la masa total del planeta ($\mathrm{M_p}$), la masa del núcleo ($\mathrm{M_c}$) y la masa de la envoltura gaseosa ($\mathrm{M_g}$) en función del tiempo (Caso $J1\sqrt{10}$). Notemos que la mitad del tiempo de formación está destinada a la formación del núcleo sólido el cual, cuando alcanza una masa de alrededor de 10 $\mathrm{M_\oplus}$, comienza a acretar gas de una manera más significativa. Después de que la masa del núcleo iguala a la de la envoltura comienza el crecimiento runaway del gas, donde el planeta acreta la mayor parte de su masa en muy poco tiempo. El segundo panel muestra la evolución de la luminosidad en función del tiempo, donde $\mathrm{L_T}$ representa la luminosidad total y $\mathrm{L_{Pl}}$ la que proviene de la acreción de los planetesimales. En el tercer panel se muestran la tasa de acreción de sólidos y de gas, $\log(\mathrm{d\,M_c/dt})$ y $\log(\mathrm{d\,M_g/dt})$, respectivamente; y en el cuarto se representa la evolución de la densidad superficial de sólidos, $\Sigma$, en la zona de alimentación del planeta.

Para el caso que estamos analizando, $M_{\rm cross} = 16,4\ {\rm M}_\oplus$ y $t_{\rm cross} = 6,6 \times 10^6$ años. Cuando se supera la masa de cruce, el crecimiento de la envoltura se exacerba y comienza el llamado "runaway de gas", que lleva a que el planeta adquiera la mayor parte de su masa en un corto intervalo de tiempo. Siendo la masa del planeta cada vez más grande, la autogravedad del planeta no puede ser balanceada por la energía térmica, lo cual provoca el colapso de las capas de gas. Es importante notar que, a pesar del marcado cambio en la tasa de acreción gaseosa, este proceso sigue siendo de carácter hidrostático. Durante el crecimiento runaway del gas, el planeta aumenta su masa muy rápidamente, lo cual hace que la zona de alimentación se expanda (recordemos que la zona de alimentación es proporcional al radio de Hill el cual, a su vez, es proporcional a $M_{\rm p}^{1/3}$). Esto incrementa notablemente la cantidad de sólidos disponibles, aumentando la tasa de acreción de planetesimales y, consecuentemente, la masa del núcleo. Sin embargo, la energía que aportan estos planetesimales no es suficiente para contener el crecimiento de la envoltura.

La última etapa de la formación del planeta, que involucra la finalización de la acreción, no está incluida en nuestro modelo, ni tampoco ha sido estudiada en profundidad por otros autores en el marco del modelo de inestabilidad nucleada. De acuerdo con Lissauer & Stevenson (2007), de las simulaciones hidrodinámicas con modelos tridimensionales realizadas para el estudio de la interacción entre un planeta que se encuentra acretando gas y el disco protoplanetario, se encuentra que, mientras la masa del planeta es relativamente pequeña ($M_{\rm p} \lesssim 10\,{\rm M}_\oplus$), el disco permanece inalterado y puede suministrar gas al planeta sin restricciones. Sin embargo, cuando la masa del planeta aumenta se producen torques en el disco que alejan al gas nebular del planeta, abriéndose una brecha entre ambos. Los límites hidrodinámicos para la tasa de acreción de gas son del orden de $10^{-2}\ {\rm M}_\oplus$/año para planetas de entre 50 y 100 ${\rm M}_\oplus$. En cuanto a la masa final del núcleo, por otra parte, hay que tener en cuenta que una vez iniciado el runaway de gas, la expansión de la zona de alimentación hacia regiones del disco que, en nuestro modelo, conservan los valores de la densidad inicial, sobredimensiona la tasa de acreción de sólidos. Tengamos presente que nuestro modelo no considera la migración de los planetesimales en el disco ni la presencia de otros embriones que también pueden estar acretando material. De hecho, la migración de los planetesimales puede reducir considerablemente la cantidad de material disponible para ser acretado (Thommes, Duncan & Levison 2003). Además, cuando el protoplaneta se vuelve muy masivo provoca también la eyección de parte de los planetesimales hacia afuera de su zona de alimentación.

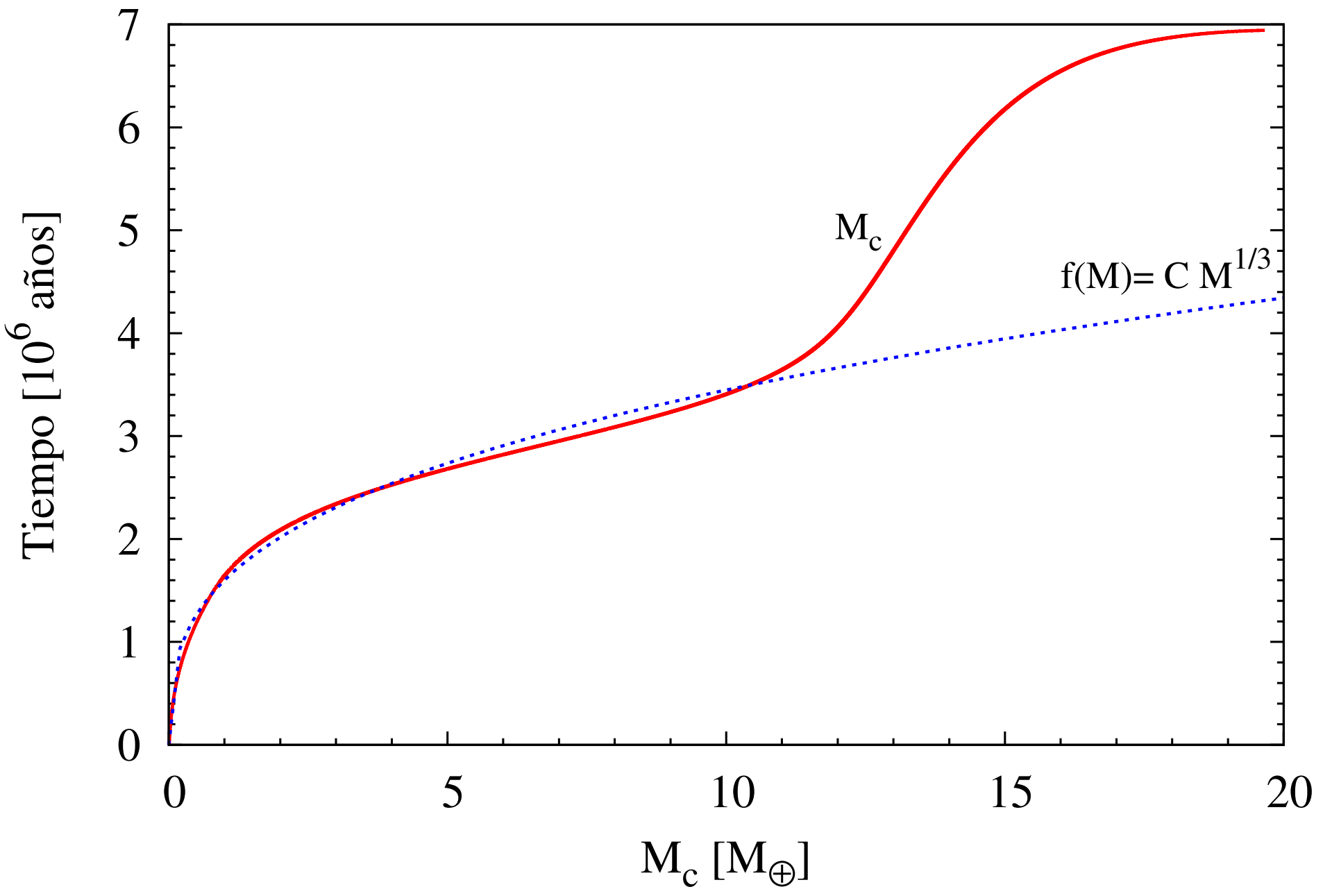

**Figura 6.2.** $J1\sqrt{10}$. Esta figura muestra la masa del núcleo ($\mathrm{M_c}$) en la abscisa y el tiempo en la ordenada. Como se mencionó en § 4.2.2, el tiempo característico del crecimiento oligárquico satisface: $T_{grow} \propto M^{1/3}$. En línea roja se muestra el resultado de nuestra simulación, mientras que la línea punteada representa una función proporcional a $M^{1/3}$. La diferencia entre ambas curvas es ínfima hasta que el núcleo alcanza aproximadamente $10\,\mathrm{M_\oplus}$. Luego ambas curvas comienzan a separarse debido a que el crecimiento del núcleo se acelera por la presencia de la envoltura. Cuando comienza el runaway de gas ($M_\mathrm{g} = M_\mathrm{c} \simeq 16\,\mathrm{M_\oplus}$), el núcleo aumenta su masa en forma notable porque la zona de alimentación del planeta se expande considerablemente y se puebla, nuevamente, de gran cantidad de planetesimales.

Por estos motivos, de nuestras simulaciones no se pueden obtener los valores finales precisos para la masa del núcleo ni para el tiempo de formación. Resultan sí , mucho más representativas, las magnitudes de la masa y del tiempo de cruce. Tomaremos entonces, en lo que sigue, a ambas variables como referencia a la hora de comparar los resultados de distintas simulaciones.

Siguiendo con el análisis de nuestro ejemplo $J1\sqrt{10}$, notamos de la figura 6.3 que el radio de acreción define el borde del planeta desde el inicio de la formación y hasta después de comenzado el runaway de gas. Dado que tanto el radio de acreción como el radio de Hill dependen solo de la masa total, y no de la masa de la envoltura, el momento en el cual se produce el cambio entre uno y otro para definir el borde no está relacionado con la etapa del proceso que esté atravesando el planeta. El hecho que el radio de acreción domine al principio de la formación tiene que ver con la ubicación del planeta en el disco y con las condiciones de la nebulosa (que definen la velocidad del sonido para el radio de acreción). Después de que ambos radios se igualan, el radio de Hill resulta ser el menor entre los dos por su dependencia con la raíz cúbica de la masa, lo que hace que el crecimiento del radio de Hill sea mucho más lento que el del radio de acreción, el cual crece linealmente con la masa. Es por esto que, una vez que el radio de Hill satisface la condición para ser el borde del planeta, lo seguirá siendo hasta el final de la formación, siempre y cuando, claro está, no haya eventos de pérdida de masa.

En la bibliografía pueden encontrarse autores que definen el borde del planeta, en todo momento, como igual al radio de Hill (por ejemplo, Wuchterl 1995). El argumento para no considerar al radio de acreción en la definición del radio del planeta es que, dado que el planeta está rodeado del gas nebular, por más que las moléculas tengan la energía térmica como para escapar del campo gravitatorio del planeta, la distribución aleatoria de velocidades hará que otras moléculas se muevan en la dirección contraria, no habiendo entonces un efecto neto de material que "escape" de sus límites gravitatorios. De todos modos, dado que las capas externas del planeta contienen muy poca masa, en las simulaciones que hemos hecho no encontramos cambios significativos en los resultados si consideramos, o no, el radio de acreción en la definición del radio del planeta. Hemos adoptado la definición clásica para el borde del planeta porque nos permite comparar más limpiamente nuestros resultados con los de otros autores y porque, como dijimos, la inclusión del radio de acreción no afecta sustancialmente los resultados.

Para estudiar lo que ocurre en el interior del planeta durante su formación elegiremos uno de los modelos de la secuencia de esta simulación para hacer el análisis (llamamos *modelo* al conjunto de variables que describen, para un determinado instante $t$, el estado de la envoltura del planeta, desde el borde externo hasta la base del núcleo sólido). Tomamos el modelo que corresponde a una masa total de $47\,\mathrm{M}_\oplus$ ($t \simeq 6,8 \times 10^6$ años), donde la masa de la envoltura es, aproximadamente, $30\,\mathrm{M}_\oplus$, el radio del planeta es $\log(R_\mathrm{p}) = 12,4$, en centímetros, y el radio del núcleo es $\log(R_\mathrm{p}) = 9,3$. En la figura 6.4 se grafican cuatro paneles que muestran la masa, la luminosidad, la energía depositada por los planetesimales y la velocidad de los planetesimales en función del radio (en realidad, en todos los casos se consideró el logaritmo de estas cantidades). Se eligió el radio como referencia porque, dado que el planeta presenta una estructura muy extendida, esta coordenada permite apreciar en detalle las variaciones de las cantidades que se quieren estudiar. Del primer panel podemos ver que prácticamente la totalidad de la masa está contenida en $\log(R) < 10,4$. Entre este valor y el borde del planeta la densidad del gas es muy baja y el aporte que hacen estas capas a la masa total es inferior al 10 %. Sin embargo, en parte de esas capas ($10,5 < \log(R) < 11$) está el aporte más importante a la luminosidad debido a la contracción del gas. En el panel C esto se ve reflejado en la diferencia entre la luminosidad total y la correspondiente a la acreción de los planetesimales. La energía cinética que los planetesimales pierden en forma de calor por el efecto de la viscosidad del gas es:

$$\frac{dE}{dr} \propto \rho_\mathrm{g} v^2, \tag{6.10}$$

como lo establece la ecuación (4.61). Luego, la energía que absorba cada capa de gas dependerá de la densidad $\rho_\mathrm{g}$ y de la velocidad $v$. Cuando los planetesimales ingresan al planeta lo hacen con velocidad $v_\mathrm{rel}$ (Ec. (4.16)). La densidad del gas en las capas exteriores del planeta es baja, y la ecuación de movimiento (Ec. 4.56) está dominada por la aceleración gravitatoria, lo que hace que la velocidad de los planetesimales aumente (panel D). Como la energía que pierden los planetesimales depende de la velocidad de los mismos, si ésta aumenta, también lo hace la energía que ceden a la envoltura. Cuando los planetesimales ingresan en regiones más densas, la fricción es mayor. Pero, además, la fracción de masa del planeta que atraviesan disminuye, por lo tanto la aceleración gravitatoria es menor. Entonces, la viscosidad comienza a jugar un papel importante en la ecuación de movimiento, con lo cual la velocidad decrece y la energía liberada también. Sin embargo, llegando a la base de la envoltura, la energía comienza a aumentar nuevamente. Esto ocurre cuando el frenado producido por la viscosidad compensa a la aceleración gravitatoria y la velocidad se vuelve casi constante. La energía liberada aumenta ahora por su dependencia con la densidad, ya que el planetesimal atraviesa las regiones donde el gas es más denso.

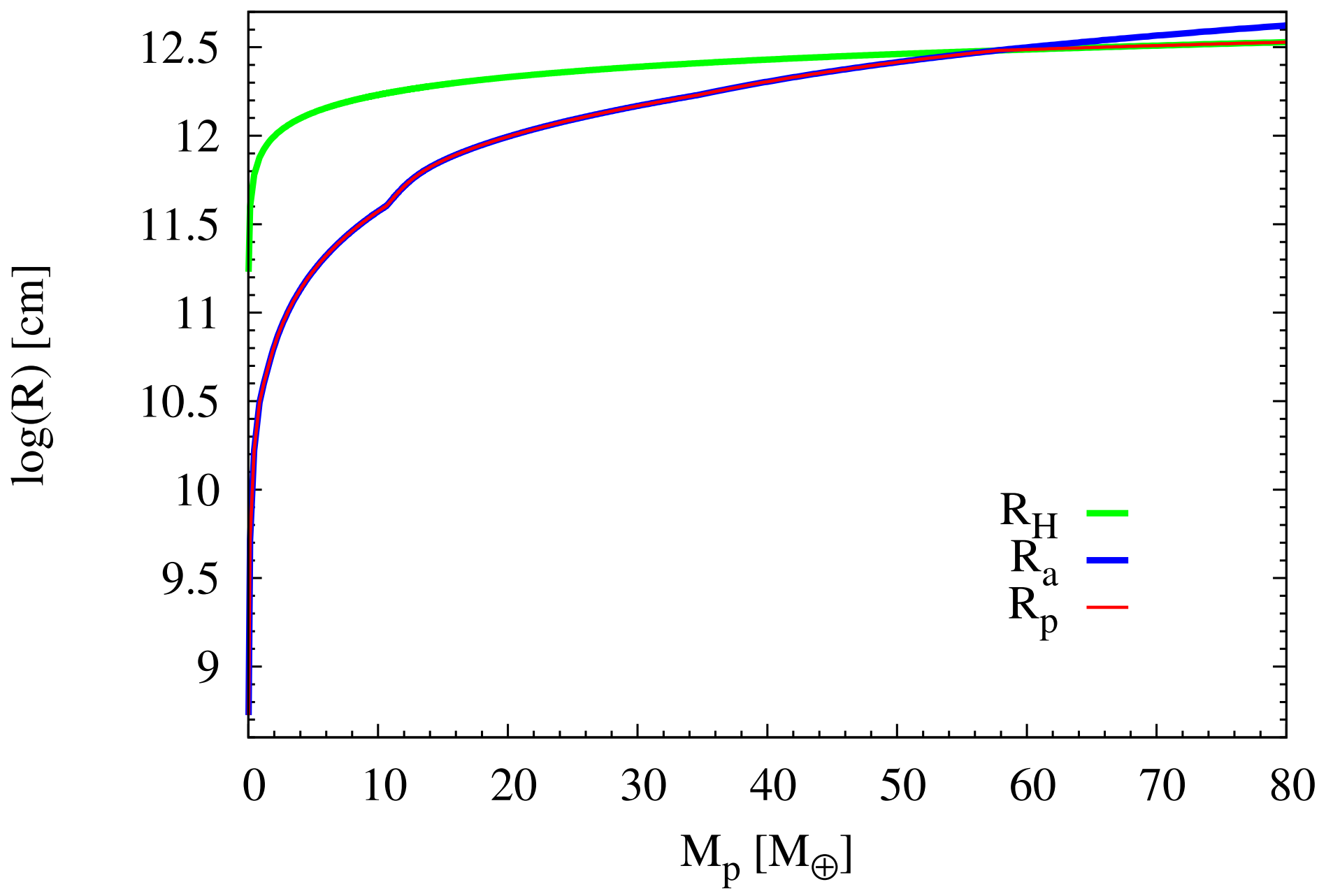

**Figura 6.3.** $J1\sqrt{10}$. Evolución del radio con la masa del planeta. En línea verde se muestra el radio de Hill, en azul el radio de acreción y en rojo el radio del planeta, el cual se define como: $R_p = \min[R_a, R_H]$. El radio de acreción representa el límite externo durante casi todo el proceso de formación del planeta, salvo cuando éste se vuelve muy masivo y provoca que el borde pase a estar definido por el radio de Hill. Esto ocurre después de comenzado el runaway, cuando la masa de la envoltura domina la masa total del planeta.

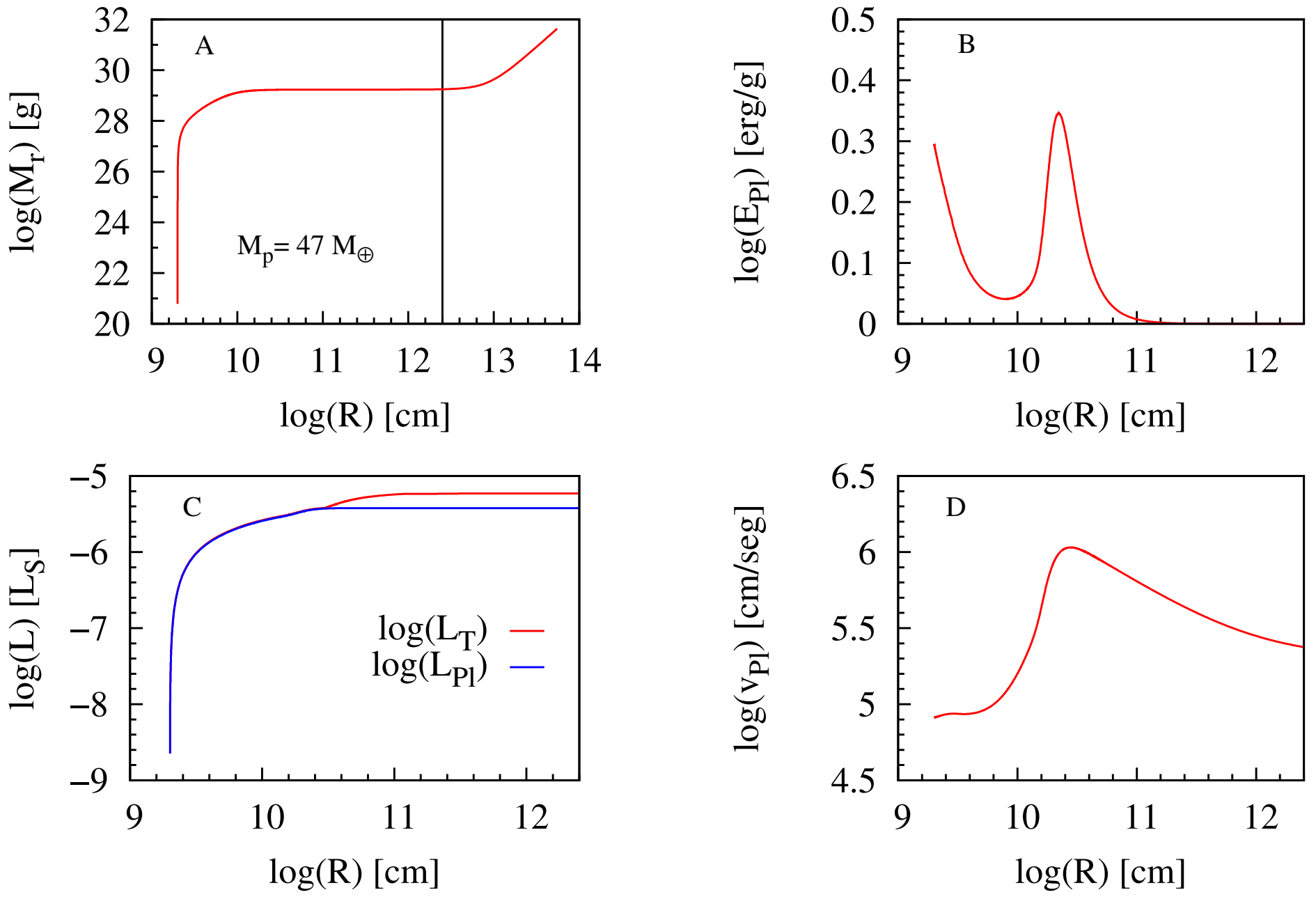

**Figura 6.4.** Los cuatro paneles que componen esta figura muestran el perfil radial de la masa ($\mathrm{M_r}$), la energía que aportan los planetesimales acretados, la luminosidad total y la velocidad de los planetesimales en el interior del modelo correspondiente a $\mathrm{M_p} = 47\,\mathrm{M_\oplus}$, que forma parte del ejemplo $J1\sqrt{10}$. La abscisa representa el logaritmo de la coordenada radial. En el panel A vemos la distribución de la masa de la envoltura en función del radio. La línea vertical establece la ubicación del radio del planeta. Para radios mayores, el gas no forma parte del planeta sino de la nebulosa protoplanetaria. Notemos que la masa del planeta se mantiene casi constante desde el borde del planeta hacia el centro en una extensión radial que involucra una variación de $R$ de dos órdenes de magnitud. Entonces, si bien el planeta presenta una estructura gaseosa muy extendida, la mayor parte de la masa se concentra en las capas interiores. El panel B representa el logaritmo de la energía que liberan los planetesimales a medida que atraviesan las capas de la envoltura en su recorrido hacia el núcleo del planeta (la energía liberada cuando los planetesimales llegan a la superficie del núcleo no se muestra). Desde el exterior hacia el centro vemos que la energía tiene dos máximos, el primero para log(R)=10,34 y el segundo sobre la base del núcleo. El primer máximo está relacionado con el aumento de la velocidad (panel D): dado que la velocidad se incrementa, de la Ec. (4.61) vemos que también lo hace la energía que pierden los planetesimales por efecto de la viscosidad. Pero la velocidad luego decrece, lo cual hace que también lo haga la energía. Sin embargo, en las cercanías del núcleo, la velocidad se mantiene casi constante pero la densidad sigue en aumento y tomando los valores más altos, con lo cual la energía liberada por los planetesimales se incrementa nuevamente. El panel C muestra el logaritmo de la luminosidad total, log($\mathrm{L_T}$), y la luminosidad proveniente de la acreción de planetesimales, log($\mathrm{L_{Pl}}$), ambas en unidades de la luminosidad solar. Notar que son las capas externas las que contribuyen al aumento de la luminosidad respecto de la debida a la acreción de los planetesimales. Por último, el panel D muestra los cambios en la evolución de la velocidad de los planetesimales, gobernada por la Ec. (4.56).

Para estudiar la evolución de las variables termodinámicas en el interior del planeta, en la figura 6.5 se muestran los perfiles de temperatura, presión, luminosidad y densidad para cinco modelos ($M_\mathrm{p} = 15$, 32, 47, 57 y $65\,\mathrm{M}_\oplus$), donde el de menor masa está dominado por la masa del núcleo, el segundo corresponde al momento de cruce (cuando $M_\mathrm{g} = M_\mathrm{c}$), y los tres restantes se encuentran en la etapa de crecimiento runaway del gas. Mientras que la temperatura, la presión y la densidad son monótonamente decrecientes con el radio, la luminosidad es creciente. Los valores de las tres primeras en el borde del planeta corresponden a las condiciones de borde que caracterizan el estado de la nebulosa solar. A medida que la masa de la envoltura aumenta, cada una de estas variables también aumenta a radio fijo. Notemos, sin embargo, que un valor de radio constante no representa al mismo elemento de masa en dos modelos distintos. Las curvas que muestran el perfil de temperatura presentan una región achatada (en el caso, por ejemplo, de $M_\mathrm{p} = 15\,\mathrm{M}_\oplus$ corresponde a $10 < \log(R_\mathrm{c}) < 10,5$) que se corresponde con la aparición de la primera zona de transporte radiativo en el interior (figura 6.6). El gradiente de temperatura (4.26) depende del tipo de transporte que tenga lugar en al región, si el transporte es radiativo el gradiente será proporcional a $\nabla_\mathrm{rad}$ (Ec. 4.28), mientras que si es convectivo se considera proporcional al gradiente adiabático ($\nabla_\mathrm{ad}$). El tipo de transporte de energía que domina la estructura del planeta depende de la masa que tenga la envoltura. En todos los casos, la región más interna es convectiva. Luego aparece la región radiativa que recién mencionábamos. Siguiendo hacia afuera se desarrollan pequeñas zonas convectivas que crecen a medida que la masa del planeta crece. En el caso del modelo de menor masa considerado aquí , el borde externo es radiativo mientras que cuando la masa de la envoltura se encuentra en pleno crecimiento runaway, la región exterior se vuelve completamente convectiva. Ésto tiene que ver con que el gradiente de temperatura se vuelve cada vez más agudo a medida que la masa del planeta aumenta, superando al gradiente adiabático, lo cual hace que se instale el transporte convectivo. En cuanto a la luminosidad, ésta aumenta con la masa. La luminosidad impuesta por la condición de borde en la superficie del núcleo es que ésta sea cero. Sin embargo, la luminosidad en la primera capa de gas no es cero debido a la energía que dejan los planetesimales. En el caso de los modelos menos masivos, como el correspondiente a $M_\mathrm{p} = 15\,\mathrm{M}_\oplus$, la luminosidad proviene casi exclusivamente de la acreción de los planetesimales. A medida que la masa de la envoltura aumenta, comienza a aparecer un punto de quiebre en el perfil de la luminosidad. Éste está relacionado con la contracción de las capas externas que, como mencionamos en un párrafo anterior, frente al decrecimiento de la acreción de sólidos y al aumento de la masa del planeta, incrementan el valor de la luminosidad.

Por último, la figura 6.7 muestra la evolución de los valores de la densidad y la temperatura de la última capa de gas (que se encuentra sobre la superficie del núcleo) durante toda la formación del planeta. Si bien la figura se extiende sobre diversos valores de la temperatura y la densidad, es importante destacar que la zona verdaderamente representativa para el análisis del estado del gas en la adyacencia con el núcleo del planeta, comienza para $\log(T) > 3,8$ ($M_\mathrm{g} > 0,01\,\mathrm{M}_\oplus$). Para valores “centrales” menores, la masa de gas ligado es despreciable. Mencionaremos, sin embargo, que en la región donde $3,1 < \log(T) < 3,4$, la opacidad de los granos disminuye abruptamente, lo que hace que en esa zona del interior del planeta el transporte sea siempre radiativo. Por eso es que aparece, en esa región, el cambio de pendiente en la curva. Cuando los valores centrales se encuentran en este rango de temperaturas, se libera, relativamente, mucha energía y las capas se contraen considerablemente, hasta que las condiciones en región más interior del planeta salen de la zona de baja opacidad. Sin embargo, dado que cuando estas son las condiciones presentes en el planeta la masa de la envoltura es ínfima, este hecho no tiene una repercusión impor-tante en la formación del planeta. En la figura se incluyó también el diagrama de fases del hidrógeno, correspondiente a la EOS de SCVH, y la curva que representa el estado del interior del modelo de 47 $M_\oplus$. Dada la monotonía de la temperatura y la densidad, los valores "centrales" son los más altos que se registran para cada modelo. Vemos entonces que, como la curva no atraviesa la PPT, ningún modelo de los que componen la secuencia de esta simulación atravesará esta discontinuidad. Sin embargo, el interior del planeta sí atravesará dos transiciones: del hidrógeno molecular al atómico, y del atómico al ionizado, como puede observarse en la curva de 47 $M_\oplus$.

### 6.1.2. Variación de los resultados frente al cambio de las condiciones de borde externas

Cualitativamente, el proceso de formación de un planeta gigante, en base a nuestras simulaciones, puede ser descripto como lo hicimos en la sección previa, sin verse notoriamente afectado por variaciones en las condiciones de borde impuestas por la nebulosa, por cambios en su ubicación en el disco protoplanetario o por el tamaño de los planetesimales acretados. Sin embargo, tanto la masa como el tiempo de cruce, son muy sensibles a la elección de estos valores.

Consideremos primero el caso de un planeta ubicado en la posición actual de Júpiter ($a = 5,2$ UA). En la tabla 6.1 se muestran los resultados de nuestras simulaciones cuando la densidad del disco protoplanetario corresponde al rango de 6 a 10 veces NSMM. Para cada uno de estos cinco casos se consideraron dos posibles valores para el radio de los planetesimales acretados: 10 y 100 km. Como ya mencionamos, tomaremos como valores representativos para cada una de las simulaciones los resultados de $M_{\rm cross}$ y $t_{\rm cross}$. Los valores de la densidad del disco fueron seleccionados de modo que el tiempo de cruce, en todos los casos, fuera inferior a $10^7$ años (este valor en general es el que se toma como característico para la vida media de los discos, ver Haisch 2001). Notemos que considerar la densidad del disco como la correspondiente a $n$ veces la NSMM no necesariamente significa que la masa total del disco sea $n$ veces la masa estándar de la nebulosa solar. Dado que nos estamos focalizando en una determinada órbita, la cual es además fija, los valores de la densidad altos pueden deberse a enriquecimientos locales (Dodson-Robinson 2008a) y no necesariamente extenderse uniformemente en todo el disco. Entonces, el asociar la densidad para una dada posición a un valor de la NSMM es solo a modo de referencia y no implica que estemos considerando un disco que se corresponda con esa masa. En cuanto al tamaño de los planetesimales, ambos valores son los que frecuentemente se encuentran en la literatura cuando se estudia este tipo de problemas.

**Tabla 6.1.** Masa y tiempo de cruce para diez simulaciones en las cuales se variaron la densidad del disco en la ubicación del planeta ($a = 5,2$ UA), y el radio de los planetesimales acretados ($r_m = 10$, $100\,\mathrm{km}$). En la primera columna se muestra la densidad del disco protoplanetario en función del número de veces que representa respecto de la nebulosa estándar. La segunda y cuarta columna listan los resultados para el tiempo de cruce, en millones de años, y en la tercera y quinta columna se muestran las masas de cruce en unidades de masas de la Tierra.

| Masa del disco [MMSN] | Radio de los planetesimales $r_m = 100\,\mathrm{km}$ | | Radio de los planetesimales $r_m = 10\,\mathrm{km}$ | |
|---|---|---|---|---|
| | Tiempo de cruce [My] | Masa de cruce [$M_\oplus$] | Tiempo de cruce [My] | Masa de cruce [$M_\oplus$] |
| 6 | $7,20$ | $25,50$ | 2,43 | $31,80$ |
| 7 | $5,58$ | $27,90$ | 1,97 | $34,80$ |
| 8 | $4,60$ | $29,30$ | 1,49 | $36,80$ |
| 9 | $3,78$ | $30,90$ | 1,32 | $38,60$ |
| 10 | $3,20$ | $31,90$ | 1,10 | $39,90$ |

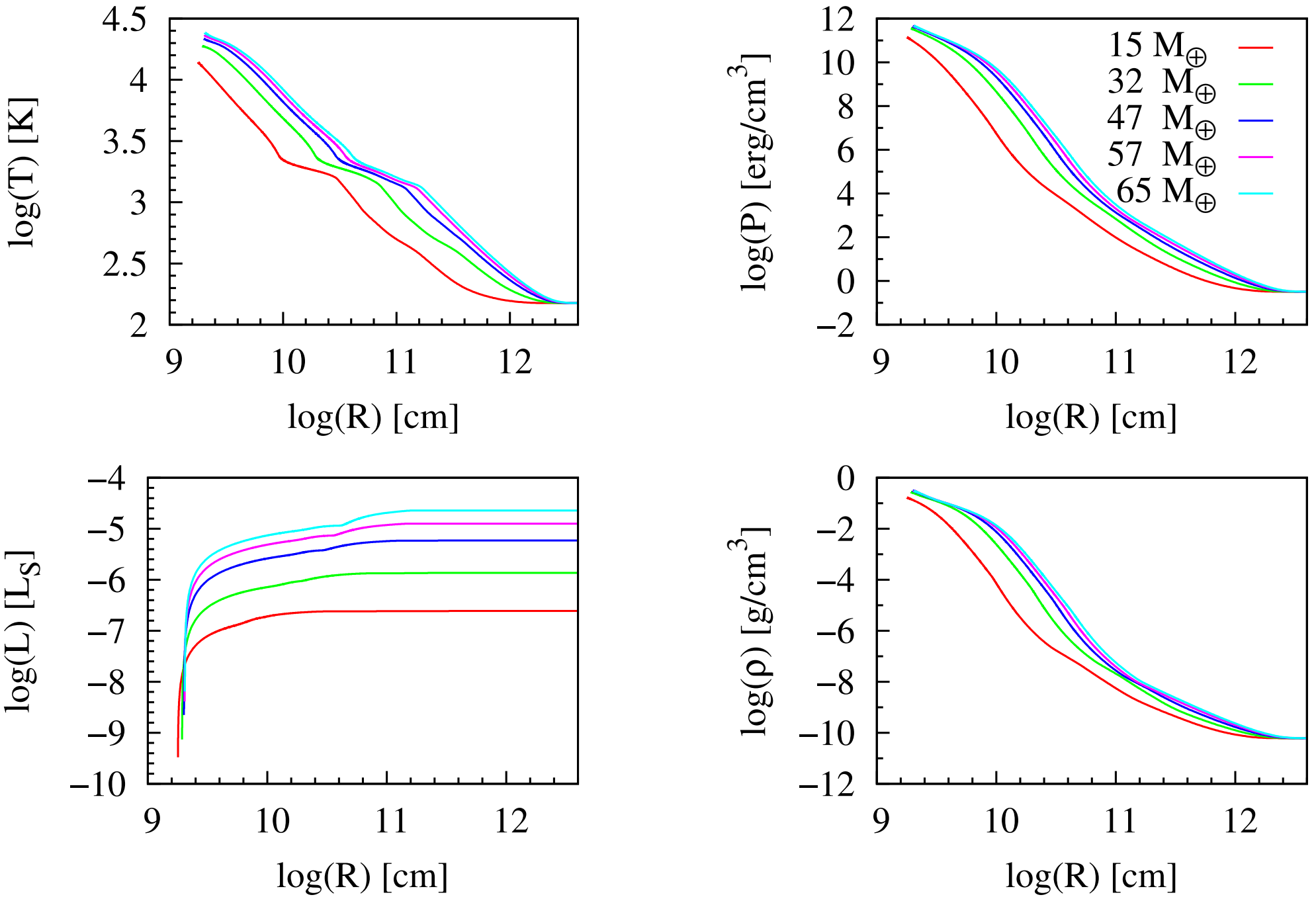

**Figura 6.5.** $J1\sqrt{10}$. Perfiles de distintas variables termodinámicas en el interior del planeta en formación cuando su masa total es: $15\,M_\oplus$, $32\,M_\oplus$, $47\,M_\oplus$, $57\,M_\oplus$, y $65\,M_\oplus$. La abscisa representa el logaritmo de la coordenada radial. El primer panel muestra el perfil de temperatura, el segundo el de la presión, el tercero la luminosidad (en unidades solares) y el cuarto el de la densidad. El límite exterior del planeta se encuentra, según el orden creciente en la masa, en: 11,9, 12,2, 12,4, 12,45, 12,5; en todos los casos este valor corresponde al $\log(R_p)$.

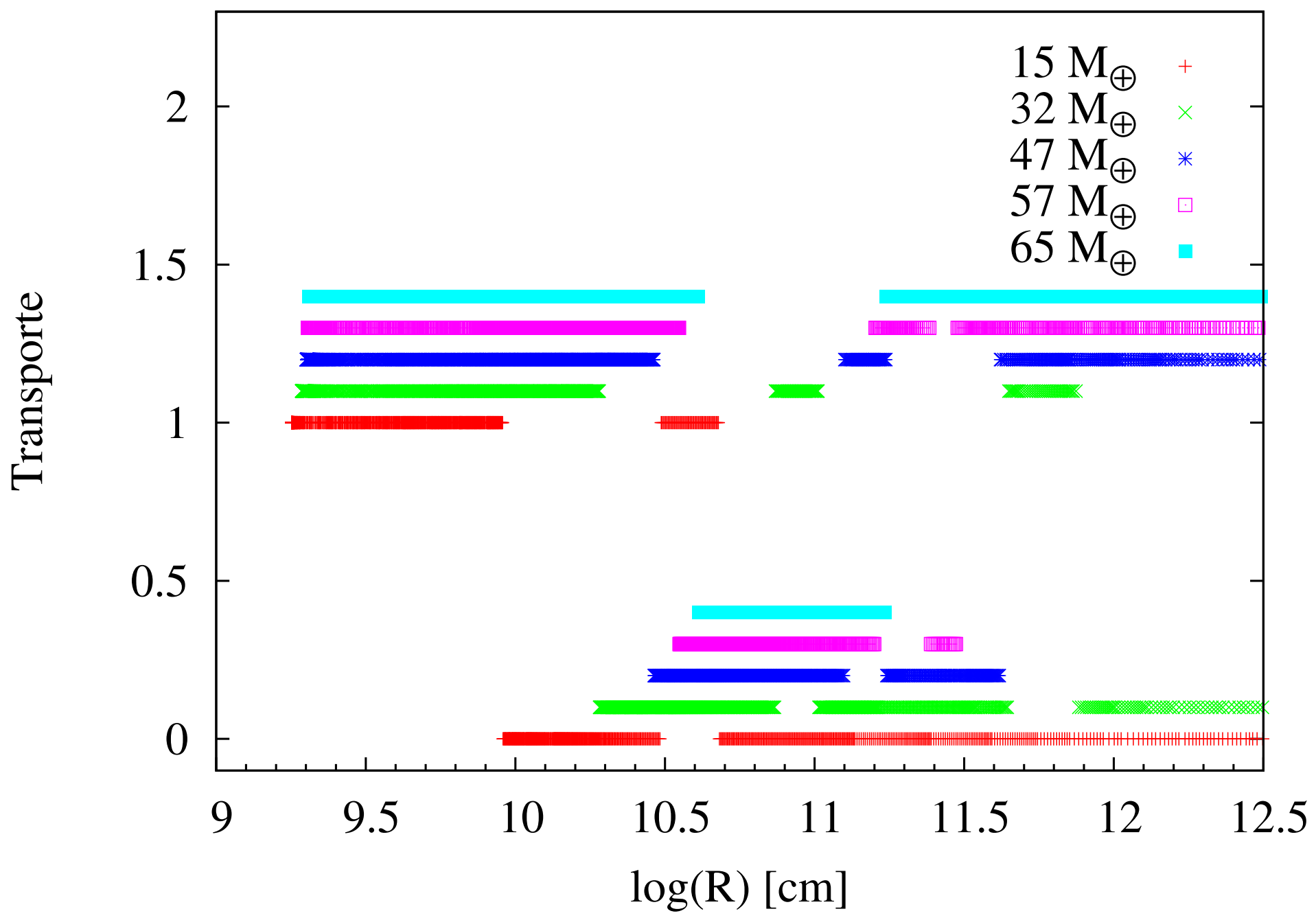

**Figura 6.6.** $J1\sqrt{10}$. Para los mismos casos que el la figura 6.5, mostramos aquí el tipo de transporte que se establece en las diferentes capas del interior del planeta. Los valores entre 0 y 0,5 corresponden a regiones de transporte radiativo, mientras que entre 1 y 1,5 al transporte convectivo (los distintos casos fueron espaciados en 0,1 unidades para su mejor visualización). El transporte radiativo domina la estructura en los estadíos de menor masa. Cuando la masa de gas ligada supera a la masa del núcleo, la convección se hace más importante, tanto en el interior profundo como en las capas externas.

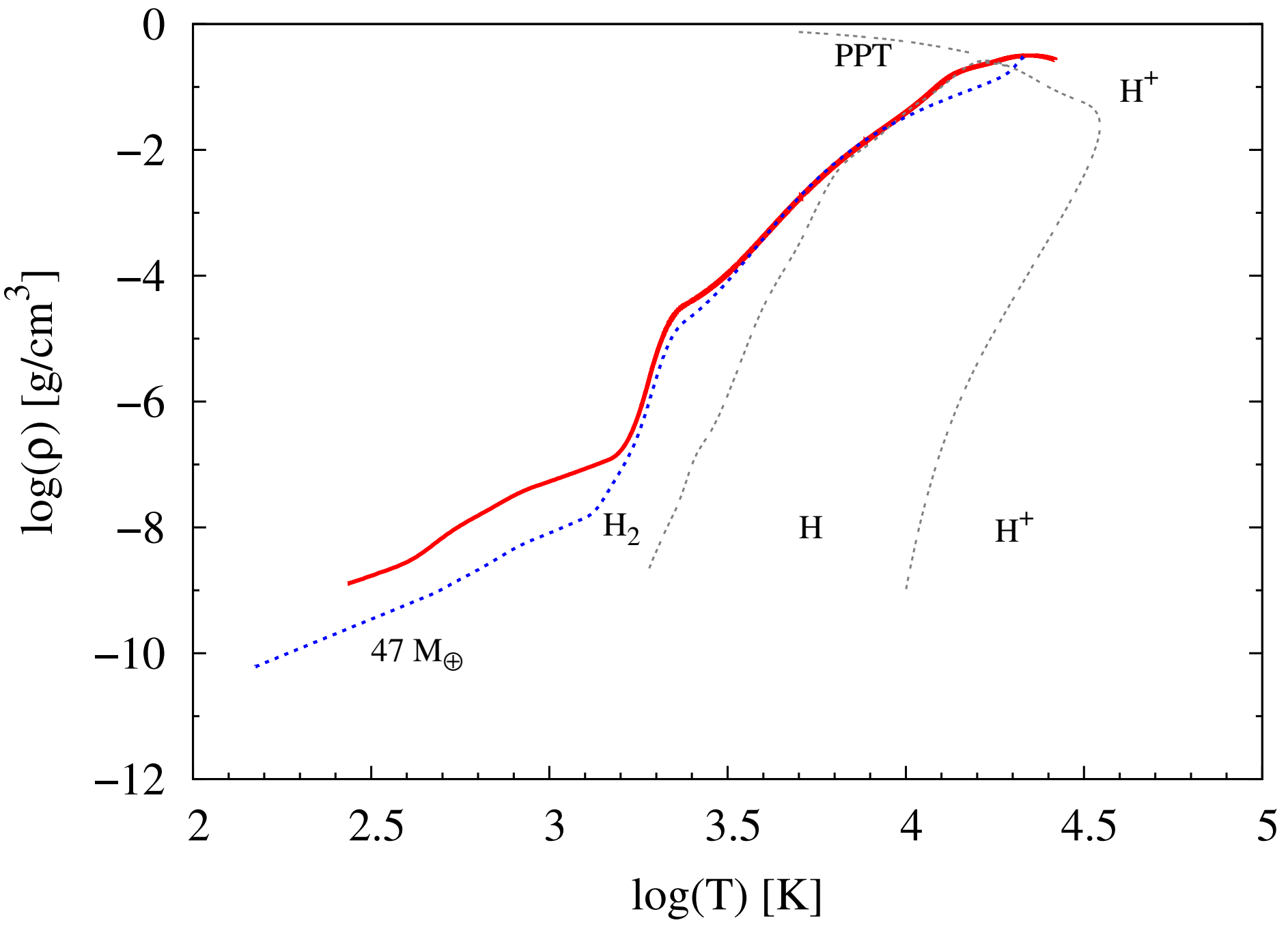

**Figura 6.7.** $J1\sqrt{10}$. El eje x corresponde al logaritmo de la temperatura y el eje y al logaritmo de la densidad. La curva roja muestra la evolución de estas dos cantidades en la base de la envoltura, durante toda la formación del planeta. A medida que el planeta incrementa su masa, las capas de gas del interior se contraen, aumentando su densidad y su temperatura. La curva en línea de puntos azules representa el interior del modelo de 47 $M_{\oplus}$. El valor mínimo responde a las condiciones nebulares, mientras que el máximo corresponde a uno de los puntos de la curva roja. En este gráfico se añadió el diagrama de fases del hidrógeno de la EOS de SCVH. En línea punteada gris se muestra la ubicación teórica de la Plasma Phase Transition (PPT), y las regiones de hidrógeno molecular, atómico e ionizado.

Hay dos características fundamentales que surgen de analizar los datos a radio fijo: cuando aumenta la densidad del disco, el tiempo de cruce disminuye y la masa de cruce aumenta. De la ecuación (4.14), se ve que la tasa de acreción es proporcional a la densidad del disco, por lo que alcanzar una determinada masa para el núcleo será más rápido cuanto más sólidos haya disponibles para ser acretados. El hecho que la masa de cruce no sea en todos los casos la misma sino que aumente con la densidad tiene que ver con que, a mayor acreción, mayor es también la energía que suministran los planetesimales a la envoltura, lo cual retarda el inicio del colapso de las capas de gas. Sin embargo, la reducción porcentual del tiempo de cruce es mucho mayor que el aumento en la masa de cruce. Tanto para el caso de $r_m = 100\,\text{km}$, como para $r_m = 10\,\text{km}$, si comparamos los dos casos de los extremos (nos referimos a las simulaciones correspondientes a 6 y 10 NMMS), al aumentar la densidad el tiempo de cruce se reduce en un factor 2,25 mientras que la masa de cruce solo aumenta un 25 %. Un resultado interesante es que, si nos restringimos al rango de densidades que estamos considerando, y pensamos a la masa de cruce como función del tiempo de cruce, se puede ajustar una recta entre los puntos para cada uno de los casos considerados (ver figura 6.8). Esto es, si:

$$M_{\rm cross} = \mathrm{a}\, t_{\rm cross} + \mathrm{b}, \tag{6.11}$$

obtenemos que para planetesimales de radio $r_m = 100\,\text{km}$, a = -1,6 y b = 36,9; mientras que para $r_m = 10\,\text{km}$, a = -5,95 y b = 46,25. Es importante mencionar que este ajuste no puede extrapolarse para densidades mucho mayores o menores a las consideradas para estos casos. Claramente, para densidades suficientemente grandes nos encontraríamos con tiempo negativos, y para densidades suficientemente pequeñas con masas de cruce también negativas. Sin embargo, por otra parte, el rango de densidades donde vale el ajuste es bastante amplio. Discos con densidades superiores a 10 NSMM serían muy masivos y, probablemente, poco frecuentes. Por otra parte, para densidades bajas el tiempo de formación sería demasiado largo.

Si ahora comparamos los resultados para una misma densidad del disco pero para los diferentes radios de los planetesimales considerados encontramos que para los planetesi-males más pequeños el tiempo de cruce se reduce notablemente pero la masa de cruce aumenta. En este caso, lo que influye en la tasa de crecimiento del núcleo (Ec. (4.14)) para que esto sea así es el radio efectivo de captura, que depende de la velocidad relativa entre los planetesimales y el embrión. De la ecuación (4.16), se ve que $v_{\rm rel}$ depende de la excentricidad, la cual depende a su vez de la masa (y por ende del radio) de los planetesimales (ver Ec. (4.19)). Luego, la ecuación (4.14) depende, aproximadamente, del inverso de la velocidad relativa (esto se ve de la Ec. (4.15), que define el radio efectivo en ausencia de gas). Los planetesimales más pequeños tienen una velocidad relativa más baja puesto que se ven más afectados por la viscosidad nebular, lo cual implica un aumento del radio efectivo de captura del planeta. Esto lleva a que la tasa de acreción aumente, lo que conlleva una reducción en las escalas de tiempo. Al igual que en el caso anterior, el hecho de incorporar más planetesimales por unidad de tiempo implica que la energía que liberan en la envoltura produce que el colapso de las capas de gas ocurra para núcleos más masivos. La reducción en el tiempo de formación al considerar planetesimales de menor radio es inclusive más marcada que cuando se considera un aumento en la densidad nebular para planetesimales de 100 km. Para una misma densidad nebular el tiempo de formación disminuye en un factor 3 si comparamos los resultados para planetesimales de 10 y 100 km. En cuanto al aumento en la masa de cruce para esta misma comparación encontramos que es de un 25 %. Esto implica que si los planetesimales presentes en la nebulosa son en su mayoría de radio menor a los 100 km (o, incluso, a los 10 km), se podrían considerar densidades más bajas para el disco, lo cual llevaría a una reducción en la masa del núcleo, con tiempos de formación todavía dentro de los límites impuestos por las observaciones.

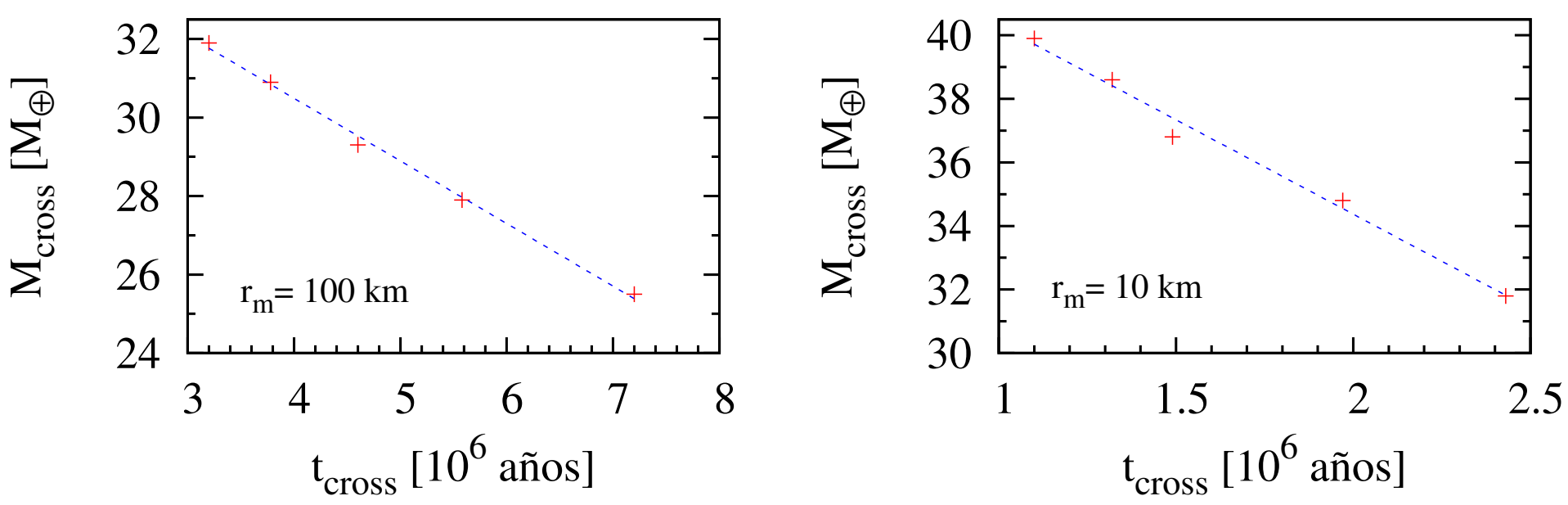

**Figura 6.8.** Ajuste de mínimos cuadrados para los datos de la tabla 6.1. El primer panel corresponde a los casos donde el radio de los planetesimales acretados es de 100 km, y el segundo cuando el radio de 10 km. En los dos casos se ajustó una recta (línea a rayas) entre los valores de la masa de cruce y el tiempo de cruce (cruces rojas) para las condiciones del disco consideradas.

Consideremos ahora dos posibles valores para la densidad del disco (3 NSMM y 7 NSMM), dos radios para los planetesimales acretados ($r_m$ = 10, 100 km) y cinco semiejes (9,55, 8, 6, 5,2, y 4 UA), entre los cuales se incluyen los correspondientes a Júpiter y

Saturno. En la tabla 6.2 se listan los resultados para la masa y el tiempo de cruce para los casos donde el tiempo de cruce sea inferior a 20 millones de años. Como era esperable, el tiempo de cruce disminuye cuando nos acercamos al Sol, puesto que la densidad aumenta en esa dirección. Notemos que ninguno de los casos considerados se ve afectado por la discontinuidad en la línea de hielo. Similarmente, la masa de cruce aumenta en dirección al Sol. Resultados análogos a los anteriores se repiten al aumentar la densidad del disco y al considerar que el tamaño de los planetesimales se reduce. En todos los casos se puede ver el amplio espectro de los tiempos de cruce, el cual se ve seriamente afectado por la distancia al Sol (y siempre teniendo en cuenta la restricción establecida por la vida media de los discos). Como se ve de estos resultados, parece difícil la formación de todos los planetas gigantes del Sistema Solar bajo las condiciones de una nebulosa solar estándar. Tengamos en cuenta que en estas simulaciones no se consideraron los semiejes correspondientes a Urano y Neptuno, mucho más alejados del Sol que Saturno. A pesar de eso, es evidente que el tiempo de formación resultaría extremadamente largo, aún para discos muy masivos.

**Notas relativas a la ecuación de estado y su impacto en los resultados**

En nuestro trabajo, Fortier, Benvenuto & Brunini (2007), presentamos gran parte de los resultados mencionados arriba. Sin embargo, meses después de su publicación, descubrimos errores de tipeo en el artículo original de Saumon, Chabrier & Van Horn (1995). Estos errores involucran a las fórmulas de interpolación de la entropía de mezcla, necesarias para calcular la ecuación de estado de un gas compuesto por hidrógeno y helio. En § 4.4.2 mencionábamos cuáles eran los errores en las ecuaciones. Aquí presentaremos los resultados obtenidos con la EOS sin corregir y los compararemos con la EOS corregida (Fortier, Benvenuto & Brunini 2009). Es interesante analizar el impacto que estos errores tienen en los resultados y sirven para poner de manifiesto la sensibilidad del proceso de formación de un planeta gigante a variaciones en la EOS.

**Tabla 6.2.** Masa y tiempo de cruce variando la ubicación del planeta, la densidad del disco y el radio de los planetesimales acretados. La masa de cruce, $M_{\rm cross}$, está en unidades de masas de la Tierra ($M_\oplus$) y el tiempo de cruce, $t_{\rm cross}$, en millones de años (My).

| $a$ [UA] | 3 NSMM | | | | 7 NSMM | | | |
|---|---|---|---|---|---|---|---|---|
| | $r_m = 100$ km | | $r_m = 10$ km | | $r_m = 100$ km | | $r_m = 10$ km | |
| | $t_{\rm cross}$ | $M_{\rm cross}$ | $t_{\rm cross}$ | $M_{\rm cross}$ | $t_{\rm cross}$ | $M_{\rm cross}$ | $t_{\rm cross}$ | $M_{\rm cross}$ |
| $9,55$ | $> 20$ | – | $> 20$ | – | $> 20$ | – | $8,7$ | $29,7$ |
| $8,00$ | $> 20$ | – | $> 20$ | – | $16,1$ | $24,5$ | $5,6$ | $31,0$ |
| $6,00$ | $> 20$ | – | $12,3$ | $17,0$ | $8,3$ | $27,0$ | $3,2$ | $33,3$ |
| $5,20$ | $> 20$ | – | $10,3$ | $15,9$ | $5,6$ | $27,9$ | $2,0$ | $34,8$ |
| $4,00$ | $18,2$ | $13,0$ | $9,2$ | $13,5$ | $3,3$ | $29,9$ | $1,0$ | $35,5$ |

En la tabla 6.3 mostramos los resultados correspondientes a la EOS sin corregir (tiempo y masa de cruce) para densidades del disco entre 6 y 10 NSMM, y para planetesimales con radios de 100 y 10 km. De comparar estos resultados con los presentados en la tabla 6.1 vemos que, con la EOS sin corregir los tiempos de cruce son más largos pero, fundamentalmente, las masas de cruce son considerablemente mayores, en la mayoría de los casos más del doble que cuando se considera la EOS corregida.

Los errores en la EOS afectan a la entropía de la mezcla (Ec. 4.39), pero fundamentalmente a las derivadas logarítmicas de la misma, $S_{\rm T}$ y $S_{\rm P}$ (Ecs. 4.44, 4.45). Estos valores son muy importantes puesto que su cociente define el gradiente adiabático (Ec. 4.46), el cual, a su vez, es fundamental para determinar el tipo de transporte. Cuando se comparan los valores del gradiente adiabático entre ambas tablas de la EOS, se ve que en la tabla corregida, $\nabla_{\rm ad}$ es siempre menor que en la tabla sin corregir, y bajo las condiciones presentes en el interior del protoplaneta en algunas regiones la diferencia puede llegar hasta un 50 %. Esto implica que nuestros primeros resultados fueron obtenidos sobreestimando los valores del gradiente adiabático. De acuerdo con el criterio de Schwarzschild (4.4.1), cuando $\nabla > \nabla_{\rm ad}$ el transporte se vuelve convectivo. Cuanto mayor sea $\nabla_{\rm ad}$, más difícil se vuelve el desarrollo de regiones convectivas que permitan un transporte más eficiente de la energía hacia el exterior. Esto frena la gran contracción de las capas de gas, lo cual demora el comienzo del crecimiento runaway. Se tiene como consecuencia la prolongación del tiempo para alcanzar la masa de cruce que, en la mayoría de los casos considerados resulta ser cercana a la masa de aislación (esto último es mucho más notorio en el caso de los planetesimales de 10 km que en los de 100 km). La masa de aislación de un protoplaneta es la masa total de sólidos presente en su zona de alimentación. En las simulaciones que que estamos mostrando a modo de ejemplo, la masa de cruce (con la EOS sin corregir) resulta ser muy cercana a la masa de aislación, lo cual implica que la contracción es significativa recién cuando el planeta está cercano a agotar su fuente de planetesimales. Esto quiere decir que el desplome de las capas de gas recién ocurre cuando deja de ingresar la energía suficiente para impedir el colapso, sin que tenga mucha incidencia en esto la capacidad del gas para transportar hacia afuera la energía aportada por los planetesimales.

El error en la fórmula de interpolación solo modificó el valor del gradiente adiabático dado por la EOS (también afectó, pero en forma despreciable, al calor específico a presión constante). En las regiones de interés para el planeta, la diferencia entre ambos valores puede tener máximos de hasta un 50 %, pero en promedio queda acotada a alrededor de un 15 %. Esta diferencia, sin embargo, repercute significativamente en los valores de la masa y el tiempo de cruce.

**Tabla 6.3.** Masa de cruce y tiempo de cruce para los mismos casos presentados 6.1, pero con la EOS sin corregir.

| Masa del disco [MMSN] | Resultados con la EOS sin corregir ($r_m$ = 100 km) | | Resultados con la EOS sin corregir ($r_m$ = 10 km) | | Masa de aislación [$M_\oplus$] |
|---|---|---|---|---|---|
| | Tiempo de cruce [My] | Masa de cruce [$M_\oplus$] | Tiempo de cruce [My] | Masa de cruce [$M_\oplus$] | |
| 6 | 11,40 | 44,70 | 4,95 | 46,70 | 46,52 |
| 7 | 8,75 | 55,40 | 3,53 | 58,70 | 58,60 |
| 8 | 7,00 | 66,60 | 2,65 | 71,40 | 71,60 |
| 9 | 5,70 | 77,60 | 2,00 | 84,85 | 85,00 |
| 10 | 4,80 | 88,50 | 1,65 | 99,00 | 100,00 |

## 6.2. Simulaciones considerando una población de planetesimales que sigue una distribución de tamaños

Como vimos en la sección anterior, al aumentar la densidad en el disco se reduce el tiempo de formación pero también aumenta la masa del núcleo. El mismo efecto se obtiene manteniendo fija la densidad pero disminuyendo el tamaño de los planetesimales acretados. Considerar densidades muy altas implica invocar mecanismos que permitan que ese enriquecimiento ocurra. Sin embargo, considerar planetesimales pequeños es, por el contrario, una hipótesis que se condice completamente con la realidad. De hecho, surge naturalmente pensar que los planetesimales que pueblan el disco no tienen una masa uniforme sino que siguen alguna ley de distribución de masas. Por ejemplo, para el caso de un planeta en la ubicación de Júpiter, se suele aceptar que un valor plausible para la densidad del disco sea $\Sigma = 10\,\mathrm{g\,cm^{-2}}$ (aproximadamente el correspondiente a 3 NSMM). La figura 6.9 muestra el proceso de formación del planeta considerando que los planetesimales acretados tienen un radio uniforme de 10 y 100 km respectivamente. Mientras que en el caso en el que los planetesimales tienen un radio de 100 km el tiempo de formación es de alrededor de 24 millones de años, superando ampliamente la cota impuesta por las observaciones, en el caso de planetesimales de 10 km el tiempo de formación es de 10 millones de años. En pocas palabras, el radio que se considere para los planetesimales acretados hace que, bajo las mismas condiciones nebulares, los resultados de una simulación puedan ser categorizados como "aceptables" o no en función del tiempo que demande su formación.

Kokubo & Ida (2000) estudiaron mediante simulaciones de $N$-cuerpos tridimensionales la ley de distribución de masas entre los planetesimales del disco protoplanetario. Sea $n_c$ el número acumulativo de planetesimales de masa $m$,[1] y sea $n(m)$ el número de planetesimales con masa en el intervalo $(m, m + dm)$. Kokubo & Ida encontraron que la distribución de masa entre los planetesimales resulta bien aproximada por una ley de potencias,

$$\frac{dn_{\rm c}}{dm} \propto n \propto m^{\alpha}. \tag{6.12}$$

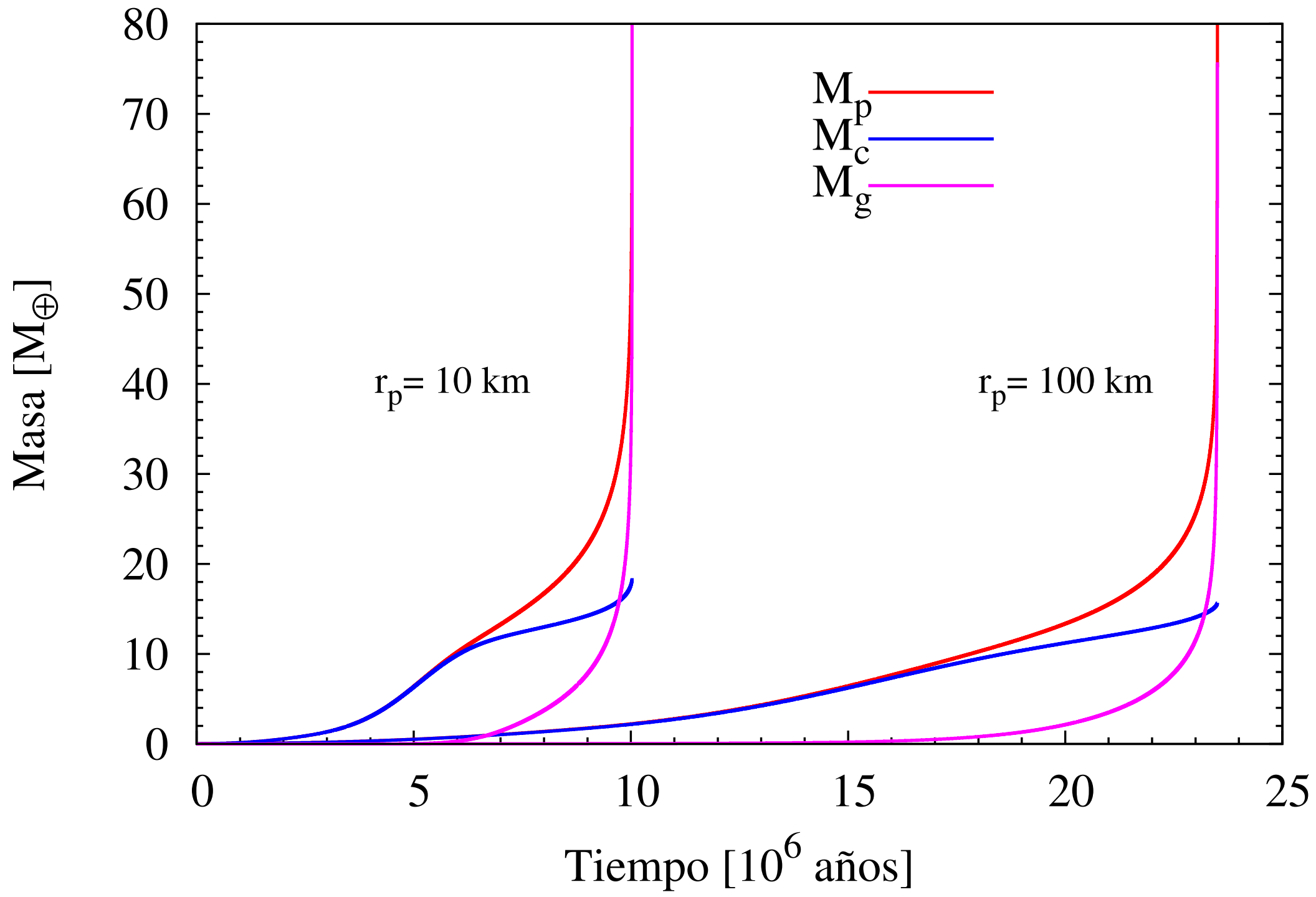

**Figura 6.9.** Evolución de la masa en función del tiempo para un planeta ubicado en $a = 5,2$ UA, bajo las condiciones de un disco protoplanetario de 3 NSMM. En línea roja se muestra la masa total, en azul la masa del núcleo y en fucsia la de la envoltura. Se consideraron dos casos: uno para planetesimales acretados de radio 10 km, y otro para 100 km. Notar que el tiempo que involucra este proceso en cada uno de los casos hace que, cuando $r_m = 100$ km se superen los $10^7$ años impuestos por la vida media de los discos. El caso de planetesimales de 10 km queda dentro de esta cota.

Dado que la masa total es:

$$\int n\, m\, dm \propto m^{\alpha+2}, \tag{6.13}$$

si $\alpha < -2$, los planetesimales más pequeños contendrán la mayor parte de la masa del sistema. Que $\alpha$ sea menor a -2 caracteriza al crecimiento runaway y post-runaway (crecimiento oligárquico), donde los planetesimales dominan la masa total del sistema mientras que los embriones son los únicos que continúan en crecimiento. En base a sus simulaciones, Kokubo & Ida encontraron que $\alpha$ decrece con el tiempo, y que toma valores ente -3 y -2, siendo el valor de referencia $\alpha = -2,5$.

En función de los resultados previos, donde la masa y el tiempo de cruce son significativamente dependientes del tamaño de los planetesimales acretados, un estudio más realista del problema lleva, entonces, a la incorporación de una distribución de tamaños para los planetesimales. Dado que en nuestro modelo la densidad de los planetesimales es constante, independientemente de su radio, transformaremos la ley de potencias para la masa en una ley de potencias para el radio. Como $m \propto r^3$, tenemos:

$$n(m)\, m\, dm \propto r^{3\,\alpha+5} dr, \tag{6.14}$$

que representa la masa contenida en planetesimales cuyo radio se encuentra en el intervalo $(r, r + dr)$.

En la práctica, un intervalo de tamaños que nos sea de interés tendrá que ser discretizado. Sea $r_{\rm max}$ y $r_{\rm min}$ los dos extremos de la distribución que estamos considerando y sea $\eta$ el número de especies en las cuales estará discretizada la distribución. Tomaremos a los valores del radio equiespaciados en escala logarítmica:

$$r_j = r_{\rm min} \left(\frac{r_{\rm max}}{r_{\rm min}}\right)^{\frac{j-1}{\eta-1}} \qquad j = 1, ..., \eta \tag{6.15}$$

donde, claramente, $r_1 = r_{\rm min}$ y $r_\eta = r_{\rm max}$. Sea $f(r) = Cr^\beta$ la ley de potencias para el radio, donde $f(r)\, dr$ es el número de planetesimales con radio en $(r, r + dr)$. Para satisfacer la ecuación (6.14), $\beta = 3\,\alpha + 5$. La masa total contenida en el intervalo $(r_{\rm min}, r_{\rm max})$ será:

$$\int_{r_{1-1/2}}^{r_{\eta+1/2}} f(r) dr = C \frac{r_1^{\beta+1}}{\beta+1} \left(\frac{r_\eta}{r_1}\right)^{-\frac{\beta+1}{2(\eta-1)}} \left[\left(\frac{r_\eta}{r_1}\right)^{\frac{\eta(\beta+1)}{\eta-1}} - 1\right]. \tag{6.16}$$

[1] $n_{\rm c}$ representa el número total de cuerpos de masa mayor a $m$.

Notemos que los límites de la integral no son $r_1$ y $r_\eta$, sino que se toman por fuera del intervalo. Ésto es así porque se está considerando que los planetesimales con radio en $(r_{j-1/2}, r_{j+1/2})$ están representados por una especie de radio $r_j$. Luego, si definimos

$$\Delta = \left(\frac{r_{\text{max}}}{r_{\text{min}}}\right)^{\frac{1}{\eta+1}}, \tag{6.17}$$

la fracción $\mathcal{F}$, respecto del total, de planetesimales con radios en $(r_{j-1/2}, r_{j+1/2})$ es:

$$\mathcal{F} = \frac{\Delta^{(j-1)(\beta+1)}[\Delta^{\beta+1} - 1]}{\Delta^{\eta(\beta+1)} - 1}. \tag{6.18}$$

En el tratamiento numérico, cada bin de tamaño se trabaja en forma independiente. Se calcula el radio efectivo para cada una de las especies de planetesimales, se establece la tasa de acreción de sólidos en cada caso y se elimina la masa de planetesimales acretada por cada uno de los bines. En cada capa de la envoltura se deposita la energía total que dejan los planetesimales, que es la suma de la energía que pierde por viscosidad cada una de las especies. Hechos estos cálculos, se procede a la resolución del modelo de la manera usual.

### 6.2.1. El modelo de Niza

Las teorías de formación del Sistema Solar indican que los planetas gigantes se formaron en órbitas circulares y coplanares. Sin embargo, esa no es, estrictamente hablando, la configuración que presentan actualmente los planetas gigantes del Sistema Solar (ver tabla 1.1). Tsiganis et al. (2005) mostraron que un sistema planetario donde las órbitas iniciales sean coplanares y cuasi circulares puede evolucionar hasta el estado actual, siempre y cuando, Júpiter y Saturno crucen la resonancia de movimientos medios[2] 1:2.

Una vez que los planetas gigantes culminaron su formación y se disipó la nebulosa solar, los planetas migraron debido al intercambio de momento angular con el disco de planetesimales. En este proceso, Júpiter debió migrar hacia el interior mientras que Saturno, Urano y Neptuno lo hicieron en el sentido contrario. Durante la migración, las excentricidades e inclinaciones de los planetas se amortiguaron debido a la fricción dinámica con las partículas del disco. Si las órbitas originales de los planetas estaban lo suficientemente próximas entre sí , es probable que hayan tenido que atravesar una resonancia de movimientos medios de bajo orden, lo cual excitaría las excentricidades de los planetas involucrados. Tsiganis et al. (2005) realizaron simulaciones numéricas considerando los siguientes semiejes iniciales: $a_J = 5,45$ UA, $a_S = 8,65$ UA (en el interior de la resonancia 1:2 con Júpiter), y $a_N, a_U \in [11, 17]$ UA (con $a_J$, $a_S$, $a_N$ y $a_U$ los semiejes de Júpiter, Saturno, Neptuno y Urano respectivamente), y la presencia de un disco de planetesimales (de entre 30 y 50

[2]Dos planetas se encuentran en resonancia de movimientos medios si existe una relación de conmensurabilidad entre sus períodos orbitales.

$M_\oplus$), que se extiende hasta 30-35 UA. Tsiganis et al. encontraron que Júpiter y Saturno cruzan la resonancia 1:2 durante su migración, lo cual excita sus excentricidades hasta valores compatibles con los que se miden actualmente. Este cambio en las excentricidades tiene un efecto drástico en todo el sistema planetario, ya que fuerza el incremento de las excentricidades de Urano y Neptuno. Las órbitas de los planetas se intersectan, lo que provoca cambios en las inclinaciones y que los gigantes de hielo migren hacia el exterior, adentrándose en el disco. De hecho, en la mitad de sus simulaciones, Tsiganis et al. obtienen que Urano y Neptuno intercambian posiciones. Que Neptuno se haya formado en una órbita interior a la de Urano explicaría la mayor abundancia de sólidos en su interior. Luego, la fricción dinámica estabiliza al sistema y la migración termina cuando los planetesimales son eyectados del disco. La configuración final del sistema, en la mayoría de sus simulaciones, muestra un excelente acuerdo con los valores actuales. Este modelo se conoce como *modelo de Niza*. El modelo de Niza consigue explicar las características orbitales más importantes de los planetas gigantes del Sistema Solar; esto es, los semiejes, excentricidades e inclinaciones. Además, es consistente con la existencia de los satélites regulares, de los Troyanos de Júpiter (y, probablemente también, con los de Neptuno), y no contradice la distribución que presenta el Cinturón Principal de Asteroides.

### 6.2.2. Resultados

En general, las simulaciones de formación de planetas gigantes dentro del marco de la inestabilidad nucleada, pueden dar cuenta de Júpiter y Saturno en tiempos razonablemente cortos (Pollack et al.1996, Hubickyj, Bodenheimer & Lissauer 2005, Dodson-Robinson et al. 2008b). Sin embargo, tanto las estimaciones analíticas como las numéricas fracasan a la hora de explicar la formación de Urano y Neptuno en escalas de tiempo acordes a la vida media de la nebulosa (Lissauer 1987, Thommes, Duncan & Levison 2003, Hillenbrand 2005, Lissauer & Stevenson 2007). La clave está en que, en las ubicaciones actuales de estos planetas, la densidad superficial de sólidos de la nebulosa primordial tendría que haber sido muy baja, lo cual extendería el proceso de acreción. De hecho, que la masa de gas de Urano y Neptuno sea bastante inferior a la masa de su núcleo podría tomarse como un indicador de que estos planetas no hayan alcanzado nunca el runaway de gas, probablemente porque la nebulosa se habría disipado antes de que esto ocurriera. Sin embargo, de acuerdo con el modelo de Niza, Urano y Neptuno se formaron mucho más cerca del Sol de lo que se encuentran actualmente. De haber sido esto así , estos planetas habrían tenido mayor cantidad material disponible, lo cual habría facilitado la acreción, acelerando el proceso.

Es importante notar que, si la configuración inicial del Sistema Solar exterior era más compacta que la actual, la nebulosa solar estándar, por construcción, resulta inconsistente con este modelo. Basándose en argumentos similares a los de Weidenschilling (1977) y Hayashi (1981), pero considerando una arquitectura primordial compatible con el modelo de Niza, Desch (2007) calculó el perfil de densidad nebular correspondiente a este caso. Al igual que para la nebulosa de masa mínima, Desch dividió al disco protoplanetario en anillos centrados en las órbitas de cada uno de los planetas (pero ahora, según lo establecido por el modelo de Niza) y distribuyó uniformemente la masa total de sólidos que ellos contienen para poder estimar la densidad del disco protoplanetario. Dado que no todo el material sólido presente en el disco es incorporado por los planetas, ya que diversos mecanismos lo remueven de la zona de alimentación de los planetas antes de ser acretados (migración, fotoevaporación, etc.), la nebulosa solar estará enriquecida respecto de la masa total de sólidos estimada hoy en día. De acuerdo a estas hipótesis, Desch encontró que el perfil de densidad de la nebulosa compatible con el modelo de Niza decrece con la distancia en forma más abrupta que en el caso de la nebulosa estándar, aunque siguiendo también una ley de potencias, análogo a la ecuación (4.2). Entonces, $\Sigma = \Sigma_0 \, a^{-p}$, con $p = 2{,}168$. La relación entre la masa de sólidos y de gas se establece en 1 en 100, en correspondencia con la abundancia solar.

En base al modelo de Niza y al perfil nebular calculado por Desch, realizamos simulaciones para estudiar la formación de Júpiter, Saturno, Urano y Neptuno. Los radios orbitales adoptados fueron: $a_J = 5,5$ UA, $a_S = 8,3$ UA, $a_U = 14$ UA y $a_N = 11$ UA. Notemos que estamos aceptando que, después de su formación y durante su migración hacia el exterior del Sistema Solar, Urano y Neptuno intercambiaron lugares. El estudio de la formación de cada planeta se hizo en forma independiente, lo cual significa que no están contempladas las interacciones entre ellos, ni cómo la presencia de cada uno podría afectar la formación del resto. Como ya mencionamos en el Capítulo 4, nuestro modelo no considera la migración planetaria, sino que la formación es calculada *in situ*. Thommes, Matsumura & Rasio (2008) calcularon la evolución de sistemas planetarios compatibles con el modelo de Niza y encontraron que, de sus simulaciones, surgen análogos al Sistema Solar siempre y cuando, durante su formación, los planetas gigantes no hayan sufrido una migración empinada y se mantuvieran en órbitas circulares. Esto le da validez a nuestra hipótesis. Es importante hacer énfasis en que la migración de los planetas gigantes predicha por el modelo de Niza ocurriría una vez disipada la nebulosa solar, y después de que los planetas han alcanzado su masa final.

Hemos considerado que en el disco de sólidos la densidad superficial dada por:

$$\Sigma = 11 \left( \frac{a}{5{,}5} \right)^{-2{,}168} \mathrm{g\,cm^{-2}}. \tag{6.19}$$

La densidad superficial de gas nebular sigue el mismo perfil de la densidad de sólidos, y ambas reproducen la abundancia solar. La temperatura de la nebulosa, en el caso de Júpiter, se tomó en $\sim$ 130 K, mientras que para los tres planetas restantes se adoptó la mínima compatible con las tablas de la EOS, $T = 125$ K.

En cuanto al tamaño de los planetesimales, aceptaremos que siguen una distribución de masa según lo propuesto por Kokubo & Ida (2000), con $n(m) \propto m^{\alpha}$. Tomaremos como valor de referencia $\alpha = -2,5$. Consideraremos que la población de planetesimales está representada por 9 especies, con $r_{min} = 0,03$ km y $r_{max} = 100$ km. La distribución de tamaños la hacemos según lo explicado en la ecuación (6.18). Dado que los planetesimales sufren un decaimiento orbital debido a la viscosidad de la nebulosa, no todos podrán ser acretados en forma eficiente. En particular, los más pequeños son los que se verán más afectados por este efecto. Si se estiman las escalas de tiempo de migración de los planetesimales y de acreción por parte del protoplaneta para diversos tamaños, se puede obtener en límite inferior para el radio de los planetesimales a partir del cual son válidas las hipótesis de nuestro planteo. De esta estimación es que surge como valor adecuado un radio mínimo de 30 m. Para el caso de los resultados que presentaremos a continuación, hemos simulado la presencia de otros embriones creciendo en la vecindad del planeta del cual estamos calculando la formación, poniéndole una restricción a la tasa de acreción de sólidos. Al igual que hicieron Hubickyj, Bodenheimer & Polack (2005), luego que se alcanza el máximo de la acreción de sólidos, inhibimos el aumento de la masa de sólidos que puede acretar el planeta. Esto es, se considera que la zona de alimentación no se repuebla, sino que a partir de este máximo el planeta solo tiene posibilidad de acretar el material que había presente hasta ese momento en su zona de alimentación.

Sobre la base del modelo que acabamos de delinear realizamos las simulaciones para la formación de Júpiter, Saturno, Urano y Neptuno. Las condiciones iniciales nebulares en las ubicaciones de los cuatro planetas gigantes y los resultados de nuestras simulaciones se listan en la tabla 6.4. En el caso de Júpiter y Saturno, las simulaciones involucran el cálculo del runaway de gas, mientras que para Urano y Neptuno los cálculos se terminan artificialmente cuando alcanzan su masa actual. Como se puede observar, en todos los casos el tiempo de formación es inferior a los 10 millones de años que imponen las observaciones de los discos circumestelares. En la figura 6.10 se muestra la evolución de la masa total y de la masa del núcleo en cada uno de los casos. Es evidentemente notorio el acuerdo entre las masas finales de los núcleos que surgen de las simulaciones y las estimaciones más recientes que surgen de los modelos del interior planetario. En el Capítulo 2 hicimos mención del estado actual del conocimiento de la estructura de los planetas gigantes. Las incertezas en la EOS, y la falta de datos observacionales más precisos (sobre todo en el caso de Urano y Neptuno), hacen que la masa del núcleo de estos planetas pueda ser estimada muy pobremente. En el caso de Júpiter, Saumon & Guillot (2004), utilizando la EOS de SCVH, encontraron que sus modelos de estructura son compatibles con una masa para el núcleo que se encuentre entre 0 y 12 $M_\oplus$. Recientemente, Nettelmann et al. (2008) y Militzer et al. (2008) hicieron nuevas estimaciones, independientes entre sí , utilizando ecuaciones de estado derivadas a partir de primeros principios. Sin embargo, sus estimaciones no son concordantes. Mientras que Nettelmann et al. ubican a la masa del núcleo en el intervalo [0,7] $M_\oplus$, Militzer et al. lo hacen en [14,18] $M_\oplus$. Militzer & Hubbard (2008) argumentaron que esta diferencia se debe al tratamiento que ambos grupos le dieron a la transición entre el hidrógeno molecular e ionizado. Si bien nuestras simulaciones fueron realizadas utilizando la EOS de SCVH, por cuanto lo más coherente sería limitarnos a comparar nuestros resultados con los de Saumon & Guillot (2004), la controversia que existe sobre cuál es la masa de los núcleos de los planetas gigantes pone de manifiesto que los valores que se manejan actualmente solo pueden tomarse como una referencia aproximada. En el caso de Saturno, las estimaciones más recientes son las de Saumon & Guillot (2004), que establecen que la masa del núcleo podría tomar valores entre 9 y 22 $M_\oplus$. Los casos de Urano y Neptuno son los más inciertos debido a la falta de datos precisos provenientes de sondas espaciales. Las cotas superiores para la masa del núcleo, de acuerdo con Guillot (2005) son 14 y 16,6 $M_\oplus$ para Urano y Neptuno respectivamente, mientras que las cotas inferiores son 9 y 12,4 $M_\oplus$ según las estimaciones de Podolak, Podolak & Marley (2000).

**Tabla 6.4.** Condiciones nebulares y resultados para la formación de los planetas gigantes del Sistema Solar. El radio orbital, $a$, y la densidad se sólidos inicial, $\Sigma_0$, surgen como consecuencia del modelo de Niza. La masa de aislación para cada uno de los casos, $M_{\rm iso}$, se lista con fines comparativos. La masa del núcleo y el tiempo de formación que surge de nuestras simulaciones se encuentran en muy buen acuerdo con los límites observaciones en los cuatro casos. Dado que Urano y Neptuno no llegan al runaway de gas, pues alcanzan su masa final antes, presentamos aquí los resultados finales para la masa del núcleo y el tiempo de formación, y no la masa y el tiempo de cruce como en los ejemplos anteriores.

| Planeta | $a$ [UA] | $\Sigma_0$ [g cm$^{-2}$] | $M_{\rm iso}$ [M$_\oplus$] | $M_{\rm c}$ [M$_\oplus$] | $t$ [$10^6$ años] |
|---|---|---|---|---|---|
| Júpiter | 5,5 | 11,0 | 15,3 | 13,34 | 0,52 |
| Saturno | 8,3 | 4,5 | 13,9 | 12,24 | 1,71 |
| Neptuno | 11,0 | 2,4 | 12,8 | 11,48 | 4,00 |
| Urano | 14,0 | 1,4 | 12,3 | 11,00 | 6,65 |

El logro de formar a los cuatro planetas gigantes en tiempos razonablemente cortos y con masas para los núcleos compatibles con las estimaciones actuales se debe al hecho de considerar que los planetesimales del disco siguen una distribución donde los más pequeños son los más abundantes. Como vimos en la sección previa, considerar planetesimales chicos produce una aceleración significativa del proceso de formación. Esto nos permite mantener baja la densidad superficial de sólidos, dado que la masa del núcleo no superará, en general, la masa de aislación. El hecho quizá más relevante es que, con mismo perfil de densidad nebular y sin la necesidad de hipótesis extra, se logra dar cuenta de la formación de los cuatro planetas en menos $10^7$ años. Este tipo de resultados no habían sido obtenidos hasta el momento y forman parte de un artículo recientemente aceptado para su publicación (Benvenuto, Fortier & Brunini 2009).

El efecto de frenado por la viscosidad del gas favorece la captura de los planetesimales más pequeños. Esto provoca que la tasa de acreción sea más alta para los planetesimales en el extremo inferior de la distribución, pero también implica que son los primeros en agotarse. En la figura 6.11 vemos la evolución de la densidad superficial de sólidos correspondiente a cada una de las especies consideradas. Los planetesimales de 30 m de radio se agotan en $2 \times 10^5$ años, mientras que la población de los más grandes, de 100 km, prácticamente no varía en ese intervalo de tiempo. Notemos, además, que al principio de la formación, la densidad superficial se mantiene constante y que, al cabo de cierto tiempo (distinto para cada una de las especies), la pendiente de la curva comienza a variar rápidamente. Esto ocurre cuando la viscosidad comienza a ser efectiva, y aumenta el radio de captura (Ec. 4.15). Dado que el efecto que tiene la fricción gaseosa en el frenado de los planetesimales es inversamente proporcional a su radio, los más pequeños son los primeros en experimentarlo, por cuanto el decaimiento en la densidad superficial de sólidos ocurre antes en estos casos.

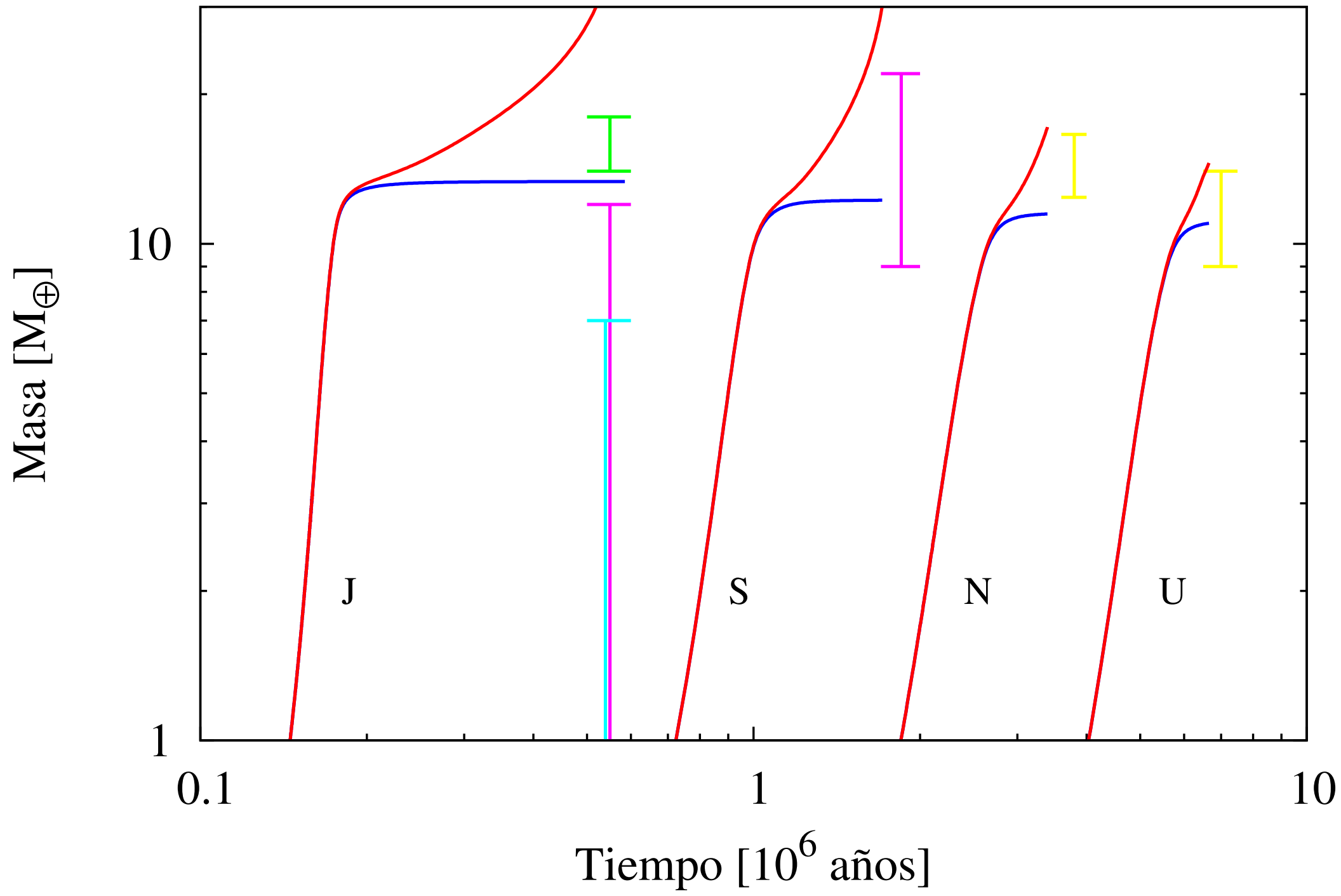

**Figura 6.10.** Evolución de la masa durante la formación de Júpiter (J), Saturno (S), Urano (U) y Neptuno (N). En línea roja se muestra la masa total y en azul la masa del núcleo. Las simulaciones se hicieron en base a las condiciones de borde que surgen del modelo de Niza, y considerando que los planetesimales acretados siguen una distribución de tamaños. Las líneas verticales representan distintas estimaciones de la masa del núcleo para los cuatro planetas: en verde la de Militzer et al. (2008), en turquesa la de Nettelmann et al. (2008), en fucsia la de Saumon & Guillot (2004) y en amarillo la de Guillot (2005) para los límites superiores y Podolak, Podolak & Marley (2000) para los inferiores.

El valor de $\alpha$ fue elegido igual a -2,5, pero en realidad la restricción para $\alpha$, según Kokubo & Ida (2000), es que esté acotado al intervalo $[-3, -2]$. Para estudiar la sensibilidad de los resultados con la potencia de la distribución realizamos las simulaciones para los casos de Júpiter, Saturno, Urano y Neptuno para otros 4 valores de $\alpha$ ($\alpha = -2,17, -2,33, -2,67, -2,83$). En la tabla 6.5 mostramos la fracción, respecto de la masa total, asociada a cada una de las 9 especies de planetesimales consideradas. Cuanto más chico es el valor de $\alpha$, más pronunciada es la distribución, con lo cual la fracción de planetesimales pequeños es mayor. Eso hace que la formación sea más rápida y que la masa del núcleo sea más grande (6.6). A medida que $\alpha$ se acerca a -3, la variación de los resultados es menos importante. Esto se debe a que los planetesimales más chicos de la distribución contienen casi el total de la masa. Notemos que, mientras que para $\alpha = -2,17$ solo el 40 % de la masa está contenida en planetesimales de un radio de 30 m, para $\alpha = -2,83$ estos mismos planetesimales contienen el 92 %. Solo en el caso de $\alpha = -2,17$, el tiempo de formación de Urano supera los $10^7$ años. Para el resto de las simulaciones, los tiempos de formación de todos los planetas son inferiores a esta cota.

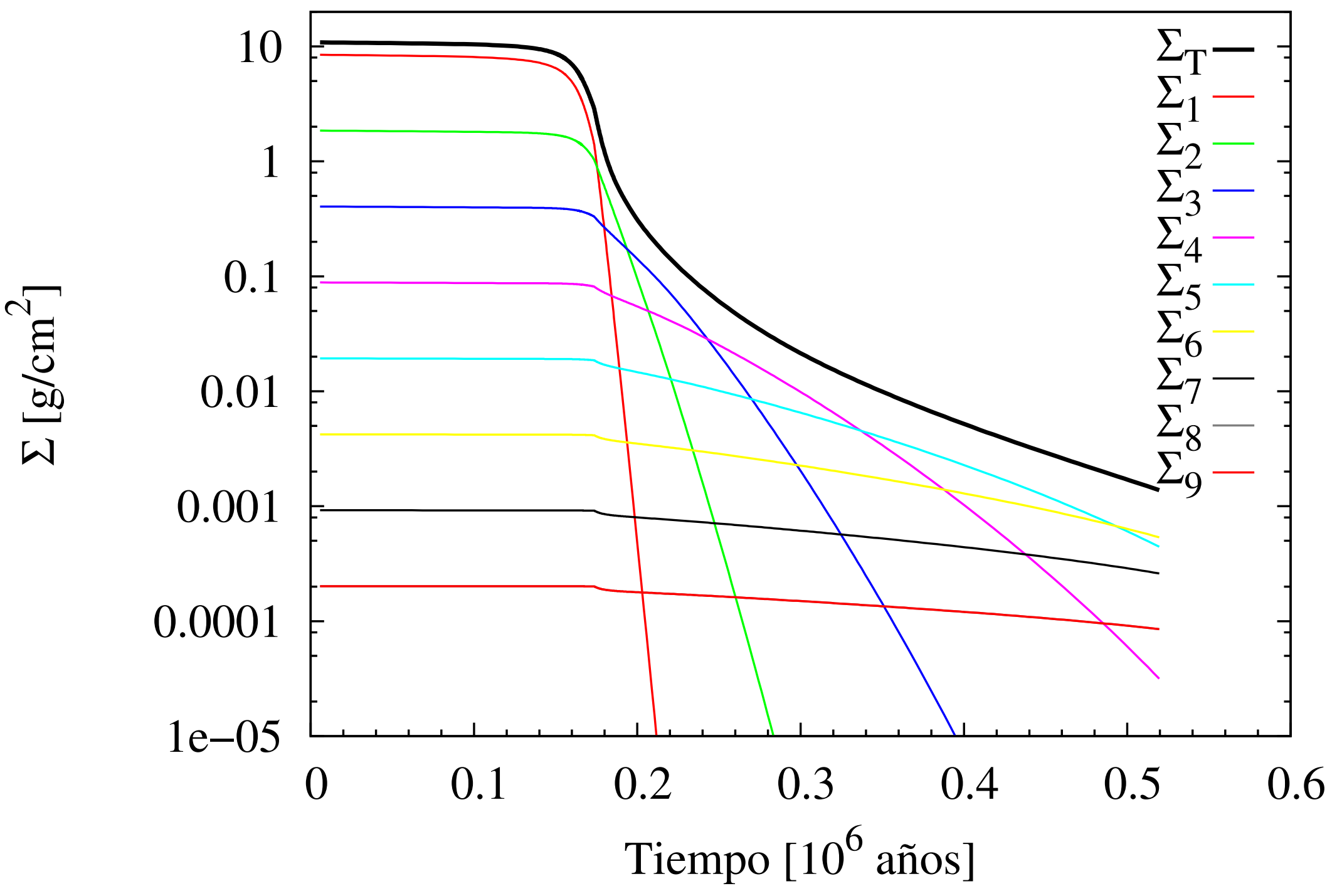

**Figura 6.11.** Evolución de la densidad superficial de sólidos durante la formación de Júpiter. La densidad total, $\Sigma_{\mathrm{T}}$, es la suma de las densidades correspondientes a cada uno de los tamaños de los planetesimales considerados. La población de planetesimales está representada por nueve especies cuyos radios, de menor a mayor, son: $r_1 = 0,03$ km, $r_2 = 0,08$ km, $r_3 = 0{,}23$ km, $r_4 = 0,63$ km, $r_5 = 1,73$ km, $r_6 = 4,77$ km, $r_7 = 13,16$, $r_8 = 36,27$ km y $r_9 = 100,00$ km. En el gráfico, $\Sigma_i = \Sigma(r_i)$.

Por último, vamos a analizar la dependencia del tiempo de formación del planeta con el tamaño del planetesimal más pequeño de la distribución. Consideramos nuevamente que $\alpha = -2,5$, pero ahora la población de planetesimales estará compuesta por 7 especies, con $r_{\rm min} = 0,1$ km y $r_{\rm max} = 100$ km. Los resultados son, para Júpiter, $t_f = 0,89$ My; para Saturno, $t_f = 3,01$ My; para Neptuno, $t_f = 6,39$ My; y para Urano, $t_f = 12,6$ My. El tiempo de formación se extiende en un 70 % para Júpiter y en casi un 90 % para Urano comparado con los casos de 9 especies. Esto pone de manifiesto, una vez más, la importancia del tamaño de los planetesimales acretados, y cómo la especie más pequeña de la distribución regula la escala del tiempo de formación.

## 6.3. Resultados inesperados: eventos cuasi periódicos de variación de la masa de la envoltura

Durante el desarrollo de todo nuestro trabajo nos hemos encontrado con simulaciones que mostraban alteraciones en el proceso de acreción de gas respecto de la evolución espe-rada, que es la que hemos descripto en las secciones previas. En los casos que presentamos antes, la masa de la envoltura crece monótonamente durante la formación del planeta. Sin embargo, esto no siempre es así , como podemos observar de la figura 6.12. Existen situaciones donde la envoltura se vuelve inestable, provocando eventos cuasi periódicos de variación de su masa, involucrando en cada uno de los "períodos" la expulsión y subsiguiente acreción de gas (Benvenuto, Brunini & Fortier 2007). La figura 6.12 muestra dos casos, a priori cualitativamente distintos, de estos eventos y de su impacto en la formación de un planeta. El primer panel corresponde a la formación de un planeta cuyo radio orbital es $a = 5,2$ UA, considerando la densidad superficial para un disco de 3 NSMM y donde los planetesimales son todos del mismo tamaño, en este caso, de 1 km de radio. Como se puede observar, las inestabilidades en la masa de gas ocurren cuando el núcleo ha, prácticamente, completado su formación, y la envoltura comienza a ser una fracción no despreciable de la masa total. La amplitud de la variación en la masa de gas aumenta conforme pasa el tiempo. La finalización abrupta de esta simulación tiene que ver con la convergencia de nuestro código, la cual se ve dificultada o impedida después de algunos "períodos" para este tipo de casos. Sin embargo, la simulación presentada en el segundo panel muestra una situación diferente. Si bien también la envoltura presenta sucesivos eventos de pérdida de gas, la inestabilidad termina siendo amortiguada y la evolución del planeta continúa de la forma habitual. En lo que sigue llamaremos al primer ejemplo de inestabilidad InC (Inestabilidad Continua), y al segundo, InD (Inestabilidad Dampeada).

**Tabla 6.5.** Fracción respecto de la masa total para cada especie de planetesimales, con $\alpha$ entre -2 y -3. La primera columna muestra los valores de los radios representantes de cada una de las 9 especies de la distribución. Para 5 valores de $\alpha$, las entradas de la tabla corresponden a la fracción de planetesimales asociada a cada una de las especies.

| $r$ [km] | $\alpha = -2,17$ | $\alpha = -2,33$ | $\alpha = -2,5$ | $\alpha = -2,67$ | $\alpha = -2,83$ |
|---|---|---|---|---|---|
| $0,03$ | $4,01 \times 10^{-1}$ | $6,37 \times 10^{-1}$ | $7,81 \times 10^{-1}$ | $8,68 \times 10^{-1}$ | $9,20 \times 10^{-1}$ |
| $0,08$ | $2,42 \times 10^{-1}$ | $2,31 \times 10^{-1}$ | $1,70 \times 10^{-1}$ | $1,14 \times 10^{-1}$ | $7,29 \times 10^{-1}$ |
| $0,23$ | $1,45 \times 10^{-1}$ | $8,38 \times 10^{-2}$ | $3,73 \times 10^{-2}$ | $1,50 \times 10^{-2}$ | $5,78 \times 10^{-3}$ |
| $0,63$ | $8,78 \times 10^{-2}$ | $3,04 \times 10^{-2}$ | $8,15 \times 10^{-3}$ | $1,97 \times 10^{-3}$ | $4,58 \times 10^{-4}$ |
| $1,73$ | $5,28 \times 10^{-2}$ | $1,10 \times 10^{-2}$ | $1,78 \times 10^{-3}$ | $2,60 \times 10^{-4}$ | $3,63 \times 10^{-5}$ |
| $4,77$ | $3,18 \times 10^{-2}$ | $4,00 \times 10^{-3}$ | $3,89 \times 10^{-4}$ | $3,42 \times 10^{-5}$ | $2,88 \times 10^{-6}$ |
| $13,16$ | $1,91 \times 10^{-2}$ | $1,45 \times 10^{-3}$ | $8,50 \times 10^{-5}$ | $4,51 \times 10^{-6}$ | $2,28 \times 10^{-7}$ |
| $36,27$ | $1,15 \times 10^{-2}$ | $5,27 \times 10^{-4}$ | $1,85 \times 10^{-5}$ | $5,93 \times 10^{-7}$ | $1,81 \times 10^{-8}$ |
| $100,00$ | $6,96 \times 10^{-3}$ | $1,91 \times 10^{-4}$ | $4,06 \times 10^{-6}$ | $7,81 \times 10^{-8}$ | $1,43 \times 10^{-9}$ |

**Tabla 6.6.** Tiempo de formación y masa del núcleo en simulaciones considerando 9 especies de planetesimales con radios entre 0,03 y 100 km. Los resultados corresponden a cada uno de los 4 valores para $\alpha$ adoptados en la distribución de tamaños. El caso para $\alpha = -2,5$ aparece en la tabla 6.4.

| Planeta | $\alpha = -2,17$ | | $\alpha = -2,33$ | | $\alpha = -2,67$ | | $\alpha = -2,83$ | |
|---|---|---|---|---|---|---|---|---|
| | $t_f$ [My] | $M_c$ [$M_\oplus$] | $t_f$ [My] | $M_c$ [$M_\oplus$] | $t_f$ [My] | $M_c$ [$M_\oplus$] | $t_f$ [My] | $M_c$ [$M_\oplus$] |
| Júpiter | 1,70 | 11,78 | 0,80 | 12,90 | 0,44 | 13,61 | 0,38 | 13,73 |
| Saturno | 3,93 | 10,90 | 2,03 | 12,00 | 1,52 | 12,42 | 1,45 | 12,52 |
| Neptuno | 7,16 | 10,00 | 4,15 | 11,00 | 3,11 | 11,67 | 2,95 | 11,72 |
| Urano | 13,70 | 9,55 | 8,15 | 10,60 | 6,73 | 11,32 | 6,35 | 11,30 |

En general, en la mayoría de las simulaciones en que encontramos inestabilidades de la envoltura, corresponden a casos análogos al InC. Lo que tienen en común todos los casos inestables es que ocurren cuando la masa del núcleo alcanza valores muy cercanos a la masa de aislación. Si bien no es una condición suficiente haber agotado el material que rodea al planeta para que se produzca la inestabilidad (como lo hemos comprobado en casos previos), sí parece ser una condición necesaria.

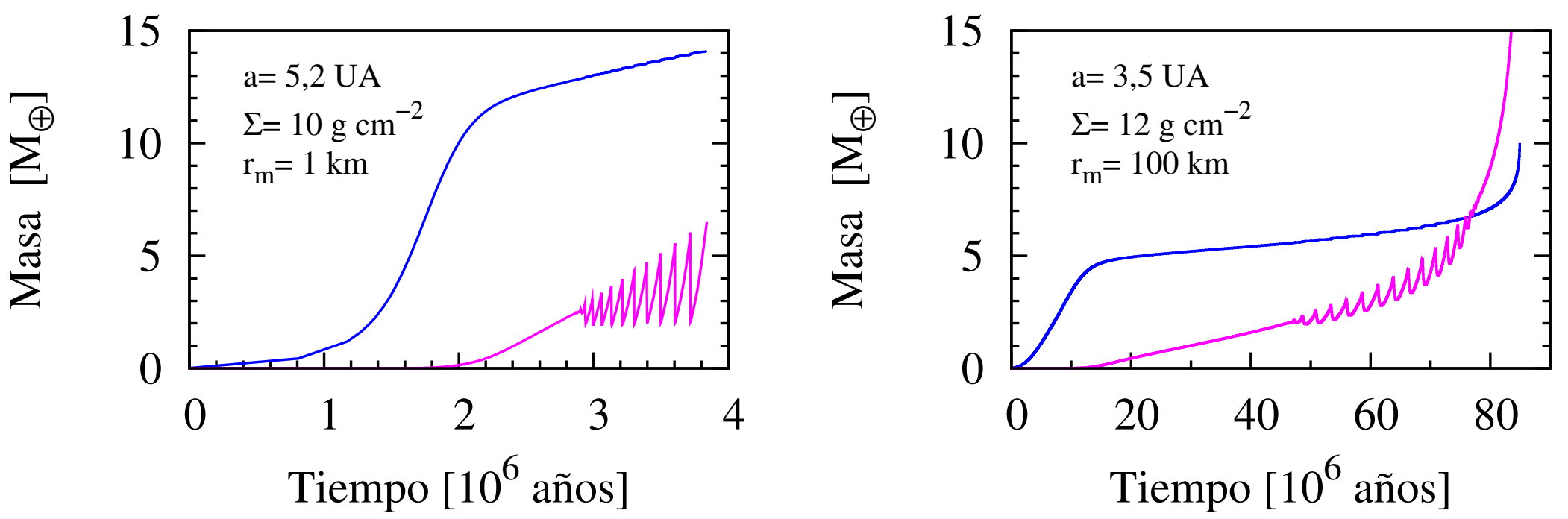

**Figura 6.12.** Evolución de la masa del núcleo (azul) y de la envoltura (fucsia) para dos casos donde se producen inestabilidades en la envoltura. El primer panel representa el caso de un protoplaneta ubicado en la posición de Júpiter ($a = 5,2$ UA), donde la densidad superficial de sólidos corresponde a 3 NSMM y los planetesimales acretados tienen un radio $r_m = 1$ km (caso InC). El segundo panel muestra la formación de un protoplaneta cuyo radio orbital es $a = 3,5$ UA, tomando la densidad de un disco de 2 NSMM y planetesimales de radio $r_m = 100$ km. En este caso (InD) el tiempo de formación supera ampliamente la vida media de un disco protoplanetario. Sin embargo, dejando de lado este hecho y analizando la curva en forma cualitativa, notamos que, a diferencia del caso del primer panel, la inestabilidad se amortigua y el planeta continúa su formación en el modo usual.

Las inestabilidades están, por cierto, asociadas a la tasa de acreción de sólidos. Cuando los planetesimales de la zona de alimentación comienzan a agotarse, el planeta pierde una fuente fundamental de energía para compensar la contracción de las capas de gas. De este modo, la tasa de acreción de gas aumenta. Si este proceso lleva a que se incorporen grandes cantidades de gas en poco tiempo, la masa total del planeta aumenta y la zona de alimentación sufre una expansión brusca, poblándose con planetesimales de las regiones del disco vecinas al planeta. Esto provoca un gran aumento en la tasa de acreción de sólidos, con el consiguiente aporte de energía por parte de los planetesimales. Sin embargo, la envoltura no logra transportar eficientemente esta energía hacia la superficie, y ésta resulta absorbida por el interior, provocando una expansión en las capas de gas (ver figura 6.13). Eventualmente, como consecuencia de esta dilatación, la masa de gas contenida en la esfera de radio $R_{\rm p}$ disminuye, lo cual significa que la masa total del planeta disminuye. Esto impide el crecimiento de la zona de alimentación que, mientras la masa de gas disminuye, se mantiene constante con lo cual, los sólidos disponibles para ser acretados entran nuevamente en una escala descendente. El proceso se retroalimenta hasta que la tasa de acreción de planetesimales alcanza de nuevo valores suficientemente bajos como para permitir que la tasa de acreción de gas cambie de signo, y vuelva a ser positiva. Es importante mencionar que en todo momento, tanto durante la acreción como durante la eyección de masa, la envoltura se encuentra en equilibrio hidrostático, por cuanto estos eventos no son de carácter hidrodinámico. Notablemente, la inestabilidad no es una pulsación libre radial como la que se encuentra en algunas estrellas variables, como las Cefeidas. Mientras que la escala de tiempo de una pulsación radial es del orden de $R_{\rm p}/c_{\rm s}$, que en estas casos sería de tan solo unas horas, los períodos que encontramos aquí son varios órdenes de magnitud mayores. Esto se debe a que estas inestabilidades son oscilaciones forzadas cuya escala de tiempo está asociada al proceso de acreción.

Como mencionamos más arriba, las inestabilidades de la envoltura no son periódicas, con lo cual hay que ser cuidadoso a la hora de estudiar los mecanismos de excitación y amortiguamiento de las oscilaciones. Cuando se estudian pulsaciones es habitual calcular la integral del trabajo que realizan las capas de gas al cabo de un ciclo (Clayton 1968),

$$W = \int dM_r \oint P dV. \tag{6.20}$$

En el plano $P - V$, si un elemento de masa a lo largo de un ciclo evoluciona en el sentido de las agujas del reloj, excita la inestabilidad. A la inversa, si va en sentido contrario a las agujas del reloj, la amortigua. En estos casos encontramos que, mientras que las capas más internas tienden a excitar la inestabilidad, las externas tienden a amortiguarla. Que las capas internas exciten la inestabilidad es esperable puesto que allí es donde liberan los planetesimales la mayor cantidad de energía. Sin embargo, cuando se calcula el trabajo total sobre todas las capas de gas del planeta se encuentra que es negativo, con lo cual el efecto neto es que el planeta tiende a amortiguar la inestabilidad.

Si bien las curvas que caracterizan cada uno de los casos muestran diferencias en su forma (sobre todo en el caso de la luminosidad), no estamos seguros de que los dos tipos de inestabilidad sean esencialmente distintos. Puede que, en los casos como el InC, simplemente ocurra que no lo hemos podido dejar evolucionar completamente, y que si lo hiciéramos, en algún momento llegaría a alcanzar las condiciones para amortiguar la inestabilidad. Lo que sí es claro en ambos casos es que este fenómeno, cuando se desencadena, tiene como principal efecto demorar la formación de un planeta.

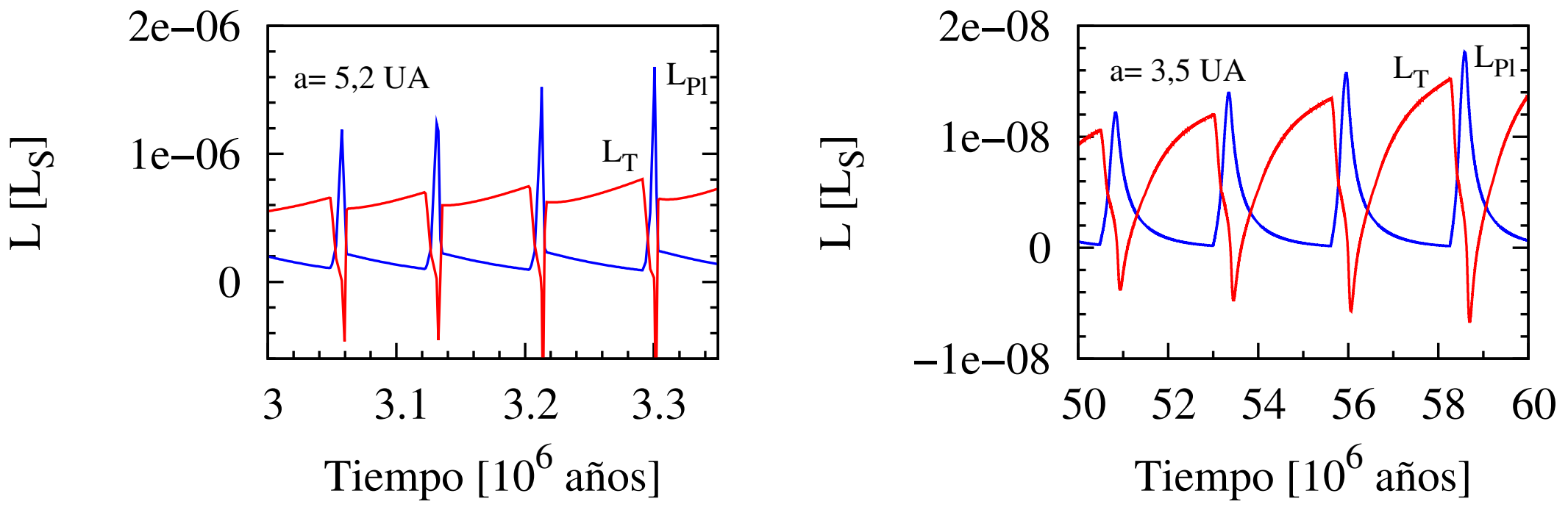

**Figura 6.13.** Para cada uno de los casos presentados en la figura 6.12 (InC e InD), mostramos la luminosidad total del planeta (rojo) y la producida por la acreción de planetesimales (azul), para cuatro "períodos" en la acreción de gas. Notemos como ambas cantidades están en contrafase.

El fenómeno de la inestabilidad de la envoltura es sumamente complejo y todavía no lo hemos comprendido en profundidad. Más trabajo en relación a este tema queda por delante.

# Capítulo 7

# Discusión

En el capítulo anterior presentamos los resultados más importantes obtenidos con nuestro modelo de formación de planetas gigantes. Otros autores (Bodenheimer & Pollack 1986, Wuchterl 1991, Pollack et al. 1996, Hubickyj, Bodenheimer & Lissauer 2005, Alibert et al. 2005, Dodson-Robinson et al. 2008) han realizado cálculos bajo hipótesis similares a las adoptadas para este trabajo. En lo que sigue, discutiremos nuestros resultados en el marco del estado actual del entendimiento de este problema.

Bodenheimer & Pollack (1986) fueron los primeros en realizar modelos numéricos de formación de planetas gigantes, de acuerdo a la hipótesis de inestabilidad nucleada, resolviendo en forma autoconsistente las ecuaciones diferenciales de la estructura gaseosa. Sus simulaciones, si bien con muchas simplificaciones, formaron parte del primer trabajo que estudiaba en forma integral, y considerando los aspectos físicos más importantes, el proceso de formación de un planeta con un núcleo sólido y una extendida envoltura gaseosa. Diez años más tarde, Pollack et al. (1996) actualizaron el código de Bodenheimer & Pollack, presentando un modelo mucho más detallado de formación de planetas gigantes, que hoy sigue teniendo vigencia, y que es considerado referente en el área por el grado de aproximación con el que se trataron ciertos aspectos físicos del problema. A continuación, describiremos sus resultados para poder realizar una posterior comparación con los obtenidos con nuestro modelo.

El modelo de Pollack et al. (1996), en adelante P96, para el cálculo de la formación de planetas gigantes contiene tres ingredientes fundamentales: un modelo para el cálculo de la componente gaseosa del planeta, la incorporación de una tasa de acreción de sólidos dependiente del tiempo y un modelo para el intercambio energético entre los planetesimales ingresantes y las capas de la envoltura, el cual tiene en cuenta la ablación que sufren los planetesimales al atravesar la envoltura. Las diferencias más importantes entre nuestro modelo y el de P96 están en la elección de la tasa de acreción de sólidos y el tratamiento que se hace cuando los planetesimales atraviesan las capas de gas de la envoltura en su trayectoria hacia el núcleo del planeta.

Nos focalizaremos primero en el modelo de acreción de planetesimales de P96 ya que este es el ingrediente que marca la diferencia entre los resultados de P96 y los nuestros. En el capítulo 4 hemos discutido largamente el proceso de crecimiento de un embrión sólido, puntualizando que, en el caso de los núcleos de los planetas gigantes, se atraviesan dos etapas: las correspondientes al régimen runaway y al régimen oligárquico. El régimen runaway es de corta duración comparado con las escalas involucradas en la formación de los planetas gigantes (Ida & Makino 1993, Kokubo & Ida 2000), y sólo sería efectivo hasta que los embriones alcanzan una masa comparable a la lunar. Para masas superiores, la dispersión de velocidades que provoca el embrión sobre los planetesimales que lo rodean lentificaría su crecimiento en forma considerable. Sin embargo, P96 incorporaron a su modelo una tasa de acreción de sólidos que favorece el rápido crecimiento de los embriones y que, sin ser estrictamente la correspondiente al caso runaway, tiene características que lo aproximan más a este tipo de crecimiento que al régimen oligárquico. P96 fundamentaron la elección de este tipo de acreción en que les permitiría establecer cotas inferiores para el tiempo de formación de un planeta gigante y, fundamentalmente, en la diferencia sustancial que esta incorporación establecía respecto de los modelos previos donde la tasa de acreción se consideraba constante. El modelo de acreción empleado por P96 es el de Lissauer (1987), y podemos resumirlo de la siguiente manera. De acuerdo con su ecuación (1), la tasa de acreción del núcleo se puede escribir como,

$$\frac{dM_{\rm c}}{dt} = \pi \mathcal{R}^2 \Sigma \Omega_{\rm k} F_{\rm g} \tag{7.1}$$

donde $\mathcal{R}$ es el radio de captura, $\Omega_{\rm k}$ es la velocidad angular kepleriana, y $F_{\rm g}$ es el factor que tiene en cuenta el enfocamiento gravitatorio. $F_{\rm g}$ depende de los valores cuadrático medios de la excentricidad e inclinación reducida de los planetesimales, como también del radio reducido de captura:

$$e_{{\rm h},m} \equiv \frac{a}{R_{\rm H}} e \qquad i_{{\rm h},m} \equiv \frac{a}{R_{\rm H}} i \qquad d_c \equiv \frac{\mathcal{R}}{R_{\rm H}}.$$

Las fórmulas analíticas para $F_{\rm g}$ fueron obtenidas por Greenzweig & Lissauer (1992) a partir de simulaciones numéricas de $N$-cuerpos. En su modelo, Greenzweig & Lissauer proponen que las inclinaciones de los planetesimales solo dependen de las interacciones planetesimal-planetesimal (no se considera la presencia del embrión). En este caso, la inclinación puede aproximarse por:

$$i_{{\rm h},m} = \frac{v_{{\rm e},m}}{\sqrt{3}\Omega_{\rm k} R_{\rm H}} \tag{7.2}$$

donde $v_{{\rm e},m}$ es la velocidad de escape de la superficie de un planetesimal. Claramente, la inclinación, $i = i_{{\rm h},m} R_{\rm H}/a$, permanece constante en el tiempo mientras que la inclinación reducida, $i_{{\rm h},m}$, decrece (dado que $R_{\rm H}$ aumenta con la masa y $a$ está fijo). En cuanto a la excentricidad, se supone que está afectada tanto por la presencia del embrión como por el resto de los planetesimales. El valor de la excentricidad reducida corresponde a:

$$e_{{\rm h},m} = \max(2 i_{{\rm h},m}, 2). \tag{7.3}$$

Esto significa que, si $e_{\mathrm{h},m} = 2i_{\mathrm{h},m}$, el embrión crece de acuerdo al régimen runaway, ya que en ese caso la excentricidad y la inclinación de los planetesimales acretados resultan independientes de la masa del embrión. Por otra parte, si $e_{\mathrm{h},m} = 2$, la excentricidad de los planetesimales estará afectada por la presencia del embrión. Esta condición corresponde a la dispersión que sufren los planetesimales cuando se encuentran dominados por la cizalladura kepleriana (*shear-dominated regime*). Sin embargo, en base a sus simulaciones de $N$-cuerpos, Ida & Makino (1993) mostraron que la dispersión protoplaneta-planetesimal bajo las condiciones de este régimen solo dura unos miles de años, luego de los cuales los planetesimales son fuertemente perturbados por la presencia del protoplaneta, con lo cual el régimen de crecimiento post-runaway del embrión pertenece al régimen donde domina la dispersión (*dispersion-dominated regime*). En este régimen, las excentricidades e inclinaciones satisfacen $e_{\mathrm{h},m}/i_{\mathrm{h},m} \simeq 2$ y $e_{\mathrm{h},m} > 2$. Luego, cuando se considera que los planetesimales se encuentran en el *shear-dominated regime*, sus excentricidades e inclinaciones se mantienen bajas, aún en la vecindad del embrión. Adoptar esta hipótesis durante todo el proceso de formación de un planeta significa considerar una tasa de crecimiento para los sólidos muy alta, o sea, un régimen de acreción considerablemente más efectivo que el correspondiente al crecimiento oligárquico.

Además de la elección de distintos regímenes de acreción de sólidos, nuestro modelo y el de P96 difieren en que P96 consideran un modelo detallado para el intercambio energético entre los planetesimales acretados y la envoltura del planeta, donde se contempla la ablación que sufren los planetesimales en este proceso. Si bien considerar la ablación es tener en cuenta un efecto importante para el cálculo de la sección eficaz de captura de un planeta (Benvenuto & Brunini 2008), este efecto se hace realmente apreciable cuando el núcleo supera, aproximadamente, las 10 $\mathrm{M}_\oplus$. Como veremos a continuación, la diferencia entre nuestros resultados y los de P96 no pueden deberse a la ablación puesto que se manifiestan mucho antes de que el embrión tenga suficiente gas ligado como para que este efecto sea importante.

Para evaluar los efectos de adoptar al crecimiento oligárquico como regulador de la formación del núcleo de los planetas gigantes en lugar de un régimen más cercano al runaway, como en el caso de P96, hicimos una serie de simulaciones bajo las mismas condiciones que estos autores. P96 utilizan como caso de referencia para sus resultados (Caso J1) una simulación caracterizada de la siguiente manera. Un embrión sólido del tamaño de Marte ($\sim 0,1\,\mathrm{M}_\oplus$), ubicado en la posición actual de Júpiter ($a = 5,2$ UA), crece acretando planetesimales de 100 km de radio y de $1,39\,\mathrm{g\,cm^{-3}}$ de densidad. La densidad del núcleo se acepta constante en $3,2\,\mathrm{g\,cm^{-3}}$. Las condiciones nebulares se caracterizan por: $T = 150$ K, $\rho = 5 \times 10^{-11}\mathrm{g\,cm^{-3}}$ y $\Sigma = 10\,\mathrm{g\,cm^{-2}}$. Los resultados de P96 para la evolución de la masa del planeta y el comportamiento de la tasa de acreción de sólidos y gas pueden observarse en la figura 7.1, la cual corresponde al artículo original de estos autores y se muestra aquí para la mejor comparación de los resultados. Bajo estas circunstancias, P96 encuentran que la masa de cruce es $M_{\mathrm{cross}} \simeq 16,2\,\mathrm{M}_\oplus$ y, el tiempo de cruce, $t_{\mathrm{cross}} = 7,6 \times 10^6$ años. Tanto en este, como en los otros resultados que estos autores muestran en su artículo, ellos distinguen que el proceso de formación atraviesa tres fases. La primera ("fase 1") involucra el principio del proceso y corresponde básicamente a la formación del núcleo. En el Caso J1, esta etapa abarca los primeros $6 \times 10^5$ años, que es el tiempo que le demanda al núcleo alcanzar unas $10\,\mathrm{M}_\oplus$. En este período, la tasa de acreción de sólidos alcanza su pico máximo, luego de lo cual decrece sensiblemente. En este caso, el embrión alcanza su masa de aislación casi en ausencia de la envoltura. Debido al vaciamiento de la zona de alimentación, pocos planetesimales ingresan en la envoltura después que esto ocurre, lo que permite el comienzo de la acreción de gas de una manera más significativa. P96 definen la finalización de la "fase 1" y el comienzo de la "fase 2" cuando la tasa de acreción de sólidos se iguala a la de gas. La "fase 2" es la etapa que regula el tiempo de formación del planeta puesto que corresponde al proceso de acreción de gas para la formación de la envoltura, y finaliza con el comienzo del runaway gaseoso. La marcada longitud de la "fase 2" se debe a que, si bien la tasa de acreción de sólidos se reduce considerablemente, la energía que aportan los planetesimales acretados durante esta etapa es lo suficientemente alta como para impedir la rápida contracción de las capas de gas. Entre el inicio y la finalización de la "fase 2" transcurren $7 \times 10^6$ años. Durante este tiempo, la relación entre la tasa de acreción de gas y de sólidos es constante, siendo la tasa de acreción de gas superior en un factor entre 2 y 3. La "fase 2" concluye cuando se alcanza la masa de cruce y comienza el runaway de gas, este último define la "fase 3". Esta última etapa caracteriza la finalización del proceso de formación, que ocurre en menos de $5 \times 10^5$ años.

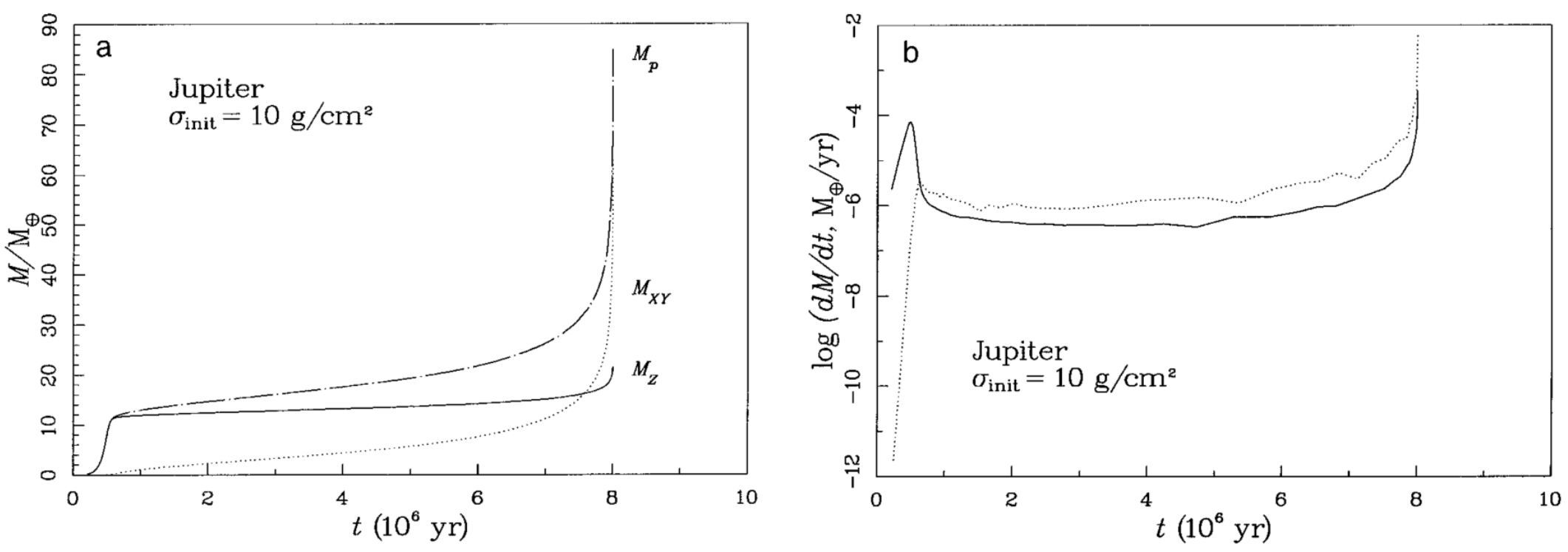

**Figura 7.1.** Esta figura corresponde a los dos primeros paneles de la Fig. 1 del artículo de Pollack et al. (1996) y se muestra aquí para facilitar su comparación con nuestros resultados (ver figura 7.2). La simulación corresponde al Caso J1, donde la densidad superficial de sólidos es $\sigma_{\rm init} = 10\ \mathrm{g\,cm^{-2}}$ en la posición del planeta ($a = 5{,}2$ UA). El panel **a** muestra la masa acumulada de gas, $M_{\rm XY}$, de sólidos, $M_{\rm Z}$, y la masa total, $M_{\rm p}$. El panel **b** muestra el logaritmo de la tasa de acreción de sólidos (línea llena) y de gas (línea punteada) en función del tiempo. P96 definen tres fases en la formación del planeta: la "fase 1", que comprende desde el principio de la formación del núcleo hasta que $\dot{M}_{\rm Z} = \dot{M}_{\rm XY}$, extendiéndose solo por $6 \times 10^5$ años; la "fase 2" que domina la formación y está caracterizada por $\dot{M}_{\rm Z}/\dot{M}_{\rm XY} \simeq$ constante, y que finaliza cuando se alcanza la masa de cruce; y, por último, la "fase 3" que corresponde al runaway de gas.

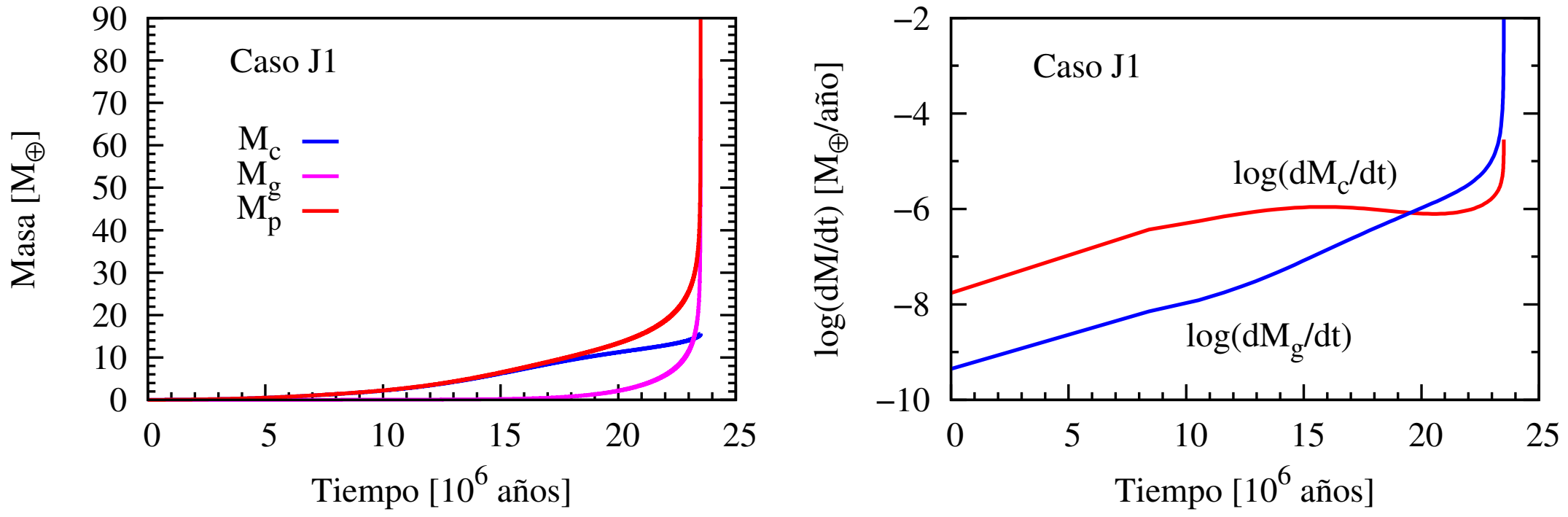

**Figura 7.2.** Nuestros resultados de la evolución de la masa del planeta con el tiempo (primer panel) y de las tasas de acreción (segundo panel) para las mismas condiciones del caso J1 de P96. $M_\mathrm{c}$ representa la masa del núcleo, $M_\mathrm{g}$ la masa de la envoltura y $M_\mathrm{p}$ la masa total del planeta.

Teniendo en cuenta las mismas condiciones que caracterizan el Caso J1 de P96, hicimos una simulación utilizando nuestro código. Como podemos observar en la figura 7.2, nuestros resultados son muy diferentes a los de P96. En nuestro caso, el tiempo de cruce es $\sim 23$ millones de años (Tabla 7.1), más de tres veces el obtenido por P96 y superior a la cota de $10^7$ años que surge de las observaciones de discos circumestelares. De hecho, ni siquiera la masa del núcleo llega a alcanzar $10\,\mathrm{M}_\oplus$ en menos de $10^7$ años (Haisch 2001). Sin embargo, la masa de cruce es muy similar en las dos simulaciones. En ambos casos ésto ocurre cuando el núcleo alcanza aproximadamente la masa de aislación (para la densidad de sólidos considerada, $M_\mathrm{iso} \simeq 16,45\,\mathrm{M}_\oplus$ para el caso en que $M_\mathrm{g} = M_\mathrm{c}$). Por otra parte, no resulta evidente en nuestros resultados la necesidad de distinguir entre la "fase 1" y la "fase 2". El proceso de acreción de gas se presenta en nuestra simulación como monótonamente creciente y con pendiente distinta de cero. En ningún momento encontramos una etapa donde la tasa de acreción de gas y de sólidos sean aproximadamente constantes ni proporcionales entre sí .

La diferencia evidente entre ambos resultados son las escalas de tiempo que involucra cada uno. Mientras que la formación del núcleo en el caso de P96 toma alrededor de medio millón de años, en nuestro caso supera los 10 millones. Las largas escalas de tiempo relacionadas con la formación de embriones sólidos está asociada al régimen de crecimiento oligárquico. Si bien P96 contemplan la ablación de los planetesimales, lo cual favorece la aceleración de la formación, este efecto es importante una vez formado el núcleo, cuando el planeta comienza a tener una atmósfera más masiva, por cuanto no opera en forma eficiente en la primera etapa, con lo cual esta diferencia entre ambos modelos no puede ser responsable de la escala de tiempo tan corta que caracteriza la formación del núcleo en el modelo de P96.

P96 comienzan sus simulaciones con un embrión sólido de $0,1\,\mathrm{M}_\oplus$. Nosotros lo hacemos con uno que es más de un orden de magnitud menor ($0,0075\,\mathrm{M}_\oplus$), el cual se encontraría ya en pleno crecimiento oligárquico. De acuerdo a nuestra simulación del Caso J1, a nuestro embrión le toma 2 millones de años alcanzar la masa inicial de P96. Con lo cual, el tiempo que insume el crecimiento oligárquico de nuestro embrión para llegar a la condición inicial de P96 es cuatro veces mayor que el tiempo en que P96 llegan a la masa de cruce con un núcleo de $11\,\mathrm{M}_\oplus$.

La existencia de la mencionada "fase 2", en el sentido en que P96 la definen, fue estudiada por Shiraishi & Ida (2008) con un modelo semi-analítico. Estos autores encuentran que la manifestación de tan largo período de acreción de gas es debida a la sobreestimación de la tasa de acreción de sólidos. En el modelo de referencia de P96 (Caso J1), la tasa de acreción de sólidos tiene un máximo del orden de $10^{-4}\,\mathrm{M}_\oplus$/ año, y durante la "fase 2" se mantiene en, aproximadamente, $10^{-6}\,\mathrm{M}_\oplus$/ año. Según los cálculos de Shiraishi & Ida, si la tasa de acreción de sólidos es ineficiente (como en el caso del crecimiento oligárquico), después que el núcleo alcanza la masa de aislación, la tasa de acreción, para este caso, sería del orden de $10^{-7}\,\mathrm{M}_\oplus$/año. En nuestra simulación, el máximo de la tasa de acreción es $10^{-6}\,\mathrm{M}_\oplus$/ año, y después de alcanzar la masa de aislación, en lo que podría tomarse como nuestra "fase 2", si usamos la misma definición que P96, la tasa de acreción es, aproximadamente, $7 \times 10^{-7}\,\mathrm{M}_\oplus$/año. El cálculo de Shiraishi & Ida tiene en cuenta que, después que el núcleo llega a unas $10\,\mathrm{M}_\oplus$, de abre una brecha con el disco de planetesimales. Esta brecha ocurre por la combinación de dos efectos: la excitación que provoca el embrión en los planetesimales que lo rodean y el decaimiento orbital que sufren por la viscosidad del gas nebular. Esto resulta en un truncamiento en la tasa de acreción de sólidos. En nuestro modelo no está contemplada la posibilidad de la apertura de una brecha de estas características. Sin embargo, aún considerando solo la dispersión que sufren los planetesimales en la zona de alimentación del planeta debido al efecto gravitatorio del embrión, se ve claramente de nuestros resultados que la definición de la "fase 2" no tiene mucha relevancia puesto que no se registra un cambio notorio en el crecimiento del planeta ni en el tipo de acreción después que éste alcanza la masa de aislación.

**Tabla 7.1.** Cuadro comparativo de los resultados obtenidos para $t_{\mathrm{cross}}$ y $M_{\mathrm{cross}}$ en los casos J1, J3 and J7 de Pollack et al. (1996). Las columnas 2 y 3 corresponden a nuestros resultados, mientras que las 4 y 5 a los de Pollack et al. U significa que la envoltura es inestable.

| Caso | $t_{\mathrm{cross}}$ [My] | $M_{\mathrm{cross}}$ [$\mathrm{M}_\oplus$] | $t_{\mathrm{cross}}^{\mathrm{P96}}$ [My] | $M_{\mathrm{cross}}^{\mathrm{P96}}$ [$\mathrm{M}_\oplus$] |
|---|---|---|---|---|
| J1 | 23,25 | 15,75 | 7,58 | 16,17 |
| J3 | 13,20 | 21,4 | 1,51 | 29,61 |
| J7 | U | U | 6,94 | 16,18 |

P96 calcularon un caso donde la densidad superficial de sólidos es superior a la del Caso J1 ($\Sigma = 15\,\mathrm{g\,cm^{-2}}$). Este aumento en la densidad lleva a que el tiempo de cruce se reduzca a 1,5 millones de años, mientras que la masa de cruce resulta, aproximadamente, 30 $\mathrm{M_\oplus}$ (Caso J3, tabla 7.1). De acuerdo a nuestras simulaciones, el tiempo de cruce sería un orden de magnitud superior (13 millones de años), mientras que la masa de cruce resultaría inferior, aproximadamente $21,5\,\mathrm{M_\oplus}$. En el caso de P96 el aumento de la densidad representa una reducción en el tiempo de formación de un factor 5, sin embargo, en nuestro caso este mismo incremento solo conduce a que el tiempo de formación se reduzca a la mitad.

La tasa de acreción de sólidos depende inversamente de la velocidad relativa de los planetesimales respecto del protoplaneta, la cual está regulada por la excentricidad e inclinación de los planetesimales que serán acretados (Ec. 4.16). Para el Caso J3, la figura 7.4 muestra el cociente entre la excentricidad (inclinación) de nuestro modelo y el de P96. Para este caso, $i_{\mathrm{h},m}$ es desde el comienzo inferior a 1 y, como es una función decreciente de la masa del protoplaneta, $e_{\mathrm{h},m}$ siempre es igual a 2. De este modo, nuestras excentricidades siempre son mayores a las de P96 en un factor 4. Además, la inclinación es constante en el caso de P96 ($i \simeq 0,004$), mientras que en nuestro modelo es $i \simeq e/2$. Esto impacta en las velocidades relativas y en el cálculo de la altura de escala del disco de planetesimales, $h$. La tasa de acreción de sólidos es inversamente proporcional a $h$, con lo cual si $h$ es más grande, porque los planetesimales están más dispersos, la tasa de acreción es menor. De este modo, siendo nuestras inclinaciones y excentricidades mucho mayores a las adoptadas en P96, es lógico que nuestros tiempos de acreción sean mucho más largos.

Thommes, Duncan & Levison (2003) estimaron el tiempo la formación de un protoplaneta sólido en crecimiento oligárquico (sin envoltura gaseosa) en un disco protoplanetario donde la densidad corresponde a 1 y a 10 NSMM. En el caso de 1 NSMM, para la posición de Júpiter, al cabo de $10^7$ de años solo llega a formarse un embrión de 1 $\mathrm{M_\oplus}$. En tanto, para 10 NSMM, un embrión de 10 $\mathrm{M_\oplus}$ se forma en, aproximadamente, $\lesssim 5 \times 10^6$ años. El Caso J1 corresponde a 3 NSMM, con lo cual no es esperable conseguir la formación de un núcleo de $10\,\mathrm{M_\oplus}$ en tiempos razonablemente cortos.

Dado que previamente habíamos encontrado que la acreción es muy sensible al tamaño de los planetesimales acretados, volvimos a considerar las condiciones del Caso J3, pero para planetesimales de 1 y 10 km de radio (ver figura 7.3). En ambos casos, el tiempo de formación resulta bastante inferior a $10^7$ años. Cuando $r_m = 10$ km, el tiempo de cruce es 3,7 millones de años y la masa de cruce es $25,5\,\mathrm{M_\oplus}$, mientras que para el caso de planetesimales de 1 km de radio, el tiempo de cruce se reduce a $1,4$ millones de años y la masa de cruce a $29\,\mathrm{M_\oplus}$. Como podemos notar, considerando planetesimales de 1 km de radio podemos reproducir los resultados de P96 para el Caso J3 (donde el radio de los planetesimales es de 100 km). El tamaño típico de los planetesimales no se conoce todavía. De hecho, como hemos mencionado previamente, la formación y consecuente distribución de tamaños de los planetesimales primordiales está todavía bajo estudio y es uno de los temas fundamentales que debe resolver las Ciencias Planetarias.

Si repetimos este procedimiento de reducir el tamaño de los planetesimales pero ahora para el Caso J1, encontramos que, para $r_m = 10$ km, el tiempo de cruce es inferior a los 10 millones de años. El caso $J1\sqrt{10}$ presentado en el capítulo 6 fue realizado bajo las mismas hipótesis del Caso J1 pero con planetesimales de radio 3,17 km. Sin embargo, no siempre reducir el tamaño de los planetesimales acretados lleva a favorecer la formación de un planeta. Si para el Caso J1 consideramos el radio de los planetesimales en 1 km obtenemos el comportamiento oscilatorio que presentamos en el capítulo anterior (figura 6.12, primer panel). Mientras que P96 vuelven a encontrar una solución estable análoga a las anteriores (Caso J7), nosotros encontramos que la envoltura se vuelve inestable cuando la masa del núcleo es $12\,\mathrm{M}_\oplus$ y la de la envoltura es $2,43\,\mathrm{M}_\oplus$. Para estos valores, la masa del núcleo es, al orden de la aproximación del cálculo de la masa de aislación, igual a la masa de aislación para un objeto con esa cantidad de gas ligado. Si bien desconocemos el motivo por el cual nosotros encontramos esta inestabilidad y otros autores no, una posibilidad es que este hecho esté ligado al tratamiento de la acreción de sólidos. Dado que nuestra acreción es más lenta, para un mismo valor de la masa del núcleo, la masa de la envoltura asociada a nuestros modelos es mayor que en la de P96. Por ejemplo, en el Caso J7, cuando la masa del núcleo de P96 es 11,4 $\mathrm{M}_\oplus$, la masa de la envoltura es $0,14\,\mathrm{M}_\oplus$, mientras que en nuestro caso, para esa misma masa del núcleo, la masa de gas es tres veces mayor, $M_{\rm g} \simeq 0,46\,{\rm M}_\oplus$. De este modo, nosotros llegamos a la masa de aislación con una envoltura más masiva, lo cual la vuelve más propensa a desarrollar este tipo de inestabilidad.

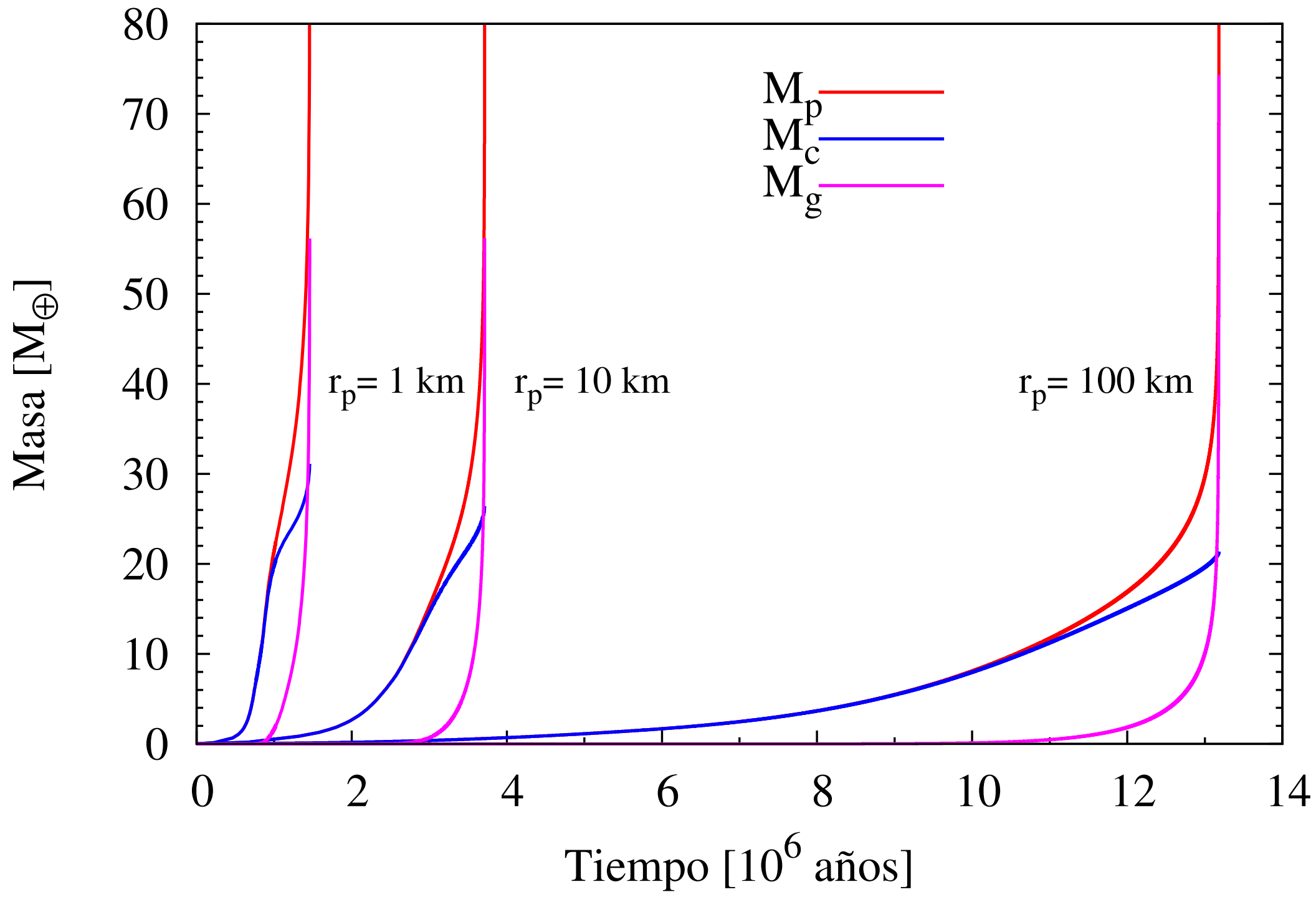

**Figura 7.3.** Tres simulaciones de la formación de Júpiter considerando solo variaciones en el tamaño de los planetesimales ($r_m = 1, 10, 100$ km). El caso correspondiente a $r_m = 100$ km fue calculado adoptando los mismos parámetros que Pollack et al. (1996) para su caso J3.

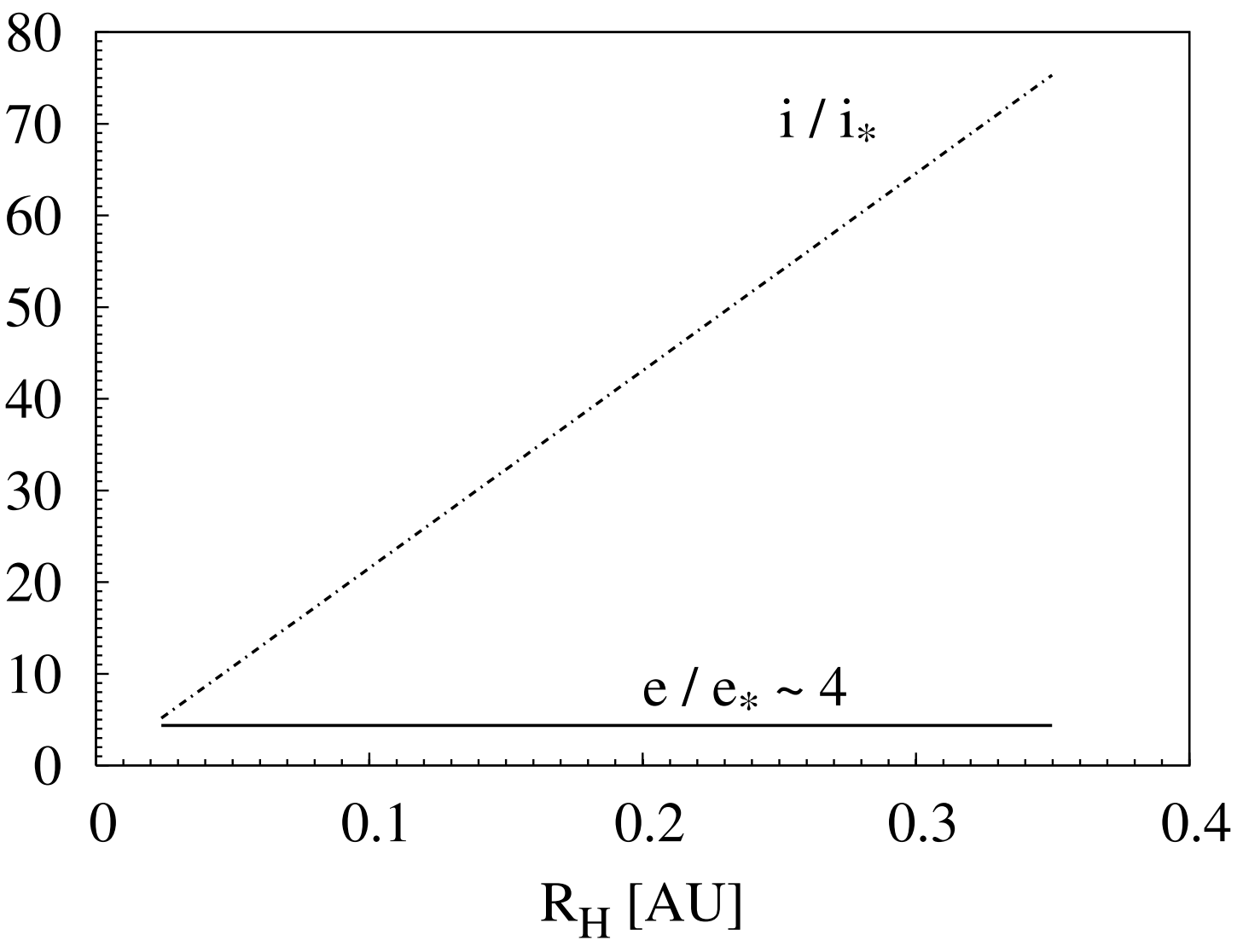

**Figura 7.4.** El impacto del modelo elegido para el crecimiento del núcleo. En línea llena se muestra el cociente entre las excentricidades $e/e_*$, donde $e$ corresponde a la calculada según la ecuación (4.19) y $e_*$ de acuerdo al modelo adoptado por Pollack et al. (1996). La línea punteada muestra el cociente entre las inclinaciones ($i/i_*$). Mientras que en nuestro modelo $i$ depende de $e$, según $i = e/2$, Pollack et al. consideran a $i_*$ constante durante toda la simulación. En la abscisa consideramos al radio de Hill para poder hacer una comparación que no involucre las escalas de tiempo.

Es importante enfatizar que los planetesimales de mayor tamaño tienden a favorecer la estabilidad de la envoltura. Hemos ya mencionado que la tasa de acreción de sólidos depende del tamaño de los planetesimales. Dado que en el crecimiento oligárquico las velocidades relativas de los planetesimales más grandes son mayores, el radio efectivo en estos casos es menor, lo cual significa que la luminosidad por acreción es también menor. Sin embargo, son los planetesimales más pequeños los más abundantes en el disco, los cuales son también más eficientemente acretados, con lo cual favorecen el desencadenamiento de la inestabilidad. El hecho que no hayamos registrado este efecto en nuestras simulaciones considerando una distribución de tamaños está relacionado con que, si se dieran las condiciones para el desarrollo de la inestabilidad, la limitación que le hemos puesto a la tasa de acreción de sólidos las inhibirían. Sin embargo, si calculamos el caso J7, sin limitar la acreción y con una distribución de tamaños (análoga a la empleada en los ejemplos del capítulo 6) encontramos que la inestabilidad está igualmente presente. Por otra parte, cuanto más densa es la nebulosa y/o mayor el radio orbital del embrión, menor será la tendencia a que se produzca la inestabilidad. Esto se debe a que en estos casos la masa de aislación es mayor y el runaway de gas se alcanza antes de llegar a las condiciones necesarias para que surjan las oscilaciones.

Varios mecanismos podrían inhibir la aparición de la inestabilidad de la envoltura. Esencialmente, la inestabilidad aparece cuando el planeta alcanza la masa de aislación, con lo cual las capas de gas sufren una abrupta contracción, se acreta gas en consecuencia y la zona de alimentación se expande incorporando una gran cantidad de planetesimales. Ahora bien, cuántos planetesimales ingresan en la zona de alimentación cuando ésta se expande dependerá del estado del disco en ese momento. En nuestro modelo no hemos incluido la evolución propia del disco protoplanetario, por cuanto cuando se expande la zona de alimentación, las nuevas regiones que se le suman se encuentran en su estado inicial. Es esperable, sin embargo, que el número de planetesimales sea menor que el original. Debido a la viscosidad de la nebulosa, los planetesimales sufren un decaimiento orbital, espiralando en dirección al Sol. Los más pequeños serán, además, los más afectados. Además, la escala de tiempo de migración de los planetesimales es del mismo orden que la de la formación del planeta (Thommes, Duncan & Levison 2003), por cuanto es esperable que este hecho tenga una incidencia no despreciable. Otro efecto que debería tenerse en cuenta es la eyección de planetesimales fuera de la zona de alimentación. Dado que esto ocurre cuando el planeta es suficientemente masivo no retrasaría la acreción inicial para formar el núcleo, con lo cual no repercutiría en el tiempo de formación. Sin embargo, operaría en la dirección de disminuir la densidad superficial de sólidos en la vecindad del planeta cuando éste tiene una masa apreciable.

En la bibliografía relativa a modelos de formación de planetas gigantes no hay registro de inestabilidades como las encontradas en nuestras simulaciones. Wuchterl (1991, 1995), con un código completamente hidrodinámico, encuentra que cuando el planeta está cercano a alcanzar la masa de cruce, atraviesa una inestabilidad dinámica debido a la cual eyecta gran parte de su envoltura. Sin embargo, la inestabilidad presente en nuestras simulaciones no es de carácter hidrodinámico, sino que puede ser descripta por una secuencia de modelos en equilibrio hidrostático debido a que la velocidad del material es suficientemente baja. Además, esta inestabilidad no está asociada al transporte de energía, sino a la interacción entre el proceso de acreción y la respuesta de la envoltura frente a cambios en la luminosidad de acreción de sólidos.

Otros escenarios se han tenido en cuenta para acelerar la formación de los planetas gigantes. Hubickyj, Bodenheimer & Lissauer (2005) analizaron la posibilidad de que la opacidad de la envoltura fuera mucho menor que la correspondiente al medio interestelar. El modelo de estos autores es esencialmente igual al de P96, con la actualización de la EOS como principal diferencia (P96 no utilizaban la EOS de SCVH sino una tabla previa). Recalcularon el Caso J1, obteniendo $M_{\rm cross} \simeq 16,2\,\mathrm{M}_{\oplus}$ (igual que P96), pero con una reducción en el tiempo de cruce ($t_{\rm cross} \simeq 6$ millones de años). Esta misma simulación fue realizada suponiendo que la opacidad de los granos era tan solo un 2 % de la opacidad interestelar, de acuerdo con los cálculos de Podolak (2003). Sus resultados muestran que el tiempo de cruce se ve significativamente afectado por la opacidad, ya que en este caso $t_{\text{cross}} \simeq 2,2$ millones de años. La masa de cruce, por otra parte, se mantiene en el mismo valor. Actualmente, no existen tablas de opacidades calculadas en forma autoconsistente con la formación del planeta. En cualquier caso, si la opacidad es menor que la considerada en nuestras simulaciones, el efecto que tendría sobre nuestros resultados sería el acortamiento de las escalas de tiempo, lo cual favorecería la formación de los planetas gigantes.

Recientemente, Dodson-Robinson et al. (2008b) estudiaron la formación *in situ* de Júpiter y Saturno. El modelo y el código empleado para los cálculos es el de Hubickyj, Bodenheimer & Lissauer (2005). El trabajo de Dodson-Robinson et al. se basa en cálculos detallados de la densidad superficial de sólidos realizados por Dodson-Robinson et al. (2008a). Para estimar la densidad superficial de sólidos, ellos combinan la evolución viscosa de un disco protosolar con un modelo cinético de formación de hielos, lo cual les permite obtener un perfil de densidad en función del tiempo y de la distancia heliocéntrica. Entre sus resultados encuentran que en la región trans-saturniana habría abundantes cantidades de amoníaco en forma de hielos, lo cual implicaría un enriquecimiento en un factor 6 respecto de la NSMM en la región de Saturno. Dodson-Robinson et al. (2008b) encontraron que Saturno se forma en $3,5 \times 10^6$ años, sin llegar a agotar nunca su zona de alimentación. Tanto en el caso donde la opacidad se considera reducida como cuando se asumen los valores originales de la tabla, el tiempo de formación les da prácticamente igual. Sin embargo, la masa del núcleo que obtienen es 44 $M_\oplus$ en el caso de las opacidades de los granos reducidas, y 54 $M_\oplus$ en el otro caso. Estos valores son mucho mayores que los estimados para la masa del núcleo de Saturno, e incluso superarían los valores aceptados para explicar enriquecimiento registrado en su atmósfera. Otra consideración a tener en cuenta es que ellos comienzan sus simulaciones con un embrión de 1 $M_\oplus$. Hemos mostrado ya que, el tiempo de formación de un embrión de esta masa no es para nada despreciable.

Por su parte, Alibert et al. (2005a) tuvieron en cuenta la migración del planeta durante su formación y la evolución del gas nebular. Su modelo es de características similares al de P96, con el agregado que consideran que el disco protoplanetario no es estático, y que el planeta puede migrar en el disco debido a su interacción con el gas. Sus simulaciones muestran que el tiempo de formación de los planetas gigantes (al menos hasta que alcanzan el crecimiento runaway), se acelera notablemente cuando de tiene en cuenta la migración del planeta. Esto se debe fundamentalmente a que desaparece la "fase 2" puesto que la zona de alimentación del planeta no llega nunca a vaciarse ya que el planeta atraviesa siempre nuevas regiones en el disco. La migración suprime la necesidad de esperar a que la envoltura alcance una masa significativa para que el planeta llegue a la masa de cruce. Para simulaciones bajo condiciones análogas a las de P96 en el Caso J1, Alibert et al. encuentran que el planeta llega a la masa de cruce en una escala de tiempo del orden del millón de años. Para conseguir esto, el embrión debe comenzar el proceso de acreción y migración a una distancia del Sol de entre 7 y 15 UA. Como mencionamos en el capítulo 3, las teorías de migración de los planetas en formación debido a su intercambio de momento angular con la nebulosa protoplanetaria tienen su origen a partir de las observaciones de los sistemas extrasolares, donde en muchos casos los planetas se encuentran muy cercanos a su estrella central. Sin embargo, todavía existen muchas dudas sobre cómo opera realmente este mecanismo. De hecho, Alibert et al. deben reducir artificialmente la tasa de migración del embrión en, al menos, un factor 100 para impedir que los protoplanetas sean engullidos por el Sol.

En un trabajo posterior, Alibert et al. (2005b), estudiaron la formación de Júpiter y Saturno considerando migración. Calcularon primero una serie de simulaciones, variando ciertos parámetros, para la formación de Júpiter. Una vez encontrados los parámetros para los cuales el resultado final concordaba con las características actuales de Júpiter (masa total, masa de sólidos, distancia al Sol, etc.), fijaban estos parámetros y buscaban la posición inicial y el retardo necesario para el comienzo de la formación de Saturno. Alibert et al. encontraron que, si Júpiter comenzaba su formación en $a \lesssim 10$ UA, y Saturno en $11,9$ UA y 0,2 millones de años más tarde, la configuración final se correspondía con la actual. El tiempo de formación de ambos planetas era inferior a $3 \times 10^6$ años. Es importante mencionar que la formación de Júpiter y Saturno no fueron calculadas en forma simultánea, por cuanto no está contemplado en este modelo la interacción entre ambos protoplanetas. Alibert et al. obtuvieron muy buenos ajustes entre su resultado final y los datos actualmente aceptados para estos planetas. Sin embargo, el escenario de formación que ellos sugieren contradice al modelo de Niza. Como mencionamos antes, el modelo de Niza consigue explicar numerosas características del Sistema Solar actual, por cuanto se lo considera robusto y confiable.

En el trabajo donde Desch (2008) calcula el perfil de densidad de la nebulosa protosolar compatible con el modelo de Niza, también realiza estimaciones para la formación de los núcleos de los cuatro planetas gigantes del Sistema Solar. Considerando el crecimiento oligárquico de los embriones, pero en ausencia de una envoltura de gas (que aumenta el radio efectivo de captura), Desch encuentra que los tiempo de formación de los núcleos son: para Júpiter, $\sim 0,5 \times 10^6$ años, para Saturno, $\sim 1,5 - 2 \times 10^6$ años, para Neptuno $\sim 5,5 - 6 \times 10^6$ años y para Urano, $\sim 10 \times 10^6$ años. El tamaño considerado para los planetesimales es $r_m = 0,1$ km. Si bien el modelo de Desch es mucho más simple que el nuestro, si comparamos estos resultados con los que obtuvimos nosotros para el caso de 7 especies de planetesimales (capítulo 6), donde el radio mínimo considerado para la distribución es 0,1 km, vemos que las escalas de tiempo involucradas en cada caso son del mismo orden.

Es importante puntualizar que nuestro modelo no considera la evolución de la nebulosa protoplanetaria. Desch (2008) estimó los tiempos de formación de los núcleos de los planetas si se permite la evolución viscosa de la nebulosa. Mientras que, dada la rápida formación de Júpiter, este hecho no tendría un impacto significativo en este caso, y para Saturno el tiempo de formación de su núcleo sería de 3 millones de años, los tiempos se verían considerablemente prolongados en el caso del resto de los planetas: para Neptuno sería de 15 millones de años y para Urano superaría ampliamente los 20 millones de años.

# Capítulo 8

# Conclusiones y perspectivas

El objetivo de esta Tesis es el estudio de la formación *in situ* de los planetas gigantes del Sistema Solar en el marco del modelo de inestabilidad nucleada. Para ello se trabajó con un modelo teórico y numérico que permite el cálculo de la tasa de acreción de sólidos y de gas en forma autoconsistente, las cuales son necesarias para un estudio realista de este problema. Si bien cálculos de esta naturaleza fueron realizados ya por otros autores (Bodenheimer & Pollack 1986; Wuchterl 1991, 1995; Pollack et al. 1996; Alibert et al. 2005, Hubickyj, Bodenheimer & Lissauer, 2005), son todavía muchas las incertezas entorno a este tema.

Existen dos factores derivados de los datos observacionales que imponen restricciones a los resultados teóricos: la vida media de los discos circumestelares (estimada en menos de $10^7$ años) y las masas de los núcleos de los planetas gigantes del Sistema Solar (en general, entre 10 y 20 $M_\oplus$). Es cierto que los valores de estas cotas no son estrictos y que hay lugar para cierta flexibilidad, pero muy probablemente puedan ser tomados como buenas aproximaciones. Encontrar soluciones a los modelos de formación, que se mantengan dentro de estos límites es una tarea que continúa entre los investigadores del área, ya que no se ha conseguido explicar en forma completamente satisfactoria la formación de Júpiter, Saturno, Urano y Neptuno en el marco de un modelo integral de evolución de un sistema planetario. El problema de la formación de los gigantes gaseosos es tan vasto, que no puede ser estudiado, por ahora, en conjunto y en forma completamente autoconsistente, con la evolución del disco protoplanetario. Sin embargo, puede ser aislado y atacado desde diversos ángulos que permitan la obtención de resultados parciales, los cuales, a su vez, den lugar a aproximaciones realistas de la solución. Teniendo presente este hecho es que hemos formulado nuestro modelo.

Los aportes más importantes de nuestro trabajo son la incorporación de la tasa de crecimiento oligárquico para el núcleo sólido y las condiciones nebulares compatibles con el modelo de Niza. En estudios previos, la tasa de acreción de sólidos considerada era cercana a la correspondiente al régimen runaway, mucho más eficiente en la captura de planetesimales pero probablemente altamente sobreestimada. De hecho, en muchas de las simulaciones donde se considera este tipo de crecimiento, la formación del núcleo ocurre en una escala de tiempo despreciable comparada con la correspondiente a la formación completa del planeta. Nuestro modelo es el primero en incorporar una tasa de acreción de sólidos realista.

Las conclusiones más importantes, en base a los resultados de nuestras simulaciones y a la comparación con modelos previos, son:

- El crecimiento oligárquico del núcleo tiene impacto directo en la formación de un planeta gigante dado que regula, no solo la escala de tiempo de formación del núcleo, sino también la tasa de acreción de sólidos. El tiempo involucrado en la formación de un núcleo a partir del cual el planeta comienza a acretar gas en forma significativa representa, aproximadamente, la mitad del tiempo necesario para alcanzar la masa de cruce. En estudios realizados por otros autores, donde la acreción de planetesimales se considera mucho más efectiva, el tiempo de formación del núcleo puede, en la mayoría de los casos, considerarse despreciable.

- La tasa de acreción de sólidos en el crecimiento oligárquico es mucho más baja debido a la dispersión de velocidades de los planetesimales, provocada por la presencia del embrión. Este hecho tiene incidencia, a su vez, en la tasa de acreción de gas. Mientras que asociado al crecimiento runaway (o cuasi runaway) de sólidos está la existencia de la "fase 2" en la formación del planeta, la cual regula la escala de tiempo de toda la formación, esta fase está prácticamente ausente si se considera el crecimiento oligárquico del núcleo. Cuando se adopta este régimen de crecimiento, la tasa de acreción de gas resulta ser siempre estrictamente creciente.

- Debido a la dependencia de la tasa de acreción de sólidos con la densidad superficial en la zona de alimentación, cuando se aumenta la densidad, los tiempos de formación se acortan y las masas de los núcleos aumentan. Un resultado análogo surge cuando se consideran variaciones en las distancias heliocéntricas. Dado que la densidad decrece hacia el exterior del disco, el tiempo de formación de un planeta aumenta en esa dirección.

- El tiempo de formación decrece sensiblemente cuando se reduce el tamaño de los planetesimales acretados por el planeta. Dado que la viscosidad del gas que constituye la envoltura es más efectiva sobre los cuerpos más pequeños, los planetesimales que la atraviesan sufren una mayor desaceleración por efecto del frenado viscoso y disminuyen sus velocidades relativas, lo cual facilita su acreción. En otras palabras, la sección eficaz de captura de los planetesimales de menor radio es mayor que la de los planetesimales más grandes, razón por la cual la formación del núcleo ocurre más rápidamente cuando la población de planetesimales en la zona de alimentación del planeta está dominada por cuerpos pequeños.

- Considerar una distribución de tamaños para los planetesimales acretados, donde los más pequeños contienen la mayor parte la masa del disco favorece la formación de los planetas gigantes. Sin embargo, el tiempo de formación depende significativamente de la elección del radio del planetesimal más pequeño.

- El proceso de formación de los planetas gigantes del Sistema Solar es compatible con el modelo de Niza. El hecho que los cuatro planetas conformaran originalmente un sistema más compacto, con radios orbitales menores a los actuales, favorece el proceso de acreción.

- La formación de un planeta gigante se puede ver seriamente afectada si la envoltura se vuelve inestable y desarrolla procesos cuasi periódicos de pérdida de masa. La aparición de estas inestabilidades está asociada a situaciones de acreción donde el núcleo llega a alcanzar la masa de aislación. Las densidades superficiales bajas y/o los planetesimales pequeños y/o las distancias heliocéntricas cortas favorecen la aparición de la inestabilidad.

- Variaciones en la ecuación de estado pueden tener serias implicancias en el tiempo de formación y en el valor de la masa del núcleo.

El resultado fundamental de nuestro trabajo es haber encontrado que el régimen de crecimiento oligárquico no es incompatible con la formación de los planetas gigantes. De hecho, aún siendo éste un escenario poco favorable, hemos podido demostrar, por primera vez y bajo hipótesis razonables, que Júpiter, Saturno, Urano y Neptuno pueden formarse en menos de 10 millones de años. Además, la masa de los núcleos resulta consistente con las estimaciones actuales.

Por otra parte, la posibilidad de la manifestación de inestabilidades en la envoltura podría dificultar la formación de los planetas. Sin embargo, otros mecanismos podrían inhibir el desarrollo de la inestabilidad. En el futuro se debería incluir un modelo de evolución del disco protoplanetario que contemple tanto la dependencia con el tiempo de la densidad del gas nebular como el decaimiento orbital de los planetesimales por el efecto de la viscosidad. Además, habría que tener en cuenta la disminución de los planetesimales en la zona de alimentación del planeta por eyección de los mismos como consecuencia de las interacciones gravitatorias con el planeta.

La sensibilidad que presentan los modelos, tanto de formación como los correspondientes a la estructura interior de los planetas gigantes, a variaciones en la ecuación de estado pone de manifiesto la necesidad de contar con tablas más detalladas. Resulta fundamental, a la hora de mejorar los modelos, poder contar con una ecuación de estado confiable. En el mismo sentido, el cálculo de la opacidad en la envoltura debería ser autoconsistente con el proceso de acreción. La ablación de los planetesimales y el cálculo detallado de la energía que estos depositan en las capas de gas es uno de los principales objetivos en la continuación de este trabajo.

Dos cuestiones interesantes para explorar en relación con este problema están relacionadas con las condiciones de borde, tanto internas como externas. Por un lado, en los estudios como el presentado en esta Tesis se considera que el núcleo es inerte y que no participa energéticamente en el proceso. Esta hipótesis debería ser evaluada en más detalle. Nosotros hemos considerado que el núcleo no absorbe ni emite energía. Si, por el contrario, existiera un flujo de energía neto en alguna de las dos direcciones, el proceso de formación se vería seriamente afectado. Si hubiera un flujo neto de energía hacia el núcleo la formación sería más rápida dado que el colapso de las capas de gas se produciría antes justamente por no contar éstas con la presión necesaria para contrarrestar el "peso" de las capas exteriores. Si, en cambio, el núcleo fuera una fuente de energía con un flujo neto hacia el exterior el proceso de colapso se retrasaría. Creemos que lo más probable es que el flujo neto de energía se dirige hacia el núcleo, lo cual favorecería al proceso de formación. Este es un punto a explorar en nuestros próximos trabajos. En el otro extremo, la condición de borde en el límite entre la nebulosa y el planeta debería ser evaluada con más cuidado. Fundamentalmente, cuando se desata el runaway de gas, la velocidad a la cual el disco puede suministrar material al planeta seguramente regulará el estadío final de la formación. La condición de borde externa determinará la masa final del planeta. Queda pendiente de estudio si este proceso modifica sustancialmente la escala de tiempo de acreción de gas.

Por último, no hay que perder de vista que los planetas no se formaron como entidades aisladas en el disco sino que el proceso tuvo lugar durante la formación simultánea con otros embriones de su vecindad. Las interacciones entre ellos, y entre ellos con el disco puede ser determinante, sobre todo si, como se espera, los planetas no se mantuvieron en órbitas fijas durante su formación sino que sufrieron variaciones en sus semiejes.

# Bibliografía

# Símbolos y unidades

$a$ = radio de la órbita del planeta entorno a la estrella central
$c$ = velocidad de la luz
$c_{\mathrm{s}}$ = velocidad del sonido
$C_{\mathrm{D}}$ = coeficiente (adimensional) de amortiguamiento de la componente gaseosa del disco
$\mathrm{CH_4}$ = metano
CO = monóxido de carbono
$d$ = distancia entre el observador y el objeto observado
$\epsilon_{\mathrm{ac}}$ = luminosidad por acreción de planetesimales
$e$ = excentricidad de la órbita
$e_m$ = excentricidad de los planetesimales
$E$ = energía
$E_{\mathrm{k},m}$ = energía cinética de un planetesimal de masa $m$
EOS= ecuación de estado (de sus siglas en inglés Equation Of State)
$\phi$ = potencial gravitatorio
$G$ = Constante de Gravitación Universal ($G = 6{,}67259 \times 10^{-8}$ dyn cm$^2$g$^{-1}$)
$h$ = altura de escala del disco de planetesimales
$H$ = altura de escala de la componente gaseosa del disco protoplanetario
H = átomo de hidrógeno
$\mathrm{H^+}$ = ion positivo de hidrógeno
$\mathrm{H_2}$ = molécula de hidrógeno
He = átomo de helio
$\mathrm{He^+}$ = átomo de helio una vez ionizado
$\mathrm{He^{++}}$ = átomo de helio dos veces ionizado
$i$ = inclinación de los planetesimales respecto del plano fundamental
$J_i$ = momentos gravitatorios
$\kappa$ = opacidad media de Rosseland
$L$ = luminosidad
$L_{\mathrm{S}} = L_{\odot}$ = luminosidad solar ($L_{\odot} = 3,839 \times 10^{33}$) erg/s)
$m$ = masa de un planetesimal
$M_{\mathrm{c}}$ = masa del núcleo sólido del planeta
$M_{\mathrm{cross}}$ = masa de cruce ($M_{\mathrm{cross}} \equiv M_{\mathrm{c}} = M_{\mathrm{g}}$)
$M_*$ = masa de la estrella central
$M_J$ = masa de Júpiter ($M_J = 1,9 \times 10^{30}$ g)

$M_{\rm g}$ = masa de la envoltura gaseosa
$M_r$ = masa contenida en una esfera de radio $r$
$\rm M_\odot$ = masa del Sol ($\rm M_\odot = 1{,}989 \times 10^{33}$g)
$\rm M_\oplus$ = masa de la Tierra ($\rm M_\oplus = 5{,}97 \times 10^{27}$g)
$M_{\rm p}$ = masa del planeta
$\nabla$ = gradiente adimensional de temperatura ($\nabla \equiv \frac{d\ln T}{d\ln P}$)
$\nabla_{\rm ad}$ = gradiente adiabático
$\nabla_{\rm rad}$ = gradiente radiativo
$NH_3$ = amoníaco
NSMM= Nebulosa Solar de Masa Mínima
$\Omega_{\rm k}$ = velocidad angular kepleriana
pc= parsec (1 pc= $3 \times 10^8$ cm = 206.265 UA = 3,262 años luz)
$P$ = presión
$P_{\rm c}$ = presión del gas de la envoltura en la superficie del núcleo
PPT= Plasma Phase Transition. Región de la EOS del hidrógeno donde se produce la ionización por presión, la cual ocurre a través de una transición de fase de primer orden.
$r$ = coordenada radial
$r_m$ = radio de un planetesimal de masa $m$
$R_{\rm a}$ = radio de acreción $\left(R_{\rm a} = \frac{GM_r}{c_{\rm s}^2}\right)$
$R_{\rm c}$ = radio del núcleo sólido del planeta
$R_{\rm ec}$ = radio ecuatorial del planeta
$R_{\rm eff}$ = radio efectivo de captura del planeta
$R_{\rm H}$ = radio de Hill $\left(R_{\rm H} = a\left(\frac{M_{\rm p}}{3M_*}\right)^{1/3}\right)$
$R_*$ = radio de la estrella central
$R_{\rm p}$ = radio del planeta
$\rm R_\oplus$ = radio de la Tierra ($\rm R_\oplus = 6,38 \times 10^8$ cm)
$\rm R_\odot$ = radio del Sol ($\rm R_\odot = 6,9 \times 10^{10}$ cm)
$\rho$ = densidad volumétrica de gas de la nebulosa protoplanetaria
$\rho_{\rm c}$ = densidad del gas de la envoltura en la superficie del núcleo
$\rho_{\rm g}$ = densidad volumétrica de gas de la envoltura del protoplaneta
$\rho_m$ = densidad en masa de un planetesimal
$\Sigma$ = densidad superficial de sólidos en el disco
$S$ = entropía por unidad de masa
$S_{\rm mix}$ = entropía de mezcla
SCVH= ecuación de estado de Saumon, Chabrier & Van Horn (1995)
$t$ = coordenada temporal
$t_{\rm cross}$ = tiempo de cruce
$T$ = temperatura
$T_{\rm c}$ = temperatura del gas de la envoltura en la superficie del núcleo
$U$ = energía interna por unidad de masa
UA= Unidad Astronómica (1 UA = $1,495 \times 10^{13}$ cm)
$v$ = velocidad

$v_{\text{esc}}$ = velocidad de escape
$v_{\text{k}}$ = velocidad kepleriana de un objeto entorno al Sol en una órbita circular
$v_{\text{rel}}$ = velocidad relativa entre el protoplaneta y los planetesimales
$X$ = fracción en masa de hidrógeno
$Y$ = fracción en masa de helio
$Z$ = fracción en masa de "elementos pesados"